\documentclass[12pt]{article}
\usepackage{amssymb,amsmath,latexsym}
\usepackage{graphicx}
\usepackage[final]{pdfpages}
\usepackage{verbatim}
\usepackage{color}
\usepackage{tikz}
\usetikzlibrary{arrows.meta}
\usepackage{float}
\usepackage{hyperref}

\definecolor{GR}{rgb}{.35,.7,.35}

\AtBeginDocument{
	\label{CorrectFirstPageLabel}
	
}

\newtheorem{theo}{Theorem}
\newtheorem{defini}[theo]{Definition}
\newtheorem{proposi}[theo]{Proposition}
\newtheorem{lemma}[theo]{Lemma}
\newtheorem{coro}[theo]{Corollary}
\newtheorem{rem}[theo]{Remark}

\newtheorem{hyp}[theo]{Hypothesis}

\numberwithin{equation}{section}
\numberwithin{figure}{section}

\newcommand{\ulc}{\mathfrak{x}}
\newcommand{\mlc}{\mathfrak{y}}
\newcommand{\llc}{\mathfrak{z}}
\newcommand{\ugl}{\mathfrak{a}}
\newcommand{\mgl}{\mathfrak{b}}
\newcommand{\lgl}{\mathfrak{c}}

\newcommand{\tick}[1]{\filldraw[black!100,line width=0.5mm,fill=none] {#1} ellipse (0.2 and 0.0)}
\newcommand{\subdivision}[6]{
	\coordinate (lt) at (-3,3);
	\tick{(lt)};
	\coordinate[label=left:{${#1}$}, label=right:{${#2}$}] (argt1) at (-3,2.5);
	\coordinate (lm1) at (-3,2);
	\tick{(lm1)};
	\coordinate[label=left:{${#3}$}] (argm1) at (-3,0);
	\coordinate (lm2) at (-3,-2);
	\tick{(lm2)};
	\coordinate[label=left:{${#4}$}, label=right:{${#5}$}] (argb1) at (-3,-2.5);
	\coordinate (lb) at (-3,-3);
	\tick{(lb)};
	\draw[-,color=black,line width=0.5mm] (lt)--(lb);
	\coordinate (rt) at (3,3);
	\tick{(rt)};
	\coordinate[label=left:{${#1}_{#6}$}, label=right:{${#2}_{#6}$}] (rm1) at (3,2.5);
	\coordinate (rm2) at (3,2.15);
	\coordinate (rm3) at (3,2);
	\tick{(rm3)};
	\coordinate[label=left:{${#3}_{#6}$}] (argm2) at (3,0);
	\coordinate (rm4) at (3,-2);
	\tick{(rm4)};
	\coordinate (rm5) at (3,-2.15);
	\coordinate[label=left:{${#4}_{#6}$}, label=right:{${#5}_{#6}$}] (rm6) at (3,-2.5);
	\coordinate (rb) at (3,-3);
	\tick{(rb)};
	\draw [-,color=black,line width=0.5mm] (rt)--(rm1);
	\draw [-,color=black,dotted,line width=0.5mm] (rm1)--(rm2);
	\draw [-,color=black,line width=0.5mm] (rm2)--(rm5);
	\draw [-,color=black,dotted,line width=0.5mm] (rm5)--(rm6);
	\draw [-,color=black,line width=0.5mm] (rm6)--(rb);
	\coordinate (rl1) at (2.2,2.15);
	\coordinate (rl2) at (2.2,-2.15);
	\coordinate (rr1) at (3.8,2.15);
	\coordinate (rr2) at (3.8,-2.15);
	\draw[-, fill=none] (rl1)--(rl2)--(rr2)--(rr1)--cycle}

\begin{document}
\allowdisplaybreaks

\title{Partially hyperbolic random dynamics on Grassmannians}

\author{Joris De Moor, Florian Dorsch, Hermann Schulz-Baldes$^*$
\\
\\
{\small Friedrich-Alexander-Universit\"at Erlangen-N\"urnberg, Department Mathematik}
\\
{\small Cauerstr. 11, D-91058 Erlangen, Germany}
\\
{\small $^*$ Email: schuba@mi.uni-erlangen.de}
}

\date{ }

\maketitle

\begin{abstract}
A sequence of invertible matrices given by a small random perturbation around a fixed diagonal partially hyperbolic matrix induces a random dynamics on the Grassmann manifolds. Under suitable weak conditions it is known to have a unique invariant (Furstenberg) measure. The main result gives concentration bounds on this measure showing that with high probability the random dynamics stays in the vicinity of stable fixed points of the unperturbed matrix, in a regime where the strength of the random perturbation dominates the local hyperbolicity of the diagonal matrix. As an application, bounds on sums of Lyapunov exponents are obtained.
\end{abstract}

\section{Overview}

Invertible matrices naturally map subspaces onto subspaces of the same dimension. If the matrices are drawn randomly, one hence obtains a random dynamical system on the corresponding Grassmannian manifold.  A well-known theorem of Furstenberg states that, provided that the distribution of the random matrices satisfies the relatively mild conditions of strong irreducibility and contractivity,  there is a unique invariant probability measure on the Grassmannian (see the monographs \cite{BQ,BL}). This paper provides information on the weight distribution of this Furstenberg measure in the particular perturbative situation where the random matrices of size $\mathsf{L}\times\mathsf{L}$ are of the form 
\begin{align}
\label{def-T}
\mathcal{T}_n
\;=\;
e^{\lambda\mathcal{P}_n}\mathcal{R}\;\in\;\textnormal{GL}(\mathsf{L},\mathbb{C})
\end{align}
where $\mathcal{R}$ is a fixed positive diagonal matrix
\begin{align}
\label{def-R}
\mathcal{R}
\;=\;
\operatorname{diag}(\kappa_{\mathsf{L}},\ldots,\kappa_1)
\,,
\qquad\qquad\kappa_1\geq\dots\geq\kappa_{\mathsf{L}}>0\,,
\end{align}
and the i.i.d. random matrices $\mathcal{P}_n$ are drawn from the Lie algebra $\textnormal{gl}(\mathsf{L},\mathbb{C})$ and all have their operator norm bounded by $1$, and finally $\lambda>0$ is a small coupling constant. The deterministic matrix $\mathcal{R}$ leads to a partially hyperbolic dynamics which will be characterized by relative gaps 
\begin{align}\label{def-eta}
\eta(\mathsf{I},\mathsf{J})
\;:=\;
1\,-\,\frac{\kappa_{\mathsf{J}}^2}{\kappa_{\mathsf{I}}^2}\;\in\; [0,1]\,,
\qquad
1\leq \mathsf{I}< \mathsf{J}\leq \mathsf{L}\;.
\end{align}
The $\eta(\mathsf{I},\mathsf{I}+1)$ will be referred to as {\it microscopic} relative gaps, while $\eta(\mathsf{I},\mathsf{J})$ for $\mathsf{J}-\mathsf{I}=\mathcal{O}(\mathsf{L})$ as {\it macroscopic} gaps. Intermediate relative gaps will also play a role and will be called {\it mesoscopic}.
The focus will be on the regime of intermediate disorder strength in which $\lambda$ is larger than the microscopic relative gaps, but smaller than the macroscopic ones. In this regime the randomness dominates the hyperbolicity of $\mathcal{R}$ on a local, but not a global scale (a precise description of the regime will be given below).  The main result of the paper, see Section~\ref{sec-main}, provides a quantitative concentration bound on the Furstenberg measure $\mu_{\lambda,\mathsf{q}}$  on the Grassmanian manifold of subspaces of dimension $\mathsf{q}\in\{1,\ldots,\mathsf{L}\}$, under suitable hypotheses. Apart from the intrinsic interest for this particular random dynamical system, these results also imply quantitative bounds on the sum of the first $\mathsf{q}$ Lyapunov exponents and are potentially useful for the study of random Schr\"odinger operators.

\vspace{.2cm}

The same set-up as above is considered in the prior work \cite{DS}. On first sight, also the main results look similar. Let us stress crucial differences already at this point. First of all, in \cite{DS} only the case $\mathsf{q}=1$ was dealt with which only allows to access the top Lyapunov exponent. Second of all, even for the case $\mathsf{q}=1$ the bound on the Furstenberg measure is improved (see Remark~\ref{rem-Compare} below). Third of all, the distribution of the $\mathcal{P}_n$ was supposed to be rotationally invariant, whereas here much less stringent conditions are imposed. For example, this allows to deal with Bernoulli distributions and only sparsely filled random matrices (see the second example in Section~\ref{sec-toy-models} in which $\mathcal{P}_n$ only contains $\mathsf{L}$ independent scalar random variables).  In particular in view of potential applications, these are all substantial generalizations which, in our opinion, also justifies the considerable technical effort in the proofs.

\vspace{.2cm}

This overview first gives an intuitive description of the random dynamics, starting with the one-dimensional Grassmannian (Section~\ref{sec-one-dimensional}), before passing to the case of the high-dimensional Grassmannian (Section~\ref{sec-more-dimensional}).  Then follows a precise statement of the main result (Section~\ref{sec-main}) as well as an illustration on how to use it in concrete situations (Section~\ref{sec-toy-models}). As an application, bounds on sums of Lyapunov exponents are stated (Section~\ref{sec-Lyapunov-exponents}). Finally, our motivation for the study is briefly laid out and the limitations of the present analysis and potential improvements are discussed  (Section~\ref{sec-motivation}).

\subsection{Heuristics for random dynamics on vectors}
\label{sec-one-dimensional}

For the intuitive description of the random dynamics, it is particularly instructive to start out with the action on  $\mathbb{S}_{\mathbb{C}}^{\mathsf{L}-1}:=\{v\in \mathbb{C}^{\mathsf{L}}\,:\,\|v\|=1\}$ which is the $\mbox{U}(1)$-cover of the complex one-dimensional projective space (this cover is irrelevant in the present context). The random dynamics on $\mathbb{S}_{\mathbb{C}}^{\mathsf{L}-1}$ is given by
\begin{equation}
\label{dyn-sphere-1}
v_n
\;=\;
\mathcal{T}_n\circ v_{n-1}
\;,
\end{equation}
where
\begin{equation}
\label{dyn-sphere-2}
\mathcal{T}\circ v\;=\; 
\frac{\mathcal{T} v}{\|\mathcal{T}v\|}
\;,
\qquad
v\in\mathbb{S}_{\mathbb{C}}^{\mathsf{L}-1}
\;,
\end{equation}
and $v_0\in \mathbb{S}_{\mathbb{C}}^{\mathsf{L}-1}$ is some initial condition.  For $\mathcal{T}$ given by~\eqref{def-T}, the Furstenberg measure $\mu_\lambda$ on $\mathbb{S}_{\mathbb{C}}^{\mathsf{L}-1}$ is then characterized \cite{BL} by
$$
\mathbb{E} \int \mu_\lambda(dv)\,f(\mathcal{T}\circ v)\;=\;\int \mu_\lambda(dv)\,f(v)
\;,
\qquad
f\in C(\mathbb{S}_{\mathbb{C}}^{\mathsf{L}-1})
\;.
$$
For $\lambda=0$, there is no random term and the deterministic dynamics $\mathcal{R}\circ$ given by $v_n=(\mathcal{R}^n)\circ v_0$ is fairly elementary to analyze. Let us choose a labeling of the standard basis vectors adapted to the above ordering of the diagonal entries $\kappa_1,\ldots,\kappa_{\mathsf{L}}$ of $\mathcal{R}$:
\begin{equation}
\label{def-basis-vectors}
e_1
\;=\;
\begin{pmatrix}
0 \\ \vdots \\ 0 \\ 0 \\ 1 
\end{pmatrix}
\;,
\quad
e_2
\;=\;
\begin{pmatrix}
0 \\ \vdots \\ 0 \\ 1 \\ 0
\end{pmatrix}
\;,
\quad
\ldots\;,
\quad
e_{\mathsf{L}}
\;=\;
\begin{pmatrix}
1 \\ 0 \\ \vdots \\ 0 \\ 0
\end{pmatrix}
\;.
\end{equation}
Then each $e_{\mathsf{I}}$ is a fixed point of $\mathcal{R}\circ$. However, $e_{\mathsf{I}}$ is an unstable fixed point when $\eta({\mathsf{J}},{\mathsf{I}})>0$ for some ${\mathsf{J}}<{\mathsf{I}}$, as then $\kappa_{\mathsf{I}}<\kappa_{\mathsf{J}}$. Hence the unit vectors form a cascade $e_1, e_2,\ldots,e_{\mathsf{L}}$ of fixed points with decreasing stability. In the maximally hyperbolic situation where $\eta({\mathsf{I}},{\mathsf{I}}+1)>0$ for all ${\mathsf{I}}$, these are the only fixed points, with $e_1$ being the only stable one. Moreover, one can then readily show that $(\mathcal{R}^n)\circ v_0\to e_{\mathsf{I}}$ where ${\mathsf{I}}$ is the smallest index such that the scalar product $\langle e_{\mathsf{I}}|v_0\rangle$ does not vanish. In general, there may be many more fixed points and then also the limiting behavior is more cumbersome to write out. Indeed, if $\eta({\mathsf{I}},{\mathsf{J}})=0$ for some ${\mathsf{I}}<{\mathsf{J}}$, then all vectors in the span of $e_{\mathsf{I}},\ldots,e_{\mathsf{J}}$ are fixed points and within this span no direction is privileged. 

\vspace{.2cm}

Next let us consider the case of $\lambda>0$. If $\lambda\ll \eta(1,2)$ and the distribution of the perturbations $\mathcal{P}_n$ couples all directions (having an absolutely continuous component is more than sufficient), the random dynamics leaves any unstable fixed point and is driven to the vicinity of the stable fixed point $e_1$ in which it then remains. Thus in this case the Furstenberg invariant measure $\mu_{\lambda}$ on $\mathbb{S}_{\mathbb{C}}^{\mathsf{L}-1}$ is supported on a small neighborhood of the only stable fixed point. More generally, if $\lambda \ll\eta(\mathsf{I}-1,\mathsf{I})$ for some $\mathsf{I}$, then $\mu_{\lambda}$ is supported by a small neighborhood of $\{0\}^{\mathsf{L}-\mathsf{I}}\times\mathbb{S}_{\mathbb{C}}^{\mathsf{I}-1}\subset\mathbb{S}_{\mathbb{C}}^{\mathsf{L}-1}$.  Proofs of these facts were provided in \cite{DS}.

\vspace{.2cm}

The main focus of this paper is on the regime of intermediate disorder strength $\lambda$. In particular, it will be supposed that $\lambda>\eta(\mathsf{I},\mathsf{I}+1)$ for all $\mathsf{I}$, or at least for a relevant fraction of all possible $\mathsf{I}$. In this situation and under suitable coupling assumptions on the distribution of the $\mathcal{P}_n$, the random dynamics can explore the full phase space $\mathbb{S}_{\mathbb{C}}^{\mathsf{L}-1}$. Such paths have been constructed explicitly in the proof of the last claim of Theorem~1.1 in \cite{DS} (see, in particular, Lemma~2.7 therein). As it is helpful to understand the strategy of the proof of the main result of this paper, let us describe how this is possible. Suppose that at some time $N$ one has  $v_N=e_{\mathsf{I}}$. Let us now show how it is then possible for the random dynamics to ascend to $e_{\mathsf{I}+1}$. Set $\eta=\eta(\mathsf{I},\mathsf{I}+1)$ and suppose that $\lambda$ and $\eta$ are both small, with $\lambda>\eta$, even though possibly not much larger. Suppose $e^{\lambda\mathcal{P}}$ acts as a rotation by $\pm\lambda$ in the two-dimensional subspace spanned  by $e_{\mathsf{I}+1}$ and $e_{\mathsf{I}}$. For that purpose, decompose $v=xe_{\mathsf{I}+1}+ye_{\mathsf{I}}$ with $x,y\in [-1,1]$ and $x^2+y^2=1$. Then one finds for the two steps of the dynamics on the two-dimensional subspace 
$$
\mathcal{R}\circ v
\;=\;
\begin{pmatrix}
\kappa_{\mathsf{I}+1} & 0 \\ 0 & \kappa_{\mathsf{I}}
\end{pmatrix}
\circ
\binom{x}{y}
\;=\;
x(1-\tfrac{\eta}{2}y^2)e_{\mathsf{I}+1}
\;+\;
y(1+\tfrac{\eta}{2}x^2)e_{{\mathsf{I}}}
\,+\,\mathcal{O}(\eta^2)
$$
and
$$
e^{\pm\lambda\mathcal{P}}\circ v
\;=\;
\begin{pmatrix}
\cos(\lambda) & \pm\sin(\lambda) \\ \mp\sin(\lambda) & \cos(\lambda)
\end{pmatrix}
\circ
\binom{x}{y}
\;=\;
(x\pm \lambda y)e_{\mathsf{I}+1}
\;+\;
(y\mp \lambda x)e_{{\mathsf{I}}}
\,+\,\mathcal{O}(\lambda^2)
\;,
$$
so that
$$
(e^{\pm\lambda\mathcal{P}}\mathcal{R}) \circ v
\;=\;
(x\pm \lambda y-\tfrac{\eta}{2}xy^2)e_{\mathsf{I}+1}
\;+\;
(y\mp \lambda x+\tfrac{\eta}{2}x^2y)e_{{\mathsf{I}}}
\,+\,\mathcal{O}(\eta^2,\eta\lambda,\lambda^2)
\;.
$$
As $\lambda > \eta$, one can deduce that $e^{+\lambda\mathcal{P}}\mathcal{R}\circ$ corresponds (up to $\mathcal{O}(\lambda^2)$) to a rotation of at least $\frac{3}{4}\lambda$. Hence if $v_N=e_{\mathsf{I}}$ so that $x_N=0$ and $y_N=1$, always choosing the sign $+$, the vectors $v_{N+n}$ rotate after $\mathcal{O}(\lambda^{-1})$ iterations  towards $e_{\mathsf{I}+1}$, even though $\mathcal{R}\circ$ produces a deterministic drift to the bottom vector $e_{\mathsf{I}}$. Of course, this is a rare event because the signs $\pm$ both appear with equal probability since $\mathcal{P}$ is centered. Nevertheless, there are such rare realizations. Moreover, once $e_{\mathsf{I}+1}$ is reached, it is possible to proceed to $e_{\mathsf{I}+2}$ by a similar procedure. If the random perturbation also has non-vanishing couplings $\langle e_{\mathsf{I}}|\mathcal{P} e_{\mathsf{I}+2}\rangle$, such realizations can enable the even stronger drift from  $e_{\mathsf{I}+2}$ to $e_{\mathsf{I}}$ which may then dominate all possible random terms. After many iterations and on a very unlikely path of realizations,  one can attain the most unstable fixed point $e_{\mathsf{L}}$ in this manner. Of course, it is also possible to move from $e_{\mathsf{I}}$ directly to $e_{\mathsf{J}}$ for some $\mathsf{J}>\mathsf{I}$ as long as  $\lambda>\eta(\mathsf{I},\mathsf{J})$, but it is not feasible to overcome macroscopic relative gaps and typically requires  successive elections of suitable realizations of $\mathcal{P}$ with small probability.  We will refer to the scenario just described as the {\it ascension of the ladder}, which is schematically depicted in Fig~\ref{fig-ascension} on a microscopic level. For a macroscopic ascension from $e_{\mathsf{I}}$ to $e_{\mathsf{J}}$ with $\mathsf{J}-\mathsf{I}=\mathcal{O}(\mathsf{L})$ one may have a higher-dimensional pyramid in mind. Let us stress again that then couplings of $\mathcal{P}$ from $e_{\mathsf{J}}$ to $e_{\mathsf{I}}$ for some $\mathsf{I}<\mathsf{J}$ are even more effective in impeding the ascension of the ladder. 

\begin{figure}[H]
\begin{center}
\begin{tikzpicture}[line join = round, line cap = round]
\coordinate [label=above:{$\|e_{\mathsf{I+2}}\|=1$}] (A) at (0,5);
\coordinate [label=right:{$\|e_{\mathsf{I}+1}\|=1$}] (B) at (5,0);
\coordinate [label=left:{$\|e_{\mathsf{I}}\|=1$}] (C) at (-5,0);
\draw[-, fill=white!30, opacity=.5] (A) --(B)--(C)--cycle;
\coordinate  (a) at (-0.5,4.5);
\coordinate (b) at (3.5,0.5);
\coordinate (c) at (-4.5,0.5);
\draw[-, fill=gray!100, opacity=.5] (a) --(b)--(c)--cycle;
\draw [<-,color=black,line width=1mm](0.5-1.5,1.5) -- (1.5-1.5,2.5) ;
\draw [->,color=black,line width=0.3mm](-0.7,0.25) -- (0.7,0.25) ;
\draw [->,color=black,line width=0.3mm](3.0,1.5) -- (2.0,2.5) ;
\end{tikzpicture}
\end{center}
\caption{{\it Schematic illustration of the ascension from $e_{\mathsf{I}}$ via $e_{\mathsf{I}+1}$ to $e_{\mathsf{I}+2}$ in the case where $\eta(\mathsf{I},\mathsf{I}+1)<\lambda < \eta(\mathsf{I},\mathsf{I}+2)$ and $\eta(\mathsf{I}+1,\mathsf{I}+2)<\lambda$. Each point in the triangle corresponds to a unit vector in the span of the three vectors $e_{\mathsf{I}}$ via $e_{\mathsf{I}+1}$ and $e_{\mathsf{I}+2}$. In the shaded region, the deterministic dynamics $\mathcal{R}\circ$ dominates and immediately drives also the random dynamics away from $e_{\mathsf{I}+2}$. The only possible way to ascend to $e_{\mathsf{I}+2}$ is by following the thin arrows in the white region, which under appropriate conditions is shown to be a very unlikely event for the random dynamics.}}
\label{fig-ascension}
\end{figure}

\vspace{.2cm}

In conclusion, it is very unlikely that the random dynamics~\eqref{dyn-sphere-1} of vectors leads to an orbit attaining the most unstable fixed point $e_{\mathsf{L}}$. Actually, one even expects that the dynamics stays in the vicinity of the stable fixed point $e_{1}$. Section~\ref{sec-main} states the main result of this paper in this respect which confirms that expectation. The result provides an upper bound on the expected value of a suitable notion of distance on the Grassmannian. It also covers the case of higher-dimensional subspaces which is described in the next section.

\subsection{Dynamics and distance on higher-dimensional Grassmannian}
\label{sec-more-dimensional}

For $\mathsf{q},\mathsf{L}\in\mathbb{N}$ such that $\mathsf{q}\leq\mathsf{L}$, let $\mathbb{G}_{\mathsf{L},\mathsf{q}}$ denote the Grassmannian manifold of $\mathsf{q}$-dimensional subspaces of $\mathbb{C}^{\mathsf{L}}$. Here it will be convenient to identify such a subspace with a $\mathsf{q}$-dimensional orthogonal projection on $\mathbb{C}^{\mathsf{L}}$, namely we will choose the concrete representation
$$
\mathbb{G}_{\mathsf{L},\mathsf{q}}
\;:=\;
\left\{Q\in\mathbb{C}^{\mathsf{L}\times\mathsf{L}}:\quad Q=Q^2=Q^*\,,\quad \operatorname{tr}(Q)=\mathsf{q}\right\}
\;.
$$
Let us also use the notation $\mathbb{G}_{\mathsf{L}}:=\bigcup\limits_{\mathsf{w}=0}^{\mathsf{L}} \mathbb{G}_{\mathsf{L},\mathsf{w}}$ for the collection of all Grassmannians in $\mathbb{C}^{\mathsf{L}}$. Again an invertible matrix $\mathcal{T}\in\mathbb{C}^{\mathsf{L}\times\mathsf{L}}$ naturally acts on the Grassmannian $\mathbb{G}_{\mathsf{L},\mathsf{q}}$ by simply mapping a $\mathsf{q}$-dimensional subspace to its image under $\mathcal{T}$. One can write out this action 
$$
\cdot : \textnormal{GL}(\mathsf{L},\mathbb{C})\times \mathbb{G}_{\mathsf{L},\mathsf{q}}\rightarrow\mathbb{G}_{\mathsf{L},\mathsf{q}}
$$ 
as
$$
\mathcal{T}\cdot Q
\;=\;
\mathcal{T} Q \mathcal{T}^*
(\mathcal{T} Q \mathcal{T}^*)^{-2}
\mathcal{T} Q \mathcal{T}^*
\;,
$$
where the inverse in the middle is understood as a map defined on the range $\operatorname{Ran}(\mathcal{T} Q \mathcal{T}^*)$. Alternatively, one can express $Q=\Phi\Phi^*$ in terms of a frame $\Phi$ from the $\mbox{U}(\mathsf{q})$-cover of $\mathsf{q}$-frames of the Grassmannian $\mathbb{G}_{\mathsf{L},\mathsf{q}}$ given by
$$
\mathbb{F}_{\mathsf{L},\mathsf{q}}
\;:=\;
\left\{\Phi\in\mathbb{C}^{\mathsf{L}\times\mathsf{q}}: \Phi^*\Phi=\mathbf{1}_{\mathsf{q}}\right\}
\;,
$$
and then
\begin{align}\label{def-action}
\mathcal{T}\cdot Q
\;=\; 
\mathcal{T}\Phi\left(\Phi^*\mathcal{T}^*\mathcal{T}\Phi\right)^{-1}\Phi^*\mathcal{T}^*
\;.
\end{align}
Clearly $\cdot$ is a group action, namely for all $\mathcal{T}_2,\mathcal{T}_1\in\textnormal{GL}(\mathsf{L},\mathbb{C})$ and $Q\in\mathbb{G}_{\mathsf{L},\mathsf{q}}$ one has
$$
\mathcal{T}_2\cdot (\mathcal{T}_1\cdot Q)
\;=\;
(\mathcal{T}_2\mathcal{T}_1) \cdot Q\,.
$$

Now let $\mathcal{T}_n=e^{\lambda\mathcal{P}_n}\mathcal{R}\in\textnormal{GL}(\mathsf{L},\mathbb{C})$ be an i.i.d. sequence given as in~\eqref{def-T}. Then one obtains a random dynamics on $\mathbb{G}_{\mathsf{L},\mathsf{q}}$ by setting
\begin{align}\label{dyn-grassmanian}
Q_n
\;:=\;
\mathcal{T}_n\cdot Q_{n-1}\,,\qquad Q_0\in\mathbb{G}_{\mathsf{L},\mathsf{q}}\,, \qquad n\in\mathbb{N}\,.
\end{align}
Extrapolating the arguments of Section~\ref{sec-one-dimensional}, one expects that $Q_n$ is close to the $\mathsf{q}$-dimensional projection given by the span of the $\mathsf{q}$  directions $e_1,\ldots,e_{\mathsf{q}}$ which are the most expanding directions of the unperturbed dynamics $\mathcal{R}\cdot$. This span is the stable fixed point (provided that $\kappa_{\mathsf{q}}>\kappa_{\mathsf{q}+1}$) of the dynamics $\mathcal{R}\cdot$ and it is possible to construct a partial order of fixed points of $\mathsf{q}$-dimensional projections, analogous to the total order given by the standard basis vectors~\eqref{def-basis-vectors} in the case $\mathsf{q}=1$. Hence one can expect $Q_n$ to align with the stable $\mathsf{q}$-dimensional projection. A somewhat weaker statement is that $Q_n$ is almost orthogonal to a projection on unstable directions which is of a dimension that is smaller than $\mathsf{L}-\mathsf{q}$. In order to introduce a quantitative measure of this orthogonality, let us decompose $\mathsf{L}=\mathsf{L}_{\ugl}+\mathsf{L}_{\mgl}+\mathsf{L}_{\lgl}$, where $\mathsf{L}_{\ugl},\mathsf{L}_{\mgl},\mathsf{L}_{\lgl}\in\mathbb{N}$, and subdivide vectors $v\in\mathbb{S}_{\mathbb{C}}^{\mathsf{L}-1}$ into
$$
v=\begin{pmatrix}
\ugl(v)\\\mgl(v)\\\lgl(v)
\end{pmatrix}\,,
$$
where $\ugl(v)$, $\mgl(v)$ and $\lgl(v)$ are of lengths $\mathsf{L}_{\ugl}$, $\mathsf{L}_{\mgl}$ and $\mathsf{L}_{\lgl}$ respectively. For a complementary partial vector, we will write $\lgl^c(v) = \binom{\ugl(v)}{\mgl(v)}$. Closely related to the given partition are three projections $\hat{P}_{{\ugl}}$, $\hat{P}_{{\mgl}}$ and $\hat{P}_{{\lgl}}$ of rank $\mathsf{L}_{\ugl}$, $\mathsf{L}_{\mgl}$ and $\mathsf{L}_{\lgl}$ respectively, given by
$$
\hat{P}_{{\ugl}}
\;=\;
\operatorname{diag}(
\mathbf{1}_{\mathsf{L}_{\ugl}},0,0)
\;,
\qquad
\hat{P}_{{\mgl}}
\;=\;
\operatorname{diag}(
0,\mathbf{1}_{\mathsf{L}_{\mgl}},0)
\;,
\qquad
\hat{P}_{{\lgl}}
\;=\;
\operatorname{diag}(
0,0,\mathbf{1}_{\mathsf{L}_{\lgl}})
\;.
$$
Note that these objects carry a hat, which here designates them as reference projections. For a fixed partition $\mathsf{L}=\mathsf{L}_{\ugl}+\mathsf{L}_{\mgl}+\mathsf{L}_{\lgl}$ and $\mathsf{q}\leq \mathsf{L}-\mathsf{L}_{\ugl}$, let us introduce a function $\mathsf{d}:\mathbb{G}_{\mathsf{L},\mathsf{q}}\longrightarrow [0,\mathsf{q}]$ by
\begin{equation}
\label{def-d}
\mathsf{d}(Q)
\;:=\;
\operatorname{tr}\big(\hat{P}_{{\ugl}}Q\hat{P}_{{\ugl}}\big)
\;=\;
\operatorname{tr}\big(\Phi^*\hat{P}_{{\ugl}}\Phi\big)
\;,
\qquad
Q\;=\;\Phi\Phi^*
\;.
\end{equation}
The quantity $\mathsf{d}(Q)$ will play a central role in this work. It is a quantitative measure of the orthogonality of $Q$ with the reference projection $ \hat{P}_{{\ugl}}$. Alternatively, one may view  $\mathsf{d}(Q)$ as a measure of how well the range of $Q$ is covered by the range of $ \hat{P}_{{\mgl}}+ \hat{P}_{{\lgl}}$. Other than the notation may suggest, $\mathsf{d}(Q)$ is {\sl not} a metric distance between $Q$ and $ \hat{P}_{{\ugl}}$, except if $\mathsf{q}= \mathsf{L}-\mathsf{L}_{\ugl}$. Indeed, in the latter case one has $\mathsf{d}(Q)=\mathsf{d}(\mathbf{1}-\hat{P}_{{\ugl}},Q)$ where $\mathsf{d}(P,Q):=\operatorname{tr}(P-PQP)$ is a metric on $\mathbb{G}_{\mathsf{L},\mathsf{q}}$ which can be shown to be equivalent of the Riemannian metric on $\mathbb{G}_{\mathsf{L},\mathsf{q}}$ as defined in \cite{Won}. In this paper, the focus will, however, be on the case $\mathsf{q}\ll \mathsf{L}-\mathsf{L}_{\ugl}$.  Let us stress that on top of being a natural measure of distance, the quantity $\mathsf{d}(Q)$ can be effectively used in the analysis of the Lyapunov exponents associated with the random matrices~\eqref{def-T}, as explained in detail in Section~\ref{sec-Lyapunov-exponents}.

\subsection{Main result}
\label{sec-main}

In this section, it is supposed that the partition $\mathsf{L}=\mathsf{L}_{\ugl}+\mathsf{L}_{\mgl}+\mathsf{L}_{\lgl}$ is fixed and is such that $\mathsf{q}\leq \mathsf{L}_{\lgl}$.  Let us now state the assumptions for the main result below. We decided to be explicit about the constants even though their values can certainly be optimized in other ways, because this stresses that the method of proof provides quantitative bounds. 

\begin{hyp}[Macroscopic relative gap for $\mathcal{R}$]\label{hyp-R}
$\pmb{\eta}:=\eta(\mathsf{L}_{\lgl},\mathsf{L}_{\mgl}+\mathsf{L}_{\lgl})$ satisfies $\pmb{\eta}>0$.
\end{hyp}

\begin{hyp}[Coupling assumption on $\mathcal{P}$]\label{hyp-P}
The distribution of the random matrix  $\mathcal{P}$ is centered: $\mathbb{E}(\mathcal{P})=0$.
The support $\textnormal{supp}(\mathcal{P})$ is contained in $\mathfrak{P}:=\left\{P\in\mathbb{C}^{\mathsf{L}\times \mathsf{L}}: \|{P}\|\leq 1\right\}$. Further,
$$
\beta
\;:=\;
\inf\left\{\mathbb{E}\,\|\lgl((\mathbf{1}-W)\mathcal{P}v)\|^2:  \quad v\in\mathbb{S}_{\mathbb{C}}^{\mathsf{L}-1}\,,\quad \lgl(v)=0\,,\quad W\in\mathbb{G}_{\mathsf{L},\mathsf{q}-1}\,,\quad W\leq\hat{P}_{{\lgl}}
\right\}
$$
is a strictly positive quantity. Note that $\beta$ depends on $\mathsf{q}$ and $\mathsf{L}_\lgl$ and when it is necessary to stress this dependence, we will also write $\beta=\beta(\mathsf{q},\mathsf{L}_\lgl)$.
\end{hyp}

\begin{hyp}[Small coupling constant]\label{hyp-lambda}
$\lambda \in \left(0,2^{-13}\right)$ satisfies $\vartheta\lambda \leq 2^{-17}\beta^{\frac{8}{3}}\pmb{\eta}^{-\frac{1}{3}}$, using the abbreviation $\vartheta := \log(2^{-\frac{54}{5}}\lambda^{-1}) \geq 1$, depending on $\lambda$ {\rm (}although this is not explicit in the notation{\rm )}.
\end{hyp}

\begin{hyp}[Condition on the dimension $\mathsf{q}$]\label{hyp-q}
$\mathsf{q} \leq 2^{-\frac{36}{5}}\beta^{\frac{1}{5}}\pmb{\eta}^{\frac{3}{5}}\vartheta^{-\frac{1}{5}}\lambda^{-\frac{1}{5}}$.
\end{hyp}

\begin{hyp}[Dominated microscopic gaps]\label{hyp-eta} $\eta(\mathsf{I},\mathsf{I}+1)< 2^4\,\lambda$ for all $\mathsf{I}\in \{\mathsf{L}_{\lgl},\dots,\mathsf{L}_{\mgl}+\mathsf{L}_{\lgl}\}$.
\end{hyp}

Hypothesis~\ref{hyp-R} is equivalent to $\kappa_{\mathsf{L}_{\mgl}+\mathsf{L}_{\lgl}}<\kappa_{\mathsf{L}_{\lgl}}$ and hence only concerns $\mathcal{R}$. Hypothesis~\ref{hyp-P} is a quantitative measure of how effective the random perturbation is when it comes to moving vectors $v$ which are in the span of the unstable fixed points (namely $\lgl(v)=0$) into the span of the complementary stable fixed points (namely into the $\lgl$-part, even after the elimination of $\mathsf{q}-1$ directions in the $\lgl$-part by the projection $W^\perp=\mathbf{1}-W$). Note that the condition $\beta>0$ is independent of $\lambda$ and it will be shown in Remark~\ref{rem-beta-monotone} that it is decreasing in $\mathsf{q}$. Hypothesis~\ref{hyp-lambda} is a condition on the size of the coupling constant. Hypothesis~\ref{hyp-q} is then a restriction on the dimension $\mathsf{q}$ of subspaces that can be controlled in the results below. Finally Hypothesis~\ref{hyp-eta} is also about the interplay of $\mathcal{R}$ and $\lambda$. It requires the microscopic gaps (between neighboring diagonal entries of $\mathcal{R}$) to be small compared to the coupling constant $\lambda$ of the random perturbation. One can show that if Hypothesis~\ref{hyp-eta} does not hold, the ascension of the ladder is excluded. However, Hypothesis~\ref{hyp-eta} is a crucial element in one technical step of the proof (more precisely, see Section~\ref{sec-subdivision}), but nevertheless we believe that this condition is not necessary for Theorem~\ref{theorem-main} to hold. In the interesting regime of intermediate $\lambda$  Hypothesis~\ref{hyp-eta} is satisfied in  applications  (see, in particular, the second toy model in Section~\ref{sec-toy-models}). Let us briefly indicate that a rescaling $\mathcal{P}\mapsto r\mathcal{P}$ and $\lambda\mapsto r^{-1}\lambda$ with some $r\in(0,1]$ affects both $\mathfrak{P}$ and $\beta$ in Hypothesis~\ref{hyp-P} which then makes  Hypothesis~\ref{hyp-lambda}  and~\ref{hyp-q} more restrictive and leads to a worse bound in Theorem~\ref{theorem-main} below, but may allow to satisfy Hypothesis~\ref{hyp-eta}. Let us now formulate the main result of this work.

\begin{theo}\label{theorem-main}
Under {\rm Hypotheses~1} to {\rm 5},
all $Q_0\in\mathbb{G}_{\mathsf{L},\mathsf{q}}$ and $T\geq T_0:=4\beta^{-1}\mathsf{q}^2\vartheta\lambda^{-2}$ obey
\begin{align}\label{ineq-main}
\mathbb{E}\,\mathsf{d}(Q_T)\;\leq\; 10\,\pmb{\eta}^{-1}\mathsf{q}\,\lambda^2\,.
\end{align}
\end{theo}

\begin{rem}
\label{rem-Compare}
{\rm
Reference \cite{DS} proved the weaker bound $\mathbb{E}\,\mathsf{d}(Q_T)\leq C\lambda^{\frac{2\mathsf{L}_{\mathfrak{c}}}{2+\mathsf{L}_{\mathfrak{c}}}}$ only for the case $\mathsf{q}=1$ under considerably stronger hypotheses on the distribution of the $\mathcal{P}_n$ (in particular, its rotational invariance).
\hfill $\diamond$
}
\end{rem}

\begin{rem}
{\rm
The scaling of the upper bound~\eqref{ineq-main} in $\pmb{\eta}^{-1}$, $\mathsf{q}$ and $\lambda^2$ is as expected. In fact, the $\lambda^2$ follows from the assumption that the distribution of $\mathcal{P}$ is centered, hence there is no contribution to the expected value of $\mathsf{d}$ that is linear in $\lambda$. The fact that $\mathsf{d}$ is a quantity giving information on $\mathsf{q}$ linearly independent directions (\textit{i.e.}, $Q \in \mathbb{G}_{\mathsf{L},\mathsf{q}}$) justifies the scaling with this parameter. Also the $\pmb{\eta}$ dependence is sensible: a smaller gap allows for a larger average contribution to $\mathsf{d}$.

\vspace{.1cm}

The bound $T_0$ on the equilibration time scales with $\mathsf{q}^2$ and $\beta^{-1}\lambda^{-2}$ (up to a correction that is logarithmic in $\lambda$). The factor $\beta^{-1}\lambda^{-2}$ results from the diffusion described in Hypothesis~\ref{hyp-P}: on average, every iteration transfers a mass of the order $\mathcal{O}(\beta\,\lambda^2)$ out of the upper part by the perturbation. On the other hand, we believe that the factor $\mathsf{q}^2$ is an artifact of the inductive technique of proof techniques and expect that it does not reflect the correct behavior.
\hfill $\diamond$
}
\end{rem}

\begin{rem}
{\rm
With Markov's inequality, Theorem~\ref{theorem-main} (using the same notation and conditions) implies that
$$
\mathbb{P}\left[\mathsf{d}(Q_T) \geq \varepsilon\right] \;\leq\; \frac{10\,\mathsf{q}\,\lambda^2}{\pmb{\eta}\,\varepsilon}
$$
for arbitrary $\varepsilon > 0$.\hfill $\diamond$
}
\end{rem}

\begin{rem}
{\rm
Let $\mu_{\lambda,\mathsf{q}}$ be an invariant measure on $\mathbb{G}_{\mathsf{L},\mathsf{q}}$ for the Markov process~\eqref{dyn-grassmanian}. Provided that Hypotheses~1 to 5 hold, Theorem~\ref{theorem-main} implies that
$$
\int_{\mathbb{G}_{\mathsf{L},\mathsf{q}}} \mu_{\lambda,\mathsf{q}}(dQ)\;\mathsf{d}(Q)\;\leq \;10\,\pmb{\eta}^{-1}\mathsf{q}\,\lambda^2
\;.
$$
Under suitable weak assumptions ($\mathsf{q}$-strong irreducibility and $\mathsf{q}$-contractibility also called proximality, see Section~\ref{sec-Lyapunov-exponents}), it can be shown that there is a unique invariant measure called the Furstenberg measure \cite{BL,BQ}.
}
\hfill $\diamond$
\end{rem}

\begin{rem}
{\rm
It is possible to deduce Theorem~\ref{theorem-main} for general $\mathsf{q}$ from the special case $\mathsf{q}=1$ by describing $\mathsf{q}$-dimensional subspaces by $\mathsf{q}$-fold wedge product which are then one-dimensional vectors in $\Lambda^{\mathsf{q}}\mathbb{C}^{\mathsf{L}}$, see \cite{BL,CL}. However, the (second quantization) representation $d \Lambda^{\mathsf{q}} (\mathcal{P})$ of the random Lie algebra element $\mathcal{P}$ then can have a norm of order $\mathsf{q}$. If then all hypotheses hold, a naive application of Theorem~\ref{theorem-main} to the second quantization thus only provides a bound with $(\mathsf{q}\,\lambda)^2$ on the r.h.s. of~\eqref{ineq-main}, rather than $\mathsf{q}\,\lambda^2$. As will be explained in Section~\ref{sec-outline} and carried out later on, the argument leading to Theorem~\ref{theorem-main} is rather based on an iterative probabilistic treatment.}
\hfill $\diamond$
\end{rem}

\begin{rem}
\label{rem-beta-monotone}
{\rm
It is possible to rewrite $\beta$ as
$$
\beta
\;=\;
\inf\left\{\mathbb{E}(v^*\mathcal{P}^* \widetilde{W}\mathcal{P}v)\;:  \quad v\in\mathbb{S}_{\mathbb{C}}^{\mathsf{L}-1}\,,\quad \lgl(v)=0\,,\quad \widetilde{W}\in\mathbb{G}_{\mathsf{L},\mathsf{L}_{\mathfrak{c}}-\mathsf{q}+1}\,,\quad \widetilde{W}\leq\hat{P}_{{\lgl}}\right\}
\;.
$$
This follows from the identity
$$
\|\lgl((\mathbf{1}-W)\mathcal{P}v)\|^2
\;=\;
\|\lgl(W^\perp\mathcal{P}v)\|^2
\;=\;
v^*\mathcal{P}^*W^\perp \hat{P}_{{\lgl}} W^\perp \mathcal{P}v
\;=\;
v^*\mathcal{P}^*(\hat{P}_{{\lgl}} -W) \mathcal{P}v
\;,
$$
where the last step follows from $W\leq\hat{P}_{{\lgl}}$ in the definition of $\beta$. Hence setting $\widetilde{W}=\hat{P}_{{\lgl}} -W$ one obtains the alternative expression for $\beta$. It clearly shows that $\mathsf{q} \mapsto \beta(\mathsf{q},\mathsf{L}_{\lgl})$ is non-increasing.
}
\hfill $\diamond$
\end{rem}

\begin{rem} 
{\rm Yet another way to express $\beta$ in Hypothesis~\ref{hyp-P} is
 $$
\beta
\;=\;
\inf\left\{\mathbb{E}\,\|\lgl((\mathbf{1}-W)\mathcal{P}v)\|^2:  \quad v\in\mathbb{S}_{\mathbb{C}}^{\mathsf{L}-1}\,,\quad \lgl(v)=0\,,\quad W\in\mathbb{G}_{\mathsf{L},\mathsf{q}-1}\right\}
\;,
$$
notably the condition $W\leq\hat{P}_{\lgl}$ can be dropped. To verify this, let us denote the r.h.s. by $\beta^{\prime}$. Clearly one has $\beta^{\prime}\leq \beta$. On the other hand, using the notation $W^\perp =\mathbf{1}-W$ one can rewrite
$$
\beta^{\prime}
\;=\;
\inf\left\{\mathbb{E}(v^*\mathcal{P}^* W^\perp \hat{P}_{\lgl} W^\perp\mathcal{P}v)\;:  \quad v\in\mathbb{S}_{\mathbb{C}}^{\mathsf{L}-1}\,,\quad \lgl(v)=0\,,\quad W\in\mathbb{G}_{\mathsf{L},\mathsf{q}-1}\right\}
\;.
$$ 
Next since
\begin{align*}
\dim\left(\operatorname{Ker}(\hat{P}_{\mathfrak{c}}^{\perp}) \cap \operatorname{Ker}(W)\right) &
\;=\; 
\dim\left(\operatorname{Ker}(\hat{P}_{\mathfrak{c}}^{\perp})\right) + \dim\left(\operatorname{Ker}(W)\right) - \dim\left(\operatorname{Ker}(\hat{P}_{\mathfrak{c}}^{\perp}) + \operatorname{Ker}(W)\right)
\\
&\;\geq\; \mathsf{L}_{\mathfrak{c}} \,+ \,(\mathsf{L}-\mathsf{q}+1) \,-\, \mathsf{L} 
\\
&
\;=\; 
\mathsf{L}_{\mathfrak{c}}\, -\, \mathsf{q}\,+\,1\,,
\end{align*}
there exist at least $\mathsf{L}_{\mathfrak{c}} - \mathsf{q}+1$ linearly independent vectors that are in the span of both $\hat{P}_{\mathfrak{c}}$ and $W^{\perp}$. If the projection on the subspace they span is denoted by $Q$, then $Q\leq \hat{P}_{\mathfrak{c}}$, $Q\leq W^{\perp}$ and the rank of $Q$ is at least $\mathsf{L}_{\mathfrak{c}} - \mathsf{q}+1$. Thus one finds $W^{\perp}\hat{P}_{\mathfrak{c}}W^{\perp} \geq W^{\perp}QW^{\perp} = Q$ so that
$$
\beta^{\prime}
\;\geq\;
\inf\left\{\mathbb{E}(v^*\mathcal{P}^* Q\mathcal{P}v)\;:  \quad v\in\mathbb{S}_{\mathbb{C}}^{\mathsf{L}-1}\,,\quad \lgl(v)=0\,,\quad Q\in\mathbb{G}_{\mathsf{L},\mathsf{L}_{\mathfrak{c}}-\mathsf{q}+1}\,,\quad Q\leq\hat{P}_{{\lgl}}\right\}
\;.
$$
Now decompose $\hat{P}_{{\lgl}}=Q\oplus W$ with $W\in\mathbb{G}_{\mathsf{L},\mathsf{q}-1}$. Then $W\leq \hat{P}_{{\lgl}}$ and $Q=\hat{P}_{{\lgl}}-W=W^\perp\hat{P}_{{\lgl}} W^\perp$. Replacing this shows $\beta^{\prime}\geq\beta$.
}
\hfill $\diamond$
\end{rem}

\subsection{Toy models}
\label{sec-toy-models}

This brief section presents two toy models for which the crucial part $\beta>0$ of Hypothesis~\ref{hyp-P} holds. As already stated, $\beta$ measures the efficiency of $\mathcal{P}$ to spread out weight over the basis of $\mathcal{R}$. The first example, similar to the model studied in \cite{DS}, assumes that $\mathcal{P}$ contains a factor which is Haar distributed. The second example supposes that $\mathcal{P}$ is a random Toeplitz matrix, similar as for the transfer matrices in the Anderson model which will be described in Section~\ref{sec-motivation}.

\vspace{.2cm}

\noindent {\bf First example.} Let us suppose that $\mathcal{P}$ is of the form
$$
\mathcal{P}
\;=\;AUB
\;
$$
with independent random $A,B,U\in\mathbb{C}^{\mathsf{L}\times\mathsf{L}}$ where $U$ is Haar distributed on $\textnormal{U}(\mathsf{L})$, the random matrices $A$ and $B$ satisfy deterministic bounds $\|A\|\leq 1$ and $\|B\|\leq 1$ and the squares of their smallest singular values $\mu_1(AA^*)$ and $\mu_1(B^*B)$ have strictly positive expectation values $\mathbb{E}(\mu_1(AA^*))$ and $\mathbb{E}(\mu_1(B^*B))$. In order to verify $\beta>0$, let us take $(W,v)\in\mathbb{G}_{\mathsf{L},\mathsf{q}-1}\times\mathbb{S}_{\mathbb{C}}^{\mathsf{L}-1}$ and set $W^\perp=\mathbf{1}-W$. Then
\begin{align*}
\mathbb{E}\,\|\lgl(W^{\perp}\mathcal{P}v)\|^2
&
\;=\;
\mathbb{E}\,\operatorname{tr}\left[U^*\left(A^*W^{\perp}\hat{P}_{\lgl}W^{\perp}A\right)U\left(Bvv^*B^*\right)\right]\\
&
\;=\;
\mathsf{L}^{-1}\,\mathbb{E}\,\operatorname{tr}\left[A^*W^{\perp}\hat{P}_{\lgl}W^{\perp}A\right]\,\operatorname{tr}\left[Bvv^*B^*\right]\\
&
\;=\;\mathsf{L}^{-1}\,\mathbb{E}\,\operatorname{tr}\left[\hat{P}_{\lgl}W^{\perp}AA^*W^{\perp}\hat{P}_{\lgl}\right]\,\langle v|B^*Bv\rangle
\;,
\end{align*}
where in the second step Lemma~2 in~\cite{RS} was used. Now one can further bound:
\begin{align*}
\mathbb{E}\,\|\lgl(W^{\perp}\mathcal{P}v)\|^2
&
\;\geq\; 
\mathsf{L}^{-1}\,\mathbb{E}\big(\mu_1(AA^*)\big)\cdot\mathbb{E}\big(\mu_1(B^*B)\big)\cdot \operatorname{tr}\big(\hat{P}_{\lgl}W^{\perp}\hat{P}_{\lgl}\big)\\
&
\;\geq\; 
\mathsf{L}^{-1}\,\mathbb{E}\big(\mu_1(AA^*)\big)\cdot \mathbb{E}\big(\mu_1(B^*B)\big)\cdot\left(\mathsf{L}_{\lgl}- (\mathsf{q}-1)\right)
\,,
\end{align*}
the latter because
$$
\operatorname{tr}\big(\hat{P}_{\lgl}W^{\perp}\hat{P}_{\lgl}\big)
\;=\;
\mathsf{L}_{\lgl}- \operatorname{tr}\big(W\hat{P}_{\lgl}W\big)
\;\geq\;
\mathsf{L}_{\lgl}- \operatorname{tr}\big( W\big)
\;=\;
\mathsf{L}_{\lgl}- (\mathsf{q}-1)
\;.
$$
Hence one concludes that Hypothesis~\ref{hyp-P} holds with
$$
\beta{(\mathsf{q},\mathsf{L}_\lgl)}
\;\geq\;
\mathbb{E}\big(\mu_1(AA^*)\big)\cdot\mathbb{E}\big(\mu_1(B^*B)\big)\cdot \frac{\mathsf{L}_{\lgl}-\mathsf{q}+1}{\mathsf{L}}
\;>\;
0\,.
$$

\vspace{.2cm}

\noindent {\bf Second example.} This second toy model is motivated by potential applications discussed in Section~\ref{sec-motivation}. Let $S$ denote the cyclic shift on $\mathbb{C}^\mathsf{L}$ and so that $\Delta^{\mathsf{L}}_1=-(S+S^*)$
is the one-dimensional discrete Laplacian on $\mathsf{L}$ points with periodic boundary conditions. Its spectrum lies in $[-2,2]$, so that $\Delta^{\mathsf{L}}_1+ s>0$ for $s>2$. Suitable modifications of the arguments below also allow to consider the discrete Laplacian in dimension $d\geq 1$, but for sake of clarity we refrain from giving details (at least in the case in which $\mathsf{L}_{\mathfrak{c}}$ is a multiple of the periodicities). The discrete Laplacian $\Delta^{\mathsf{L}}_1$ is diagonalized by the Fourier transformation $\mathcal{F}:\mathbb{C}^{\mathsf{L}}\to \mathbb{C}^{\mathsf{L}}$ which will be written out explicitly further down. Then the model of the from~\eqref{def-T} is specified by
\begin{equation}
\label{eq-second-toy-model}
\mathcal{R}
\;=\;
\mathcal{F} (\Delta^{\mathsf{L}}_1+ s)\mathcal{F}^*
\;,
\qquad
\mathcal{P}
\;=\;
\mathcal{F} \Big(\sum_{j=1}^{\mathsf{L}}\omega_j\,|j\rangle\langle j|\Big)\mathcal{F}^*
\;\in\;
\mathbb{C}^{\mathsf{L}\times\mathsf{L}}
\;,
\end{equation}
where the numbers $\omega_i\in[-1,1]$ are i.i.d. centered random values. Here $|j\rangle$ denotes the state localized at site $j=1,\ldots,\mathsf{L}$ (these are the same vectors as the $e_j$ above, but in Fourier space). Note that $\mathcal{P}$ takes the most general form of a random Toeplitz matrix. The main aim in the following will be to show that this toy model satisfies Hypothesis~\ref{hyp-P}. Clearly this is actually independent of the particular form of $\mathcal{R}$. On the other hand, the real diagonal matrix $\mathcal{R}\in\mathbb{C}^{\mathsf{L}\times\mathsf{L}} $ written out explicitly below satisfies Hypothesis~\ref{hyp-R} and for large enough $\mathsf{L}$ also Hypothesis~\ref{hyp-eta} so that Theorem~\ref{theorem-main} applies.

\vspace{.2cm}

For the sake of simplicity, let us assume that $\mathsf{L}$ is odd. The Fourier transform $\mathcal{F}$ written in terms of the plane wave states   $c_k\in\mathbb{C}^{\mathsf{L}}$, $k \in \lbrace 1,\dots,\mathsf{L}\rbrace$, is
$$
	c_k
	\;=\;
	\frac{1}{\sqrt{\mathsf{L}}}
	\begin{pmatrix}
	\exp(\tfrac{2\pi \imath k}{\mathsf{L}})
	\\
	\exp(\tfrac{2\pi \imath 2 k}{\mathsf{L}})
	\\
	\vdots 
	\\
	\exp(\tfrac{2\pi \imath \mathsf{L}k}{\mathsf{L}})
	\end{pmatrix}
	\;, \qquad \mathcal{F}^*
	\;=\;
	\big(
	c_{\mathsf{L}},c_1,c_{\mathsf{L}-1},c_2,c_{\mathsf{L}-2},\ldots,c_{\frac{\mathsf{L}-3}{2}},c_{\frac{\mathsf{L}+3}{2}},c_{\frac{\mathsf{L}-1}{2}},c_{\frac{\mathsf{L}+1}{2}}
	\big)
	\;.
$$
Here the order of the vectors is chosen such that $\mathcal{R}$ is by construction already in the ordered form~\eqref{def-R}:
$$
	\mathcal{R}
	\;=\;
	s\,{\bf 1}\,-\,2\,\mbox{\rm diag}\big(
	1,
	\cos(\tfrac{2\pi }{\mathsf{L}}),
	\cos(\tfrac{2\pi }{\mathsf{L}}),\cos(\tfrac{2\pi 2}{\mathsf{L}}),
	\cos(\tfrac{2\pi 2}{\mathsf{L}}),
	\ldots,\cos(\tfrac{\pi (\mathsf{L}-1)}{\mathsf{L}}),
	\cos(\tfrac{\pi (\mathsf{L}-1)}{\mathsf{L}})
	\big)
	\;.
$$
Let us start out with the case $\mathsf{q} = 1$ so that the projection in the dynamics is of rank one. From the characterization of $\beta$ in Remark~\ref{rem-beta-monotone} and using the explicit form $\mathcal{P} = \sum_{j=1}^{\mathsf{L}}\omega_{\sigma(j)}\,|c_j\rangle\langle c_j|$ (for some permutation $\sigma$ of $\lbrace 1,\dots,\mathsf{L} \rbrace$ due to the reordering in $\mathcal{F}$) from~\eqref{eq-second-toy-model}, one finds
\begin{align*}
\beta{(1,\mathsf{L}_\lgl)} 
&\;=\;
	\inf\left\{\mathbb{E}(v^*\mathcal{P}^* \widetilde{W}\mathcal{P}v)\;:  \quad v\in\mathbb{S}_{\mathbb{C}}^{\mathsf{L}-1}\,,\quad \lgl(v)=0\,,\quad \widetilde{W}\in\mathbb{G}_{\mathsf{L},\mathsf{L}_{\mathfrak{c}}}\,,\quad \widetilde{W}\leq\hat{P}_{{\lgl}}\right\}\\
	&
	\;=\; \mathbb{E}(\omega^2)\inf\left\{\sum_{j=1}^{\mathsf{L}}\langle v|c_j\rangle \langle c_j|\hat{P}_{\mathfrak{c}}c_j\rangle \langle c_j|v\rangle\;:  \quad v\in\mathbb{S}_{\mathbb{C}}^{\mathsf{L}-1}\,,\quad \lgl(v)=0\right\}\\
	&= \mathbb{E}(\omega^2)\inf\left\{\sum_{j=1}^{\mathsf{L}}|\langle v|c_j\rangle|^2 \sum_{k=\mathsf{L}-\mathsf{L}_{\mathfrak{c}}+1}^{\mathsf{L}} |\langle c_j|k\rangle|^2 \;:  \quad v\in\mathbb{S}_{\mathbb{C}}^{\mathsf{L}-1}\,,\quad \lgl(v)=0\right\}\\
	&\;=\; \frac{\mathsf{L}_{\mathfrak{c}}}{\mathsf{L}}\,\mathbb{E}(\omega^2)\; >\; 0\,,
\end{align*}
as $|\langle c_j|k\rangle| = \mathsf{L}^{-\frac{1}{2}}$ for all $j,k \in \lbrace 1,\dots, \mathsf{L} \rbrace$, and since $\widetilde{W} \in \mathbb{G}_{\mathsf{L},\mathsf{L}_{\mathfrak{c}}}$ and $\widetilde{W} \leq \hat{P}_{\mathfrak{c}}$ imply that $\widetilde{W} = \hat{P}_{\mathfrak{c}}$.

\vspace{.2cm}

We are unable to prove a quantitative lower bound for general $\mathsf{q} \leq \mathsf{L}_{\mathfrak{c}}$, but can prove that $\beta{(\mathsf{q},\mathsf{L}_\lgl)}>0$ for $\mathsf{q} \leq \mathsf{L}_{\lgl}$. As $\mathsf{q}\mapsto \beta{(\mathsf{q},\mathsf{L}_\lgl)}$ is non-increasing by Remark~\ref{rem-beta-monotone}, it is sufficient to show $\beta{(\mathsf{L}_\lgl,\mathsf{L}_\lgl)}>0$, namely to consider the case $\mathsf{q} = \mathsf{L}_{\mathfrak{c}}$.  Starting from the rewriting in Remark~\ref{rem-beta-monotone}, setting $\widetilde{W} = ww^* \in \mathbb{G}_{\mathsf{L},1}$ for some $w \in \mathbb{S}_{\mathbb{C}}^{\mathsf{L}-1}$ with then $\mathfrak{c}^c(w) = 0$ from $\widetilde{W} \leq \hat{P}_{\mathfrak{c}}$, one first finds
\begin{align}
\label{eq-beta-toy-model}
\beta{(\mathsf{L}_\lgl,\mathsf{L}_\lgl)} 
&
\,=\, 
\inf\left\{\mathbb{E}(v^*\mathcal{P}^* \widetilde{W}\mathcal{P}v)\;:  \quad v\in\mathbb{S}_{\mathbb{C}}^{\mathsf{L}-1}\,,\quad \lgl(v)=0\,,\quad \widetilde{W}\in\mathbb{G}_{\mathsf{L},1}\,,\quad \widetilde{W}\leq\hat{P}_{{\lgl}}\right\}
\nonumber
\\
&
\,=\, 
\mathbb{E}(\omega^2)\,\inf\left\{\sum_{j=1}^L
	|\langle c_j|v\rangle|^2
	|\langle c_j|w\rangle|^2\;:  \quad v, w\in\mathbb{S}_{\mathbb{C}}^{\mathsf{L}-1}\,,\quad \lgl(v)=0\,,\quad \lgl^c(w)=0\right\}
	.
\end{align}
Let us now consider index sets $J = \lbrace j_1, \dots, j_{\mathsf{L}-\mathsf{L}_{\mathfrak{c}}} \rbrace$ and $K = \lbrace k_1, \dots, k_{\mathsf{L}_{\mathfrak{c}}} \rbrace$, subsets of $\lbrace 1,\dots,\mathsf{L} \rbrace$, having $\mathsf{L}-\mathsf{L}_{\mathfrak{c}}$ and $\mathsf{L}_{\mathfrak{c}}$ elements respectively. A reduction to a Vandermonde determinant shows
\begin{align*}
\det\big(\lgl^c(c_{j_1}),\dots,\lgl^c(c_{j_{\mathsf{L}-\mathsf{L}_{\mathfrak{c}}}})\big) 
&
\;=\; 
\mathsf{L}^{-\frac{\mathsf{L}-\mathsf{L}_{\mathfrak{c}}}{2}} \Big[\prod_{j \in J} e^{\tfrac{2\pi \imath  j}{\mathsf{L}}} \Big] \det\begin{pmatrix} 1 & \cdots & 1\\ e^{\tfrac{2\pi \imath  j_1}{\mathsf{L}}} & \cdots & e^{\tfrac{2\pi \imath  j_{\mathsf{L}-\mathsf{L}_{\mathfrak{c}}}}{\mathsf{L}}}\\ \Big[e^{\tfrac{2\pi \imath  j_1}{\mathsf{L}}}\Big]^2 & \cdots & \Big[e^{\tfrac{2\pi \imath  j_{\mathsf{L}-\mathsf{L}_{\mathfrak{c}}}}{\mathsf{L}}}\Big]^2\\ \vdots & \ddots & \vdots \\ \Big[e^{\tfrac{2\pi \imath  j_1}{\mathsf{L}}}\Big]^{\mathsf{L}-\mathsf{L}_{\mathfrak{c}}-1} & \cdots & \Big[e^{\tfrac{2\pi \imath  j_{\mathsf{L}-\mathsf{L}_{\mathfrak{c}}}}{\mathsf{L}}}\Big]^{\mathsf{L}-\mathsf{L}_{\mathfrak{c}}-1} \end{pmatrix}\\
&
\;=\; 
\mathsf{L}^{-\frac{\mathsf{L}-\mathsf{L}_{\mathfrak{c}}}{2}} \prod_{j \in J} \Big[\exp(\tfrac{2\pi \imath  j}{\mathsf{L}}) \prod_{j^{\prime} \in J,\,j^{\prime} \neq j} \big[\exp(\tfrac{2\pi \imath  j}{\mathsf{L}}) - \exp(\tfrac{2\pi \imath  j^{\prime}}{\mathsf{L}})\big]\Big]\\
&
\;\neq\; 0\,,
\end{align*}
(including an irrelevant sign ambiguity) and similarly
\begin{align*}
\det\big(\lgl(c_{k_1}),\dots,\lgl(c_{k_{\mathsf{L}_{\mathfrak{c}}}})\big) 
\;=\; 
\mathsf{L}^{-\frac{\mathsf{L}_{\mathfrak{c}}}{2}} \prod_{k \in K} \Big[\exp(\tfrac{2\pi \imath  \mathsf{L}_{\mathfrak{c}} k}{\mathsf{L}}) \prod_{k^{\prime} \in K,\,k^{\prime} \neq k} \big[\exp(\tfrac{2\pi \imath  k}{\mathsf{L}}) - \exp(\tfrac{2\pi \imath  k^{\prime}}{\mathsf{L}})\big]\Big] \;\neq\; 0\,.
\end{align*}
This implies that $\lbrace \hat{P}_{\mathfrak{c}}^{\perp}c_j \;:\; j \in J \rbrace$ and $\lbrace \hat{P}_{\mathfrak{c}}c_k \;:\; k \in K \rbrace$ are sets of linearly independent vectors. Now suppose that $\beta$ as characterized in~\eqref{eq-beta-toy-model} equals zero. That means that either the set $J^{\prime} = \lbrace j \in \lbrace 1,\dots,\mathsf{L} \rbrace \;:\; \langle c_j|w\rangle \neq 0 \rbrace$ contains no more than $\mathsf{L} - \mathsf{L}_{\mathfrak{c}}$ elements, or the set $K^{\prime} = \lbrace k \in \lbrace 1,\dots,\mathsf{L} \rbrace \;:\; \langle c_k|v\rangle \neq 0 \rbrace$ contains at most $\mathsf{L}_{\mathfrak{c}}$ elements. In the first case, one finds $0 = \hat{P}_{\mathfrak{c}}^{\perp}w = \hat{P}_{\mathfrak{c}}^{\perp}\sum_{j=1}^{\mathsf{L}} \langle c_j|w \rangle c_j = \sum_{j \in J^{\prime}} \langle c_j|w \rangle \hat{P}_{\mathfrak{c}}^{\perp}c_j$, which is a sum of linearly independent vectors (as there are at most $\mathsf{L} - \mathsf{L}_{\mathfrak{c}}$ summands) that equals zero. Then all coefficients must vanish, meaning that $J^{\prime} = \emptyset$. But then $0 = \sum_{j=1}^{\mathsf{L}} \langle c_j|w \rangle c_j = w \in \mathbb{S}_{\mathbb{C}}^{\mathsf{L}-1}$, a contradiction. With the second case treated similarly, one concludes that not all summands of~\eqref{eq-beta-toy-model} can vanish, hence $\beta {(\mathsf{L}_\lgl,\mathsf{L}_\lgl)} > 0$.

\subsection{Bounds on Lyapunov exponents}
\label{sec-Lyapunov-exponents}

This section uses the upper bound on $\mathbb{E}\,\mathsf{d}(Q_n)$ obtained in Theorem~\ref{theorem-main} to prove a  bounds on the average of the  $\mathsf{q}$ Lyapunov exponents associated to the sequence of random matrices $(\mathcal{T}_n)_{n\in\mathbb{N}}$ given by~\eqref{def-T}. The general theory of Lyapunov exponents is laid out in \cite{BL,CL} for the case of real matrices, while the more general case of complex matrices is covered by \cite{BQ}. There is one change of terminology in these books, namely the contracting semigroups in \cite{BL,CL} (and Hypothesis 14 below) are called proximal in \cite{BQ}. Let us begin by recalling the definition of the Lyapunov exponents which according to \cite{BQ,BL,CL} makes sense (namely the limits exist).

\begin{defini}
The Lyapunov exponents $\gamma_1,\dots,\gamma_{\mathsf{L}}\geq 0$ associated to $(\mathcal{T}_n)_{n\in\mathbb{N}}$ are defined by
\begin{align}\label{def-Lyapunov-exponents}
\sum\limits_{\mathsf{w}=1}^{\mathsf{q}}\gamma_{\mathsf{w}}
\;=\;
\lim\limits_{N\rightarrow\infty}\frac{1}{N}\,\mathbb{E}\,\log\,\left\|\Lambda^{\mathsf{q}}(\mathcal{T}_N\cdots\mathcal{T}_1)\right\|_{\Lambda^{\mathsf{q}}\mathbb{C}^{\mathsf{L}}}\,,\qquad\mathsf{q}=1,\dots,\mathsf{L}\;.
\end{align}
\end{defini}

By A.III.5 in \cite{BL}, the map $\mathsf{w}\mapsto \gamma_{\mathsf{w}}$ is non-increasing. The following three hypotheses are standard assumptions in the theory of Lyapunov exponents. They are known to hold in many situations. Hypothesis~\ref{hyp-Lyapunov-existence} is trivially satisfied in the situations considered in this work. 


\begin{hyp}\label{hyp-Lyapunov-existence}
The averages $\mathbb{E}\max\left\{\log\|\mathcal{T}_1\|,0\right\}$ and $\mathbb{E}\max\left\{\log\|\mathcal{T}_1^{-1}\|,0\right\}$ are finite.
\end{hyp}
 
\begin{hyp}\label{hyp-Lyapunov-irreducible}
The semigroup $\mathbf{S}$ generated by $\textnormal{supp}(\mathcal{T}_1)$ is $\mathsf{q}$-strongly irreducible, \textit{i.e.}, for any finite union $F$ of proper linear subspaces of $\Lambda^{\mathsf{q}}\mathbb{C}^{\mathsf{L}}$, there exists some $\mathcal{T}\in\mathbf{S}$ such that $\Lambda^{\mathsf{q}}\mathcal{T}F\not\subset F$.
\end{hyp}

\begin{hyp}\label{hyp-Lyapunov-contracting}
The semigroup $\mathbf{S}$ generated by $\textnormal{supp}(\mathcal{T}_1)$ is $\mathsf{q}$-contracting, \textit{i.e.}, there exists a sequence $(T_n)_{n\in\mathbb{N}}\subset \mathbf{S}$ for which $\Lambda^{\mathsf{q}}T_n\|\Lambda^{\mathsf{q}}T_n\|_{\Lambda^{\mathsf{q}}\mathbb{C}^{\mathsf{L}}}^{-1}$ converges to an operator of rank one.
\end{hyp}

\begin{theo}\label{theorem-lower-bound-Lyapunov}
Under {\rm Hypotheses~1} to {\rm 5} as well as {\rm Hypotheses~\ref{hyp-Lyapunov-irreducible}} and {\rm~\ref{hyp-Lyapunov-contracting}}, the average of the $\mathsf{q}$ largest Lyapunov exponents associated to $(\mathcal{T}_n)_{n\in\mathbb{N}}$ is bounded from below by
$$
\frac{1}{\mathsf{q}}\,\sum\limits_{\mathsf{w}=1}^{\mathsf{q}}\gamma_{\mathsf{w}}
\;\geq \;
\log(\kappa_{\mathsf{L}_{\mgl}+\mathsf{L}_{\lgl}})-\left[\mbox{\small $\frac{3}{2}$}-10\,\pmb{\eta}^{-1}\,\log\mbox{\small $\frac{\kappa_{\mathsf{L}}}{\kappa_{\mathsf{L}_{\mgl}+\mathsf{L}_{\lgl}}}$}\right]\lambda^2\,.
$$
\end{theo}

This bound is in accordance with the phenomenon proved in Theorem~\ref{theorem-main}. The $\mathsf{q}$ most stable directions are away from the upper part of $\mathcal{R}$ in which the diagonal entries of $\mathcal{R}$ are smaller that $\kappa_{\mathsf{L}_{\mgl}+\mathsf{L}_{\lgl}}$. Therefore the corresponding effects of the $\mathsf{q}$ most stable directions are at least of the power $\log(\kappa_{\mathsf{L}_{\mgl}+\mathsf{L}_{\lgl}})$. While detailed proofs are deferred to Appendix~\ref{app-Lyapunov}, let us outline why Theorem~\ref{theorem-lower-bound-Lyapunov} indeed follows from Theorem~\ref{theorem-main}. This is, first of all, based on the well-known fact that the Lyapunov exponents can be accessed via averages of the random dynamics~\eqref{dyn-grassmanian} with an arbitrary fixed initial condition (see {\it e.g.} Section 2.4 of \cite{SB1}).

\begin{proposi}\label{proposi-Lyapunov-rewriting}
Under {\rm Hypotheses~\ref{hyp-Lyapunov-existence},~\ref{hyp-Lyapunov-irreducible}} and {\rm~\ref{hyp-Lyapunov-contracting}}, the sum of the $\mathsf{q}$ largest Lyapunov exponents associated to $(\mathcal{T}_n)_{n\in\mathbb{N}}$ is given in terms of the dynamics~\eqref{dyn-grassmanian} by
$$
\sum\limits_{\mathsf{w}=1}^{\mathsf{q}}\gamma_{\mathsf{w}}
\;=\;
\frac{1}{2}\,\lim\limits_{N\rightarrow\infty}\frac{1}{N}\sum\limits_{n=0}^{N-1}\,\mathbb{E}\,\log\,\det\left(\Phi_n^*\mathcal{T}_{n+1}^*\mathcal{T}_{n+1}\Phi_n\right)\,,
$$
where $\Phi_n\in\mathbb{F}_{\mathsf{L},\mathsf{q}}$ is such that $Q_n=\Phi_n\Phi_n^*$.
\end{proposi}

The next purely computational lemma now bounds the r.h.s. of the statement of Proposition~\ref{proposi-Lyapunov-rewriting} by the quantity $\mathsf{d}$ defined by~\eqref{def-d}. This result combined with Theorem~\ref{theorem-main} then readily implies Theorem~\ref{theorem-lower-bound-Lyapunov}.

\begin{lemma}\label{lemma-Lyapunov-estimate}
Let $\mathtt{P}\in\mathfrak{P}$ and  $Q\in\mathbb{G}_{\mathsf{L},\mathsf{q}}$ and  $\Phi\in\mathbb{F}_{\mathsf{L},\mathsf{q}}$  be such that $Q=\Phi\Phi^*$. Then one has
\begin{align*}
\log\,\det & \left(\Phi^*(e^{\lambda\mathtt{P}}\mathcal{R})^*e^{\lambda\mathtt{P}}\mathcal{R}\Phi\right)
\\
&
\;\geq\; 
2\left[\,\mathsf{q}\,\log(\kappa_{\mathsf{L}_{\mgl}+\mathsf{L}_{\lgl}})+\mathsf{d}(Q)\,\log\mbox{\small $\frac{\kappa_{\mathsf{L}}}{\kappa_{\mathsf{L}_{\mgl}+\mathsf{L}_{\lgl}}}$}+\lambda\,e^{\frac{3}{4}\lambda^2}\,\operatorname{tr}\,\left[(\mathtt{P}+\mathtt{P}^*)(\mathcal{R}\cdot Q)\right]\right]-3\,\lambda^2\,\mathsf{q}\,.
\end{align*}
\end{lemma}

By a reflection principle, it is also possible to derive an upper bound on the $\mathsf{q}$ smallest Lyapunov exponents by considering the adjoint inverses of the sequence~\eqref{def-T}. The reflection principle stated next holds for general sequences of i.i.d. random matrices. While we suspect that it is known, we could not track down a reference.

\begin{proposi}\label{proposi-Lyapunov-reflection}
Suppose that {\rm Hypotheses~\ref{hyp-Lyapunov-existence}} to~{\rm\ref{hyp-Lyapunov-contracting}} hold for all $\mathsf{q}=1,\dots,\mathsf{L}$ with $\mathcal{T}_1$ and also with $\mathcal{T}_1^{-1}$ instead of $\mathcal{T}_1$. Then, the Lyapunov exponents $\gamma^{\prime}_1,\dots,\gamma^{\prime}_{\mathsf{L}}$ associated to $((\mathcal{T}_n^*)^{-1})_{n\in\mathbb{N}}$, given by
$$
\sum\limits_{\mathsf{w}=1}^{\mathsf{q}}\gamma_{\mathsf{w}}^{\prime}
\;=\;
\lim\limits_{N\rightarrow\infty}\frac{1}{N}\,\mathbb{E}\,\log\,\left\|\Lambda^{\mathsf{q}}((\mathcal{T}_N^*)^{-1}\cdots(\mathcal{T}_1^*)^{-1})\right\|_{\Lambda^{\mathsf{q}}\mathbb{C}^{\mathsf{L}}}\,,\qquad\mathsf{q}=1,\dots,\mathsf{L}\,,
$$
are related to the Lyapunov exponents $\gamma_1,\dots,\gamma_{\mathsf{L}}$ associated to $(\mathcal{T}_n)_{n\in\mathbb{N}}$  via
\begin{align}\label{eq-Lyapunov-reflection}
\gamma_{\mathsf{q}}
\;=\;
-\,\gamma_{\mathsf{L}-\mathsf{q}+1}^{\prime}\,,\qquad\mathsf{q}=1,\dots,\mathsf{L}\,.
\end{align}
\end{proposi}

The proof of  Proposition~\ref{proposi-Lyapunov-reflection} is also given in Appendix~\ref{app-Lyapunov} and is essentially based on the following auxiliary statement:

\begin{lemma}\label{lemma-complement-action}
Let $Q\in\mathbb{G}_{\mathsf{L},\mathsf{q}}$ and $\mathcal{T} \in \textnormal{GL}(\mathsf{L}, \mathbb{C})$. Then $(\mathcal{T} \cdot Q)^{\perp} = (\mathcal{T}^{-1})^* \cdot Q^{\perp}$.
\end{lemma}

Let us note that $(\mathcal{T}_n^*)^{-1}=e^{-\lambda\mathcal{P}_n^*}\mathcal{R}^{-1}$ is again of the form~\eqref{def-T}, as $\mathcal{R}^{-1}$ is real and diagonal, and the random matrix $\mathcal{P}$ is replaced by $-\mathcal{P}^*$. The order of the diagonal entries of $\mathcal{R}^{-1}$ is now reversed, as $\kappa_{\mathsf{L}}^{-1} \geq \dots \geq \kappa_1^{-1}> 0$. The relative gaps, however, remain invariant (see the proof of Theorem~\ref{theorem-upper-bound-Lyapunov} below). Using this adapted action (with an appropriate modification of the assumptions) and the identity~\eqref{eq-Lyapunov-reflection}, one obtains a counterpart of Theorem~\ref{theorem-lower-bound-Lyapunov}.

\begin{theo}\label{theorem-upper-bound-Lyapunov}
	Suppose that {\rm Hypotheses~\ref{hyp-R}} to~{\rm \ref{hyp-eta}} hold for $\mathcal{P}^*$ instead of $\mathcal{P}$ and for $\ugl(v)$ instead of $\lgl(v)$, and furthermore that {\rm Hypotheses~\ref{hyp-Lyapunov-irreducible}} and {\rm~\ref{hyp-Lyapunov-contracting}} hold. Then the average of the $\mathsf{q}$ smallest Lyapunov exponents $\gamma_{\mathsf{L}-\mathsf{q}+1},\dots,\gamma_{\mathsf{L}}$ associated to $(\mathcal{T}_n)_{n\in\mathbb{N}}$ is bounded from above by
	$$
	\frac{1}{\mathsf{q}}\,\sum\limits_{\mathsf{w}=\mathsf{L}-\mathsf{q}+1}^{\mathsf{L}}\gamma_{\mathsf{w}}
	\;\leq \;
	\log(\kappa_{\mathsf{L}_{\mgl}})+\left[\mbox{\small $\frac{3}{2}$}-10\,\pmb{\eta}^{-1}\,\log\mbox{\small $\frac{\kappa_{\mathsf{L}_{\mgl}}}{\kappa_{1}}$}\right]\lambda^2\,.
	$$
\end{theo}

\noindent\textbf{Proof of Theorem~\ref{theorem-upper-bound-Lyapunov}.}
Let us observe that the condition $\kappa_1 \geq \dots \geq \kappa_{\mathsf{L}} \geq 0$ on the diagonal entries of $\mathcal{R}$ implies that $\kappa_{\mathsf{L}}^{-1} \geq \dots \geq \kappa_1^{-1} \geq 0$ for those of the diagonal matrix $\mathcal{R}^{-1}$. This means that the larger entries of $\mathcal{R}^{-1}$ are now in its upper part (contrary to the lower part for $\mathcal{R}$), so the stable directions now lie in the $\ugl$-part (instead of in the $\lgl$-part). In spite of this reversal, the gap structure of $\mathcal{R}^{-1}$ and $\mathcal{R}$ remains identical, \textit{i.e.}, for $1 \leq \mathsf{I} < \mathsf{J} \leq \mathsf{L}$,
$$
1 - \frac{(\kappa_{\mathsf{I}}^{-1})^2}{(\kappa_{\mathsf{J}}^{-1})^2} 
\;=\; 
1 - \frac{\kappa_{\mathsf{J}}^2}{\kappa_{\mathsf{I}}^2} 
\;=\; 
\eta(\mathsf{I},\mathsf{J})\,.
$$
Upon adapting all Hypotheses as indicated in the statement, and using the notation $\gamma^{\prime}_1,\dots,\gamma^{\prime}_{\mathsf{L}}$ for the Lyapunov exponents associated to $((\mathcal{T}_n^*)^{-1})_{n\in\mathbb{N}}$ as introduced in Proposition~\ref{proposi-Lyapunov-rewriting}, it follows from Theorem~\ref{theorem-lower-bound-Lyapunov} that
$$
\frac{1}{\mathsf{q}}\,\sum\limits_{\mathsf{w}=1}^{\mathsf{q}}\gamma^{\prime}_{\mathsf{w}}
\;\geq \;
-\log(\kappa_{\mathsf{L}_{\mgl}})-\left[\mbox{\small $\frac{3}{2}$}-10\,\pmb{\eta}^{-1}\,\log\mbox{\small $\frac{\kappa_{\mathsf{L}_{\mgl}}}{\kappa_1}$}\right]\lambda^2\,.
$$
Here, the diagonal entries $\kappa_{\mathsf{L}_{\mgl}+\mathsf{L}_{\lgl}}$ and $\kappa_{\mathsf{L}}$ (delimiting the $\ugl$-part) of $\mathcal{R}$ are replaced by $\kappa_{\mathsf{L}_{\mgl}}^{-1}$ and $\kappa_{1}^{-1}$ of $\mathcal{R}^{-1}$ (delimiting the $\lgl$-part). Finally, inserting the result~\eqref{eq-Lyapunov-reflection} from Proposition~\ref{proposi-Lyapunov-reflection} yields the desired statement.
\hfill $\square$

\subsection{Motivation, limitations and potential improvements}
\label{sec-motivation}

Our main motivation to undertake the technical endeavor of this paper is rooted in the study of random discrete Schr\"odinger operators in high dimension $d\geq 3$ and in a weak coupling regime of the disorder. The prototypical model is the $d$-dimensional Anderson Hamiltonian $H=\Delta_d+\lambda V_d$ on $\ell^2(\mathbb{Z}^d)$ where $\Delta_d=-\sum_{j=1}^d(S_j+S_j^*)$ is the $d$-dimensional discrete Laplacian constructed from the shifts in the lattice directions and $V_d=\sum_{j\in\mathbb{Z}^d}\omega_j\,|j\rangle\langle j|$ is a random potential with i.i.d. real centered random variables $\omega_j\in[-1,1]$.  It is expected (an explanation and reference can be found in {\it e.g.} \cite{Bel} or the introduction to \cite{AW}) that in dimension $d\geq 3$ there is a non-trivial wave packet spreading for the associated quantum dynamics, more precisely a diffusive one. To provide a mathematical proof of this statement has been a challenge for decades. One way to approach the problem is to study solutions of the Schr\"odinger equation $H\phi=E\phi$ at finite volume and to show that there are many such solutions that are spread out over the whole finite sample ({\it i.e.} the eigenfunctions are roughly of constant modulus). As the solutions can satisfy a three-term recurrence relation which can be written with transfer matrices, this can be achieved by upper bounds on suitable parts of the transfer matrix across the sample. Let us show how this leads to the toy model~\eqref{eq-second-toy-model} studied in Section~\ref{sec-toy-models}. To avoid inessential algebraic difficulties (irrelevant for the set-up of the transfer matrix formalism), let us focus on the case in which the fibers are one-dimensional. Moreover, making a finite-volume approximation in these fibers, the transfer matrix at energy $E\in\mathbb{R}$ is of the form
$$
\mathcal{T}^E
\;=\;
\begin{pmatrix}
E-(\Delta^{\mathsf{L}}_1+\lambda V^{\mathsf{L}}_1) & -\mathbf{1} \\ \mathbf{1} & 0
\end{pmatrix}
\;\in\;
\mathbb{C}^{2\mathsf{L}\times 2\mathsf{L}}
\;,
$$
where $\Delta^{\mathsf{L}}_1:\mathbb{C}^{\mathsf{L}}\to \mathbb{C}^{\mathsf{L}}$ is the one-dimensional discrete Laplacian with periodic boundary conditions and $V^{\mathsf{L}}_1=\sum_{j=1}^{\mathsf{L}}\omega_j\,|j\rangle\langle j|$ the random potential. The matrices $\Delta^{\mathsf{L}}_1$ and $V^{\mathsf{L}}_1$ are exactly as in~\eqref{eq-second-toy-model}. Then $\mathcal{T}^E$ is the transfer matrix of the Anderson model on a strip of width $\mathsf{L}$ ({\it e.g.} \cite{SB1}). For larger $d$, the finite-volume fibers would consist of a Hilbert space of dimension $\mathsf{L}^{d-1}$ on which would then act fiber operators $\Delta^{\mathsf{L}}_{d-1}$ and $V^{\mathsf{L}}_{d-1}$, but the transfer matrix has the same structure and the analysis below is essentially the same as well. The transfer matrix $\mathcal{T}^E$ can be factorized  in the form~\eqref{def-T}:
$$
\mathcal{T}^E
\;=\;
\begin{pmatrix}
\mathbf{1} & -\lambda V \\ 0 & \mathbf{1} 
\end{pmatrix}
\begin{pmatrix}
E-\Delta & -\mathbf{1} \\ \mathbf{1} & 0
\end{pmatrix}
\;=\;
\exp\left(\lambda
\begin{pmatrix}
0 & - V \\ 0 & 0 
\end{pmatrix}
\right)
\begin{pmatrix}
E-\Delta & -\mathbf{1} \\ \mathbf{1} & 0
\end{pmatrix}
\;.
$$
The Fourier transformation $\mathcal{F}:\mathbb{C}^{\mathsf{L}}\to \mathbb{C}^{\mathsf{L}}$ already introduced in Section~\ref{sec-toy-models} block-diagonalizes the second matrix, while the first nilpotent factor contains a block-entry given by a Toeplitz matrix after Fourier transform, that is,
$$
\mathcal{F}\mathcal{T}^E\mathcal{F}^*
\;=\;
e^{\lambda \mathcal{P}}\mathcal{R}
\;,
$$
where
$$
\mathcal{P}
\;=\;
\begin{pmatrix}
0 & -\mathcal{F} V\mathcal{F}^* \\ 0 & 0
\end{pmatrix}
\;,
\qquad
\mathcal{R}
\;=\;
\begin{pmatrix}
E-\mathcal{F}\Delta\mathcal{F}^* & -\mathbf{1} \\ \mathbf{1} & 0
\end{pmatrix}
\;.
$$
Of course, $\mathcal{R}$ is not yet diagonal as in~\eqref{def-R}, but it is a direct sum of $2\times 2$ symplectic blocks, each of which can readily be diagonalized ({\it e.g.} \cite{SB1}). Apart from this, the essential feature of the transfer matrix $\mathcal{T}^E$ is that it is of the form~\eqref{def-T} with a random perturbation that is given by a Toeplitz matrix (once again: up to the diagonalization of the $2\times 2$ blocks). Therefore the toy model~\eqref{eq-second-toy-model} to which Theorem~\ref{theorem-main} applies has several features in common with the transfer matrices of the Anderson model. Most importantly, it has the same partial hyperbolicity. Clearly some further analysis of the symplectic structure of $\mathcal{T}^E$ is needed, but this goes beyond the scope of this work. Several technical elements in this last respect can be found in \cite{RS,SB1} and in the work of Sadel and Vir\'ag \cite{SV} whose main focus is, however, on the derivation of so-called DMPK equations in a scaling limit in which the strip width $\mathsf{L}$ scales like $\lambda^{-2}$, just as in \cite{BaR,VV}, but including hyperbolic channels.

\vspace{.2cm}

Let us now come to a discussion of the limitations of the present study and potential improvements of the main result. First of all, it requires the discrete time $T\geq T_0$ to be very large. As already explained, this allows to deduce statements on the Furstenberg measure, but really only addresses the quasi-one-dimensional limit of the Anderson model, rather than the finite volume approximations by cubes for which one needs $T_0\approx \mathsf{L}$. Let us stress, though, that the contraction arguments presented in this work use mesoscopic gaps and do allow to make (weak) statements already on this scale. 

\vspace{.2cm}

Based on the description of the Anderson model above, one realizes that in applications $\mathcal{P}$ may actually be a sparsely filled random matrix with few independent random entries. Then it may be hard to verify Hypothesis~\ref{hyp-P}. A way out is to regroup a finite number $N$ of the random blocks. This naturally leads to
$$
\mathcal{R} e^{\lambda \mathcal{P}_N}\cdots \mathcal{R} e^{\lambda \mathcal{P}_1}
\;=\;
\mathcal{R}^N
e^{\lambda\mathcal{P}^{(N)}+\lambda^2 \mathcal{Q}^{(N)}(\lambda)}
\;,
$$
with $\mathcal{P}^{(N)}=\sum_{n=1}^N \mathcal{R}^{N-n}\mathcal{P}_n\mathcal{R}^{n-1}$ and a bounded remainder $\mathcal{Q}^{(N)}(\lambda)$. Clearly $\mathcal{P}^{(N)}$ has considerably better coupling properties (see \cite{SS} where Lie algebraic hypoellipticity properties are used). On the other hand, the new r.h.s. is not of the form~\eqref{def-T}, but rather contains the supplementary higher order term $\lambda^2 \mathcal{Q}^{(N)}(\lambda)$. While this can in principle be dealt with by the techniques of the present work, it leads to considerably more involved perturbative expansions. 

\vspace{.2cm}

The next point considers Hypothesis~\ref{hyp-eta}. As already pointed out, we believe that this assumption is redundant. It is, however, the element in the present proof that leads to the iterative procedure in the dimension $\mathsf{q}$. Replacing this iterative procedure by collective contraction arguments on subspaces would constitute a considerable improvement of the argument. On the other hand, thinking of the Anderson model or the second toy model of Section~\ref{sec-toy-models} is clear that Hypothesis~\ref{hyp-eta} holds if only the size of the fibers is taken sufficiently large (because the microscopic gaps roughly scale as  $\mathsf{L}^{-(d-1)}$).

\vspace{.2cm}

\noindent {\bf Acknowledgments} We thank Andreas Knauf for many discussions as well as helpful and constructive comments. F.~D. received funding from the \textit{Studienstiftung des deutschen Volkes}. This work was also supported by the DFG grant SCHU 1358/6-2.

\section{Outline of the proof}\label{sec-outline}

As the proof of Theorem~\ref{theorem-main} is quite involved, this section lays out the main ideas and technical steps in a structured manner and defers most of the proofs and further technicalities to Section~\ref{sec-Proofs}.

\subsection{Perturbative expansion on dynamics on Grassmannian}
\label{sec-expansion}

A key tool is a quantitative estimate (which we believe to be close to optimal) on the remainder terms in the perturbation of the action on the Grassmannian. For the present purposes, it is sufficient to write out the second order in $\lambda$ and control the third order terms.

\begin{lemma}\label{lemma-expansion}
	Suppose that $\lambda\leq 2^{-6}$. Let us define two maps $\mathtt{X}, \mathtt{Y}: \mathbb{G}_{\mathsf{L}}\times\mathfrak{P}\rightarrow\mathbb{C}^{\mathsf{L}\times\mathsf{L}}$ by
\begin{align*}
& \mathtt{X}(Q,\mathtt{P})\;=\;Q^{\perp}\mathtt{P}Q+Q\mathtt{P}^*Q^{\perp}\,,
\\
&
\mathtt{Y}(Q,\mathtt{P})\;=\;Q^{\perp}\mathtt{P}Q\mathtt{P}^*Q^{\perp}-Q\mathtt{P}^*Q^{\perp}\mathtt{P}Q+\frac{1}{2}\left[Q^{\perp}\mathtt{P}(Q^{\perp}-Q)\mathtt{P}Q+Q\mathtt{P}^*(Q^{\perp}-Q)\mathtt{P}^*Q^{\perp}\right]\,.
\end{align*}
Further introduce a map $\mathtt{Z}^{(\lambda)}: \mathbb{G}_{\mathsf{L}}\times\mathfrak{P}\rightarrow\mathbb{C}^{\mathsf{L}\times \mathsf{L}}$ by
	\begin{align}\label{def-Z}	
	e^{\lambda\mathtt{P}}\cdot Q=Q+\lambda\mathtt{X}(Q,\mathtt{P})+\lambda^2\mathtt{Y}(Q,\mathtt{P})+\lambda^3\,\mathtt{Z}^{(\lambda)}(Q,\mathtt{P})\,.
	\end{align}
	Then for all $\mathtt{P} \in \mathfrak{P}$ and another $Q^{\prime}\in \mathbb{G}_{\mathsf{L}}$ satisfying $Q Q^{\prime}=\mathbf{0}$ one has
\begin{align}\label{ineq-expansion}
	\|\mathtt{X}(Q,\mathtt{P})\|&\leq 1\,,&\operatorname{rk}\big(\mathtt{X}(Q +Q^{\prime},\mathtt{P})-\mathtt{X}(Q ,\mathtt{P})\big)&\leq 2\,\operatorname{rk}(Q^{\prime})\,,\nonumber\\
	\|\mathtt{Y}(Q,\mathtt{P})\|&\leq \mbox{\small $\frac{3}{2}$}\,,&\operatorname{rk}\big(\mathtt{Y}(Q +Q^{\prime},\mathtt{P})-\mathtt{Y}(Q ,\mathtt{P})\big)&\leq 3\,\operatorname{rk}(Q^{\prime})\,,\\
	\|\mathtt{Z}^{(\lambda)}(Q,\mathtt{P})\|&\leq 20\,,&\operatorname{rk}\big(\mathtt{Z}^{(\lambda)}(Q +Q^{\prime},\mathtt{P})-\mathtt{Z}^{(\lambda)}(Q ,\mathtt{P})\big)&\leq 4\,\operatorname{rk}(Q^{\prime})\,.\nonumber
	\end{align}
\end{lemma}

\subsection{Contraction estimates}

This sections presents the bounds resulting from the hyperbolic character of $\mathcal{R}$, making the drift downwards into the $\lgl$-part quantitative. For this purpose, it is useful to introduce notations for particular frames for $\hat{P}_{{\ugl}}$ and $\hat{P}_{{\lgl}}$ and their orthogonal complements. Set
$$
\hat{\mathfrak{\alpha}}=
\mbox{\footnotesize $
\begin{pmatrix}
\mathbf{1}_{\mathsf{L}_{\ugl}}\\\mathbf{0}_{\mathsf{L}_{\ugl}\times (\mathsf{L}_{\mgl}+\mathsf{L}_{\lgl})}
\end{pmatrix}
$}
\;,
\quad
\hat{\mathfrak{\alpha}}^{\perp}=
\mbox{\footnotesize $
\begin{pmatrix}
\mathbf{0}_{\mathsf{L}_{\ugl}\times(\mathsf{L}_{\mgl}+\mathsf{L}_{\lgl})}\\\mathbf{1}_{\mathsf{L}_{\mgl}+\mathsf{L}_{\lgl}}
\end{pmatrix}
$}
\;,
\quad\hat{\mathfrak{\gamma}}=
\mbox{\footnotesize $
\begin{pmatrix}
\mathbf{0}_{(\mathsf{L}_{\ugl}+\mathsf{L}_{\mgl})\times\mathsf{L}_{\lgl}}\\\mathbf{1}_{\mathsf{L}_{\lgl}}
\end{pmatrix}
$}
\;,\quad
\hat{\mathfrak{\gamma}}^{\perp}=
\mbox{\footnotesize $
\begin{pmatrix}
\mathbf{1}_{\mathsf{L}_{\ugl}+\mathsf{L}_{\mgl}}\\\mathbf{0}_{\mathsf{L}_{\lgl}\times (\mathsf{L}_{\ugl}+\mathsf{L}_{\mgl})}
\end{pmatrix}
$}
\;.
$$
Then $\hat{\alpha}\in\mathbb{F}_{\mathsf{L},\mathsf{L}_{\ugl}}$ and one has $\hat{P}_{{\ugl}}=\hat{\alpha}\hat{\alpha}^*$ and $\mathbf{1}-\hat{P}_{{\ugl}}=\hat{\alpha}^\perp(\hat{\alpha}^\perp)^*$, and similarly for $\hat{\gamma}\in\mathbb{F}_{\mathsf{L},\mathsf{L}_{\lgl}}$. Moreover, one can rewrite~\eqref{def-d} as
\begin{equation}
\label{def-d-bis}
\mathsf{d}(Q)
\;=\;
\operatorname{tr}\left[\hat{\alpha}^*Q\hat{\alpha}\right]
\;.
\end{equation}
The norm $\|\hat{\alpha}^*Q\hat{\alpha}\|$ of the upper  left block matrix $\hat{\alpha}^*Q\hat{\alpha}$ of a projection $Q\in\mathbb{G}_{\mathsf{L},\mathsf{q}}$ can be viewed as a measure for the largest deviation of the range of $Q$ from the subspace $\{0\}^{\mathsf{L}_{\ugl}}\times \mathbb{C}^{\mathsf{L}_{\mgl}+\mathsf{L}_{\lgl}}$.
This space corresponds to the $\mathsf{L}_{\mgl}+\mathsf{L}_{\lgl}$ most stable directions in space, provided that the inequality
$$
\kappa_{\mathsf{L}_{\mgl}+\mathsf{L}_{\lgl}}
\;>\;
\kappa_{\mathsf{L}_{\mgl}+\mathsf{L}_{\lgl}+1}\,,
$$
is strict, which is equivalent to the positivity of the direct neighbor gap $\eta(\mathsf{L}_{\mgl}+\mathsf{L}_{\lgl},\mathsf{L}_{\mgl}+\mathsf{L}_{\lgl}+1)$. In this case, it is relatively elementary to show that the norm $\|\hat{\alpha}^*Q\hat{\alpha}\|$ is strictly diminished by the hyperbolic action:
\begin{lemma}\label{lemma-norm-contraction}
All $Q\in\mathbb{G}_{\mathsf{L},\mathsf{q}}$ satisfy the inequality
\begin{align}\label{ineq-norm-contraction}
\|\hat{\alpha}^*\,(\mathcal{R}\cdot Q)\,\hat{\alpha}\|
\;\leq\; 
\Big[1-\eta(\mathsf{L}_{\mgl}+\mathsf{L}_{\lgl},\mathsf{L}_{\mgl}+\mathsf{L}_{\lgl}+1)\,[1-\|\hat{\alpha}^*\,Q\,\hat{\alpha}\|]\Big]\|
\hat{\alpha}^*\,Q\,\hat{\alpha}\|\,.
\end{align}
\end{lemma}
However, the inequality~\eqref{ineq-norm-contraction} may be of little use because the next neighbor microscopic gap $\eta(\mathsf{L}_{\mgl}+\mathsf{L}_{\lgl},\mathsf{L}_{\mgl}+\mathsf{L}_{\lgl}+1)$ may be tiny compared to $\lambda$. Hence it is clearly too weak to dominate  the subsequent perturbation. In contrast, the macroscopic gap $\pmb{\eta}=\eta(\mathsf{L}_{\lgl},\mathsf{L}_{\lgl}+\mathsf{L}_{\mgl})$ introduced in Hypothesis~\ref{hyp-R} is much larger than the perturbation.
The next bound uses a macroscopic gap for the diminishment of the trace $\operatorname{tr}[\hat{\alpha}^*Q\hat{\alpha}]$ of the upper left block matrix $\hat{\alpha}^*Q\hat{\alpha}$ of a projection $Q\in\mathbb{G}_{\mathsf{L},\mathsf{q}}$, contrary to the elementary Lemma~\ref{lemma-norm-contraction} which only relies on a microscopic gap. The quantity $\operatorname{tr}[\hat{\alpha}^*Q\hat{\alpha}]$ can be viewed as a measure for the aggregate deviation of the range of $Q$ from the space $\{0\}^{\mathsf{L}_{\ugl}}\times \mathbb{C}^{\mathsf{L}_{\mgl}+\mathsf{L}_{\lgl}}$. Lemma~\ref{lemma-trace-contraction} shows how it changes under the hyperbolic action $\mathcal{R}\cdot$, and also provides an analogous bound on its counterpart, which is the trace $\operatorname{tr}[\hat{\gamma}^*Q\hat{\gamma}]$ of the lower right block matrix of $Q$.

\begin{lemma}\label{lemma-trace-contraction}
All $Q\in\mathbb{G}_{\mathsf{L},\mathsf{q}}$ satisfy the inequalities
\begin{align}\label{ineq-trace-contraction-1}
\operatorname{tr}\left[\hat{\alpha}^*(\mathcal{R}\cdot Q)\hat{\alpha}\right]
\;\leq\;
\operatorname{tr}\left[\hat{\alpha}^* Q\big(\mathbf{1}_{\mathsf{L}}-\pmb{\eta}\,\hat{\gamma}\hat{\gamma}^*\big)Q\hat{\alpha}\right]
\end{align}
and
\begin{align}\label{ineq-trace-contraction-2}
\operatorname{tr}\left[\hat{\gamma}^*(\mathcal{R}\cdot Q)\hat{\gamma}\right]
\;\geq\;
\operatorname{tr}\left[\hat{\gamma}^* Q\big(\mathbf{1}_{\mathsf{L}}+\pmb{\eta}\,\hat{\alpha}\hat{\alpha}^*\big)Q\hat{\gamma}\right]\,.
\end{align}
\end{lemma}

Due to~\eqref{def-d-bis}, one can derive a contraction inequality for $\mathsf{d}(Q)$ from~\eqref{ineq-trace-contraction-1}.

\begin{coro}\label{coro-d-contraction}
All $Q\in\mathbb{G}_{\mathsf{L},\mathsf{q}}$ satisfy the inequality
\begin{align}\label{ineq-d-contraction-bis}
\mathsf{d}(\mathcal{R}\cdot Q)
\;\leq\; 
\left(1-\pmb{\eta}\,\big(1-\|(\hat{\gamma}^{\perp})^*\,Q\,\hat{\gamma}^{\perp}\|\big)\right)\,\mathsf{d}(Q)\,.
\end{align}
\end{coro}

\subsection{The subdivision of the middle part}
\label{sec-subdivision}

The idea of proof below requires a suitable series of subdivisions of the middle part of size $\mathsf{L}_{\mgl}$ into subparts which have  corresponding relative gaps of $\mathcal{R}$ leading to desired local contraction powers. The existence of such suitable subdivisions relies on the assumption of small direct neighbor gaps of the diagonal entries of $\mathcal{R}$ corresponding to the middle part (see~Hypothesis~\ref{hyp-eta}) and is based on the following general statement:

\begin{lemma}\label{lemma-subdivision}
	Let $\phi \in (0,1)$ and $\mathsf{A},\mathsf{B},\mathsf{F}\in\mathbb{N}$ satisfy $\mathsf{A}<\mathsf{B}<\mathsf{L}$ and $\eta(\mathsf{B},\mathsf{A})>0$. Suppose that
	\begin{align}\label{ineq-subdivision-assumption}
	\eta(\mathsf{J},\mathsf{J}+1)\;\leq \;\frac{\phi}{\mathsf{F}}\,\eta(\mathsf{A},\mathsf{B})
	\end{align}
	holds for all $\mathsf{J}\in\left\{\mathsf{A},\dots,\mathsf{B}-1\right\}\,.$ Then, there exists a partition
	\begin{align}\label{ineq-subdivision-strict-partition}
	\mathsf{A}\;=\;\mathsf{I}_0\;<\; \mathsf{I}_1\;<\;\dots \;<\;\mathsf{I}_{\mathsf{F}}\;=\;\mathsf{B}
	\end{align}
	for which all $\mathsf{f}\in\{1,\dots,\mathsf{F}\}$ satisfy
	\begin{align}\label{ineq-subdivision-result}
	\eta(\mathsf{I}_{\mathsf{f}-1},\mathsf{I}_{\mathsf{f}})\;\geq \;\frac{1-\phi}{\mathsf{F}}\,\eta(\mathsf{A},\mathsf{B})\,.
	\end{align}
\end{lemma}

\vspace{.2cm}

Let us now apply Lemma~\ref{lemma-subdivision} with
$$
\mathsf{A}\;=\;\mathsf{L}_{\lgl}\,,
\qquad
\mathsf{B}\;=\;\mathsf{L}_{\mgl}+\mathsf{L}_{\lgl}\,,
\qquad
\mathsf{F}\;=\;2\,,
\qquad
\phi \;=\; 2^5\pmb{\eta}^{-1}\lambda\,,
$$
as then Hypothesis~\ref{hyp-eta} is precisely~\eqref{ineq-subdivision-assumption}.
Lemma~\ref{lemma-subdivision} then implies the existence of an integer $\mathsf{H}$ with
\begin{align}\label{ineq-global-gaps}
\mathsf{L}_{\lgl}\;<\;\mathsf{H}\;<\;\mathsf{L}_{\mgl}+\mathsf{L}_{\lgl}\,,
\qquad
\eta(\mathsf{L}_{\lgl},\mathsf{H})\geq 2^{-2}\,\pmb{\eta}\,,
\qquad
\eta(\mathsf{H},\mathsf{L}_{\lgl}+\mathsf{L}_{\mgl})\geq 2^{-1}(1-2^{-8})\,\pmb{\eta}\,.
\end{align}
since (an estimate that will be used several times in this context, implied by Hypotheses~\ref{hyp-lambda} and~\ref{hyp-q})
\begin{align}\label{ineq-subdivision-estimate}
\begin{split}
\mathsf{q}\,\lambda &\leq \mathsf{q}\,\lambda(\mathsf{q}^{-1}\mathsf{q})^2(\vartheta^{-1}\lambda^{-1}\vartheta\lambda)^{\frac{3}{5}}2^{\frac{58}{5}}\beta^{-2}\mathsf{q}\,\vartheta\\
&\leq 2^{\frac{58}{5}}\beta^{-2}\mathsf{q}^2\vartheta\lambda(\mathsf{q}^{-1}2^{-\frac{36}{5}}\beta^{\frac{1}{5}}\pmb{\eta}^{\frac{3}{5}}\vartheta^{-\frac{1}{5}}\lambda^{-\frac{1}{5}})^2(\vartheta^{-1}\lambda^{-1}2^{-17}\beta^{\frac{8}{3}}\pmb{\eta}^{-\frac{1}{3}})^{\frac{3}{5}} = 2^{-13}\pmb{\eta}\,,
\end{split}
\end{align}
which implies $1-2^5\pmb{\eta}^{-1}\lambda \geq 1-2^{12}\pmb{\eta}^{-1}\mathsf{q}\,\lambda \geq 2^{-1}$ and $1-2^5\pmb{\eta}^{-1}\lambda \geq 1-2^5\pmb{\eta}^{-1}\mathsf{q}\,\lambda \geq 1-2^{-8}$. Accordingly, let us subdivide the middle part $\mathfrak{b}(v)$ of a vector $v\in\mathbb{S}_{\mathbb{C}}^{\mathsf{L}-1}$ into 
$$
\mgl(v)\;=\;\begin{pmatrix}
\mgl_{\uparrow}(v)\\\mgl_{\downarrow}(v)
\end{pmatrix}\,,
$$
where $\mgl_{\uparrow}(v)$ and $\mgl_{\downarrow}(v)$ have length $\mathsf{L}_{\mgl}+\mathsf{L}_{\lgl}-\mathsf{H}$ and $\mathsf{H}-\mathsf{L}_{\lgl}$, respectively. The frames
$$
\hat{\gamma}_{\uparrow}\;=\;
\mbox{\footnotesize $
\begin{pmatrix}
\mathbf{0}_{(\mathsf{L}-\mathsf{H})\times\mathsf{H}}\\\mathbf{1}_{\mathsf{H}}
\end{pmatrix}
$}
\,,
\qquad\qquad
\hat{\gamma}_{\uparrow}^{\perp}\;=\;
\mbox{\footnotesize $
\begin{pmatrix}
\mathbf{1}_{\mathsf{L}-\mathsf{H}}\\\mathbf{0}_{\mathsf{H}\times (\mathsf{L}-\mathsf{H})}
\end{pmatrix}
$}
$$
are defined in correspondence with this subdivision. As in~\eqref{ineq-d-contraction-bis}, one has
\begin{align}\label{ineq-d-contraction}
\begin{split}
\mathsf{d}(\mathcal{R}\cdot Q)&\leq \left(1-\eta(\mathsf{H},\mathsf{L}_{\lgl}+\mathsf{L}_{\mgl})\,\left[1-\|(\hat{\gamma}_{\uparrow}^{\perp})^*\,Q\,\hat{\gamma}_{\uparrow}^{\perp}\|\right]\right)\,\mathsf{d}(Q)\\&\leq \left(1-2^{-1}(1-2^{-8})\,\pmb{\eta}\,\left[1-\|(\hat{\gamma}_{\uparrow}^{\perp})^*\,Q\,\hat{\gamma}_{\uparrow}^{\perp}\|\right]\right)\,\mathsf{d}(Q)\,,
\end{split}
\end{align}
where the second step follows from~\eqref{ineq-global-gaps}. Consequently, the next aim is to control the quantity $\|(\hat{\gamma}_{\uparrow}^{\perp})^*\,Q\,\hat{\gamma}_{\uparrow}^{\perp}\|$. For this, it turns out to be convenient to apply Lemma~\ref{lemma-subdivision} again, namely with
$$
\mathsf{A}\;=\;\mathsf{L}_{\lgl}\,,
\qquad
\mathsf{B}\;=\;\mathsf{H}\,,
\qquad 
\mathsf{F}\;=\;\mathsf{q}\,,
\qquad
\phi \;=\; 2^6\pmb{\eta}^{-1}\mathsf{q}\,\lambda\,.
$$
Due to the second item of~\eqref{ineq-global-gaps}, the requirement~\eqref{ineq-subdivision-assumption} is then again fulfilled by Hypothesis~\ref{hyp-eta}. Lemma~\ref{lemma-subdivision} then implies the existence of a partition
$$
\mathsf{L}_{\lgl}\;=\;\mathsf{I}_0\;<\;\mathsf{I}_1\;<\;\dots\;<\;\mathsf{I}_{\mathsf{q}}\;=\;\mathsf{H}\,,
$$
which satisfies $\eta(\mathsf{I}_{\mathsf{w}-1},\mathsf{I}_{\mathsf{w}})\geq 2^{-2}\mathsf{q}^{-1}(1-2^6\pmb{\eta}^{-1}\mathsf{q}\,\lambda)\,\pmb{\eta}$ for all $\mathsf{w} \in \lbrace 1,\dots,\mathsf{q} \rbrace$, so in particular
\begin{align}\label{ineq-local-gap}
\eta(\mathsf{I}_{\mathsf{w}-1},\mathsf{I}_{\mathsf{w}})
\;\geq\; 
2^{-3}\,\mathsf{q}^{-1}\,\pmb{\eta}\,,\qquad\qquad\qquad \mathsf{w}=1,\dots,\mathsf{q}\,,
\end{align}
as~\eqref{ineq-subdivision-estimate} implies $1-2^6\pmb{\eta}^{-1}\mathsf{q}\,\lambda \geq 1-2^{12}\pmb{\eta}^{-1}\mathsf{q}\,\lambda \geq 2^{-1}$. Accordingly, $\mgl_{\downarrow}(v)$ is subdivided further~into
$$
\mgl_{\downarrow}(v)\;=\;\begin{pmatrix}
\mgl_{\mathsf{q}}(v)\\\mgl_{\mathsf{q}-1}(v)\\\vdots\\\mgl_{\mathsf{w}}(v)\\\vdots\\\mgl_1(v)
\end{pmatrix}\,,
$$
where the $\mgl_{\mathsf{w}}$ are of the length $\mathsf{I}_{\mathsf{w}}-\mathsf{I}_{\mathsf{w}-1}$, respectively. Let us use the abbreviations
$$
\ugl_{\mathsf{w}}(v)\;=\;\begin{pmatrix}
\ugl(v)\\\mgl_{\uparrow}(v)\\\mgl_{\mathsf{q}}(v)\\\vdots\\\mgl_{{\mathsf{w}}+1}(v)
\end{pmatrix}\qquad\qquad\textnormal{and}\qquad\qquad
\lgl_{\mathsf{w}}(v)\;=\;\begin{pmatrix}
\mgl_{{\mathsf{w}}-1}(v)\\\vdots\\\mgl_1(v)\\\lgl(v)
\end{pmatrix}
$$
including $\ugl_{\mathsf{q}}(v)=\mbox{\footnotesize $\begin{pmatrix}\ugl(v)\\\mgl_{\uparrow}(v)\end{pmatrix}$}$ and $\lgl_1(v)=\lgl(v)$. For $\mathsf{w}\in\{0,\dots,\mathsf{q}\}$, let us then introduce~the~frames
$$
\hat{\alpha}_{\mathsf{w}}\;=\;
\mbox{\footnotesize $
\begin{pmatrix}
\mathbf{1}_{\mathsf{L}-\mathsf{I}_{\mathsf{w}}}\\\mathbf{0}_{\mathsf{I}_{\mathsf{w}}\times (\mathsf{L}-\mathsf{I}_{\mathsf{w}})}
\end{pmatrix}
$}
,
\quad
\hat{\alpha}_{\mathsf{w}}^{\perp}\;=\;
\mbox{\footnotesize $
\begin{pmatrix}
\mathbf{0}_{(\mathsf{L}-\mathsf{I}_{\mathsf{w}})\times\mathsf{I}_{\mathsf{w}}}\\\mathbf{1}_{\mathsf{I}_{\mathsf{w}}}
\end{pmatrix}
$},
\quad
\hat{\gamma}_{\mathsf{w}}\;=\;
\mbox{\footnotesize $
\begin{pmatrix}
\mathbf{0}_{(\mathsf{L}-\mathsf{I}_{\mathsf{w}-1})\times\mathsf{I}_{\mathsf{w}-1}}\\\mathbf{1}_{\mathsf{I}_{\mathsf{w}-1}}
\end{pmatrix}
$}
,
\quad
\hat{\gamma}_{\mathsf{w}}^{\perp}\;=\;
\mbox{\footnotesize $
\begin{pmatrix}
\mathbf{1}_{\mathsf{L}-\mathsf{I}_{\mathsf{w}-1}}\\\mathbf{0}_{\mathsf{I}_{\mathsf{w}-1}\times (\mathsf{L}-\mathsf{I}_{\mathsf{w}-1})}
\end{pmatrix}
$}\,.
$$

\begin{figure}[H]
	\begin{center}
		\begin{minipage}{0.5\textwidth}
			\begin{center}
				\begin{tikzpicture}[line join = round, line cap = round]
					\subdivision{\mathfrak{a}}{\hat{\alpha}}{\mathfrak{b}}{\mathfrak{c}}{\hat{\gamma}}{\mathsf{w}};
					\coordinate (ct) at (0,3);
					\tick{(ct)};
					\coordinate[label=left:{$\mathfrak{a}$}, label=right:{$\hat{\alpha}$}] (argt2) at (0,2.5);
					\coordinate (cm1) at (0,2);
					\tick{(cm1)};
					\coordinate[label=left:{${\mathfrak{b}}_{\uparrow}\ $}] (argm3) at (0,1.4);
					\coordinate (cm2) at (0,0.8);
					\tick{(cm2)};
					\coordinate[label=left:{${\mathfrak{b}}_{\mathsf{q}}\ $}] (argm4) at (0,0.65);
					\coordinate (cm3) at (0,0.5);
					\tick{(cm3)};
					\coordinate (cm4) at (0,0);
					\coordinate (cm5) at (0,-0.15);
					\tick{(cm5)};
					\coordinate (cm6) at (0,-0.45);
					\tick{(cm6)};
					\coordinate (cm7) at (0,-0.6);
					\coordinate (cm8) at (0,-1.1);
					\tick{(cm8)};
					\coordinate[label=left:{${\mathfrak{b}}_{3}\ $}] (argm5) at (0,-1.25);
					\coordinate (cm9) at (0,-1.4);
					\tick{(cm9)};
					\coordinate[label=left:{${\mathfrak{b}}_{2}\ $}] (argm7) at (0,-1.55);
					\coordinate (cm10) at (0,-1.7);
					\tick{(cm10)};
					\coordinate[label=left:{${\mathfrak{b}}_{1}\ $}] (argm8) at (0,-1.85);
					\coordinate (cm11) at (0,-2);
					\tick{(cm11)};
					\coordinate[label=left:{$\mathfrak{c}$}, label=right:{$\hat{\gamma}$}] (argb2) at (0,-2.5);
					\coordinate (cb) at (0,-3);
					\tick{(cb)};
					\draw [-,color=black,line width=0.5mm] (ct)--(cm3);
					\draw [-,color=black,dotted,line width=0.5mm] (cm3)--(cm4);
					\draw [-,color=black,line width=0.5mm] (cm4)--(cm7);
					\draw [-,color=black,dotted,line width=0.5mm] (cm7)--(cm8);
					\draw [-,color=black,line width=0.5mm] (cm8)--(cb);
					\coordinate (cl1) at (-0.3,0);
					\coordinate (cl2) at (-0.3,-0.6);
					\coordinate (cr1) at (0.3,0);
					\coordinate (cr2) at (0.3,-0.6);
					\draw[-, fill=none] (cl1)--(cl2)--(cr2)--(cr1)--cycle;
					\draw [-] (cr1)--(rl1);
					\draw [-] (cr2)--(rl2);
				\end{tikzpicture}
			\end{center}
		\end{minipage}
		\hspace{-0.5cm}
		=
		\hspace{-0.5cm}
		\begin{minipage}{0.5\textwidth}
			\begin{center}
				\begin{tikzpicture}[line join = round, line cap = round]
					\subdivision{\mathfrak{x}}{\hat{\chi}}{\mathfrak{y}}{\mathfrak{z}}{\hat{\zeta}}{\mathsf{m}};
					\coordinate (ct) at (0,3);
					\tick{(ct)};
					\coordinate[label=left:{$\mathfrak{x}$}, label=right:{$\hat{\chi}$}] (argt2) at (0,2.5);
					\coordinate (cm1) at (0,2);
					\tick{(cm1)};
					\coordinate[label=left:{${\mathfrak{y}}_{\mathsf{M}+1}\ $}] (argm3) at (0,1.7);
					\coordinate (cm2) at (0,1.4);
					\tick{(cm2)};
					\coordinate[label=left:{${\mathfrak{y}}_{\mathsf{M}}\ $}] (argm4) at (0,1.25);
					\coordinate (cm3) at (0,1.1);
					\tick{(cm3)};
					\coordinate (cm4) at (0,0.6);
					\coordinate (cm5) at (0,0.45);
					\tick{(cm5)};
					\coordinate (cm6) at (0,0.15);
					\tick{(cm6)};
					\coordinate (cm7) at (0,0);
					\coordinate (cm8) at (0,-0.5);
					\tick{(cm8)};
					\coordinate[label=left:{${\mathfrak{y}}_{3}\ $}] (argm5) at (0,-0.65);
					\coordinate (cm9) at (0,-0.8);
					\tick{(cm9)};
					\coordinate[label=left:{${\mathfrak{y}}_{2}\ $}] (argm6) at (0,-0.95);
					\coordinate (cm10) at (0,-1.1);
					\tick{(cm10)};
					\coordinate[label=left:{${\mathfrak{y}}_{1}\ $}] (argm7) at (0,-1.25);
					\coordinate (cm11) at (0,-1.4);
					\tick{(cm11)};
					\coordinate[label=left:{${\mathfrak{y}}_{0}\ $}] (argm8) at (0,-1.7);
					\coordinate (cm12) at (0,-2);
					\tick{(cm12)};
					\coordinate[label=left:{$\mathfrak{z}$}, label=right:{$\hat{\zeta}$}] (argb2) at (0,-2.5);
					\coordinate (cb) at (0,-3);
					\tick{(cb)};
					\draw [-,color=black,line width=0.5mm] (ct)--(cm3);
					\draw [-,color=black,dotted,line width=0.5mm] (cm3)--(cm4);
					\draw [-,color=black,line width=0.5mm] (cm4)--(cm7);
					\draw [-,color=black,dotted,line width=0.5mm] (cm7)--(cm8);
					\draw [-,color=black,line width=0.5mm] (cm8)--(cb);
					\coordinate (cl1) at (-0.3,0.6);
					\coordinate (cl2) at (-0.3,0);
					\coordinate (cr1) at (0.3,0.6);
					\coordinate (cr2) at (0.3,0);
					\draw[-, fill=none] (cl1)--(cl2)--(cr2)--(cr1)--cycle;
					\draw [-] (cr1)--(rl1);
					\draw [-] (cr2)--(rl2);
				\end{tikzpicture}
			\end{center}
		\end{minipage}
		\caption{The complete subdivision of all vector entries, shown in multiple stages. Next to each partial vector, its symbol (on the left) and corresponding frame (on the right) are indicated.}\label{fig-subdivision}
	\end{center}
\end{figure}
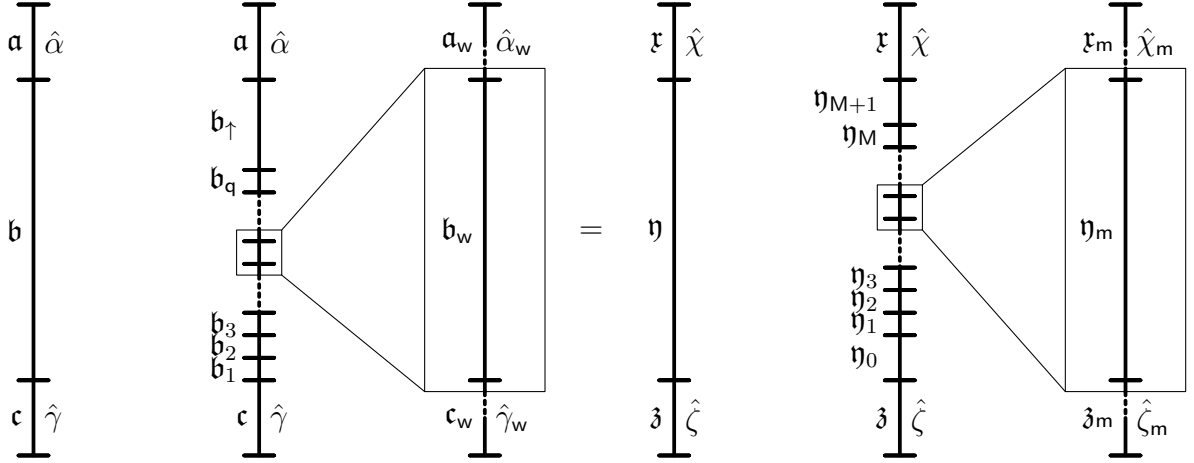

At this point, we performed the subdivision of the $\mathsf{L}$ vector entries as indicated in the left three diagrams of Figure~\ref{fig-subdivision}. The finer subdivision in the other three diagrams will only be used in Section~\ref{sec-auxiliary-sytem}.

\subsection{The inductive approach}\label{sec-induction}

In order to prove Theorem~\ref{theorem-main}, the structure constructed in Section~\ref{sec-subdivision} will be used. In Section~\ref{sec-auxiliary-sytem}, we will proceed in an inductive manner from $\mathsf{w}=0$ to $\mathsf{q}$. For this, we introduce the following.

\vspace{.2cm}

\noindent \textbf{Induction Hypothesis.}
{\sl For $W_0 \in \mathbb{G}_{\mathsf{L},\mathsf{w}}$, let $W_n=\mathcal{T}_n\cdot W_{n-1}$ be as in~\eqref{dyn-grassmanian} for all $n\in\mathbb{N}$. Then, for all $n \in \left[2^{\frac{18}{5}}\beta^{-1}\mathsf{q}\,\mathsf{w}\,\vartheta\lambda^{-2},\infty\right)$ one has } 
\begin{align*}
\mathbb{P}\left(\operatorname{tr}\left[(\hat{\alpha}_{\mathsf{w}})^*W_n\hat{\alpha}_{\mathsf{w}}\right]>2^{-\frac{21}{5}}\beta^{\frac{3}{5}}\pmb{\eta}^{-\frac{1}{5}}\vartheta^{-\frac{3}{5}}\lambda^{\frac{7}{5}}\right)
\;\leq\; 
2^4(2^{\frac{54}{5}}\lambda)^{3(\mathsf{q}-\mathsf{w}+1)}\,.
\end{align*}

\vspace{.2cm}

Obviously, the initial step for $\mathsf{w}=0$ holds, as all zero-dimensional projections vanish, \textit{i.e.}, $W_n\in \mathbb{G}_{\mathsf{L},0} = \{\mathbf{0}\}$ for all $n\in\mathbb{N}$.  It will be shown in Section~\ref{sec-auxiliary-sytem} that the induction step also holds:

\begin{lemma}\label{lemma-inductive-result}
	For all $\mathsf{w} \in \lbrace 0,\dots,\mathsf{q} \rbrace$, the statement of the above {\rm Induction Hypothesis} holds.
\end{lemma}

Lemma~\ref{lemma-inductive-result} implies, in particular, the validity of the statement of the Induction Hypothesis for $\mathsf{w}=\mathsf{q}$, which allows to prove the main theorem.

\subsection{Proof of the main result}

This section shows how the technical elements presented in Sections~\ref{sec-expansion} to~\ref{sec-induction} can be combined to complete the proof of the main result.

\vspace{.2cm}

\noindent\textbf{Proof of Theorem~\ref{theorem-main}.}
For $\mathsf{w} = \mathsf{q}$, as $\hat{\alpha}_{\mathsf{q}} = \hat{\gamma}^{\perp}_{\uparrow}$, Lemma~\ref{lemma-inductive-result} reads
\begin{align}\label{ineq-induction-stop}
\mathbb{P}\left(\operatorname{tr}\left[(\hat{\gamma}^{\perp}_{\uparrow})^*Q_n\hat{\gamma}^{\perp}_{\uparrow}\right]>2^{-\frac{21}{5}}\beta^{\frac{3}{5}}\pmb{\eta}^{-\frac{1}{5}}\vartheta^{-\frac{3}{5}}\lambda^{\frac{7}{5}}\right) \leq 2^4(2^{\frac{54}{5}}\lambda)^3\,,
\end{align}
for all $n \in \left[2^{\frac{18}{5}}\beta^{-1}\mathsf{q}^2\vartheta\lambda^{-2},\infty\right)\cap\mathbb{N}$ as we can set $W_0 = Q_0$ (so then $W_n = Q_n$ holds for all $n \in \mathbb{N}$). Let us note that since $\lambda < 2^{-13}$, in fact $\vartheta \geq 2^{\frac{3}{5}}$. Using this and Hypothesis~\ref{hyp-q} we then estimate
\begin{align}\label{ineq-induction-stop-estimate}
\begin{split}
2^3(1-2^{-7})\mathsf{q}^2\pmb{\eta}(2^{\frac{54}{5}}\lambda)^3 &\leq 2^{\frac{177}{5}}(1-2^{-7})\mathsf{q}^2\lambda^3 \leq (2^{\frac{177}{5}}-80)\mathsf{q}^2\lambda^3\\
&\leq 2^{\frac{177}{5}}\beta^{-1}\pmb{\eta}^{-3}\mathsf{q}^6\lambda^3(2^{-\frac{3}{5}}\vartheta) - 80\,\mathsf{q}\,\lambda^3 \leq 2^{-\frac{6}{5}}\mathsf{q}\,\lambda^2 - 80\,\mathsf{q}\,\lambda^3\\
&\leq \left[5 \cdot 2^{-\frac{1}{54}} - \frac{9}{2}\right]\mathsf{q}\,\lambda^2 - 80\,\mathsf{q}\,\lambda^3\,.
\end{split}
\end{align}
For all $Q \in \mathbb{G}_{\mathsf{L},\mathsf{q}}$, the matrix $\mathtt{Y}(Q,\mathcal{P})$ is self-adjoint. Therefore we can estimate
\begin{align}\label{ineq-trace-norm-rank}
\begin{split}
\operatorname{tr}\left[\hat{\alpha}^*\mathtt{Y}(Q,\mathcal{P})\hat{\alpha}\right] &= \operatorname{tr}\left[\hat{\alpha}^*\chi_{\mathbb{R}\setminus\lbrace 0 \rbrace}\left(\mathtt{Y}(Q,\mathcal{P})\right)\mathtt{Y}(Q,\mathcal{P})\chi_{\mathbb{R}\setminus\lbrace 0 \rbrace}\left(\mathtt{Y}(Q,\mathcal{P})\right)\hat{\alpha}\right]\\
&\leq \|\mathtt{Y}(Q,\mathcal{P})\| \operatorname{tr}\left[\chi_{\mathbb{R}\setminus\lbrace 0 \rbrace}\left(\mathtt{Y}(Q,\mathcal{P})\right)\hat{\alpha}\hat{\alpha}^*\chi_{\mathbb{R}\setminus\lbrace 0 \rbrace}\left(\mathtt{Y}(Q,\mathcal{P})\right)\right]\\
&\leq \|\mathtt{Y}(Q,\mathcal{P})\| \|\hat{\alpha}\hat{\alpha}^*\| \operatorname{tr}\left[\chi_{\mathbb{R}\setminus\lbrace 0 \rbrace}\left(\mathtt{Y}(Q,\mathcal{P})\right)\right] 
\\
&
= \|\mathtt{Y}(Q,\mathcal{P})\| \operatorname{rk}\left(\mathtt{Y}(Q,\mathcal{P})\right)\,.
\end{split}
\end{align}
The same calculation can be made for $\mathtt{Z}^{(\lambda)}(Q,\mathcal{P})$. From Lemma~\ref{lemma-expansion} and Hypothesis~\ref{hyp-P} we find $\mathbb{E}\mathtt{X}(Q,\mathcal{P}) = \mathbf{0}$, $\|\mathtt{Y}(Q,\mathcal{P})\| \leq \frac{3}{2}$, $\|\mathtt{Z}^{(\lambda)}(Q,\mathcal{P})\| \leq 20$, $\operatorname{rk}\left(\mathtt{Y}(Q,\mathcal{P})\right) \leq 3\mathsf{q}$ and $\operatorname{rk}\left(\mathtt{Z}^{(\lambda)}(Q,\mathcal{P})\right) \leq 4\mathsf{q}$. Combining these bounds for $Q = \mathcal{R} \cdot Q_{n-1}$ and $\mathcal{P} = \mathcal{P}_n$,~\eqref{ineq-d-contraction} and~\eqref{ineq-trace-norm-rank} yields for all $n \in \mathbb{N}$
\begin{align}\label{ineq-expectation-iteration-1}
\begin{split}
\mathbb{E}\left[\mathsf{d}(Q_n)\right] &= \mathbb{E}\left[\operatorname{tr}\left(\hat{\alpha}^*\left[e^{\lambda\mathcal{P}_n} \cdot (\mathcal{R} \cdot Q_{n-1})\right]\hat{\alpha}\right)\right]\\
&\leq \mathbb{E}\left[\operatorname{tr}\left(\hat{\alpha}^*(\mathcal{R} \cdot Q_{n-1})\hat{\alpha}\right)\right] + \frac{9}{2}\mathsf{q}\,\lambda^2 + 80\,\mathsf{q}\,\lambda^3\\
&\leq \mathbb{E}\left[\left(1-2^{-1}(1-2^{-8})\,\pmb{\eta}\left[1-\|(\hat{\gamma}^{\perp}_{\uparrow})^*Q_{n-1}\hat{\gamma}^{\perp}_{\uparrow}\|\right]\right)\operatorname{tr}\left(\hat{\alpha}^*Q_{n-1}\hat{\alpha}\right)\right] + \frac{9}{2}\mathsf{q}\,\lambda^2 + 80\,\mathsf{q}\,\lambda^3\\
&\leq \mathbb{E}\left[\mathsf{d}(Q_{n-1})\right] + \frac{9}{2}\mathsf{q}\,\lambda^2 + 80\,\mathsf{q}\,\lambda^3 - 2^{-1}(1-2^{-8})\,\pmb{\eta}\mathbb{E}\left[\left[1-\operatorname{tr}\left((\hat{\gamma}^{\perp}_{\uparrow})^*Q_{n-1}\hat{\gamma}^{\perp}_{\uparrow}\right)\right]\mathsf{d}(Q_{n-1})\right]\,.
\end{split}
\end{align}
To handle the product in the expectation value, we note the following. If $a$, $b$ and $c$ are real-valued random variables (that possibly depend on each other), and there exists some $\varepsilon \in \mathbb{R}$ such that $b(c-a) \leq \varepsilon$ a.s., it holds that $\mathbb{E}\left[ab\right] \geq \mathbb{E}\left[bc\right] - \varepsilon\mathbb{P}\left[a < c\right]$. This clearly holds if $\mathbb{P}\left[a \geq c\right] = 0$, as then $\mathbb{P}\left[a < c\right] = 1$. If $\mathbb{P}\left[a < c\right] = 0$, then $a \geq c$ holds a.s., and the inequality indeed follows. Excluding these two cases, the statement holds due to the defining condition for $\varepsilon$ and
\begin{align*}
\mathbb{E}\left[ab\right] &= \mathbb{E}\left[ab \ | \ a \geq c\right]\mathbb{P}\left[a \geq c\right] + \mathbb{E}\left[ab \ | \ a < c\right]\mathbb{P}\left[a < c\right]\\
&\geq \mathbb{E}\left[bc \ | \ a \geq c\right]\mathbb{P}\left[a \geq c\right] + \left(\mathbb{E}\left[ab \ | \ a < c\right] + \mathbb{E}\left[bc \ | \ a < c\right] - \mathbb{E}\left[bc \ | \ a < c\right]\right)\mathbb{P}\left[a < c\right]\\
&= \mathbb{E}\left[bc\right] - \mathbb{E}\left[b(c-a) \ | \ a < c\right]\mathbb{P}\left[a < c\right]\,.
\end{align*}
For $a = 1 - \operatorname{tr}\left((\hat{\gamma}^{\perp}_{\uparrow})^*Q_{n-1}\hat{\gamma}^{\perp}_{\uparrow}\right)$, $b = \mathsf{d}(Q_{n-1})$, $c = 1 - 2^{-\frac{21}{5}}\beta^{\frac{3}{5}}\pmb{\eta}^{-\frac{1}{5}}\vartheta^{-\frac{3}{5}}\lambda^{\frac{7}{5}} \geq 1 - 2^{-16} \geq \frac{2^{-\frac{1}{108}}}{1-2^{-8}}$ (which follows from~\eqref{ineq-trace-order-lambda} and Hypothesis~\eqref{hyp-lambda}) and $\varepsilon = \mathsf{q}^2$, which is indeed an upper bound for $b(c-a) = \mathsf{d}(Q_{n-1})\left[\operatorname{tr}\left((\hat{\gamma}^{\perp}_{\uparrow})^*Q_{n-1}\hat{\gamma}^{\perp}_{\uparrow}\right) - 2^{-\frac{21}{5}}\beta^{\frac{3}{5}}\pmb{\eta}^{-\frac{1}{5}}\vartheta^{-\frac{3}{5}}\lambda^{\frac{7}{5}}\right]$, one finds
\begin{align}
\label{ineq-expectation-decoupling}
\mathbb{E} & \left[\left[1-\operatorname{tr}\left((\hat{\gamma}^{\perp}_{\uparrow})^*Q_{n-1}\hat{\gamma}^{\perp}_{\uparrow}\right)\right]\mathsf{d}(Q_{n-1})\right] 
\nonumber
\\
&
\;\;\;\geq \;
\frac{\mathbb{E}\left[\mathsf{d}(Q_{n-1})\right]}{2^{\frac{1}{108}}(1-2^{-8})} - \mathsf{q}^2\,\mathbb{P}\left[\operatorname{tr}\left((\hat{\gamma}^{\perp}_{\uparrow})^*Q_{n-1}\hat{\gamma}^{\perp}_{\uparrow}\right) > 2^{-\frac{21}{5}}\beta^{\frac{3}{5}}\pmb{\eta}^{-\frac{1}{5}}\vartheta^{-\frac{3}{5}}\lambda^{\frac{7}{5}}\right]\,.
\end{align}
For $n \geq 2^{\frac{18}{5}}\beta^{-1}\mathsf{q}^2\vartheta\lambda^{-2}$, combining~\eqref{ineq-induction-stop},~\eqref{ineq-induction-stop-estimate},~\eqref{ineq-expectation-iteration-1} and~\eqref{ineq-expectation-decoupling} yields the inequality
$$
\mathbb{E}\left[\mathsf{d}(Q_n)\right] \leq \left[1-2^{-\frac{109}{108}}\pmb{\eta}\right]\mathbb{E}\left[\mathsf{d}(Q_{n-1})\right] + 5 \cdot 2^{-\frac{1}{54}}\mathsf{q}\,\lambda^2\,.
$$
We iterate from $n = \lceil2^{\frac{18}{5}}\beta^{-1}\mathsf{q}^2\vartheta\lambda^{-2}\rceil+1$ to $T^{\prime} := \lceil2^{\frac{19}{5}}\beta^{-1}\mathsf{q}^2\vartheta\lambda^{-2}\rceil$, using the following general principle that will be used a few times later on: suppose that for $n_1, n_2 \in \mathbb{N}_0$ with $n_1 < n_2$, a set $\lbrace f_n \ : \ n \in \lbrace n_1,\dots,n_2 \rbrace\rbrace \subset \mathbb{R}$, $g \in (0,1)$ and a positive real number $h$, it holds that $0 \leq f_{n+1} \leq (1-g)f_n + h$ for all $n \in \lbrace n_1,\dots,n_2-1 \rbrace$; then one can iteratively estimate
\begin{align}\label{ineq-iterative-estimate}
\begin{split}
f_{n_2} &\leq (1-g)^{n_2-n_1}f_{n_1} + h\sum_{n=n_1}^{n_2-1}(1-g)^{n_2-1-n}\\
&\leq (1-g)^{n_2-n_1}f_{n_1} + h\sum_{k \in \mathbb{N}_0}(1-g)^k\\
&\leq f_{n_1}\exp\left[-g(n_2-n_1)\right] + g^{-1}h\,.
\end{split}
\end{align}
Using Hypotheses~\ref{hyp-lambda} and~\ref{hyp-q}, the fact that $\left(\frac{5e}{6}\right)^{\frac{6}{5}}e^{-x} \leq x^{-\frac{6}{5}}$ for all $x>0$ and~\eqref{ineq-iterative-estimate}, one obtains
\begin{align}\label{ineq-expectation-iteration-2}
\begin{split}
\mathbb{E} & \left[\mathsf{d}(Q_{T^{\prime}})\right] 
\\
&\leq \exp\left[-2^{-\frac{109}{108}}\pmb{\eta}\left(\lceil2^{\frac{19}{5}}\beta^{-1}\mathsf{q}^2\vartheta\lambda^{-2}\rceil-\lceil2^{\frac{18}{5}}\beta^{-1}\mathsf{q}^2\vartheta\lambda^{-2}\rceil\right)\right]\mathbb{E}\left[\mathsf{d}(Q_{\lceil2^{\frac{18}{5}}\beta^{-1}\mathsf{q}^2\vartheta\lambda^{-2}\rceil})\right] 
\\
&
\;\;\;\;\;\;+ 5 \cdot 2^{\frac{107}{108}}\pmb{\eta}^{-1}\mathsf{q}\,\lambda^2
\\
&\leq \mathsf{q}\exp\left[-2^{-\frac{109}{108}}(2^{\frac{19}{5}}-2^{\frac{18}{5}})\pmb{\eta}\beta^{-1}\mathsf{q}^2\vartheta\lambda^{-2}+2^{-\frac{109}{108}}\pmb{\eta}\right] + 5 \cdot 2^{\frac{107}{108}}\pmb{\eta}^{-1}\mathsf{q}\,\lambda^2\\
&\leq 2^{-\frac{67}{50}}\mathsf{q}\left(\frac{5e}{6}\right)^{\frac{6}{5}}\exp\left[-2^{-\frac{1}{5}}\pmb{\eta}\beta^{-1}\mathsf{q}^2\vartheta\lambda^{-2}\right]\exp\left[2^{-\frac{109}{108}}\right] + 5 \cdot 2^{\frac{107}{108}}\pmb{\eta}^{-1}\mathsf{q}\,\lambda^2\\
&\leq 2^{-\frac{67}{50}}\mathsf{q}\left(2^{\frac{1}{5}}\pmb{\eta}^{-1}\beta\,\mathsf{q}^{-2}\vartheta^{-1}\lambda^2\right)^{\frac{6}{5}}2^{\frac{4}{5}}(\mathsf{q}^{-1}\mathsf{q})^{\frac{1}{2}}(\vartheta^{-1}\lambda^{-1}\vartheta\lambda)^{\frac{3}{10}}\beta^{-\frac{21}{10}}\vartheta^{\frac{8}{5}}\mathsf{q}^{\frac{29}{10}} + 5 \cdot 2^{\frac{107}{108}}\pmb{\eta}^{-1}\mathsf{q}\,\lambda^2\\
&\leq 2^{-\frac{3}{10}}\pmb{\eta}^{-\frac{6}{5}}\beta^{-\frac{9}{10}}\mathsf{q}^{\frac{3}{2}}\vartheta^{\frac{2}{5}}\lambda^{\frac{12}{5}}(\mathsf{q}^{-1}2^{-\frac{36}{5}}\beta^{\frac{1}{5}}\pmb{\eta}^{\frac{3}{5}}\vartheta^{-\frac{1}{5}}\lambda^{-\frac{1}{5}})^{\frac{1}{2}}(\vartheta^{-1}\lambda^{-1}2^{-17}\beta^{\frac{8}{3}}\pmb{\eta}^{-\frac{1}{3}})^{\frac{3}{10}} + 5 \cdot 2^{\frac{107}{108}}\pmb{\eta}^{-1}\mathsf{q}\,\lambda^2\\
&\leq 2^{-9}\pmb{\eta}^{-1}\mathsf{q}\,\lambda^{2} + (10-2^{-9})\pmb{\eta}^{-1}\mathsf{q}\,\lambda^2\\
&= 10\,\pmb{\eta}^{-1}\mathsf{q}\,\lambda^2\,.
\end{split}
\end{align}
Now, since the inequality~\eqref{ineq-expectation-iteration-2} holds for all starting points $Q_0\in\mathbb{G}_{\mathsf{L},\mathsf{q}}$, it also holds for all starting distributions. Thus, a start of the entire process at a later time is possible so that the upper bound in~\eqref{ineq-expectation-iteration-2} remains valid if $T^{\prime}$ is replaced by any larger integer. But $T_0\geq T^{\prime}$.
\hfill $\square$

\subsection{An auxiliary random dynamical system}\label{sec-auxiliary-sytem}

This section spells out the main steps of the inductive argument leading to Lemma~\ref{lemma-inductive-result}. In each induction step, an extra dimension is added to a projection $W$, by means of an extra vector $v$ that is orthogonal to it. The dynamics on $\mathcal{T}\cdot W$ is given by~\eqref{def-action}, that of the vector by $\mathcal{T}\circ v$, see~\eqref{dyn-sphere-2}, but clearly $\mathcal{T}\circ v$ does not need to be orthogonal to $\mathcal{T}\cdot W$ and hence $(\mathcal{T}\cdot W,\mathcal{T}\circ v) $ has to be orthogonalized. To spell this out, let us fix $\mathsf{w}\in\{0,\dots,\mathsf{q}-1\}$ and consider the space
$$
\mathfrak{W}
\;:=\;
\left\{(W,v)\in\mathbb{G}_{\mathsf{L},\mathsf{w}}\times\mathbb{S}_{\mathbb{C}}^{\mathsf{L}-1}:Wv=0\right\}
\;.
$$
Then the action $\star$ on $\mathfrak{W}$ is defined by
\begin{align}\label{def-star-action}
\star : \textnormal{GL}(\mathsf{L},\mathbb{C})\times \mathfrak{W}\rightarrow\mathfrak{W}\,,\qquad (\mathcal{T},(W,v))\longmapsto \big(\mathcal{T}\cdot W,((\mathcal{T}\cdot W)^{\perp}\mathcal{T})\circ v\big)\,,
\end{align}

The next lemma shows that this is actually well-defined.

\begin{lemma}\label{lemma-auxiliary-action}
Let $\mathcal{T}\in\textnormal{GL}(\mathsf{L},\mathbb{C})$ and $(W,v)\in\mathfrak{W}$. Then, $(\mathcal{T}\cdot W)^{\perp}\mathcal{T} v\neq 0$ and one has
\begin{align}\label{eq-auxiliary-action}
\mathcal{T}\cdot (W+vv^*)=\mathcal{T}\cdot W+\left[((\mathcal{T}\cdot W)^{\perp}\mathcal{T})\circ v\right]\left[((\mathcal{T}\cdot W)^{\perp}\mathcal{T})\circ v\right]^*\,.
\end{align}
\end{lemma}

Moreover, one can readily verify that all $\mathcal{T}_1,\mathcal{T}_2\in\textnormal{GL}(\mathsf{L},\mathbb{C})$ and $(W,v)\in\mathfrak{W}$ obey
$$
\mathcal{T}_2\star (\mathcal{T}_1\star (W,v))=(\mathcal{T}_2\mathcal{T}_1) \star (W,v)\,,
$$
namely $\star$ is a group action.

\vspace{.2cm}

After these preparations, it is natural to introduce a random dynamical system on $\mathfrak{W}$ by
\begin{align}\label{dyn-auxiliary}
(W_n,v_n):=\mathcal{T}_n\star (W_{n-1},v_{n-1})\,,\qquad\qquad (W_0,v_0)\in\mathfrak{W}\,,
\end{align}
where $\mathcal{T}_n$ is given by~\eqref{def-T}. For $n\in\mathbb{N}$, let us now pick some $\Upsilon_n\in\mathbb{F}_{\mathsf{L},\mathsf{w}}$ such that $\Upsilon_n\Upsilon_n^*=W_n$. We  set ${\Phi_n = \begin{pmatrix}\Upsilon_n & v_n\end{pmatrix}\in\mathbb{F}_{\mathsf{L},\mathsf{w}+1}}$ so that one has $\Phi_n\Phi_n^*=W_n+v_nv_n^*$. Next introduce
$$
\ulc(v):=\ugl_{\mathsf{w}+1}(v)\,,\qquad\qquad\mlc(v):=\mgl_{\mathsf{w}+1}(v)\,,\qquad\qquad\llc(v):=\lgl_{\mathsf{w}+1}(v)\,,
$$  
and
$$
\mathsf{D}:=\mathsf{I}_{\mathsf{w}}\,,\qquad\qquad\qquad\mathsf{E}:=\mathsf{I}_{\mathsf{w}+1}\,.
$$
Then, $\ulc(v)$, $\mlc(v)$ and $\llc(v)$ are of the length $\mathsf{L}-\mathsf{E}$, $\mathsf{E}-\mathsf{D}$ and $\mathsf{D}$, respectively, and, due to~\eqref{ineq-local-gap},
$$
\eta(\mathsf{D},\mathsf{E})\geq 2^{-3}\,\mathsf{q}^{-1}\,\pmb{\eta}\,.
$$
As before, we have the projection $\hat{P}_{\llc}$. Moreover, let us introduce the frames
$$
\hat{\mathfrak{\chi}}=
\mbox{\footnotesize $
	\begin{pmatrix}
	\mathbf{1}_{\mathsf{L}-\mathsf{E}}\\\mathbf{0}_{\mathsf{E}\times (\mathsf{L}-\mathsf{E})}
	\end{pmatrix}
	$}
,
\qquad\quad
\hat{\mathfrak{\chi}}^{\perp}=
\mbox{\footnotesize $
	\begin{pmatrix}
	\mathbf{0}_{(\mathsf{L}-\mathsf{E})\times\mathsf{E}}\\\mathbf{1}_{\mathsf{E}}
	\end{pmatrix}
	$},
\qquad\quad
\hat{\mathfrak{\zeta}}=
\mbox{\footnotesize $
	\begin{pmatrix}
	\mathbf{0}_{(\mathsf{L}-\mathsf{D})\times\mathsf{D}}\\\mathbf{1}_{\mathsf{D}}
	\end{pmatrix}
	$}
,
\qquad\quad
\hat{\mathfrak{\zeta}}^{\perp}=
\mbox{\footnotesize $
	\begin{pmatrix}
	\mathbf{1}_{\mathsf{L}-\mathsf{D}}\\\mathbf{0}_{\mathsf{D}\times (\mathsf{L}-\mathsf{D})}
	\end{pmatrix}
	$}
,
$$
so $\hat{P}_{\llc}=\hat{\zeta}\hat{\zeta}^*$. Now all notation is at hand to rewrite the induction step in the proof of Lemma~\ref{lemma-inductive-result}.
\begin{lemma}\label{lemma-induction-step}
	Let $(W_0, v_0) \in \mathfrak{W}$ and $(W_n, v_n) \in \mathfrak{W}$ for all $n \in \mathbb{N}$ as defined in~\eqref{dyn-auxiliary}.
	If for all $n \in \left[2^{\frac{18}{5}}\beta^{-1}\mathsf{q}\,\mathsf{w}\,\vartheta\lambda^{-2},\infty\right)\cap\mathbb{N}$ it holds that
	\begin{align}\label{ineq-induction-hypothesis}
	\mathbb{P}\left(\operatorname{tr}\left[(\hat{\zeta}^{\perp})^*W_n\hat{\zeta}^{\perp}\right]>2^{-\frac{21}{5}}\beta^{\frac{3}{5}}\pmb{\eta}^{-\frac{1}{5}}\vartheta^{-\frac{3}{5}}\lambda^{\frac{7}{5}}\right)\leq 2^4(2^{\frac{54}{5}}\lambda)^{3(\mathsf{q}-\mathsf{w}+1)}\,,
	\end{align}
	then for all $n \in \left[2^{\frac{18}{5}}\beta^{-1}\mathsf{q}(\mathsf{w}+1)\vartheta\lambda^{-2},\infty\right)\cap\mathbb{N}$ it follows that
	\begin{align}\label{ineq-induction-outcome}
	\mathbb{P}\left(\operatorname{tr}\left[\hat{\chi}^*(W_n+v_nv_n^*)\hat{\chi}\right]>2^{-\frac{21}{5}}\beta^{\frac{3}{5}}\pmb{\eta}^{-\frac{1}{5}}\vartheta^{-\frac{3}{5}}\lambda^{\frac{7}{5}}\right)\leq 2^4(2^{\frac{54}{5}}\lambda)^{3(\mathsf{q}-\mathsf{w})}\,.
	\end{align}
\end{lemma}

\noindent\textbf{Proof of Lemma~\ref{lemma-inductive-result}.} As stated before, the result will be proved by induction. The basis step for $\mathsf{w}=0$ is trivially true, as all projections vanish (that is, $W_n \in \mathbb{G}_{\mathsf{L},0} = \lbrace \mathbf{0} \rbrace$ for all $n \in \mathbb{N}$). Now suppose that the statement is true for some $\mathsf{w} \in \lbrace 0,\dots,\mathsf{q}-1 \rbrace$, and it needs to be shown for $\mathsf{w}+1$ being the rank of the projections of the considered random dynamical system, say $W^{\prime}_n \in \mathbb{G}_{\mathsf{L},\mathsf{w}+1}$ for all $n \in \mathbb{N}$. Now denote $W_0 \in \mathbb{G}_{\mathsf{L},\mathsf{w}}$ for the projection on some $\mathsf{w}$-dimensional subspace of the range of $W^{\prime}_0$. Then $W^{\prime}_0 - W_0$ is a projection of rank $1$, so one can choose a $v_0 \in \mathbb{S}_{\mathbb{C}}^{\mathsf{L}-1}$ such that $W^{\prime}_0 = W_0 + v_0v_0^*$. Now $(W_0,v_0) \in \mathfrak{W}$, so we can define $(W_n,v_n) \in \mathfrak{W}$ for all $n \in \mathbb{N}$ as in~\eqref{dyn-auxiliary}. By~\eqref{eq-auxiliary-action}, it then holds that $W^{\prime}_n = W_n + v_nv_n^*$ for all $n \in \mathbb{N}$. Recall that $\hat{\alpha}_{\mathsf{w}} = \hat{\gamma}_{\mathsf{w}+1}^{\perp} = \hat{\zeta}^{\perp}$ and $\hat{\alpha}_{\mathsf{w}+1} = \hat{\chi}$. What remains to prove is exactly the content of Lemma~\ref{lemma-induction-step}.
\hfill $\square$

\vspace{.2cm}

The aim of the remaining subsections is to outline the proof of Lemma~\ref{lemma-induction-step}.

\subsubsection{The local subdivision of the middle part}

The first step is a suitable subdivision of the local middle part $\mlc$, as indicated in Figure~\ref{fig-subdivision}. For this, we apply Lemma~\ref{lemma-subdivision} twice again.
Let us start by setting
$$
\mathsf{A}=\mathsf{D}\,,\qquad\qquad\qquad\mathsf{B}=\mathsf{E}\,,\qquad\qquad\qquad\mathsf{F}=3\,,\qquad\qquad\qquad\phi = 3 \cdot 2^7\pmb{\eta}^{-1}\mathsf{q}\,\lambda\,.
$$
As~\eqref{ineq-local-gap} holds, the requirement~\eqref{ineq-subdivision-assumption} follows from Hypothesis~\ref{hyp-eta}. Lemma~\ref{lemma-subdivision} then implies the existence of a partition
$$
\mathsf{D}=\mathsf{A}_0\;<\;\mathsf{A}_1\;<\;\mathsf{A}_{\mathsf{M}+1}\;<\;\mathsf{A}_{\mathsf{M}+2}=\mathsf{E}\,,
$$
which satisfies
\begin{align}\label{ineq-mesoscopic-gaps}
\tau_0 := \eta(\mathsf{A}_0,\mathsf{A}_1)\geq 2^{-5}\,\mathsf{q}^{-1}\,\pmb{\eta}\,,\quad\qquad\tau_{\mathsf{M}+1} := \eta(\mathsf{A}_{\mathsf{M}+1},\mathsf{A}_{\mathsf{M}+2})\geq 2^{-5}\,\mathsf{q}^{-1}\,\pmb{\eta}\,,
\end{align}
and $\eta(\mathsf{A}_1,\mathsf{A}_{\mathsf{M}+1})\geq 2^{-5}\,\mathsf{q}^{-1}\,\pmb{\eta}$, where we estimated $2^{-2} \leq 3^{-1} (1 - 3 \cdot 2^7\pmb{\eta}^{-1}\mathsf{q}\,\lambda)$ which follows from~\eqref{ineq-subdivision-estimate}, as this implies $3 \cdot 2^7\mathsf{q}\,\lambda \leq 2^{11}\mathsf{q}\,\lambda \leq 2^{-2}\pmb{\eta}$.
Next take
\begin{align}\label{def-M}
	\mathsf{A}=\mathsf{A}_1\,,\quad\quad\qquad\mathsf{B}=\mathsf{A}_{\mathsf{M}+1}\,,\quad\quad\qquad\mathsf{F}=\mathsf{M}:=\lfloor 2^{-10}\pmb{\eta}\,\mathsf{q}^{-1}\lambda^{-1}\rfloor\,,\quad\quad\qquad\phi = \frac{1}{2}\,.
\end{align}
Since $\eta(\mathsf{A}_1,\mathsf{A}_{\mathsf{M}+1})\geq 2^{-5}\,\mathsf{q}^{-1}\,\pmb{\eta}$, this time~\eqref{ineq-subdivision-assumption} is satisfied by Hypothesis~\ref{hyp-eta} and the estimate
\begin{align}\label{ineq-microscopic-partition}
\frac{\phi}{\mathsf{M}}\eta(\mathsf{A}_1,\mathsf{A}_{\mathsf{M}+1}) \geq \frac{2^{-6}\,\mathsf{q}^{-1}\,\pmb{\eta}}{\lfloor 2^{-10}\,\pmb{\eta}\,\mathsf{q}^{-1}\lambda^{-1}\rfloor} \geq \frac{2^{-6}\,\mathsf{q}^{-1}\,\pmb{\eta}}{2^{-10}\,\pmb{\eta}\,\mathsf{q}^{-1}\lambda^{-1}} = 2^4\lambda\,.
\end{align}
Lemma~\eqref{lemma-subdivision} then implies the existence of a partition
\begin{align}\label{ineq-concrete-partition}
\mathsf{A}_0\;<\;\mathsf{A}_1\;<\;\mathsf{A}_2\;<\;\dots\;<\;\mathsf{A}_{\mathsf{M}}\;<\;\mathsf{A}_{\mathsf{M}+1}\;<\;\mathsf{A}_{\mathsf{M}+2}\,,
\end{align}
such that
\begin{align}\label{ineq-microscopic-gaps}
{\tau}_{\mathsf{m}}:=\eta(\mathsf{A}_{\mathsf{m}},\mathsf{A}_{\mathsf{m}+1}) \geq 2^4\lambda\,,\qquad\qquad\qquad \mathsf{m}=1,\dots,\mathsf{M}\,.
\end{align}
The latter follows from the fact that the lower bound of~\eqref{ineq-subdivision-result} equals the left of~\eqref{ineq-microscopic-partition} for $\phi = \frac{1}{2}$.
In accordance with the partition~\eqref{ineq-concrete-partition}, we subdivide the local middle part $\mlc$ into
$$
\mlc(v)=\begin{pmatrix}
\mlc_{\mathsf{M}+1}(v)\\\mlc_{\mathsf{M}}(v)\\\vdots\\\mlc_{1}(v)\\\mlc_{0}(v)
\end{pmatrix}
$$
where the lengths of the $\mlc_{\mathsf{m}}$ are given by $\mathsf{A}_{\mathsf{m}+1}-\mathsf{A}_{\mathsf{m}}$, respectively.
Further, we use the abbreviations
$$
\ulc_{\mathsf{m}}(v):=\begin{pmatrix}
\ulc(v)\\\mlc_{\mathsf{M}+1}(v)\\\vdots\\\mlc_{\mathsf{m}+1}(v)
\end{pmatrix}\,,\qquad\qquad
\llc_{\mathsf{m}}(v):=\begin{pmatrix}
\mlc_{\mathsf{m}-1}(v)\\\vdots\\\mlc(v)\\\llc(v)
\end{pmatrix}\,,
$$
including $\ulc_{\mathsf{M}+1}(v)=\ulc(v)$ and $\llc_0(v)=\llc(v)$. For $\mathsf{m}\in\{0,\dots,\mathsf{M}+1\}$, we introduce the frames
$$
\hat{\mathfrak{\chi}}_{\mathsf{m}}=
\mbox{\footnotesize $
\begin{pmatrix}
\mathbf{1}_{\mathsf{L}-\mathsf{A}_{\mathsf{m}+1}}\\\mathbf{0}_{\mathsf{A}_{\mathsf{m}+1}\times (\mathsf{L}-\mathsf{A}_{\mathsf{m}+1})}
\end{pmatrix}
$}
,
\quad
\hat{\mathfrak{\chi}}^{\perp}_{\mathsf{m}}=
\mbox{\footnotesize $
\begin{pmatrix}
\mathbf{0}_{(\mathsf{L}-\mathsf{A}_{\mathsf{m}+1})\times\mathsf{A}_{\mathsf{m}+1}}\\\mathbf{1}_{\mathsf{A}_{\mathsf{m}+1}}
\end{pmatrix}
$},
\quad
\hat{\mathfrak{\zeta}}_{\mathsf{m}}=
\mbox{\footnotesize $
\begin{pmatrix}
\mathbf{0}_{(\mathsf{L}-\mathsf{A}_{\mathsf{m}})\times\mathsf{A}_{\mathsf{m}}}\\\mathbf{1}_{\mathsf{A}_{\mathsf{m}}}
\end{pmatrix}
$}
,
\quad
\hat{\mathfrak{\zeta}}^{\perp}_{\mathsf{m}}=
\mbox{\footnotesize $
\begin{pmatrix}
\mathbf{1}_{\mathsf{L}-\mathsf{A}_{\mathsf{m}}}\\\mathbf{0}_{\mathsf{A}_{\mathsf{m}}\times (\mathsf{L}-\mathsf{A}_{\mathsf{m}})}
\end{pmatrix}
$}
.
$$
This yields the complete subdivision as depicted in Figure~\ref{fig-subdivision}.

\subsubsection{Expansion of the perturbation for the auxiliary dynamics}

Similar as in Lemma~\ref{lemma-expansion}, one needs to expand the effect of the perturbation when applied in combination with an orthogonal projection as in the last summand of equation~\eqref{eq-auxiliary-action}.

\begin{lemma}\label{lemma-vector-expansion}
	Let $\lambda \leq 2^{-6}$. Let $\hat{\Psi} \in \bigcup\limits_{\mathsf{w}=0}^{\mathsf{L}} \mathbb{F}_{\mathsf{L},\mathsf{w}}$, and let $\mathfrak{d}(\cdot)$ be the partial vector corresponding to $\hat{\Psi}$, \textit{i.e.}, for all $v \in \mathbb{S}_{\mathbb{C}}^{\mathsf{L}-1}$ it holds that $\hat{\Psi}^*v=\mathfrak{d}(v)$. Then, all $\mathtt{P}\in\mathfrak{P}$ and $(W,v) \in \mathfrak{W}$ satisfy the bound
	\begin{align}\label{ineq-vector-expansion-1}
	\big|\|\mathfrak{d}([(e^{\lambda\mathtt{P}}\cdot W)^{\perp}e^{\lambda\mathtt{P}}]\circ v)\|^2-\|\mathfrak{d}(v)\|^2\big| \leq \mbox{\small $\frac{3}{2}$}\,\lambda\,.
	\end{align}
	Using the notation from {\rm Lemma~\ref{lemma-expansion}}, let us define the map $\mathbf{A}_{\mathfrak{d}}: \mathfrak{W}\times \mathfrak{P}\rightarrow \mathbb{R}$ by
	\begin{align*}
	\mathbf{A}_{\mathfrak{d}}(W,v,\mathtt{P})=\operatorname{tr}\left(\hat{\Psi}^*\left[\mathtt{X}(W+vv^*,\mathtt{P})-\mathtt{X}(W,\mathtt{P})\right]\hat{\Psi}\right)\,,
	\end{align*}
	which fulfills for all $\mathtt{P}\in\mathfrak{P}$ and $(W,v) \in \mathfrak{W}$ the estimates
	\begin{align}\label{ineq-vector-expansion-2}
	\big|\mathbf{A}_{\mathfrak{d}}(W,v,\mathtt{P})\big|\leq 2^{\frac{1}{2}}\,,
	\end{align}
	and
	\begin{align}\label{ineq-vector-expansion-3}
	\big|\|\mathfrak{d}([(e^{\lambda\mathtt{P}}\cdot W)^{\perp}e^{\lambda\mathtt{P}}]\circ v)\|^2-\|\mathfrak{d}(v)\|^2 - \lambda\,\mathbf{A}_{\mathfrak{d}}(W,v,\mathtt{P})\big| \leq 9\,\lambda^2 + 160\,\lambda^3\,.
	\end{align}
\end{lemma}

\subsubsection{Contraction and growth  inequalities}

Next let us analyze the effect of the hyperbolic action when applied in combination with an orthogonal projection as in the last summand of equation~\eqref{eq-auxiliary-action} on the norm of the upper and lower part of a vector.

\begin{lemma}\label{lemma-vector-contraction}
	For all $\mathsf{m}\in\{0,\dots,\mathsf{M}+1\}$ and $(W,v)\in\mathfrak{W}$, one has the symmetric inequalities 
	\begin{align}\label{ineq-vector-contraction-1}
		\| \ulc_{\mathsf{m}}\left(\left((\mathcal{R} \cdot W)^{\perp}\mathcal{R}\right) \circ v\right) \|^2&\leq \|\ulc_{\mathsf{m}}(v)\|^2\left[1-\tau_{\mathsf{m}}\,\|\llc_{\mathsf{m}}(v)\|^2\right]+2\,\operatorname{tr}\left[(\hat{\zeta}^{\perp}_{\mathsf{m}})^*W\hat{\zeta}^{\perp}_{\mathsf{m}}\right]
		\end{align}
and
	\begin{align}\label{ineq-vector-contraction-2}
		\| \llc_{\mathsf{m}}\left(\left((\mathcal{R} \cdot W)^{\perp}\mathcal{R}\right) \circ v\right) \|^2&\geq \|\llc_{\mathsf{m}}(v)\|^2\left[1+\tau_{\mathsf{m}}\,\|\ulc_{\mathsf{m}}(v)\|^2\right]-2\,\operatorname{tr}\left[(\hat{\zeta}^{\perp}_{\mathsf{m}})^*W\hat{\zeta}^{\perp}_{\mathsf{m}}\right]\,.
\end{align}
\end{lemma}	

In view of Lemma~\ref{lemma-vector-contraction}, the purely hyperbolic action (not combined with an orthogonal projection) on a vector $v$ yields a contraction of the norm upper part of $v$ and a growth of the norm of the lower part  of $v$. More precisely, the inequalities~\eqref{ineq-vector-contraction-1} and~\eqref{ineq-vector-contraction-2} with $W=0$ read
\begin{align*}
\| \ulc_{\mathsf{m}}\left(\mathcal{R} \circ v\right) \|^2\leq \|\ulc_{\mathsf{m}}(v)\|^2\left[1-\tau_{\mathsf{m}}\,\|\llc_{\mathsf{m}}(v)\|^2\right]\,,\qquad\quad\| \llc_{\mathsf{m}}\left(\mathcal{R} \circ v\right) \|^2\geq \|\llc_{\mathsf{m}}(v)\|^2\left[1+\tau_{\mathsf{m}}\,\|\ulc_{\mathsf{m}}(v)\|^2\right]\,.
\end{align*}
Of course, these inequalities are more accessible. 
However, by the induction hypothesis made in Section~\ref{sec-induction}, the terms $\operatorname{tr}\big[(\hat{\zeta}^{\perp}_{\mathsf{m}})^*W\hat{\zeta}^{\perp}_{\mathsf{m}}\big]$ are eventually small enough to be treated as perturbations~\textemdash~at least with overwhelming probability.  Therefore, we introduce a subset of $\mathfrak{W}$ in which the vector $v_n$ of the dynamics~\eqref{dyn-auxiliary} approximately fulfills a contraction and a growth inequality:

\vspace{.2cm}

\noindent\textbf{The overwhelming region.} This is defined as the set
$$
\mathfrak{Q}:=\left\{(W,v)\in\mathfrak{W}: \operatorname{tr}\left[(\hat{\zeta}^{\perp})^*W\hat{\zeta}^{\perp}\right]\leq 2^{-\frac{21}{5}}\beta^{\frac{3}{5}}\pmb{\eta}^{-\frac{1}{5}}\vartheta^{-\frac{3}{5}}\lambda^{\frac{7}{5}}\right\}\,.
$$

\vspace{.2cm}

Clearly, the analysis of the effect of the hyperbolic action is simpler if the path of the random dynamics~\eqref{dyn-auxiliary} lies in $\mathfrak{Q}$. 
The arguments below are performed precisely for such paths. Thus, it is also convenient to introduce  the condition that the path of the dynamics~\eqref{dyn-auxiliary} lies in the overwhelming region $\mathfrak{Q}$ within certain spaces of time as an event in the overall probability space:

\vspace{.2cm}

\noindent\textbf{The overwhelming event.}
For $n_1,n_2\in\mathbb{N}$ with $n_1<n_2$, this is defined as the set
$$
\mathfrak{O}_{n_1,n_2}:=\left\{\forall\,n\in\{n_1+1,\dots,n_2\}: (W_n,v_n)\in\mathfrak{Q} \right\}\,.
$$

\vspace{.2cm}

Now in the overwhelming region, the effect of the overall action when applied in combination with an orthogonal projection as in the last summand of equation~\eqref{eq-auxiliary-action} on the norm of the upper and lower part of a vector is deterministically bounded. This is stated in Corollary~\ref{lemma-deterministic-vector}, which follows from inequality~\eqref{ineq-vector-expansion-1} of Lemma~\ref{lemma-vector-expansion}, Lemma~\ref{lemma-vector-contraction} and the following observation. If $(W,v)\in\mathfrak{Q}$, the fact that $\hat{\zeta}_{\mathsf{m}}^{\perp}(\hat{\zeta}^{\perp}_{\mathsf{m}})^*\leq\hat{\zeta}^{\perp}(\hat{\zeta}^{\perp})^*$ and Hypothesis~\ref{hyp-q} imply
\begin{align}\label{ineq-trace-order-lambda}
\begin{split}
\operatorname{tr}\left[(\hat{\zeta}_{\mathsf{m}}^{\perp})^*W\hat{\zeta}^{\perp}_{\mathsf{m}}\right]&=\operatorname{tr}\left[W\hat{\zeta}_{\mathsf{m}}^{\perp}(\hat{\zeta}^{\perp}_{\mathsf{m}})^*W\right]\leq \operatorname{tr}\left[W\hat{\zeta}^{\perp}(\hat{\zeta}^{\perp})^*W\right]=\operatorname{tr}\left[(\hat{\zeta}^{\perp})^*W\hat{\zeta}^{\perp}\right]\\
&\leq 2^{-\frac{21}{5}}\beta^{\frac{3}{5}}\pmb{\eta}^{-\frac{1}{5}}\vartheta^{-\frac{3}{5}}\lambda^{\frac{7}{5}} \leq 2^{-\frac{21}{5}}\beta^{\frac{3}{5}}\pmb{\eta}^{-\frac{1}{5}}\vartheta^{-\frac{3}{5}}\lambda^{\frac{7}{5}}(\mathsf{q}^{-1}\mathsf{q})^22^{\frac{78}{5}}\beta^{-1}\pmb{\eta}^{-1}\mathsf{q}^2\vartheta\\
&\leq 2^{\frac{67}{5}}\beta^{-\frac{2}{5}}\pmb{\eta}^{-\frac{6}{5}}\mathsf{q}^2\vartheta^{\frac{2}{5}}\lambda^{\frac{7}{5}}(\mathsf{q}^{-1}2^{-\frac{36}{5}}\beta^{\frac{1}{5}}\pmb{\eta}^{\frac{3}{5}}\vartheta^{-\frac{1}{5}}\lambda^{-\frac{1}{5}})^2 = 2^{-3}\lambda\,.
\end{split}
\end{align}

\begin{coro}\label{lemma-deterministic-vector}
Let $\mathsf{m}\in\{0,\dots,\mathsf{M}+1\}$ and $(W,v)\in\mathfrak{Q}$. Then, one has the symmetric inequalities
\begin{align}\label{ineq-deterministic-vector-1}
\| \ulc_{\mathsf{m}}\left(\left(((e^{\lambda\mathtt{P}}\mathcal{R}) \cdot W)^{\perp}(e^{\lambda\mathtt{P}}\mathcal{R})\right) \circ v\right) \|^2\leq \|\ulc_{\mathsf{m}}(v)\|^2+\mbox{\small $\frac{7}{4}$}\lambda
\end{align}
and 
\begin{align}\label{ineq-deterministic-vector-2}
\| \llc_{\mathsf{m}}\left(\left(((e^{\lambda\mathtt{P}}\mathcal{R}) \cdot W)^{\perp}(e^{\lambda\mathtt{P}}\mathcal{R})\right) \circ v\right) \|^2\geq \|\llc_{\mathsf{m}}(v)\|^2-\mbox{\small $\frac{7}{4}$}\lambda\,.
\end{align}
\end{coro}

Moreover, there are subsets of the overwhelming region (specified by inequality~\eqref{ineq-ladder} below), in which the overall action when applied in combination with an orthogonal projection as in the last summand of equation~\eqref{eq-auxiliary-action} strictly decreases (increases) the norm of the upper (lower) part of a vector. This is stated in Corollary~\ref{coro-ladder}, which also follows from inequality~\eqref{ineq-vector-expansion-1} of Lemma~\ref{lemma-vector-expansion} and Lemma~\ref{lemma-vector-contraction}.

\begin{coro}\label{coro-ladder}
Let $\mathsf{m}\in\{0,\dots,\mathsf{M}+1\}$ and $(W,v)\in\mathfrak{Q}$ be such that
\begin{align}\label{ineq-ladder}
\|\ulc_{\mathsf{m}}(v)\|^2\|\llc_{\mathsf{m}}(v)\|^2\geq 2\frac{\lambda}{{\tau}_{\mathsf{m}}}
\end{align}
is satisfied. Then, one has the symmetric inequalities
\begin{align}\label{ineq-diminish-strictly}
\| \ulc_{\mathsf{m}}\left(\left(((e^{\lambda\mathtt{P}}\mathcal{R}) \cdot W)^{\perp}(e^{\lambda\mathtt{P}}\mathcal{R})\right) \circ v\right) \|^2\leq \|\ulc_{\mathsf{m}}(v)\|^2-2^{-2}\lambda
\end{align}
and 
\begin{align}\label{ineq-enlarge-strictly}
\| \llc_{\mathsf{m}}\left(\left(((e^{\lambda\mathtt{P}}\mathcal{R}) \cdot W)^{\perp}(e^{\lambda\mathtt{P}}\mathcal{R})\right) \circ v\right) \|^2\geq \|\llc_{\mathsf{m}}(v)\|^2+2^{-2}\lambda\,.
\end{align}
\end{coro}

\subsubsection{The ladder construction}

The \textit{ladder construction} is a collection of deterministic constraints on the path of the dynamics~\eqref{dyn-auxiliary} that constitutes the core of the proof. It is based on Corollary~\ref{coro-ladder} and the following preparatory remark on the assumption~\eqref{ineq-ladder} made therein.

\begin{rem}
{\rm
The assumption~\eqref{ineq-ladder} of Corollary~\ref{coro-ladder} holds if there is some $\sigma\in (0,1)$ for which 
\begin{align}\label{ineq-ladder-1}
\|\ulc_{\mathsf{m}}(v)\|^2\geq \sigma
\end{align}
and
\begin{align}\label{ineq-ladder-2}
\|\llc_{\mathsf{m}}(v)\|^2\geq \frac{2}{\sigma}\frac{\lambda}{\tau_{\mathsf{m}}}
\end{align}
hold.
}
\hfill $\diamond$
\end{rem}

Now, inequalities~\eqref{ineq-ladder-1} and~\eqref{ineq-ladder-2} motivate to introduce the following couple of notions.

\begin{defini} 
\noindent Let $\sigma,\tau\in (0,1)$. For $\mathsf{m}\in\{0,1,\dots,\mathsf{M}+1\}$,  the $\mathsf{m}$-th cones are defined as
$$
\mathfrak{C}^{\sigma}_{\mathsf{m}}:=\left\{(W,v)\in\mathfrak{W}: \|\ulc_{\mathsf{m}}(v)\|^2\leq \sigma\right\}\,,
$$
the $\mathsf{m}$-th anti-cones are
$$
\mathfrak{A}^{\sigma,\tau}_{\mathsf{m}}:=\left\{(W,v)\in\mathfrak{W}: \|\llc_{\mathsf{m}}(v)\|^2\leq \frac{2}{\sigma}\frac{\lambda}{\tau}\right\}
\;,
$$
and the $\mathsf{m}$-th steps of the ladder are 
\begin{align*}
\mathfrak{S}_{\mathsf{m}}^{\sigma,\tau}:=\mathfrak{C}^{\sigma}_{\mathsf{m}}\cap\mathfrak{A}^{\sigma,\tau}_{\mathsf{m}}
\;.
\end{align*}
Furthermore, for half-integers $\mathsf{m}^{\prime}\in\{1/2,3/2,\dots,\mathsf{M}+1/2\}$ the $\mathsf{m}^{\prime}$-th interspaces are
$$
\mathfrak{I}_{\mathsf{m}^{\prime}}^{\sigma,\tau}:=\left(\mathfrak{C}^{\sigma}_{\mathsf{m}^{\prime}+1/2}\setminus\mathfrak{C}^{\sigma}_{\mathsf{m}^{\prime}-1/2}\right)\cap\left(\mathfrak{A}^{\sigma,\tau}_{\mathsf{m}^{\prime}-1/2}\setminus\mathfrak{A}^{\sigma,\tau}_{\mathsf{m}^{\prime}+1/2}\right)\;.
$$
\end{defini}

\begin{rem} {\rm The construction is illustrated in Figure~\ref{fig-ladder}.
It follows immediately from the definition that the cones satisfy the sequence of inclusions
$$
\mathfrak{C}^{\sigma}_0\subset\mathfrak{C}^{\sigma}_1\subset\mathfrak{C}^{\sigma}_2\subset\dots\subset\mathfrak{C}^{\sigma}_{\mathsf{M}}\subset\mathfrak{C}^{\sigma}_{\mathsf{M}+1}\,,
$$
while the anti-cones satisfy 
$$
\mathfrak{A}^{\sigma,\tau}_0\supset\mathfrak{A}^{\sigma,\tau}_1\supset\mathfrak{A}^{\sigma,\tau}_2\supset\dots\supset\mathfrak{A}^{\sigma,\tau}_{\mathsf{M}}\supset\mathfrak{A}^{\sigma,\tau}_{\mathsf{M}+1}\,.
$$
If one has $\sigma+\frac{2}{\sigma}\frac{\lambda}{\tau}<1$ and $\mathsf{m}<\mathsf{n}$, then the $\mathsf{m}$-th cone $\mathfrak{C}^{\sigma}_{\mathsf{m}}$ and the $\mathsf{n}$-th anti-cone $\mathfrak{A}^{\sigma,\tau}_{{\mathsf{n}}}$ are disjoint. However, the intersection of $\mathfrak{C}^{\sigma}_{\mathsf{m}}$ with $\mathfrak{A}^{\sigma,\tau}_{{\mathsf{m}}}$ is non-empty and makes up the steps of the ladder. More explicitly, the steps of the ladder are
$$
\mathfrak{S}_{\mathsf{m}}^{\sigma,\tau}
=\left\{(W,v)\in\mathfrak{W}: \|\ulc_{\mathsf{m}}(v)\|^2\leq \sigma\,,\quad \|\llc_{\mathsf{m}}(v)\|^2\leq \frac{2}{\sigma}\frac{\lambda}{\tau} \right\}
\;,
$$
and the interspaces
$$
\mathfrak{I}_{\mathsf{m}^{\prime}}^{\sigma,\tau}
=\left\{(W,v)\in\mathfrak{W}: \|\ulc_{\mathsf{m}^{\prime}-1/2}(v)\|^2> \sigma\geq\|\ulc_{\mathsf{m}^{\prime}+1/2}(v)\|^2\,,\; \|\llc_{\mathsf{m}^{\prime}-1/2}(v)\|^2\leq \frac{2}{\sigma}\frac{\lambda}{\tau}<\|\llc_{\mathsf{m}^{\prime}+1/2}(v)\|^2 \right\}.
$$
As a matter of fact, any two different elements of $\left\{\mathfrak{S}_k^{\sigma,\tau}\right\}_{k=1}^{\mathsf{M}}\cup\left\{\mathfrak{I}_{\mathsf{m}^{\prime}}^{\sigma,\tau}\right\}_{\mathsf{m}^{\prime}=3/2}^{\mathsf{M}-1/2}$ are disjoint.
}
\hfill $\diamond$
\end{rem}

Lemma~\ref{lemma-allowed-movements} states a couple of deterministic constraints for the path of the dynamics~\eqref{dyn-auxiliary} in terms of the notions just introduced. Loosely speaking, these constraints limit the possible paths from the attractive to the repulsive region of the state space, as displayed in Figure~\ref{fig-ladder}.

\begin{figure}[H]
	\begin{center}
		\begin{tikzpicture}[line join = round, line cap = round]
			\coordinate (st) at (0,4.5);
			\coordinate (s1) at (0,2.5);
			\coordinate (i0) at (0,0);
			\coordinate (s2) at (0,-2.5);
			\coordinate (sb) at (0,-4.5);
			\coordinate (i1) at (0,1.5);
			\coordinate (i2) at (1.5,3);
			\coordinate (i3) at (4.5,0);
			\coordinate (i4) at (1.5,-3);
			\coordinate (i5) at (0,-1.5);
			\coordinate (i6) at (-1.5,-3);
			\coordinate (i7) at (-4.5,0);
			\coordinate (i8) at (-1.5,3);
			\coordinate[label=right:{$\,\mathfrak{A}^{\sigma,\tau}_{\mathsf{m}+1}$}] (l1) at (-3.5,5);
			\coordinate[label=right:{$\,\mathfrak{A}^{\sigma,\tau}_{\mathsf{m}}$}] (l2) at (-8,3.5);
			\coordinate[label=right:{$\,\mathfrak{C}^{\sigma}_{\mathsf{m}+1}$}] (l3) at (-8,-3.5);
			\coordinate[label=right:{$\,\mathfrak{C}^{\sigma}_{\mathsf{m}}$}] (l4) at (-3.5,-5);
			\coordinate (r1) at (3.5,5);
			\coordinate (r2) at (8,3.5);
			\coordinate (r3) at (8,-3.5);
			\coordinate (r4) at (3.5,-5);
			\coordinate (d1) at (0,5);
			\coordinate (d2) at (2.5,2.5);
			\coordinate (d3) at (7.5,-2.5);
			\coordinate (d4) at (3.75,-3.75);
			\coordinate (d5) at (0,-5);
			\fill[gray!20] (st)--(i2)--(i1)--(i8)--cycle;
			\fill[gray!20] (sb)--(i6)--(i5)--(i4)--cycle;
			\draw[-,color=gray!130,line width=1mm] (l1)--(i1)--(r1);
			\draw[-,color=gray!130,line width=1mm] (l2)--(sb)--(r2);
			\draw[-,color=gray!90,line width=1mm] (l3)--(st)--(r3);
			\draw[-,color=gray!90,line width=1mm] (l4)--(i5)--(r4);
			\coordinate[label=right:{$\,\mathfrak{S}^{\sigma,\tau}_{\mathsf{m}}$}] (i6) at (i6);
			\coordinate[label=right:{$\ \mathfrak{I}^{\sigma,\tau}_{\mathsf{m}+1/2}$}] (i7) at (i7);
			\coordinate[label=right:{$\,\mathfrak{S}^{\sigma,\tau}_{\mathsf{m}+1}$}] (i8) at (i8);
			\draw[arrows={-Stealth[inset=0pt,angle=15:12pt]},color=black,line width=0.7mm] (d1)--(s1);
			\draw[arrows={-Stealth[inset=0pt,angle=15:12pt]},color=black,line width=0.7mm] (s1)--(i0);
			\draw[arrows={-Stealth[inset=0pt,angle=15:12pt]},color=black,line width=0.7mm] (i0)--(s2);
			\draw[arrows={-Stealth[inset=0pt,angle=15:12pt]},color=black,line width=0.7mm] (s2)--(d5);
			\draw[arrows={-Stealth[inset=0pt,angle=15:12pt]},color=black,line width=0.7mm] (d1)--(d2);
			\draw[arrows={-Stealth[inset=0pt,angle=15:12pt]},color=black,line width=0.7mm] (d2)--(d3);
			\draw[arrows={-Stealth[inset=0pt,angle=15:12pt]},color=black,line width=0.7mm] (d3)--(d4);
			\draw[arrows={-Stealth[inset=0pt,angle=15:12pt]},color=black,line width=0.7mm] (d4)--(d5);
			\draw[arrows={-Stealth[inset=0pt,angle=15:12pt]},color=black,line width=0.7mm] (s1)--(d2);
			\draw[arrows={-Stealth[inset=0pt,angle=15:12pt]},color=black,line width=0.7mm] (d2)--(i0);
			\draw[arrows={-Stealth[inset=0pt,angle=15:12pt]},color=black,line width=0.7mm] (i0)--(d3);
			\draw[arrows={-Stealth[inset=0pt,angle=15:12pt]},color=black,line width=0.7mm] (s2)--(d3);
			\draw[arrows={-Stealth[inset=0pt,angle=15:12pt]},color=black,line width=0.7mm] (s2)--(d4);
			\draw[arrows={-Stealth[inset=0pt,angle=15:12pt]},color=black,line width=0.7mm] (-1.25,-4.25)--(-0.25,-3.25);
			\draw[arrows={-Stealth[inset=0pt,angle=15:12pt]},color=black,line width=0.7mm] (-0.25,-2.75)--(-1.25,-1.75);
			\draw[arrows={-Stealth[inset=0pt,angle=15:12pt]},color=black,line width=0.7mm] (-0.25,3.25)--(-1.25,4.25);
			\draw[arrows={-Stealth[inset=0pt,angle=15:12pt]},color=black,line width=0.7mm] (-1.25,1.75)--(-0.25,2.75);
		\end{tikzpicture}
	\end{center}
	\caption{For some $\mathsf{m} \in \lbrace 0, \dots, \mathsf{M} \rbrace$, the anticones $\mathfrak{A}^{\sigma,\tau}_{\mathsf{m}}$ and $\mathfrak{A}^{\sigma,\tau}_{\mathsf{m}+1}$ (in dark gray), the cones $\mathfrak{C}^{\sigma}_{\mathsf{m}}$ and $\mathfrak{C}^{\sigma}_{\mathsf{m}+1}$ (in medium gray), the steps $\mathfrak{S}^{\sigma,\tau}_{\mathsf{m}}$ and $\mathfrak{S}^{\sigma,\tau}_{\mathsf{m}+1}$ (the light gray regions) and the interspace $\mathfrak{I}^{\sigma,\tau}_{\mathsf{m}+1/2}$ (the central enclosed white region) are depicted schematically. All arrows shown (and those that can be composed by joining arrows having a starting and ending point in common) are the only paths that the dynamics can go in one step, according to Lemma~\ref{lemma-allowed-movements}.}\label{fig-ladder}
\end{figure}

\begin{lemma}\label{lemma-allowed-movements}
Let $\mathsf{m}\in\{0,\dots,\mathsf{M}+1\}$ and suppose that $\sigma\in (0,1)$ and $\tau\in (0,\tau_{\mathsf{m}}]$~satisfy
\begin{align}\label{ineq-sigma-tau}
\sigma+\frac{7}{4}\lambda+\frac{2}{\sigma}\frac{\lambda}{{\tau}}<1\,.
\end{align}
If the dynamics~\eqref{dyn-auxiliary} lies in $\mathfrak{Q}$ persistently,
 the following statements hold true:
\begin{enumerate}
\item If the dynamics is neither in $\mathfrak{C}^{\sigma}_{\mathsf{m}}$ nor in $\mathfrak{A}^{\sigma,\tau}_{\mathsf{m}}$, then there is a deterministic run into $\mathfrak{C}^{\sigma}_{\mathsf{m}}$ and the dynamics will not enter $\mathfrak{A}^{\sigma,\tau}_{\mathsf{m}}$ before having entered $\mathfrak{C}^{\sigma}_{\mathsf{m}}$. In particular: If the dynamics is entering $\mathfrak{A}^{\sigma,\tau}_{\mathsf{m}}$, it is entering $\mathfrak{S}^{\sigma,\tau}_{\mathsf{m}}$.
\item If the dynamics is in $\mathfrak{C}^{\sigma}_{\mathsf{m}}\setminus\mathfrak{S}^{\sigma,\tau}_{\mathsf{m}}=\mathfrak{C}^{\sigma}_{\mathsf{m}}\setminus\mathfrak{A}^{\sigma,\tau}_{\mathsf{m}}$, then it will not leave $\mathfrak{C}^{\sigma}_{\mathsf{m}}$ before having entered~$\mathfrak{A}^{\sigma,\tau}_{\mathsf{m}}$. In other words: If the dynamics is leaving $\mathfrak{C}^{\sigma}_{\mathsf{m}}$, it is leaving $\mathfrak{S}^{\sigma,\tau}_{\mathsf{m}}$.
\item Assume $\mathsf{m}\neq \mathsf{M}+1$ and $\tau\leq \tau_{\mathsf{m}+1}$. If the dynamics is entering $\mathfrak{A}^{\sigma,\tau}_{\mathsf{m}+1}$, it is leaving $\mathfrak{I}_{\mathsf{m}+1/2}^{\sigma,\tau}$.
\item If the dynamics is being in $\mathfrak{S}^{\sigma,\tau}_{\mathsf{m}}$ and leaving $\mathfrak{C}^{\sigma}_{\mathsf{m}}$ but staying in $\mathfrak{A}_{\mathsf{m}}^{\sigma,\tau}$, it is entering $\mathfrak{I}^{\sigma,\tau}_{\mathsf{m}+1/2}$.
\item Assume $\mathsf{m}\neq 0$. If the dynamics is leaving $\mathfrak{I}_{\mathsf{m}-1/2}^{\sigma,\tau}$, it is leaving $\mathfrak{A}^{\sigma,\tau}_{\mathsf{m}-1}$ or entering $\mathfrak{S}^{\sigma,\tau}_{\mathsf{m}-1}\cup\mathfrak{S}^{\sigma,\tau}_{\mathsf{m}}$.
\end{enumerate}
\end{lemma}

Remark~\ref{rem-sigma-tau} provides a rationale for the particular choice of $\sigma$ and $\tau$ as of now.

\begin{rem}\label{rem-sigma-tau}
{\rm 
The assumption~\eqref{ineq-sigma-tau} suffices for the ladder construction as proved in Lemma~\ref{lemma-allowed-movements}. It is, however, convenient to maximize the distance $1-\sigma+\frac{2}{\sigma}\frac{\lambda}{\tau}$ between $\mathfrak{C}^{\sigma}_{\mathsf{m}}$ and $\mathfrak{A}^{\sigma,\tau}_{\mathfrak{m}+1}$,~\textit{i.e.}, to choose $\sigma=(2\lambda\tau^{-1})^{\frac{1}{2}}$. In view of the assumption of Lemma~\ref{lemma-allowed-movements} that $\tau\in (0,\tau_{\mathsf{m}}]$ for all\linebreak $\mathsf{m}\in\{0,\dots,\mathsf{M}+1\}$, we want to choose $\tau$ smaller than all the $\tau_{\mathsf{m}}$. Due to~\eqref{ineq-microscopic-gaps}, one has $\tau_{\mathsf{m}}\geq 2^4\lambda$ for all $\mathsf{m}\in\{1,\dots,\mathsf{M}\}$. Further,~\eqref{ineq-subdivision-estimate} and~\eqref{ineq-mesoscopic-gaps} imply that the inequalities $\tau_0\geq 2^4\lambda$ and $\tau_{\mathsf{M}+1}\geq 2^4\lambda$ hold, as $2^{-5}\mathsf{q}^{-1}\pmb{\eta} \geq 2^4\mathsf{q}^{-1}(2^{-11}\pmb{\eta}) \geq 2^4\lambda$. Therefore, we choose
$$
\tau=\overline{\tau}:=2^4\lambda
$$
and accordingly
$$
\sigma=\overline{\sigma}:=(2\,\lambda\,\overline{\tau}^{-1})^{\frac{1}{2}}=2^{-\frac{3}{2}}\,.
$$
One then has
$$
\overline{\tau}\leq \min\limits_{\mathsf{m}=0}^{\mathsf{M}+1}\tau_{\mathsf{m}}
$$
and the distance between $\mathfrak{C}_{\mathsf{m}}^{\overline{\sigma}}$ and $\mathfrak{A}_{\mathsf{m}+1}^{\overline{\sigma},\overline{\tau}}$ is 
\begin{align}\label{eq-sigma-tau}
1-\overline{\sigma}-\frac{2}{\overline{\sigma}}\frac{\lambda}{\overline{\tau}}=1-2^{-\frac{1}{2}}\,.
\end{align}
Then Hypothesis~\ref{hyp-lambda} guarantees that~\eqref{ineq-sigma-tau} holds for $\sigma = \overline{\sigma}$ and $\tau = \overline{\tau}$.
}
\hfill $\diamond$
\end{rem}

The task is now to control the probability that the dynamics lies in $\mathfrak{A}_{\mathsf{M}+1}^{\overline{\sigma},\overline{\tau}}$ by an overwhelmingly small number after an adequate period of time. This is then used to derive the induction outcome~\eqref{ineq-induction-outcome} from the induction hypothesis~\eqref{ineq-induction-hypothesis}. The region $\mathfrak{A}_{\mathsf{M}+1}^{\overline{\sigma},\overline{\tau}}$ may be viewed as the repulsive region in the ladder construction.

\subsubsection{Diffusive bounds}

The purpose of this section is to bound the probability with which the dynamics~\eqref{dyn-auxiliary} is not contained in $24$-th anti-cone $\mathfrak{A}_{24}^{\overline{\sigma},\overline{\tau}}$ from below (again the number $24$ is merely a result of our estimates). The exterior of $\mathfrak{A}_{24}^{\overline{\sigma},\overline{\tau}}$ may be viewed as an attractive ground region from which it is unlikely to climb up the ladder in a short period of time towards the repulsive region $\mathfrak{A}_{\mathsf{M}+1}^{\overline{\sigma},\overline{\tau}}$.
The penultimate sentence is true, however, only up to the overwhelming event necessary for our analysis. More precisely, we prove a suitable upper bound for the intersection of the complementary event with the overwhelming event (see Lemma~\ref{lemma-diffusion-conclusion} below).

\vspace{.2cm}

Intuitively, the ascension of the ladder is an event of very small probability.  If the dynamics tries to ascend, but leaves $\mathfrak{A}_{24}^{\overline{\sigma},\overline{\tau}}$, then it has to start anew almost from the bottom.  

\vspace{.2cm}

To leave the anti-cone $\mathfrak{A}_0^{\overline{\sigma},\tau_0}$ also for a larger second superindex $\tau_0\gg\overline{\tau}$ turns out to be useful to achieve  the leave of $\mathfrak{A}_{24}^{\overline{\sigma},\overline{\tau}}$ itself. This has two reasons.
First of all, Lemma~\ref{lemma-deterministic-path} shows that the leave of $\mathfrak{A}_0^{\overline{\sigma},\tau_0}$, given the overwhelming event, results in a leave of $\mathfrak{A}_{24}^{\overline{\sigma},\overline{\tau}}$.

\begin{lemma}\label{lemma-deterministic-path}
If all $n\hspace{-0.5mm}\in\hspace{-0.5mm} \{0,\dots,\lfloor 4\,\lambda^{-1}\rfloor\hspace{-0.2mm}-\hspace{-0.2mm}1\}$ obey $(W_n,v_n)\hspace{-0.2mm}\in\hspace{-0.2mm}\mathfrak{Q}$, one has the deterministic implication
$$
(W_0,v_0)\not\in\mathfrak{A}_0^{\overline{\sigma},\tau_0}\qquad\Longrightarrow\qquad \big(W_{\lfloor 4\,\lambda^{-1}\rfloor},v_{\lfloor 4\,\lambda^{-1}\rfloor}\big)\not\in\mathfrak{A}_{24}^{\overline{\sigma},\overline{\tau}}\,.
$$
\end{lemma}

Secondly, the thinness of $\mathfrak{A}_0^{\overline{\sigma},\tau_0}$, which is due to the largeness of $\tau_0$, simply makes it easier to be left. In fact, we call such a leave of  $\mathfrak{A}_0^{\overline{\sigma},\tau_0}$ \textit{diffusion} for reasons that are explained next. By definition, the anti-cones  $\mathfrak{A}_0^{\overline{\sigma},\tau_0}$ is a thin region specified by the condition that the squared norm of the lower part is bounded by a term of the order $\mathcal{O}(\lambda)$. By the nature of the system, there is no hyperbolic contribution for the increase of this (squared) norm  and thus only the perturbation can cause such an increase by randomness. As mentioned above for the overall statement of the section (Lemma~\ref{lemma-diffusion-conclusion}), the diffusion statement made in the following Lemma~\ref{lemma-diffusion-probability} does not provide a lower bound for the probability of a leave of  $\mathfrak{A}_0^{\overline{\sigma},\tau_0}$, but an upper bound for the intersection of the probability of the complementary event with the overwhelming event.

\begin{lemma}\label{lemma-diffusion-probability}
If $(W_0,v_0)\in\mathfrak{Q}$, then all $N\in\mathbb{N}$ satisfy
\begin{align}\label{ineq-diffusion-probability}
\mathbb{P}\big((W_N,v_N)\in\mathfrak{A}^{\sigma,\tau}_0\,\wedge\,\mathfrak{O}_{0,N}\big)
\leq 1-\left[\frac{2^{\frac{6}{5}}}{3}\beta^{\frac{2}{5}}\pmb{\eta}^{\frac{1}{5}}\vartheta^{\frac{3}{5}}\lambda^{\frac{3}{5}}\Big(1-\exp\big[-2^{-\frac{6}{5}}\beta^{\frac{3}{5}}\pmb{\eta}^{-\frac{1}{5}}\vartheta^{-\frac{3}{5}}\lambda^{\frac{7}{5}}N\big]\Big)-\frac{1}{\sigma}\frac{\lambda}{\tau}\right]\,.
\end{align}
\end{lemma}	

Lemma~\ref{lemma-diffusion-probability} with $N-\lfloor 4\,\lambda^{-1} \rfloor$ instead of $N$ and with  $(\sigma,\tau)=(\overline{\sigma},\tau_0)$ implies the following.

\begin{coro}\label{coro-diffusion-consequence}
	Suppose that $(W_0,v_0) \in \mathfrak{Q}$ and $N\in\mathbb{N}$ fulfills
	\begin{align}\label{ineq-diffusion-N}
	3 \cdot 2^{10}\beta^{-1}\pmb{\eta}^{-1}\mathsf{q}\,\lambda^{-1}\leq N\leq 2^{-\frac{4}{5}}\beta^{-\frac{3}{5}}\pmb{\eta}^{\frac{1}{5}}\vartheta^{\frac{3}{5}}\lambda^{-\frac{7}{5}}\,.
	\end{align}
	Then, one has
	$$
	\mathbb{P}\,\Big((W_{N-\lfloor 4\,\lambda^{-1}\rfloor},v_{N-\lfloor 4\,\lambda^{-1}\rfloor}) \in \mathfrak{A}^{\overline{\sigma},\tau_0}_{0}\,\wedge\,\mathfrak{O}_{0,N-\lfloor 4\,\lambda^{-1}\rfloor}\Big)\leq 1 - 2^{-2}\,N\,\beta\,\lambda^2\,.
	$$
\end{coro}

\vspace{.2cm}

Combining the above allows to prove the desired overall statement of the section. 

\begin{lemma}\label{lemma-diffusion-conclusion}
	Suppose that $(W_0,v_0) \in \mathfrak{Q}$ and $N\in\mathbb{N}$ fulfills~\eqref{ineq-diffusion-N}.
	Then, one has
	\begin{align}
	\mathbb{P}\,\Big((W_N,v_N) \in \mathfrak{A}^{\overline{\sigma},\overline{\tau}}_{24}\,\wedge\,\mathfrak{O}_{0,N}\Big) \leq 1-2^{-\frac{17}{2}}\,N\,\beta\,\lambda^2\,.
	\end{align}
\end{lemma}

\subsubsection{The ascension of the ladder}

We now analyze the ascension of a full step of the ladder. More precisely, we prove an upper bound for the probability that the dynamics lies in $(\mathsf{m}+1)$-th anti-cone at some point in time under the condition that it started even outside of the $\mathsf{m}$-th anti-cone. Again only the probability of the intersection of the described event with the overwhelming event is considered.

\begin{lemma}\label{lemma-ascension}
	Let $\mathsf{m}\in\{0,\dots,\mathsf{M}\}$ and suppose that $N\in\mathbb{N}$ satisfies
	\begin{align}\label{ineq-ascension-assumption}
	N\leq 2^{-\frac{4}{5}}\beta^{-\frac{3}{5}}\pmb{\eta}^{\frac{1}{5}}\vartheta^{\frac{3}{5}}\lambda^{-\frac{7}{5}}\,.
	\end{align}
	Further, suppose that $(W_0,v_0)\in\mathfrak{W}$ satisfies $(W_0,v_{0})\not\in\mathfrak{A}_{\mathsf{m}}$ and $(W_0,v_0)\in\mathfrak{Q}$. Then,
	\begin{align}\label{ineq-ascension-statement}
	\mathbb{P}\Big((W_N,v_N)\in\mathfrak{A}_{\mathsf{m}+1}^{\overline{\sigma},\overline{\tau}}\,\wedge\,\mathfrak{O}_{0,N}\Big)\leq \exp\big[-2^{-6}N^{-1}\lambda^{-2}\big]\,.
	\end{align}
\end{lemma}

\subsubsection{The probability bound}

We summarize the two main intermediate results up to now (Lemmata~\ref{lemma-diffusion-conclusion} and~\ref{lemma-ascension}) by an iterative application of conditional probabilities to control the probability for the presence of the dynamics in the repulsive region $\mathfrak{A}_{\mathsf{M}+1}^{\overline{\sigma},\overline{\tau}}$ in Lemma~\ref{lemma-anticone-probability}. Again, all this is up to the overwhelming event.

\begin{lemma}\label{lemma-anticone-probability}
	If $(W_0,v_0) \in \mathfrak{Q}$ and $N \in \mathbb{N}$ fulfills~\eqref{ineq-diffusion-N}, one has
	$$
	\mathbb{P}\,\Big(\big(W_{N\mathsf{M}},v_{N\mathsf{M}}\big)\in\mathfrak{A}_{\mathsf{M}+1}^{\overline{\sigma},\overline{\tau}}\,\wedge\,\mathfrak{O}_{0,N\mathsf{M}}\Big)\leq 2^4\exp\left[-2^{-12}N\beta\,\pmb{\eta}\,\mathsf{q}^{-1}\lambda\right]\,.
	$$
\end{lemma}

The remaining task is performed in Lemma~\ref{lemma-outside-anticone}: the event to be controlled in the induction outcome is a consequence of the one controlled in Lemma~\ref{lemma-anticone-probability}~\textemdash~up to the overwhelming event. 

\begin{lemma}\label{lemma-outside-anticone} 
	With $S_0 = \lceil 2^{-\frac{9}{2}}\lambda^{-1} \rceil$, one has the deterministic implication
	$$
	(W_{0},v_{0})\in\mathfrak{Q}\setminus\mathfrak{A}_{\mathsf{M}+1}^{\overline{\sigma},\overline{\tau}}\,\wedge\,\mathfrak{O}_{0,S_0}\quad\Rightarrow\quad \operatorname{tr}\big[\hat{\chi}^*(W_{S_0}+v_{S_0}v_{S_0}^*)\hat{\chi}\big]\leq 2^{-\frac{21}{5}}\beta^{\frac{3}{5}}\pmb{\eta}^{-\frac{1}{5}}\vartheta^{-\frac{3}{5}}\lambda^{\frac{7}{5}}\,.
	$$
\end{lemma}

Lemma~\ref{lemma-anticone-probability} and Lemma~\ref{lemma-outside-anticone} finally allow to prove the induction step (Lemma~\ref{lemma-induction-step}).

\vspace{.2cm}

\noindent\textbf{Proof of Lemma~\ref{lemma-induction-step}.} Let $n \in \mathbb{N}$ such that $n \geq 2^{\frac{18}{5}}\beta^{-1}\mathsf{q}(\mathsf{w}+1)\vartheta\lambda^{-2}$. Let us set
\begin{align}\label{def-choice-of-N}
N=\left\lceil 3 \cdot 2^{12}\beta^{-1}\pmb{\eta}^{-1}\mathsf{q}^2\vartheta\lambda^{-1}\right\rceil\,,
\end{align}
which fulfills~\eqref{ineq-diffusion-N}, by Hypothesis~\ref{hyp-q} and the fact that $\lceil x \rceil \leq \frac{101}{100}x \leq \frac{2^{\frac{8}{5}}}{3}x$ for all $x \geq 100$. Now $\mathsf{M} = \lfloor 2^{-10}\,\pmb{\eta}\,\mathsf{q}^{-1}\,\lambda^{-1}\rfloor$ from~\eqref{def-M}, $S_0 = \lceil 2^{-\frac{9}{2}}\lambda^{-1} \rceil$ as introduced in Lemma~\ref{lemma-outside-anticone} and $\lambda < 2^{-12}$ imply
\begin{align}\label{ineq-induction-step-time}
\begin{split}
n - N\mathsf{M} - S_0 &= n - \left\lceil 3 \cdot 2^{12}\beta^{-1}\pmb{\eta}^{-1}\mathsf{q}^2\vartheta\lambda^{-1}\right\rceil\lfloor 2^{-10}\,\pmb{\eta}\,\mathsf{q}^{-1}\,\lambda^{-1}\rfloor - \lceil 2^{-\frac{9}{2}}\lambda^{-1} \rceil\\
&\geq n - \left(3 \cdot 2^{12}\beta^{-1}\pmb{\eta}^{-1}\mathsf{q}^2\vartheta\lambda^{-1}+1\right) \cdot 2^{-10}\,\pmb{\eta}\,\mathsf{q}^{-1}\,\lambda^{-1} - 2^{-\frac{9}{2}}\lambda^{-1} - 1\\
&> n - 3 \cdot 2^2\beta^{-1}\mathsf{q}\,\vartheta\lambda^{-2} - 2^{-10}\,\pmb{\eta}\,\mathsf{q}^{-1}\,\lambda^{-1} - 2^{-\frac{33}{2}}\lambda^{-2} + 3 - 2^{-10}\lambda^{-1}\\
&\geq n - 2^2(3+2^{-10}+2^{-\frac{33}{2}}+2^{-10})\beta^{-1}\mathsf{q}\,\vartheta\lambda^{-2} + 2\\
&\geq n - 2^{\frac{18}{5}}\beta^{-1}\mathsf{q}\,\vartheta\lambda^{-2} + 2\\
&\geq 2^{\frac{18}{5}}\beta^{-1}\mathsf{q}\,\mathsf{w}\,\vartheta\lambda^{-2} + 2\,.
\end{split}
\end{align}
In particular, $n > S_0 + N\mathsf{M}$. The contrapositive of Lemma~\ref{lemma-outside-anticone} with a time shift by $n-S_0$ reads
\begin{align*}
\operatorname{tr}\big[\hat{\chi}^*(W_n+v_nv_n^*)\hat{\chi}\big]> 2^{-\frac{21}{5}}\beta^{\frac{3}{5}}\pmb{\eta}^{-\frac{1}{5}}\vartheta^{-\frac{3}{5}}\lambda^{\frac{7}{5}} \, \Rightarrow \, (W_{n-S_0},v_{n-S_0})\not\in\mathfrak{Q}\setminus\mathfrak{A}_{\mathsf{M}+1}^{\overline{\sigma},\overline{\tau}}\,\vee\,\neg\,\mathfrak{O}_{n-S_0,n}\,.
\end{align*}
Using this and the definition of the overwhelming event, we estimate
\begin{align*}
\begin{split}
&\mathbb{P}\left(\operatorname{tr}\left[\hat{\chi}^*(W_n+v_nv_n^*)\hat{\chi}\right]>2^{-\frac{21}{5}}\beta^{\frac{3}{5}}\pmb{\eta}^{-\frac{1}{5}}\vartheta^{-\frac{3}{5}}\lambda^{\frac{7}{5}}\right)\\
&\leq \mathbb{P}\left((W_{n-S_0},v_{n-S_0})\not\in\mathfrak{Q}\setminus\mathfrak{A}_{\mathsf{M}+1}^{\overline{\sigma},\overline{\tau}}\,\vee\,\neg\,\mathfrak{O}_{n-S_0,n}\right)\\
&= \mathbb{P}\left((W_{n-S_0},v_{n-S_0})\in\mathfrak{A}_{\mathsf{M}+1}^{\overline{\sigma},\overline{\tau}}\,\vee\,(W_{n-S_0},v_{n-S_0})\not\in\mathfrak{Q}\,\vee\bigvee\limits_{n^{\prime}=n-S_0+1}^{n}(W_{n^{\prime}},v_{n^{\prime}})\not\in\mathfrak{Q}\right)\\
&= \mathbb{P}\Bigg(\left[(W_{n-S_0},v_{n-S_0})\in\mathfrak{A}_{\mathsf{M}+1}^{\overline{\sigma},\overline{\tau}}\wedge\bigwedge\limits_{n^{\prime}=n-S_0-N\mathsf{M}+1}^{n-S_0}(W_{n^{\prime}},v_{n^{\prime}})\in\mathfrak{Q}\right]\,\vee\bigvee\limits_{n^{\prime}=n-S_0}^{n}(W_{n^{\prime}},v_{n^{\prime}})\not\in\mathfrak{Q}\\
&\qquad\quad\vee\,\left[(W_{n-S_0},v_{n-S_0})\in\mathfrak{A}_{\mathsf{M}+1}^{\overline{\sigma},\overline{\tau}}\wedge\bigvee\limits_{n^{\prime}=n-S_0-N\mathsf{M}+1}^{n-S_0}(W_{n^{\prime}},v_{n^{\prime}})\not\in\mathfrak{Q}\right]\Bigg)\\
&\leq \mathbb{P}\left((W_{n-S_0},v_{n-S_0})\in\mathfrak{A}_{\mathsf{M}+1}^{\overline{\sigma},\overline{\tau}}\wedge
\bigwedge\limits_{n^{\prime}=n-S_0-N\mathsf{M}+1}^{n-S_0}
(W_{n^{\prime}},v_{n^{\prime}})\in\mathfrak{Q}\right) + \mathbb{P}\left(\bigvee\limits_{n^{\prime}=n-S_0-N\mathsf{M}+1}^{n}(W_{n^{\prime}},v_{n^{\prime}})\not\in\mathfrak{Q}\right)\\
&\leq \mathbb{P}\left((W_{n-S_0},v_{n-S_0})\in\mathfrak{A}_{\mathsf{M}+1}^{\overline{\sigma},\overline{\tau}}\wedge\mathfrak{O}_{n-S_0-N\mathsf{M},n-S_0}\right) + \sum_{n^{\prime}=n-S_0-N\mathsf{M}+1}^{n} \mathbb{P}\left((W_{n^{\prime}},v_{n^{\prime}})\not\in\mathfrak{Q}\right)\\
&\leq \mathbb{P}\left(\left[(W_{n-S_0},v_{n-S_0})\in\mathfrak{A}_{\mathsf{M}+1}^{\overline{\sigma},\overline{\tau}}\wedge\mathfrak{O}_{n-S_0-N\mathsf{M},n-S_0}\right]\wedge(W_{n-S_0-N\mathsf{M}},v_{n-S_0-N\mathsf{M}})\in\mathfrak{Q}\right)\\
&\qquad + \sum_{n^{\prime}=n-S_0-N\mathsf{M}}^{n} \mathbb{P}\left((W_{n^{\prime}},v_{n^{\prime}})\not\in\mathfrak{Q}\right)\\
&\leq 2^4\exp\left[-2^{-12}N\beta\,\pmb{\eta}\,\mathsf{q}^{-1}\lambda\right]\mathbb{P}\left((W_{n-S_0-N\mathsf{M}},v_{n-S_0-N\mathsf{M}})\in\mathfrak{Q}\right) + \sum_{n^{\prime}=n-S_0-N\mathsf{M}}^{n} \mathbb{P}\left((W_{n^{\prime}},v_{n^{\prime}})\not\in\mathfrak{Q}\right)\\
\end{split}
\end{align*}
in which we combined Lemma~\ref{lemma-anticone-probability} and the following general fact in the ultimate step:
if we have shown that $\mathfrak{N} \Rightarrow \mathbb{P}(\mathfrak{M}) \leq m$ for some events $\mathfrak{M}$ and $\mathfrak{N}$ with $m \in [0,1]$, then
$$
\mathbb{P}(\mathfrak{M} \wedge \mathfrak{N}) \leq m \,\mathbb{P}(\mathfrak{N})\,.
$$
Indeed, this estimate holds trivially when $\mathbb{P}(\mathfrak{N}) = 0$, and when $\mathbb{P}(\mathfrak{N}) > 0$ it is proved by
$$\mathbb{P}(\mathfrak{M} \wedge \mathfrak{N}) = \mathbb{P}(\mathfrak{M} \, | \, \mathfrak{N})\mathbb{P}(\mathfrak{N}) \leq m \,\mathbb{P}(\mathfrak{N})\,.$$
Inserting~\eqref{def-choice-of-N}, one deduces
\begin{align}\label{ineq-induction-step-1}
\begin{split}
&\mathbb{P}\left(\operatorname{tr}\left[\hat{\chi}^*(W_n+v_nv_n^*)\hat{\chi}\right]>2^{-\frac{21}{5}}\beta^{\frac{3}{5}}\pmb{\eta}^{-\frac{1}{5}}\vartheta^{-\frac{3}{5}}\lambda^{\frac{7}{5}}\right)\\
&\leq 2^4(2^{\frac{54}{5}}\lambda)^{3\mathsf{q}} + \sum_{n^{\prime}=n-S_0-N\mathsf{M}}^{n} \mathbb{P}\left(\operatorname{tr}\left[\hat{\chi}^*W_{n^{\prime}}\hat{\chi}\right]>2^{-\frac{21}{5}}\beta^{\frac{3}{5}}\pmb{\eta}^{-\frac{1}{5}}\vartheta^{-\frac{3}{5}}\lambda^{\frac{7}{5}}\right)\,.
\end{split}
\end{align}
If $\mathsf{w} = 0$, all probabilities in the final sum vanish, and therefore the statement~\eqref{ineq-induction-outcome} is proved. For the case that $\mathsf{w} \geq 1$,~\eqref{ineq-induction-step-time} allows us to insert the assumption~\eqref{ineq-induction-hypothesis} and to estimate
\begin{align}\label{ineq-induction-step-2}
\begin{split}
2^4(2^{\frac{54}{5}}\lambda)^{3\mathsf{q}} + \sum_{n^{\prime}=n-S_0-N\mathsf{M}}^{n} &\mathbb{P}\left(\operatorname{tr}\left[\hat{\chi}^*W_{n^{\prime}}\hat{\chi}\right]>2^{-\frac{21}{5}}\beta^{\frac{3}{5}}\pmb{\eta}^{-\frac{1}{5}}\vartheta^{-\frac{3}{5}}\lambda^{\frac{7}{5}}\right)\\
&\leq 2^4(2^{\frac{54}{5}}\lambda)^{3\mathsf{q}} + 2^4(S_0 + N\mathsf{M} + 1)(2^{\frac{54}{5}}\lambda)^{3(\mathsf{q}-\mathsf{w}+1)}\\
&\leq 2^4(S_0 + N\mathsf{M} + 2)(2^{\frac{54}{5}}\lambda)^{3(\mathsf{q}-\mathsf{w}+1)}\\
&\leq 2^4(n - 2^{\frac{18}{5}}\beta^{-1}\mathsf{q}\,\mathsf{w}\,\vartheta\lambda^{-2})(2^{\frac{54}{5}}\lambda)^{3(\mathsf{q}-\mathsf{w}+1)}\\
&\leq 2^{\frac{38}{5}}\beta^{-1}\mathsf{q}\,\vartheta\lambda^{-2}(\mathsf{q}^{-1}\mathsf{q})^5(2^{\frac{54}{5}}\lambda)^{3(\mathsf{q}-\mathsf{w}+1)}\pmb{\eta}^{-3}\mathsf{q}^4\\
&\leq 2^{\frac{38}{5}}\beta^{-1}\pmb{\eta}^{-3}\mathsf{q}^5\vartheta\lambda^{-2}(\mathsf{q}^{-1}2^{-\frac{36}{5}}\beta^{\frac{1}{5}}\pmb{\eta}^{\frac{3}{5}}\vartheta^{-\frac{1}{5}}\lambda^{-\frac{1}{5}})^5(2^{\frac{54}{5}}\lambda)^{3(\mathsf{q}-\mathsf{w}+1)}\\
&= 2^4(2^{\frac{54}{5}}\lambda)^{3(\mathsf{q}-\mathsf{w})}\,,
\end{split}
\end{align}
where we used~\eqref{ineq-induction-step-time} in the third and Hypothesis~\ref{hyp-q} in the fifth step. Combining~\eqref{ineq-induction-step-1} and~\eqref{ineq-induction-step-2} yields~\eqref{ineq-induction-outcome}.
\hfill $\square$

\section{Proofs}
\label{sec-Proofs}

\noindent\textbf{Proof of Lemma~\ref{lemma-expansion}.} Let $\mathtt{P} \in \mathfrak{P}$, $Q\in\mathbb{G}_{\mathsf{L},\mathsf{w}}$ and $\Phi\in\mathbb{F}_{\mathsf{L},\mathsf{w}}$ such that $Q=\Phi\Phi^*$.
Let us formulate three auxiliary inequalities:
\begin{align}
\|\mathtt{A}-\mathtt{B}\|&\leq \max\left\lbrace\|\mathtt{A}\|,\|\mathtt{B}\|\right\rbrace\,,\qquad\qquad \forall\,\mathtt{A},\mathtt{B}\in\mathbb{C}^{\mathsf{L}\times\mathsf{L}}\,,\quad \mathtt{A},\mathtt{B}\geq 0\,,\label{ineq-norm-1}\\
\|Q\mathtt{C}Q+Q^{\perp}\mathtt{D}Q^{\perp}\|&\leq\max\left\lbrace\|\mathtt{C}\|,\|\mathtt{D}\|\right\rbrace\,,\qquad\qquad \forall\,\mathtt{C},\mathtt{D}\in\mathbb{C}^{\mathsf{L}\times\mathsf{L}}\,,\label{ineq-norm-2}\\
\|Q^{\perp}\mathtt{E}Q+Q\mathtt{F}Q^{\perp}\|&\leq\max\left\lbrace\|\mathtt{E}\|,\|\mathtt{F}\|\right\rbrace\,,\qquad\qquad \forall\,\mathtt{E},\mathtt{F}\in\mathbb{C}^{\mathsf{L}\times\mathsf{L}}\,.\label{ineq-norm-3}
\end{align}
The estimate $-\|\mathtt{B}\|\leq -\mathtt{B}\leq \mathtt{A}-\mathtt{B}\leq\mathtt{A}\leq \|\mathtt{A}\|$ implies~\eqref{ineq-norm-1}. Further, inequality~\eqref{ineq-norm-2} follows from
\begin{align}\label{ineq-norm-2-proof}
\begin{split}
\|Q\mathtt{C}Q+Q^{\perp}\mathtt{D}Q^{\perp}\|&=\sup\limits_{v\in\mathbb{S}_{\mathbb{C}}^{\mathsf{L}-1}}\|(Q\mathtt{C}Q+Q^{\perp}\mathtt{D}Q^{\perp})v\|\\
&= \sup\limits_{v\in\mathbb{S}_{\mathbb{C}}^{\mathsf{L}-1}}\left[\|Q\mathtt{C}Qv\|^2+\|Q^{\perp}\mathtt{D}Q^{\perp}v\|^2\right]^{\frac{1}{2}}\\
&= \sup\limits_{\alpha\in [0,\frac{\pi}{2}]}\sup\limits_{v_1,v_2\in\mathbb{S}_{\mathbb{C}}^{\mathsf{L}-1}\atop Qv_2=Q^{\perp}v_1=\mathbf{0}}\left[\cos(\alpha)^2\,\|Q\mathtt{C}Qv_1\|^2+\sin(\alpha)^2\,\|Q^{\perp}\mathtt{D}Q^{\perp}v_2\|^2\right]^{\frac{1}{2}}\\
&= \max\bigg\{\sup\limits_{v_1\in\mathbb{S}_{\mathbb{C}}^{\mathsf{L}-1}\atop Q^{\perp}v_1=\mathbf{0}}\|Q\mathtt{C}Qv_1\|,\sup\limits_{v_2\in\mathbb{S}_{\mathbb{C}}^{\mathsf{L}-1}\atop Qv_2=\mathbf{0}}\|Q^{\perp}\mathtt{D}Q^{\perp}v_2\|\bigg\}\\
&=\max\{\|Q\mathtt{C}Q\|,\|Q^{\perp}\mathtt{D}Q^{\perp}\|\}\,.
\end{split}
\end{align}
To show inequality~\eqref{ineq-norm-3}, we apply the identity~\eqref{ineq-norm-2-proof} for $\mathtt{C}=\mathtt{E}^*Q^{\perp}\mathtt{E}$ and $\mathtt{D}=\mathtt{F}^*Q\mathtt{F}$ and obtain 
\begin{align*}
\|Q^{\perp}\mathtt{E}Q+Q\mathtt{F}Q^{\perp}\|&=\|(Q^{\perp}\mathtt{E}Q+Q\mathtt{F}Q^{\perp})^*(Q^{\perp}\mathtt{E}Q+Q\mathtt{F}Q^{\perp})\|^{\frac{1}{2}}\\
&=\|Q\mathtt{E}^*Q^{\perp}\mathtt{E}Q+Q^{\perp}\mathtt{F}^*Q\mathtt{F}Q^{\perp}\|^{\frac{1}{2}}\\
&=\max\left\{\|Q\mathtt{E}^*Q^{\perp}\mathtt{E}Q\|^{\frac{1}{2}},\|Q^{\perp}\mathtt{F}^*Q\mathtt{F}Q^{\perp}\|^{\frac{1}{2}}\right\}\,.
\end{align*}
Now the bound for $\|\mathtt{X}(Q,\mathtt{P})\|$ follows directly from~\eqref{ineq-norm-3}. The bound for $\|\mathtt{Y}(Q,\mathtt{P})\|$ is obtained by an application of~\eqref{ineq-norm-1} to the first two terms, as well as~\eqref{ineq-norm-1} and~\eqref{ineq-norm-3} to the remainder.
As for $\|\mathtt{Z}^{(\lambda)}(Q,\mathtt{P})\|$, we define the maps $\mathtt{U}$ and $\mathtt{V}$ by
\begin{align*}
\mathtt{U}:\mathfrak{P}\rightarrow\mathbb{C}^{\mathsf{L}\times\mathsf{L}}&:\mathtt{P}\mapsto\mathtt{P}+\mathtt{P}^*\,,\\
\mathtt{V}: \mathbb{G}_{\mathsf{L}}\times\mathfrak{P}\rightarrow\mathbb{C}^{\mathsf{L}\times\mathsf{L}}&:(Q,\mathtt{P})\mapsto\mathtt{P}^*Q^{\perp}\mathtt{P}-\mathtt{P}Q\mathtt{P}^*+\mbox{\small $\frac{1}{2}$}\left[\mathtt{P}\left(Q^{\perp}-Q\right)\mathtt{P}+\mathtt{P}^*\left(Q^{\perp}-Q\right)\mathtt{P}^*\right]\,.
\end{align*}
Then we have for all $\mathtt{P}\in\mathfrak{P}$ and $Q\in\mathbb{G}_{\mathsf{L}}$ that $\|\mathtt{U}(\mathtt{P})\|\leq 2$ and (by applying~\eqref{ineq-norm-1} three times) $\|\mathtt{V}(Q,\mathtt{P})\|\leq 2$. Clearly, the matrix $\mathtt{T}^{(\lambda)}(\mathtt{P})$, characterized by the equation
\begin{align}\label{def-exponential}
e^{\lambda\mathtt{P}}=\mathbf{1}+\lambda\mathtt{P}+\mbox{\small $\frac{1}{2}$}\lambda^2\mathtt{P}^2+\lambda^3\mathtt{T}^{(\lambda)}(\mathtt{P})\,,
\end{align}
is uniformly bounded in $\lambda$ and, more precisely, satisfies $\|\mathtt{T}^{(\lambda)}(\mathtt{P})\|\leq \frac{1}{5}$. Combining this fact with~\eqref{def-exponential} allows to verify that also $\mathtt{S}^{(\lambda)}(\Phi,\mathtt{P})$, characterized~by
\begin{align}\label{def-S}
\Phi^*(e^{\lambda\mathtt{P}})^*e^{\lambda\mathtt{P}}\Phi&=\mathbf{1}+\Phi^*\big[\lambda \mathtt{U}(\mathtt{P})+\lambda^2\mathtt{V}(\mathbf{0},\mathtt{P})\big]\Phi+\lambda^3\mathtt{S}^{(\lambda)}(\Phi,\mathtt{P})\,,
\end{align}
is uniformly bounded in $\lambda$ and $\lambda^3\|\mathtt{S}^{(\lambda)}(\Phi,\mathtt{P})\|$ is bounded from above by the sum of all terms of
\begin{align*}
\left[1+\lambda\|\mathtt{P}\|+\mbox{\small $\frac{1}{2}$}\lambda^2\|\mathtt{P}^2\|+\lambda^3\|\mathtt{T}^{(\lambda)}(\mathtt{P})\|\right]^2\leq\left[1+\lambda+\mbox{\small $\frac{1}{2}$}\,\lambda^2+\mbox{\small $\frac{1}{5}$}\,\lambda^3\right]^2
\end{align*}
that are beyond the second order in $\lambda$, \textit{i.e.}, one has
\begin{align}\label{ineq-S}
\|\mathtt{S}^{(\lambda)}(\Phi,\mathtt{P})\|\leq\lambda^{-3}\left[\mbox{\small $\frac{7}{5}$}\,\lambda^3+\mbox{\small $\frac{13}{20}$}\,\lambda^4+\mbox{\small $\frac{1}{5}$}\,\lambda^5+\mbox{\small $\frac{1}{25}$}\,\lambda^6\right]\leq \mbox{\small $\frac{3}{2}$}\,.
\end{align}
Next, we verify for all $\mathtt{G},\mathtt{H},\mathtt{I}\in\mathbb{C}^{\mathsf{L}\times\mathsf{L}}$ such that $\mathbf{1}+\lambda\mathtt{G}+\lambda^2\mathtt{H}+\lambda^3\mathtt{I}$ is invertible the identity
\begin{align}\label{eq-inverse}
\left[\mathbf{1}+\lambda\mathtt{G}+\lambda^2\mathtt{H}+\lambda^3\mathtt{I}\right]^{-1}=\mathbf{1}-\lambda\mathtt{G}-\lambda^2\left[\mathtt{H}-\mathtt{G}^2\right]+\lambda^3\mathtt{R}\,,
\end{align}
in which the norms of these matrices are uniformly bounded in $\lambda$, and 
\begin{align*}
\mathtt{R}=\left[\mathbf{1}+\lambda\mathtt{G}+\lambda^2\mathtt{H}+\lambda^3\mathtt{I}\right]^{-1}\left(\left[\mathtt{G}+\lambda\mathtt{H}+\lambda^2\mathtt{I}\right]\left[\mathtt{H}-\mathtt{G}^2\right] + \left[\mathtt{H}+\lambda\mathtt{I}\right]\mathtt{G} - \mathtt{I}\right)\,.
\end{align*}
Now setting $\mathtt{G}=\Phi^*\mathtt{U}(\mathtt{P})\Phi$, $\mathtt{H}=\Phi^*\mathtt{V}(\mathbf{0},\mathtt{P})\Phi$, $\mathtt{I}=\mathtt{S}^{(\lambda)}(\Phi,\mathtt{P})$ allows us to apply~\eqref{eq-inverse}
to the matrix inverse of~\eqref{def-S} and then use the identity $$\mathtt{H}-\mathtt{G}^2=\Phi^*\left[\mathtt{V}(\mathbf{0},\mathtt{P})-\mathtt{U}(\mathtt{P})Q\mathtt{U}(\mathtt{P})\right]\Phi=\Phi^*\mathtt{V}(Q,\mathtt{P})\Phi$$
to compute the r.h.s. of~\eqref{eq-inverse}. This proves that $\mathtt{R}^{(\lambda)}(\Phi,\mathtt{P})$, characterized by
\begin{align}\label{def-inverse-implicit}
\left[\Phi^*(e^{\lambda\mathtt{P}})^*e^{\lambda\mathtt{P}}\Phi\right]^{-1}&=\mathbf{1}-\Phi^*\big[\lambda\mathtt{U}(\mathtt{P})+\lambda^2\mathtt{V}(Q,\mathtt{P})\big]\Phi+\lambda^3\,\mathtt{R}^{(\lambda)}(\Phi,\mathtt{P})\,,
\end{align}
is uniformly bounded in $\lambda$ and is explicitly given by
\begin{align}\label{def-inverse-explicit}
\begin{split}
\mathtt{R}^{(\lambda)}(\Phi,\mathtt{P})&=\left[\Phi^*(e^{\lambda\mathtt{P}})^*e^{\lambda\mathtt{P}}\Phi\right]^{-1}\Big(\lambda^{-1}\left[\Phi^*(e^{\lambda\mathtt{P}})^*e^{\lambda\mathtt{P}}\Phi-\mathbf{1}\right]\Phi^*\mathtt{V}(Q,\mathtt{P})\Phi\\
&\qquad\qquad\qquad\qquad\qquad+\left[\Phi^*\mathtt{V}(\mathbf{0},\mathtt{P})\Phi+\lambda\,\mathtt{S}^{(\lambda)}(\Phi,\mathtt{P})\,\right]\Phi^*\mathtt{U}(\mathtt{P})\Phi-\mathtt{S}^{(\lambda)}(\Phi,\mathtt{P})\Big)\,.
\end{split}
\end{align}
Due to~\eqref{def-inverse-explicit}, the bounds $\|\mathtt{U}(\mathtt{P})\|\leq 2$, $\|\mathtt{V}(Q,\mathtt{P})\|\leq 2$,~\eqref{ineq-S} and $(e^{\lambda\mathtt{P}})^*(e^{\lambda\mathtt{P}}) \geq e^{-2\lambda}\mathbf{1}$ then yield
\begin{align}\label{ineq-inverse}
\|\mathtt{R}^{(\lambda)}(\Phi,\mathtt{P})\|\leq e^{2\lambda}\left(2\lambda^{-1}\left[e^{2\lambda}-1\right] + 2\left[2+\mbox{\small $\frac{3}{2}$}\,\lambda\right]+\mbox{\small $\frac{3}{2}$}\right)\leq 10\,.
\end{align}
Using~\eqref{def-Z},~\eqref{def-exponential},~\eqref{def-inverse-implicit} and~\eqref{ineq-inverse}, it follows that $\lambda^3\,\|\mathtt{Z}^{(\lambda)}(Q,\mathtt{P})\|$ is bounded by all terms of 
\begin{align*}
\begin{split}
&\big[1+\lambda\|\mathtt{P}\|+\mbox{\small $\frac{1}{2}$}\lambda^2\|\mathtt{P}^2\|+\lambda^3\|\mathtt{T}^{(\lambda)}(\mathtt{P})\|\big]^2\big[1+\lambda\|\mathtt{U}(\mathtt{P})\|+\lambda^2\|\mathtt{V}(Q,\mathtt{P})\|+\lambda^3\|\mathtt{R}^{(\lambda)}(\Phi,\mathtt{P})\|\big]\\
&\leq\big[1+\lambda+\mbox{\small $\frac{1}{2}$}\,\lambda^2+\mbox{\small $\frac{1}{5}$}\,\lambda^3\big]^2\left[1+2\,\lambda+2\,\lambda^2+10\,\lambda^3\right]
\end{split}
\end{align*}
that are beyond the second order in $\lambda$, which then proves the last norm bound of~\eqref{ineq-expansion}:
$$
\|\mathtt{Z}^{(\lambda)}(Q,\mathtt{P})\|\leq\lambda^{-3}\mbox{\small $\left[\frac{97}{5}\,\lambda^3+\frac{549}{20}\,\lambda^4+\frac{243}{10}\,\lambda^5+\frac{787}{50}\,\lambda^6+\frac{349}{50}\,\lambda^7+\frac{52}{25}\,\lambda^8+\frac{2}{5}\,\lambda^9\right]$}\leq 20\,.
$$
To obtain the bounds on the ranks indicated in~\eqref{ineq-expansion}, we use the facts that the rank of a sum of matrices is bounded by the sum of the ranks of the summands and the rank of a product of matrices is smaller than the minimum of the ranks of the factors. The reader can verify that 
\begin{align}\label{eq-order-differences}
\begin{split}
\mathtt{X}(Q +Q^{\prime},\mathtt{P})-\mathtt{X}(Q ,\mathtt{P})&=Q^{\prime}\mathtt{K}(Q ,Q^{\prime},\mathtt{P})+\mathtt{K}(Q ,Q^{\prime},\mathtt{P})^*Q^{\prime}\,,\\
\mathtt{Y}(Q +Q^{\prime},\mathtt{P})-\mathtt{Y}(Q ,\mathtt{P})&=Q^{\prime}\mathtt{L}(Q ,Q^{\prime},\mathtt{P})+\mathtt{L}(Q ,Q^{\prime},\mathtt{P})^*Q^{\prime}+(Q ^{\perp}\mathtt{P}-Q \mathtt{P}^*)Q^{\prime}(\mathtt{P}^*Q ^{\perp}-\mathtt{P}Q )\,,
\end{split}
\end{align}
for some matrices $\mathtt{K}(Q ,Q^{\prime},\mathtt{P})$ and $\mathtt{L}(Q ,Q^{\prime},\mathtt{P})$. This implies $\operatorname{rk}[\mathtt{X}(Q +Q^{\prime},\mathtt{P})-\mathtt{X}(Q ,\mathtt{P})]\leq 2\,\operatorname{rk}(Q^{\prime})$ and
$\operatorname{rk}[\mathtt{Y}(Q +Q^{\prime},\mathtt{P})-\mathtt{Y}(Q ,\mathtt{P})]\leq 3\,\operatorname{rk}(Q^{\prime})$. As for the last inequality of~\eqref{ineq-expansion}, we first note that $e^{\lambda\mathtt{P}}\cdot Q  \leq e^{\lambda\mathtt{P}}\cdot(Q +Q^{\prime})$, hence $\operatorname{rk}[e^{\lambda\mathtt{P}}\cdot(Q +Q^{\prime})-e^{\lambda\mathtt{P}}\cdot Q ]=\operatorname{rk}(Q^{\prime})$. Recall that multiplication by a nonzero scalar does not change the rank of a matrix. This, together with~\eqref{def-Z} and~\eqref{eq-order-differences} proves
\begin{align*}
&\operatorname{rk}\left[\mathtt{Z}^{(\lambda)}(Q +Q^{\prime},\mathtt{P})-\mathtt{Z}^{(\lambda)}(Q ,\mathtt{P})\right]\\
&\leq\operatorname{rk}\big[e^{\lambda\mathtt{P}}\cdot(Q +Q^{\prime})-e^{\lambda\mathtt{P}}\cdot Q \big]\\
&\quad+\operatorname{rk}\big[Q +Q^{\prime}-Q +\lambda\,\mathtt{X}(Q +Q^{\prime},\mathtt{P})-\lambda\,\mathtt{X}(Q ,\mathtt{P})+\lambda^2\,\mathtt{Y}(Q +Q^{\prime},\mathtt{P})-\lambda^2\,\mathtt{Y}(Q ,\mathtt{P})\big]\\
&=\operatorname{rk}(Q^{\prime})+\operatorname{rk}\Big(Q^{\prime}\big[\mathbf{1}+\lambda\,\mathtt{K}(Q ,Q^{\prime},\mathtt{P})+\lambda^2\,\mathtt{L}(Q ,Q^{\prime},\mathtt{P})\big]+\lambda^2\,(Q ^{\perp}\mathtt{P}-Q \mathtt{P}^*)Q^{\prime}(\mathtt{P}^*Q ^{\perp}-\mathtt{P}Q )\\
&\qquad\qquad\qquad\quad+\big[\lambda\,\mathtt{K}(Q ,Q^{\prime},\mathtt{P})^*+\lambda^2\,\mathtt{L}(Q ,Q^{\prime},\mathtt{P})^*\big]Q^{\prime}\Big)\\
&\leq 4\,\operatorname{rk}(Q^{\prime})\,,
\end{align*}
which concludes the proof.
\hfill $\square$

\vspace{.2cm}

\noindent\textbf{Proof of Lemma~\ref{lemma-norm-contraction}.} Let $\Phi\in\mathbb{F}_{\mathsf{L},\mathsf{q}}$ such that $Q=\Phi\Phi^*$. If $\|\hat{\alpha}^*\,Q\,\hat{\alpha}\|=1$,~\eqref{ineq-norm-contraction} is trivially satisfied, as its r.h.s. equals $1$. Hence, we may assume that $\|\Phi^*\hat{\alpha}\hat{\alpha}^*\Phi\|=\|\hat{\alpha}^*\,Q\,\hat{\alpha}\|<1$. This implies that $\Phi^*\hat{\alpha}^{\perp}(\hat{\alpha}^{\perp})^*\Phi=\mathbf{1}_{\mathsf{q}}-\Phi^*\hat{\alpha}\hat{\alpha}^*\Phi$ is invertible. 
Under this condition, one~can~estimate
\begin{align*}
\left\|\hat{\alpha}^*(\mathcal{R}\cdot Q)\hat{\alpha}\right\|&=\left\|\hat{\alpha}^*\mathcal{R}\Phi[\Phi^*\mathcal{R}^2\Phi]^{-1}\Phi^*\mathcal{R}\hat{\alpha}\right\|\\
&=\left\|\hat{\alpha}^*\mathcal{R}\Phi\left[\Phi^*\left(\mathcal{R}\hat{\alpha}\hat{\alpha}^*\mathcal{R}+\mathcal{R}\hat{\alpha}^{\perp}(\hat{\alpha}^{\perp})^*\mathcal{R}\right)\Phi\right]^{-1}\Phi^*\mathcal{R}\hat{\alpha}\right\|\\
&\leq \left\|\hat{\alpha}^*\mathcal{R}\Phi\left[\Phi^*\left(\mathcal{R}\hat{\alpha}\hat{\alpha}^*\mathcal{R}+\kappa_{\mathsf{L}_{\mgl}+\mathsf{L}_{\lgl}}^2\hat{\alpha}^{\perp}(\hat{\alpha}^{\perp})^*\right)\Phi\right]^{-1}\Phi^*\mathcal{R}\hat{\alpha}\right\|\\
&=\bigg\|\left[\Phi^*\left(\mathcal{R}\hat{\alpha}\hat{\alpha}^*\mathcal{R}+\kappa_{\mathsf{L}_{\mgl}+\mathsf{L}_{\lgl}}^2\hat{\alpha}^{\perp}(\hat{\alpha}^{\perp})^*\right)\Phi\right]^{-\frac{1}{2}}\Phi^*\mathcal{R}\hat{\alpha}\hat{\alpha}^*\mathcal{R}\Phi\times\\
&\qquad\qquad\qquad\qquad\qquad\times\left[\Phi^*\left(\mathcal{R}\hat{\alpha}\hat{\alpha}^*\mathcal{R}+\kappa_{\mathsf{L}_{\mgl}+\mathsf{L}_{\lgl}}^2\hat{\alpha}^{\perp}(\hat{\alpha}^{\perp})^*\right)\Phi\right]^{-\frac{1}{2}}\bigg\|\\
&=\bigg\|\mathbf{1}_{\mathsf{q}}-\kappa_{\mathsf{L}_{\mgl}+\mathsf{L}_{\lgl}}^2\left[\Phi^*\left(\mathcal{R}\hat{\alpha}\hat{\alpha}^*\mathcal{R}+\kappa_{\mathsf{L}_{\mgl}+\mathsf{L}_{\lgl}}^2\hat{\alpha}^{\perp}(\hat{\alpha}^{\perp})^*\right)\Phi\right]^{-\frac{1}{2}}\times\\
&\qquad\times\Phi^*\hat{\alpha}^{\perp}(\hat{\alpha}^{\perp})^*\Phi\left[\Phi^*\left(\mathcal{R}\hat{\alpha}\hat{\alpha}^*\mathcal{R}+\kappa_{\mathsf{L}_{\mgl}+\mathsf{L}_{\lgl}}^2\hat{\alpha}^{\perp}(\hat{\alpha}^{\perp})^*\right)\Phi\right]^{-\frac{1}{2}}\bigg\|\\
&=1-\kappa_{\mathsf{L}_{\mgl}+\mathsf{L}_{\lgl}}^2\bigg\|\bigg(\left[\Phi^*\left(\mathcal{R}\hat{\alpha}\hat{\alpha}^*\mathcal{R}+\kappa_{\mathsf{L}_{\mgl}+\mathsf{L}_{\lgl}}^2\hat{\alpha}^{\perp}(\hat{\alpha}^{\perp})^*\right)\Phi\right]^{-\frac{1}{2}}\times\\
&\qquad\times\,\Phi^*\hat{\alpha}^{\perp}(\hat{\alpha}^{\perp})^*\Phi\,\left[\Phi^*\left(\mathcal{R}\hat{\alpha}\hat{\alpha}^*\mathcal{R}+\kappa_{\mathsf{L}_{\mgl}+\mathsf{L}_{\lgl}}^2\hat{\alpha}^{\perp}(\hat{\alpha}^{\perp})^*\right)\Phi\right]^{-\frac{1}{2}}\bigg)^{-1}\bigg\|^{-1}\\
&=1-\kappa_{\mathsf{L}_{\mgl}+\mathsf{L}_{\lgl}}^2\bigg\|\left[\Phi^*\left(\mathcal{R}\hat{\alpha}\hat{\alpha}^*\mathcal{R}+\kappa_{\mathsf{L}_{\mgl}+\mathsf{L}_{\lgl}}^2\hat{\alpha}^{\perp}(\hat{\alpha}^{\perp})^*\right)\Phi\right]^{\frac{1}{2}}\times\\
&\qquad\times\left[\Phi^*\hat{\alpha}^{\perp}(\hat{\alpha}^{\perp})^*\Phi\right]^{-1}\left[\Phi^*\left(\mathcal{R}\hat{\alpha}\hat{\alpha}^*\mathcal{R}+\kappa_{\mathsf{L}_{\mgl}+\mathsf{L}_{\lgl}}^2\hat{\alpha}^{\perp}(\hat{\alpha}^{\perp})^*\right)\Phi\right]^{\frac{1}{2}}\bigg\|^{-1}\\
&=1-\kappa_{\mathsf{L}_{\mgl}+\mathsf{L}_{\lgl}}^2\bigg\|\left[\Phi^*\hat{\alpha}^{\perp}(\hat{\alpha}^{\perp})^*\Phi\right]^{-\frac{1}{2}}\times\\
&\qquad\times\left[\Phi^*\left(\mathcal{R}\hat{\alpha}\hat{\alpha}^*\mathcal{R}+\kappa_{\mathsf{L}_{\mgl}+\mathsf{L}_{\lgl}}^2\hat{\alpha}^{\perp}(\hat{\alpha}^{\perp})^*\right)\Phi\right]\left[\Phi^*\hat{\alpha}^{\perp}(\hat{\alpha}^{\perp})^*\Phi\right]^{-\frac{1}{2}}\bigg\|^{-1}\\
&\leq 1-\kappa_{\mathsf{L}_{\mgl}+\mathsf{L}_{\lgl}}^2\bigg\|\left[\Phi^*\hat{\alpha}^{\perp}(\hat{\alpha}^{\perp})^*\Phi\right]^{-\frac{1}{2}}\times\\
&\qquad\times\left[\Phi^*\left(\kappa_{\mathsf{L}_{\mgl}+\mathsf{L}_{\lgl}+1}^2\hat{\alpha}\hat{\alpha}^*+\kappa_{\mathsf{L}_{\mgl}+\mathsf{L}_{\lgl}}^2\hat{\alpha}^{\perp}(\hat{\alpha}^{\perp})^*\right)\Phi\right]\left[\Phi^*\hat{\alpha}^{\perp}(\hat{\alpha}^{\perp})^*\Phi\right]^{-\frac{1}{2}}\bigg\|^{-1}\\
&= 1-\bigg\|\left[\Phi^*\hat{\alpha}^{\perp}(\hat{\alpha}^{\perp})^*\Phi\right]^{-\frac{1}{2}}\times\\
&\qquad\times\left[\Phi^*\left(\kappa_{\mathsf{L}_{\mgl}+\mathsf{L}_{\lgl}+1}^2\kappa_{\mathsf{L}_{\mgl}+\mathsf{L}_{\lgl}}^{-2}\hat{\alpha}\hat{\alpha}^*+\hat{\alpha}^{\perp}(\hat{\alpha}^{\perp})^*\right)\Phi\right]\left[\Phi^*\hat{\alpha}^{\perp}(\hat{\alpha}^{\perp})^*\Phi\right]^{-\frac{1}{2}}\bigg\|^{-1}\\
&= 1-\bigg\|\left[\Phi^*\left(\kappa_{\mathsf{L}_{\mgl}+\mathsf{L}_{\lgl}+1}^2\kappa_{\mathsf{L}_{\mgl}+\mathsf{L}_{\lgl}}^{-2}\hat{\alpha}\hat{\alpha}^*+\hat{\alpha}^{\perp}(\hat{\alpha}^{\perp})^*\right)\Phi\right]^{\frac{1}{2}}\left[\Phi^*\hat{\alpha}^{\perp}(\hat{\alpha}^{\perp})^*\Phi\right]^{-1}\times\\
&\qquad\times\left[\Phi^*\left(\kappa_{\mathsf{L}_{\mgl}+\mathsf{L}_{\lgl}+1}^2\kappa_{\mathsf{L}_{\mgl}+\mathsf{L}_{\lgl}}^{-2}\hat{\alpha}\hat{\alpha}^*+\hat{\alpha}^{\perp}(\hat{\alpha}^{\perp})^*\right)\Phi\right]^{\frac{1}{2}}\bigg\|^{-1}\\
&= 1-\bigg\|\bigg(\left[\Phi^*\left(\kappa_{\mathsf{L}_{\mgl}+\mathsf{L}_{\lgl}+1}^2\kappa_{\mathsf{L}_{\mgl}+\mathsf{L}_{\lgl}}^{-2}\hat{\alpha}\hat{\alpha}^*+\hat{\alpha}^{\perp}(\hat{\alpha}^{\perp})^*\right)\Phi\right]^{-\frac{1}{2}}\Phi^*\hat{\alpha}^{\perp}(\hat{\alpha}^{\perp})^*\Phi\times\\
&\qquad\times\left[\Phi^*\left(\kappa_{\mathsf{L}_{\mgl}+\mathsf{L}_{\lgl}+1}^2\kappa_{\mathsf{L}_{\mgl}+\mathsf{L}_{\lgl}}^{-2}\hat{\alpha}\hat{\alpha}^*+\hat{\alpha}^{\perp}(\hat{\alpha}^{\perp})^*\right)\Phi\right]^{-\frac{1}{2}}\bigg)^{-1}\bigg\|^{-1}\\
&= \bigg\|\mathbf{1}_{\mathsf{q}}-\left[\Phi^*\left(\kappa_{\mathsf{L}_{\mgl}+\mathsf{L}_{\lgl}+1}^2\kappa_{\mathsf{L}_{\mgl}+\mathsf{L}_{\lgl}}^{-2}\hat{\alpha}\hat{\alpha}^*+\hat{\alpha}^{\perp}(\hat{\alpha}^{\perp})^*\right)\Phi\right]^{-\frac{1}{2}}\Phi^*\hat{\alpha}^{\perp}(\hat{\alpha}^{\perp})^*\Phi\times\\
&\qquad\times\left[\Phi^*\left(\kappa_{\mathsf{L}_{\mgl}+\mathsf{L}_{\lgl}+1}^2\kappa_{\mathsf{L}_{\mgl}+\mathsf{L}_{\lgl}}^{-2}\hat{\alpha}\hat{\alpha}^*+\hat{\alpha}^{\perp}(\hat{\alpha}^{\perp})^*\right)\Phi\right]^{-\frac{1}{2}}\bigg\|\\
&= \bigg\|\mathbf{1}_{\mathsf{q}}-\left[\Phi^*\left(\kappa_{\mathsf{L}_{\mgl}+\mathsf{L}_{\lgl}}^2\kappa_{\mathsf{L}_{\mgl}+\mathsf{L}_{\lgl}+1}^{-2}\hat{\alpha}\hat{\alpha}^*+\hat{\alpha}^{\perp}(\hat{\alpha}^{\perp})^*\right)\Phi\right]^{-\frac{1}{2}}\Phi^*\hat{\alpha}^{\perp}(\hat{\alpha}^{\perp})^*\Phi\times\\
&\qquad\times\left[\Phi^*\left(\kappa_{\mathsf{L}_{\mgl}+\mathsf{L}_{\lgl}+1}^2\kappa_{\mathsf{L}_{\mgl}+\mathsf{L}_{\lgl}}^{-2}\hat{\alpha}\hat{\alpha}^*+\hat{\alpha}^{\perp}(\hat{\alpha}^{\perp})^*\right)\Phi\right]^{-\frac{1}{2}}\bigg\|\\
&= \bigg\|\left[\Phi^*\left(\hat{\alpha}\hat{\alpha}^*+\kappa_{\mathsf{L}_{\mgl}+\mathsf{L}_{\lgl}}^2\kappa_{\mathsf{L}_{\mgl}+\mathsf{L}_{\lgl}+1}^{-2}\hat{\alpha}^{\perp}(\hat{\alpha}^{\perp})^*\right)\Phi\right]^{-\frac{1}{2}}\Phi^*\hat{\alpha}\hat{\alpha}^*\Phi\times\\
&\qquad\times\left[\Phi^*\left(\hat{\alpha}\hat{\alpha}^*+\kappa_{\mathsf{L}_{\mgl}+\mathsf{L}_{\lgl}}^2\kappa_{\mathsf{L}_{\mgl}+\mathsf{L}_{\lgl}+1}^{-2}\hat{\alpha}^{\perp}(\hat{\alpha}^{\perp})^*\right)\Phi\right]^{-\frac{1}{2}}\bigg\|\\
&= \bigg\|\hat{\alpha}^*\Phi\left[\Phi^*\left(\hat{\alpha}\hat{\alpha}^*+\kappa_{\mathsf{L}_{\mgl}+\mathsf{L}_{\lgl}}^2\kappa_{\mathsf{L}_{\mgl}+\mathsf{L}_{\lgl}+1}^{-2}\hat{\alpha}^{\perp}(\hat{\alpha}^{\perp})^*\right)\Phi\right]^{-1}\Phi^*\hat{\alpha}\bigg\|\\
&= \bigg\|\hat{\alpha}^*\Phi\left[\mathbf{1}_{\mathsf{q}}-(1-\kappa_{\mathsf{L}_{\mgl}+\mathsf{L}_{\lgl}}^2\kappa_{\mathsf{L}_{\mgl}+\mathsf{L}_{\lgl}+1}^{-2})\Phi^*\hat{\alpha}^{\perp}(\hat{\alpha}^{\perp})^*\Phi\right]^{-1}\Phi^*\hat{\alpha}\bigg\|\\
&\leq \Big\|\left[\mathbf{1}_{\mathsf{q}}-(1-\kappa_{\mathsf{L}_{\mgl}+\mathsf{L}_{\lgl}}^2\kappa_{\mathsf{L}_{\mgl}+\mathsf{L}_{\lgl}+1}^{-2})\Phi^*\hat{\alpha}^{\perp}(\hat{\alpha}^{\perp})^*\Phi\right]^{-1}\Big\|\,\|\hat{\alpha}^*Q\hat{\alpha}\|\,.
\end{align*}
Using the inequality $[1-(1-y)x]^{-1}\leq 1-(1-y^{-1})x$ for all $x\in [0,1]$ and $y>0$, one deduces
\begin{align*}
\left\|\left[\mathbf{1}_{\mathsf{q}}-(1-\kappa_{\mathsf{L}_{\mgl}+\mathsf{L}_{\lgl}}^2\kappa_{\mathsf{L}_{\mgl}+\mathsf{L}_{\lgl}+1}^{-2})\Phi^*\hat{\alpha}^{\perp}(\hat{\alpha}^{\perp})^*\Phi\right]^{-1}\right\|&\leq \left\|\mathbf{1}_{\mathsf{q}}-\eta(\mathsf{L}_{\mgl}+\mathsf{L}_{\lgl},\mathsf{L}_{\mgl}+\mathsf{L}_{\lgl}+1)\,\Phi^*\hat{\alpha}^{\perp}(\hat{\alpha}^{\perp})^*\Phi\right\|\\
&=\left\|\mathbf{1}_{\mathsf{q}}-\eta(\mathsf{L}_{\mgl}+\mathsf{L}_{\lgl},\mathsf{L}_{\mgl}+\mathsf{L}_{\lgl}+1)\,[\mathbf{1}_{\mathsf{q}}-\Phi^*\hat{\alpha}\hat{\alpha}^*\Phi]\right\|\\
&=1-\eta(\mathsf{L}_{\mgl}+\mathsf{L}_{\lgl},\mathsf{L}_{\mgl}+\mathsf{L}_{\lgl}+1)\,[1-\|\Phi^*\hat{\alpha}\hat{\alpha}^*\Phi\|]\\
&=1-\eta(\mathsf{L}_{\mgl}+\mathsf{L}_{\lgl},\mathsf{L}_{\mgl}+\mathsf{L}_{\lgl}+1)\,[1-\|\hat{\alpha}^*Q\hat{\alpha}\|]\,.
\end{align*}
This then implies~\eqref{ineq-norm-contraction}.
\hfill $\square$

\vspace{.2cm}

\noindent\textbf{Proof of Lemma~\ref{lemma-trace-contraction}.} Let $\Phi\in\mathbb{F}_{\mathsf{L},\mathsf{q}}$ such that $Q=\Phi\Phi^*$. 
 We introduce the frame
$$
\hat{\theta}=\begin{pmatrix}
\mathbf{0}_{\mathsf{L}_{\ugl}\times\mathsf{L}_{\mgl}}\\\mathbf{1}_{\mathsf{L}_{\mgl}}\\\mathbf{0}_{\mathsf{L}_{\lgl}\times\mathsf{L}_{\mgl}}
\end{pmatrix}\,,
$$
for which one has $\mathbf{1}=\hat{\alpha}\hat{\alpha}^*+\hat{\theta}\hat{\theta}^*+\hat{\gamma}\hat{\gamma}^*$.
 This allows to estimate
\begin{align*}
&\operatorname{tr}\left[\hat{\alpha}^*(\mathcal{R}\cdot Q)\hat{\alpha}\right]\\
&=\operatorname{tr}\left[\hat{\alpha}^*\mathcal{R}\Phi[\Phi^*\mathcal{R}^2\Phi]^{-1}\Phi^*\mathcal{R}\hat{\alpha}\right]\\
&=\operatorname{tr}\left[\hat{\alpha}^*\mathcal{R}\Phi\left[\Phi^*\left(\mathcal{R}\hat{\alpha}\hat{\alpha}^*\mathcal{R}+\mathcal{R}\hat{\theta}\hat{\theta}^*\mathcal{R}+\mathcal{R}\hat{\gamma}\hat{\gamma}^*\mathcal{R}\right)\Phi\right]^{-1}\Phi^*\mathcal{R}\hat{\alpha}\right]\\
&\leq \operatorname{tr}\left[\hat{\alpha}^*\mathcal{R}\Phi\left[\Phi^*\left(\mathcal{R}\hat{\alpha}\hat{\alpha}^*\mathcal{R}+\kappa_{\mathsf{L}_{\mgl}+\mathsf{L}_{\lgl}}^2\hat{\theta}\hat{\theta}^*+\kappa_{\mathsf{L}_{\lgl}}^2\hat{\gamma}\hat{\gamma}^*\right)\Phi\right]^{-1}\Phi^*\mathcal{R}\hat{\alpha}\right]\\
&=\operatorname{tr}\bigg[\left[\Phi^*\left(\mathcal{R}\hat{\alpha}\hat{\alpha}^*\mathcal{R}+\kappa_{\mathsf{L}_{\mgl}+\mathsf{L}_{\lgl}}^2\hat{\theta}\hat{\theta}^*+\kappa_{\mathsf{L}_{\lgl}}^2\hat{\gamma}\hat{\gamma}^*\right)\Phi\right]^{-\frac{1}{2}}\Phi^*\mathcal{R}\hat{\alpha}\hat{\alpha}^*\mathcal{R}\Phi\times\\
&\qquad\qquad\qquad\qquad\qquad\times\left[\Phi^*\left(\mathcal{R}\hat{\alpha}\hat{\alpha}^*\mathcal{R}+\kappa_{\mathsf{L}_{\mgl}+\mathsf{L}_{\lgl}}^2\hat{\theta}\hat{\theta}^*+\kappa_{\mathsf{L}_{\lgl}}^2\hat{\gamma}\hat{\gamma}^*\right)\Phi\right]^{-\frac{1}{2}}\bigg]\\
&=\operatorname{tr}\bigg[\mathbf{1}_{\mathsf{q}}-\left[\Phi^*\left(\mathcal{R}\hat{\alpha}\hat{\alpha}^*\mathcal{R}+\kappa_{\mathsf{L}_{\mgl}+\mathsf{L}_{\lgl}}^2\hat{\theta}\hat{\theta}^*+\kappa_{\mathsf{L}_{\lgl}}^2\hat{\gamma}\hat{\gamma}^*\right)\Phi\right]^{-\frac{1}{2}}\times\\
&\qquad\times\Phi^*\left(\kappa_{\mathsf{L}_{\mgl}+\mathsf{L}_{\lgl}}^2\hat{\theta}\hat{\theta}^*+\kappa_{\mathsf{L}_{\lgl}}^2\hat{\gamma}\hat{\gamma}^*\right)\Phi\left[\Phi^*\left(\mathcal{R}\hat{\alpha}\hat{\alpha}^*\mathcal{R}+\kappa_{\mathsf{L}_{\mgl}+\mathsf{L}_{\lgl}}^2\hat{\theta}\hat{\theta}^*+\kappa_{\mathsf{L}_{\lgl}}^2\hat{\gamma}\hat{\gamma}^*\right)\Phi\right]^{-\frac{1}{2}}\bigg]\\
&=\mathsf{q}-\operatorname{tr}\bigg[\left[\Phi^*\left(\mathcal{R}\hat{\alpha}\hat{\alpha}^*\mathcal{R}+\kappa_{\mathsf{L}_{\mgl}+\mathsf{L}_{\lgl}}^2\hat{\theta}\hat{\theta}^*+\kappa_{\mathsf{L}_{\lgl}}^2\hat{\gamma}\hat{\gamma}^*\right)\Phi\right]^{-\frac{1}{2}}\times\\
&\qquad\times\Phi^*\left(\kappa_{\mathsf{L}_{\mgl}+\mathsf{L}_{\lgl}}^2\hat{\theta}\hat{\theta}^*+\kappa_{\mathsf{L}_{\lgl}}^2\hat{\gamma}\hat{\gamma}^*\right)\Phi\left[\Phi^*\left(\mathcal{R}\hat{\alpha}\hat{\alpha}^*\mathcal{R}+\kappa_{\mathsf{L}_{\mgl}+\mathsf{L}_{\lgl}}^2\hat{\theta}\hat{\theta}^*+\kappa_{\mathsf{L}_{\lgl}}^2\hat{\gamma}\hat{\gamma}^*\right)\Phi\right]^{-\frac{1}{2}}\bigg]\\
&=\mathsf{q}-\operatorname{tr}\bigg[\left[\Phi^*\left(\kappa_{\mathsf{L}_{\mgl}+\mathsf{L}_{\lgl}}^2\hat{\theta}\hat{\theta}^*+\kappa_{\mathsf{L}_{\lgl}}^2\hat{\gamma}\hat{\gamma}^*\right)\Phi\right]^{\frac{1}{2}}\times\\
&\qquad\times\left[\Phi^*\left(\mathcal{R}\hat{\alpha}\hat{\alpha}^*\mathcal{R}+\kappa_{\mathsf{L}_{\mgl}+\mathsf{L}_{\lgl}}^2\hat{\theta}\hat{\theta}^*+\kappa_{\mathsf{L}_{\lgl}}^2\hat{\gamma}\hat{\gamma}^*\right)\Phi\right]^{-1}\left[\Phi^*\left(\kappa_{\mathsf{L}_{\mgl}+\mathsf{L}_{\lgl}}^2\hat{\theta}\hat{\theta}^*+\kappa_{\mathsf{L}_{\lgl}}^2\hat{\gamma}\hat{\gamma}^*\right)\Phi\right]^{\frac{1}{2}}\bigg]\\
&\leq\mathsf{q}-\operatorname{tr}\bigg[\left[\Phi^*\left(\kappa_{\mathsf{L}_{\mgl}+\mathsf{L}_{\lgl}}^2\hat{\theta}\hat{\theta}^*+\kappa_{\mathsf{L}_{\lgl}}^2\hat{\gamma}\hat{\gamma}^*\right)\Phi\right]^{\frac{1}{2}}\times\\
&\qquad\times\left[\Phi^*\left(\kappa_{\mathsf{L}_{\mgl}+\mathsf{L}_{\lgl}}^2\hat{\alpha}\hat{\alpha}^*+\kappa_{\mathsf{L}_{\mgl}+\mathsf{L}_{\lgl}}^2\hat{\theta}\hat{\theta}^*+\kappa_{\mathsf{L}_{\lgl}}^2\hat{\gamma}\hat{\gamma}^*\right)\Phi\right]^{-1}\left[\Phi^*\left(\kappa_{\mathsf{L}_{\mgl}+\mathsf{L}_{\lgl}}^2\hat{\theta}\hat{\theta}^*+\kappa_{\mathsf{L}_{\lgl}}^2\hat{\gamma}\hat{\gamma}^*\right)\Phi\right]^{\frac{1}{2}}\bigg]\\
&=\mathsf{q}-\operatorname{tr}\bigg[\left[\Phi^*\left(\hat{\theta}\hat{\theta}^*+\kappa_{\mathsf{L}_{\lgl}}^2\kappa_{\mathsf{L}_{\mgl}+\mathsf{L}_{\lgl}}^{-2}\hat{\gamma}\hat{\gamma}^*\right)\Phi\right]^{\frac{1}{2}}\times\\
&\qquad\times\left[\Phi^*\left(\hat{\alpha}\hat{\alpha}^*+\hat{\theta}\hat{\theta}^*+\kappa_{\mathsf{L}_{\lgl}}^2\kappa_{\mathsf{L}_{\mgl}+\mathsf{L}_{\lgl}}^{-2}\hat{\gamma}\hat{\gamma}^*\right)\Phi\right]^{-1}\left[\Phi^*\left(\hat{\theta}\hat{\theta}^*+\kappa_{\mathsf{L}_{\lgl}}^2\kappa_{\mathsf{L}_{\mgl}+\mathsf{L}_{\lgl}}^{-2}\hat{\gamma}\hat{\gamma}^*\right)\Phi\right]^{\frac{1}{2}}\bigg]\\
&=\mathsf{q}-\operatorname{tr}\bigg[\left[\Phi^*\left(\hat{\alpha}\hat{\alpha}^*+\hat{\theta}\hat{\theta}^*+\kappa_{\mathsf{L}_{\lgl}}^2\kappa_{\mathsf{L}_{\mgl}+\mathsf{L}_{\lgl}}^{-2}\hat{\gamma}\hat{\gamma}^*\right)\Phi\right]^{-\frac{1}{2}}\times\\
&\qquad\times\Phi^*\left(\hat{\theta}\hat{\theta}^*+\kappa_{\mathsf{L}_{\lgl}}^2\kappa_{\mathsf{L}_{\mgl}+\mathsf{L}_{\lgl}}^{-2}\hat{\gamma}\hat{\gamma}^*\right)\Phi\left[\Phi^*\left(\hat{\alpha}\hat{\alpha}^*+\hat{\theta}\hat{\theta}^*+\kappa_{\mathsf{L}_{\lgl}}^2\kappa_{\mathsf{L}_{\mgl}+\mathsf{L}_{\lgl}}^{-2}\hat{\gamma}\hat{\gamma}^*\right)\Phi\right]^{-\frac{1}{2}}\bigg]\\
&=\operatorname{tr}\bigg[\mathbf{1}_{\mathsf{q}}-\left[\Phi^*\left(\hat{\alpha}\hat{\alpha}^*+\hat{\theta}\hat{\theta}^*+\kappa_{\mathsf{L}_{\lgl}}^2\kappa_{\mathsf{L}_{\mgl}+\mathsf{L}_{\lgl}}^{-2}\hat{\gamma}\hat{\gamma}^*\right)\Phi\right]^{-\frac{1}{2}}\times\\
&\qquad\times\Phi^*\left(\hat{\theta}\hat{\theta}^*+\kappa_{\mathsf{L}_{\lgl}}^2\kappa_{\mathsf{L}_{\mgl}+\mathsf{L}_{\lgl}}^{-2}\hat{\gamma}\hat{\gamma}^*\right)\Phi\left[\Phi^*\left(\hat{\alpha}\hat{\alpha}^*+\hat{\theta}\hat{\theta}^*+\kappa_{\mathsf{L}_{\lgl}}^2\kappa_{\mathsf{L}_{\mgl}+\mathsf{L}_{\lgl}}^{-2}\hat{\gamma}\hat{\gamma}^*\right)\Phi\right]^{-\frac{1}{2}}\bigg]\\
&=\operatorname{tr}\bigg[\left[\Phi^*\left(\hat{\alpha}\hat{\alpha}^*+\hat{\theta}\hat{\theta}^*+\kappa_{\mathsf{L}_{\lgl}}^2\kappa_{\mathsf{L}_{\mgl}+\mathsf{L}_{\lgl}}^{-2}\hat{\gamma}\hat{\gamma}^*\right)\Phi\right]^{-\frac{1}{2}}\Phi^*\hat{\alpha}\hat{\alpha}^*\Phi\times\\
&\qquad\qquad\qquad\qquad\qquad\times\left[\Phi^*\left(\hat{\alpha}\hat{\alpha}^*+\hat{\theta}\hat{\theta}^*+\kappa_{\mathsf{L}_{\lgl}}^2\kappa_{\mathsf{L}_{\mgl}+\mathsf{L}_{\lgl}}^{-2}\hat{\gamma}\hat{\gamma}^*\right)\Phi\right]^{-\frac{1}{2}}\bigg]\\
&=\operatorname{tr}\bigg[\hat{\alpha}^*\Phi\left[\Phi^*\left(\hat{\alpha}\hat{\alpha}^*+\hat{\theta}\hat{\theta}^*+\kappa_{\mathsf{L}_{\lgl}}^2\kappa_{\mathsf{L}_{\mgl}+\mathsf{L}_{\lgl}}^{-2}\hat{\gamma}\hat{\gamma}^*\right)\Phi\right]^{-1}\Phi^*\hat{\alpha}\bigg]\\
&=\operatorname{tr}\bigg[\hat{\alpha}^*\Phi\left[\mathbf{1}_{\mathsf{q}}-(1-\kappa_{\mathsf{L}_{\lgl}}^2\kappa_{\mathsf{L}_{\mgl}+\mathsf{L}_{\lgl}}^{-2})\Phi^*\hat{\gamma}\hat{\gamma}^*\Phi\right]^{-1}\Phi^*\hat{\alpha}\bigg]\\
\end{align*}
Now one can verify that for all $x\in [0,1]$ and $y>0$ the inequality $[1-(1-y)x]^{-1}\leq 1-(1-y^{-1})x$ holds,
which then implies~\eqref{ineq-trace-contraction-1} due to
\begin{align*}
\hat{\alpha}^*\Phi\left[\mathbf{1}_{\mathsf{q}}-(1-\kappa_{\mathsf{L}_{\lgl}}^2\kappa_{\mathsf{L}_{\mgl}+\mathsf{L}_{\lgl}}^{-2})\Phi^*\hat{\gamma}\hat{\gamma}^*\Phi\right]^{-1}\Phi^*\hat{\alpha}
& 
\leq 
\hat{\alpha}^*\Phi\left[\mathbf{1}_{\mathsf{q}}-\pmb{\eta}^{\prime}\,\Phi^*\hat{\gamma}\hat{\gamma}^*\Phi\right]\Phi^*\hat{\alpha}
\\
&
=
\hat{\alpha}^*Q\left[\mathbf{1}_{\mathsf{L}}-\pmb{\eta}^{\prime}\,\hat{\gamma}\hat{\gamma}^*\right]Q\hat{\alpha}\,.
\end{align*}
One can now estimate
\begin{align*}
&\operatorname{tr}\left[\hat{\gamma}^*(\mathcal{R}\cdot Q)\hat{\gamma}\right]\\
&=\operatorname{tr}\left[\hat{\gamma}^*\mathcal{R}\Phi[\Phi^*\mathcal{R}^2\Phi]^{-1}\Phi^*\mathcal{R}\hat{\gamma}\right]\\
&=\operatorname{tr}\left[\hat{\gamma}^*\mathcal{R}\Phi\left[\Phi^*\left(\mathcal{R}\hat{\alpha}\hat{\alpha}^*\mathcal{R}+\mathcal{R}\hat{\theta}\hat{\theta}^*\mathcal{R}+\mathcal{R}\hat{\gamma}\hat{\gamma}^*\mathcal{R}\right)\Phi\right]^{-1}\Phi^*\mathcal{R}\hat{\gamma}\right]\\
&\geq \operatorname{tr}\left[\hat{\gamma}^*\mathcal{R}\Phi\left[\Phi^*\left(\kappa_{\mathsf{L}_{\mgl}+\mathsf{L}_{\lgl}}^2\hat{\alpha}\hat{\alpha}^*+\kappa_{\mathsf{L}_{\lgl}}^2\hat{\theta}\hat{\theta}^*+\mathcal{R}\hat{\gamma}\hat{\gamma}^*\mathcal{R}\right)\Phi\right]^{-1}\Phi^*\mathcal{R}\hat{\gamma}\right]\\
&=\operatorname{tr}\bigg[\left[\Phi^*\left(\kappa_{\mathsf{L}_{\mgl}+\mathsf{L}_{\lgl}}^2\hat{\alpha}\hat{\alpha}^*+\kappa_{\mathsf{L}_{\lgl}}^2\hat{\theta}\hat{\theta}^*+\mathcal{R}\hat{\gamma}\hat{\gamma}^*\mathcal{R}\right)\Phi\right]^{-\frac{1}{2}}\Phi^*\mathcal{R}\hat{\gamma}\hat{\gamma}^*\mathcal{R}\Phi\times\\
&\qquad\qquad\qquad\qquad\qquad\times\left[\Phi^*\left(\kappa_{\mathsf{L}_{\mgl}+\mathsf{L}_{\lgl}}^2\hat{\alpha}\hat{\alpha}^*+\kappa_{\mathsf{L}_{\lgl}}^2\hat{\theta}\hat{\theta}^*+\mathcal{R}\hat{\gamma}\hat{\gamma}^*\mathcal{R}\right)\Phi\right]^{-\frac{1}{2}}\bigg]\\
&=\operatorname{tr}\bigg[\mathbf{1}_{\mathsf{q}}-\left[\Phi^*\left(\kappa_{\mathsf{L}_{\mgl}+\mathsf{L}_{\lgl}}^2\hat{\alpha}\hat{\alpha}^*+\kappa_{\mathsf{L}_{\lgl}}^2\hat{\theta}\hat{\theta}^*+\mathcal{R}\hat{\gamma}\hat{\gamma}^*\mathcal{R}\right)\Phi\right]^{-\frac{1}{2}}\times\\
&\qquad\times\Phi^*\left(\kappa_{\mathsf{L}_{\mgl}+\mathsf{L}_{\lgl}}^2\hat{\alpha}\hat{\alpha}^*+\kappa_{\mathsf{L}_{\lgl}}^2\hat{\theta}\hat{\theta}^*\right)\Phi\left[\Phi^*\left(\kappa_{\mathsf{L}_{\mgl}+\mathsf{L}_{\lgl}}^2\hat{\alpha}\hat{\alpha}^*+\kappa_{\mathsf{L}_{\lgl}}^2\hat{\theta}\hat{\theta}^*+\mathcal{R}\hat{\gamma}\hat{\gamma}^*\mathcal{R}\right)\Phi\right]^{-\frac{1}{2}}\bigg]\\
&=\mathsf{q}-\operatorname{tr}\bigg[\left[\Phi^*\left(\kappa_{\mathsf{L}_{\mgl}+\mathsf{L}_{\lgl}}^2\hat{\alpha}\hat{\alpha}^*+\kappa_{\mathsf{L}_{\lgl}}^2\hat{\theta}\hat{\theta}^*+\mathcal{R}\hat{\gamma}\hat{\gamma}^*\mathcal{R}\right)\Phi\right]^{-\frac{1}{2}}\times\\
&\qquad\times\Phi^*\left(\kappa_{\mathsf{L}_{\mgl}+\mathsf{L}_{\lgl}}^2\hat{\alpha}\hat{\alpha}^*+\kappa_{\mathsf{L}_{\lgl}}^2\hat{\theta}\hat{\theta}^*\right)\Phi\left[\Phi^*\left(\kappa_{\mathsf{L}_{\mgl}+\mathsf{L}_{\lgl}}^2\hat{\alpha}\hat{\alpha}^*+\kappa_{\mathsf{L}_{\lgl}}^2\hat{\theta}\hat{\theta}^*+\mathcal{R}\hat{\gamma}\hat{\gamma}^*\mathcal{R}\right)\Phi\right]^{-\frac{1}{2}}\bigg]\\
&=\mathsf{q}-\operatorname{tr}\bigg[\left[\Phi^*\left(\kappa_{\mathsf{L}_{\mgl}+\mathsf{L}_{\lgl}}^2\hat{\alpha}\hat{\alpha}^*+\kappa_{\mathsf{L}_{\lgl}}^2\hat{\theta}\hat{\theta}^*\right)\Phi\right]^{\frac{1}{2}}\times\\
&\qquad\times\left[\Phi^*\left(\kappa_{\mathsf{L}_{\mgl}+\mathsf{L}_{\lgl}}^2\hat{\alpha}\hat{\alpha}^*+\kappa_{\mathsf{L}_{\lgl}}^2\hat{\theta}\hat{\theta}^*+\mathcal{R}\hat{\gamma}\hat{\gamma}^*\mathcal{R}\right)\Phi\right]^{-1}\left[\Phi^*\left(\kappa_{\mathsf{L}_{\mgl}+\mathsf{L}_{\lgl}}^2\hat{\alpha}\hat{\alpha}^*+\kappa_{\mathsf{L}_{\lgl}}^2\hat{\theta}\hat{\theta}^*\right)\Phi\right]^{\frac{1}{2}}\bigg]\\
&\geq\mathsf{q}-\operatorname{tr}\bigg[\left[\Phi^*\left(\kappa_{\mathsf{L}_{\mgl}+\mathsf{L}_{\lgl}}^2\hat{\alpha}\hat{\alpha}^*+\kappa_{\mathsf{L}_{\lgl}}^2\hat{\theta}\hat{\theta}^*\right)\Phi\right]^{\frac{1}{2}}\times\\
&\qquad\times\left[\Phi^*\left(\kappa_{\mathsf{L}_{\mgl}+\mathsf{L}_{\lgl}}^2\hat{\alpha}\hat{\alpha}^*+\kappa_{\mathsf{L}_{\lgl}}^2\hat{\theta}\hat{\theta}^*+\kappa_{\mathsf{L}_{\lgl}}^2\hat{\gamma}\hat{\gamma}^*\right)\Phi\right]^{-1}\left[\Phi^*\left(\kappa_{\mathsf{L}_{\mgl}+\mathsf{L}_{\lgl}}^2\hat{\alpha}\hat{\alpha}^*+\kappa_{\mathsf{L}_{\lgl}}^2\hat{\theta}\hat{\theta}^*\right)\Phi\right]^{\frac{1}{2}}\bigg]\\
&=\mathsf{q}-\operatorname{tr}\bigg[\left[\Phi^*\left(\kappa_{\mathsf{L}_{\mgl}+\mathsf{L}_{\lgl}}^2\kappa_{\mathsf{L}_{\lgl}}^{-2}\hat{\alpha}\hat{\alpha}^*+\hat{\theta}\hat{\theta}^*\right)\Phi\right]^{\frac{1}{2}}\times\\
&\qquad\times\left[\Phi^*\left(\kappa_{\mathsf{L}_{\mgl}+\mathsf{L}_{\lgl}}^2\kappa_{\mathsf{L}_{\lgl}}^{-2}\hat{\alpha}\hat{\alpha}^*+\hat{\theta}\hat{\theta}^*+\hat{\gamma}\hat{\gamma}^*\right)\Phi\right]^{-1}\left[\Phi^*\left(\kappa_{\mathsf{L}_{\mgl}+\mathsf{L}_{\lgl}}^2\kappa_{\mathsf{L}_{\lgl}}^{-2}\hat{\alpha}\hat{\alpha}^*+\hat{\theta}\hat{\theta}^*\right)\Phi\right]^{\frac{1}{2}}\bigg]\\
&=\mathsf{q}-\operatorname{tr}\bigg[\left[\Phi^*\left(\kappa_{\mathsf{L}_{\mgl}+\mathsf{L}_{\lgl}}^2\kappa_{\mathsf{L}_{\lgl}}^{-2}\hat{\alpha}\hat{\alpha}^*+\hat{\theta}\hat{\theta}^*+\hat{\gamma}\hat{\gamma}^*\right)\Phi\right]^{-\frac{1}{2}}\times\\
&\qquad\times\Phi^*\left(\kappa_{\mathsf{L}_{\mgl}+\mathsf{L}_{\lgl}}^2\kappa_{\mathsf{L}_{\lgl}}^{-2}\hat{\alpha}\hat{\alpha}^*+\hat{\theta}\hat{\theta}^*\right)\Phi\left[\Phi^*\left(\kappa_{\mathsf{L}_{\mgl}+\mathsf{L}_{\lgl}}^2\kappa_{\mathsf{L}_{\lgl}}^{-2}\hat{\alpha}\hat{\alpha}^*+\hat{\theta}\hat{\theta}^*+\hat{\gamma}\hat{\gamma}^*\right)\Phi\right]^{-\frac{1}{2}}\bigg]\\
&=\operatorname{tr}\bigg[\mathbf{1}_{\mathsf{q}}-\left[\Phi^*\left(\kappa_{\mathsf{L}_{\mgl}+\mathsf{L}_{\lgl}}^2\kappa_{\mathsf{L}_{\lgl}}^{-2}\hat{\alpha}\hat{\alpha}^*+\hat{\theta}\hat{\theta}^*+\hat{\gamma}\hat{\gamma}^*\right)\Phi\right]^{-\frac{1}{2}}\times\\
&\qquad\times\Phi^*\left(\kappa_{\mathsf{L}_{\mgl}+\mathsf{L}_{\lgl}}^2\kappa_{\mathsf{L}_{\lgl}}^{-2}\hat{\alpha}\hat{\alpha}^*+\hat{\theta}\hat{\theta}^*\right)\Phi\left[\Phi^*\left(\kappa_{\mathsf{L}_{\mgl}+\mathsf{L}_{\lgl}}^2\kappa_{\mathsf{L}_{\lgl}}^{-2}\hat{\alpha}\hat{\alpha}^*+\hat{\theta}\hat{\theta}^*+\hat{\gamma}\hat{\gamma}^*\right)\Phi\right]^{-\frac{1}{2}}\bigg]\\
&=\operatorname{tr}\bigg[\left[\Phi^*\left(\kappa_{\mathsf{L}_{\mgl}+\mathsf{L}_{\lgl}}^2\kappa_{\mathsf{L}_{\lgl}}^{-2}\hat{\alpha}\hat{\alpha}^*+\hat{\theta}\hat{\theta}^*+\hat{\gamma}\hat{\gamma}^*\right)\Phi\right]^{-\frac{1}{2}}\Phi^*\hat{\gamma}\hat{\gamma}^*\Phi\times\\
&\qquad\qquad\qquad\qquad\qquad\times\left[\Phi^*\left(\kappa_{\mathsf{L}_{\mgl}+\mathsf{L}_{\lgl}}^2\kappa_{\mathsf{L}_{\lgl}}^{-2}\hat{\alpha}\hat{\alpha}^*+\hat{\theta}\hat{\theta}^*+\hat{\gamma}\hat{\gamma}^*\right)\Phi\right]^{-\frac{1}{2}}\bigg]\\
&=\operatorname{tr}\left[\hat{\gamma}^*\Phi\left[\Phi^*\left(\kappa_{\mathsf{L}_{\mgl}+\mathsf{L}_{\lgl}}^2\kappa_{\mathsf{L}_{\lgl}}^{-2}\hat{\alpha}\hat{\alpha}^*+\hat{\theta}\hat{\theta}^*+\hat{\gamma}\hat{\gamma}^*\right)\Phi\right]^{-1}\Phi^*\hat{\gamma}\right]\\
&=\operatorname{tr}\left[\hat{\gamma}^*\Phi\left[\mathbf{1}_{\mathsf{q}}-\pmb{\eta}\,\Phi^*\hat{\alpha}\hat{\alpha}^*\Phi\right]^{-1}\Phi^*\hat{\gamma}\right]\\
&\geq\operatorname{tr}\left[\hat{\gamma}^*\Phi\left[\mathbf{1}_{\mathsf{q}}+\pmb{\eta}\,\Phi^*\hat{\alpha}\hat{\alpha}^*\Phi\right]\Phi^*\hat{\gamma}\right]\\
&=\operatorname{tr}\left[\hat{\gamma}^*Q\left[\mathbf{1}_{\mathsf{L}}+\pmb{\eta}\,\hat{\alpha}\hat{\alpha}^*\right]Q\hat{\gamma}\right]\,,
\end{align*}
completing the proof of ~\eqref{ineq-trace-contraction-2}.
\hfill $\square$

\vspace{.2cm}

\noindent\textbf{Proof of Corollary~\ref{coro-d-contraction}.}
Let $\Phi\in\mathbb{F}_{\mathsf{L},\mathsf{q}}$ be such that $Q=\Phi\Phi^*$. Now,  using~\eqref{ineq-trace-contraction-1},
\begin{align*}
\mathsf{d}(\mathcal{R}\cdot Q)&=\operatorname{tr}\left[\hat{\alpha}^*(\mathcal{R}\cdot Q)\hat{\alpha}\right]\\
&\leq\operatorname{tr}\left[\hat{\alpha}^* Q\left[\mathbf{1}_{\mathsf{L}}-\pmb{\eta}\,\hat{\gamma}\hat{\gamma}^*\right]Q\hat{\alpha}\right]\\
&=\operatorname{tr}\left[\hat{\alpha}^* \Phi\left[\mathbf{1}_{\mathsf{q}}-\pmb{\eta}\,\left(\mathbf{1}_{\mathsf{q}}-\Phi^*\hat{\gamma}^{\perp}(\hat{\gamma}^{\perp})^*\Phi\right)\right]\Phi^*\hat{\alpha}\right]\\
&\leq\left\|\mathbf{1}_{\mathsf{q}}-\pmb{\eta}\,\left(\mathbf{1}_{\mathsf{q}}-\Phi^*\hat{\gamma}^{\perp}(\hat{\gamma}^{\perp})^*\Phi\right)\right\|\,\operatorname{tr}\left[\hat{\alpha}^* \Phi\Phi^*\hat{\alpha}\right]\\
&=\left(1-\pmb{\eta}\,\left[1-\|(\Phi^*\hat{\gamma}^{\perp}(\hat{\gamma}^{\perp})^*\Phi\|\right]\right)\,\operatorname{tr}\left[\hat{\alpha}^* \Phi\Phi^*\hat{\alpha}\right]\\
&=\left(1-\pmb{\eta}\,\left[1-\|(\hat{\gamma}^{\perp})^*\,Q\,\hat{\gamma}^{\perp}\|\right]\right)\,\mathsf{d}(Q)\,,
\end{align*}
which is~\eqref{ineq-d-contraction-bis}.
\hfill $\square$

\vspace{.2cm}

\noindent\textbf{Proof of Lemma~\ref{lemma-subdivision}.}
We begin by exploiting~\eqref{ineq-subdivision-assumption} in order to estimate for all $\mathsf{J}\in\left\{\mathsf{A}+1,\dots,\mathsf{B}\right\}$
\begin{align}\label{ineq-kappa-differences}
\begin{split}
\kappa_{\mathsf{J}-1}^2-\kappa_{\mathsf{J}}^2=\kappa_{\mathsf{J}-1}^2\,\eta({\mathsf{J}-1},\mathsf{J})\leq {\kappa_{\mathsf{J}-1}^2}\,\frac{\phi}{\mathsf{F}}\,\eta(\mathsf{A},\mathsf{B})= {\kappa_{\mathsf{J}-1}^2}\kappa^{-2}_{\mathsf{A}}\,\frac{\phi}{\mathsf{F}}\,\big[\kappa^{2}_{\mathsf{A}}-\kappa^{2}_{\mathsf{B}}\big]\leq \frac{\phi}{\mathsf{F}}\,\big[\kappa^{2}_{\mathsf{A}}-\kappa^{2}_{\mathsf{B}}\big]
\end{split}
\end{align}
and also
\begin{align}\label{ineq-sufficient-length}
\begin{split}
\mathsf{F}&<\phi^{-1}\mathsf{F}=\phi^{-1}\mathsf{F}\,\eta(\mathsf{A},\mathsf{B})^{-1}\,\kappa_{\mathsf{A}}^{-2}\,\big[\kappa^2_{\mathsf{A}}-\kappa^2_{\mathsf{B}}\big]
=\phi^{-1}\mathsf{F}\,\eta(\mathsf{A},\mathsf{B})^{-1}\,\kappa_{\mathsf{A}}^{-2}\,\sum\limits_{\mathsf{J}=\mathsf{A}+1}^{\mathsf{B}}\,\big[\kappa^2_{\mathsf{J}-1}-\kappa^2_{\mathsf{J}}\big]\\
&=\phi^{-1}\mathsf{F}\,\eta(\mathsf{A},\mathsf{B})^{-1}\,\kappa_{\mathsf{A}}^{-2}\,\sum\limits_{\mathsf{J}=\mathsf{A}+1}^{\mathsf{B}}\,\kappa^2_{\mathsf{J}-1}\,\eta({\mathsf{J}-1},{\mathsf{J}})
\leq \phi^{-1}\mathsf{F}\,\eta(\mathsf{A},\mathsf{B})^{-1}\,\kappa_{\mathsf{A}}^{-2}\,\sum\limits_{\mathsf{J}=\mathsf{A}+1}^{\mathsf{B}}\,\kappa^2_{\mathsf{J}-1}\,\frac{\phi}{\mathsf{F}}\,\eta(\mathsf{A},\mathsf{B})\\
&= \sum\limits_{\mathsf{J}=\mathsf{A}+1}^{\mathsf{B}}\,\kappa_{\mathsf{A}}^{-2}\,\kappa^2_{\mathsf{J}-1}
\leq \sum\limits_{\mathsf{J}=\mathsf{A}+1}^{\mathsf{B}}\,1
=\mathsf{B}-\mathsf{A}\,.
\end{split}
\end{align}
Next, we define iteratively from $\mathsf{f}=1$ to $\mathsf{F}$ the numbers
\begin{align}\label{def-sub-partition}
\mathsf{J}_{\mathsf{f}}:=\min\left\{\mathsf{J}\in\{\mathsf{J}_{\mathsf{f}-1}+1,\dots,\mathsf{B}-\mathsf{F}+\mathsf{f}\}:\quad {\kappa_{\mathsf{J}_{\mathsf{f}-1}}^2-\kappa_{\mathsf{J}}^2}\geq\frac{1-\phi}{\mathsf{F}}\,\big[{\kappa_{\mathsf{A}}^2-\kappa^2_{\mathsf{B}}}\big]\right\}\,,\qquad \mathsf{J}_{0}:=\mathsf{A}\,.
\end{align}
\noindent \underline{\textit{Step 1.}} \textit{The numbers $\mathsf{J}_{1},\mathsf{J}_{2},\dots,\mathsf{J}_{\mathsf{F}}$ are well-defined.}\\
\noindent We show this by induction. First of all, $\mathsf{J}_0$ is explicitly defined. We now assume that also the numbers $\mathsf{J}_1,\mathsf{J}_2,\dots,\mathsf{J}_{\mathsf{f}-1}$ are well-defined for some $\mathsf{f}\in\{1,\dots,\mathsf{F}\}$. The inequality 
$$
\mathsf{J}_{\mathsf{f}-1}\leq \mathsf{B}-\mathsf{F}+\mathsf{f}-1\,.
$$
holds for $\mathsf{f}=1$ due to~\eqref{ineq-sufficient-length} and $\mathsf{J}_0=\mathsf{A}$, and for $\mathsf{f}\in\{2,\dots,\mathsf{F}\}$ by the construction~\eqref{def-sub-partition}.
Thus, the set  $\{\mathsf{J}_{\mathsf{f}-1}+1,\dots,\mathsf{B}-\mathsf{F}+\mathsf{f}\}$ is non-empty and contains $\mathsf{B}-\mathsf{F}+\mathsf{f}$. Hence, it suffices to show~that 
\begin{align}\label{ineq-well-definedness-1}
{\kappa_{\mathsf{J}_{\mathsf{f}-1}}^2-\kappa_{\mathsf{B}-\mathsf{F}+\mathsf{f}}^2}\geq\frac{1-\phi}{\mathsf{F}}\,\big[{\kappa_{\mathsf{A}}^2-\kappa^2_{\mathsf{B}}}\big]\,.
\end{align}
Now, by construction, all $\mathsf{g}=1,\dots,\mathsf{f}-1$ satisfy
\begin{align}\label{ineq-well-definedness-2}
{\kappa_{\mathsf{J}_{\mathsf{g}-1}}^2-\kappa_{{\mathsf{J}_{\mathsf{g}}-1}}^2}<\frac{1-\phi}{\mathsf{F}}\big[{\kappa_{\mathsf{A}}^2-\kappa^2_{\mathsf{B}}}\big]\,.
\end{align}
Combining~\eqref{ineq-kappa-differences} with~\eqref{ineq-well-definedness-2} for $\mathsf{J}=\mathsf{J}_{\mathsf{g}}$ yields for all $\mathsf{g}=1,\dots,\mathsf{f}-1$ the~upper~bound
\begin{align*}
\kappa_{\mathsf{J}_{\mathsf{g}-1}}^2-\kappa_{{\mathsf{J}_{\mathsf{g}}}}^2=\big[\kappa_{\mathsf{J}_{\mathsf{g}-1}}^2-\kappa_{{\mathsf{J}_{\mathsf{g}}-1}}^2\big]+\big[\kappa_{\mathsf{J}_{\mathsf{g}}-1}^2-\kappa_{{\mathsf{J}_{\mathsf{g}}}}^2\big]<\frac{1-\phi}{\mathsf{F}}\,\big[\kappa^{2}_{\mathsf{A}}-\kappa^{2}_{\mathsf{B}}\big]+\frac{\phi}{\mathsf{F}}\,\big[\kappa^{2}_{\mathsf{A}}-\kappa^{2}_{\mathsf{B}}\big]=\frac{1}{\mathsf{F}}\,\big[\kappa^{2}_{\mathsf{A}}-\kappa^{2}_{\mathsf{B}}\big]\,.
\end{align*}
Summing this from $\mathsf{g}=1$ to $\mathsf{f}-1$ reads
\begin{align}\label{ineq-well-definedness-3}
\kappa^2_{\mathsf{A}}-\kappa^2_{\mathsf{J}_{\mathsf{f}-1}}=\sum\limits_{\mathsf{g}=1}^{\mathsf{f}-1}\big[\kappa_{\mathsf{J}_{\mathsf{g}-1}}^2-\kappa_{{\mathsf{J}_{\mathsf{g}}}}^2\big]\leq\frac{\mathsf{f}-1}{\mathsf{F}}\,\big[\kappa^{2}_{\mathsf{A}}-\kappa^{2}_{\mathsf{B}}\big]\,.
\end{align}
Moreover, the sum of the inequality~\eqref{ineq-kappa-differences} for $\mathsf{J}=\mathsf{B}-\mathsf{g}+1$ from $\mathsf{g}=1$ to $\mathsf{F}-\mathsf{f}$ reads
\begin{align}\label{ineq-well-definedness-4}
\kappa^2_{\mathsf{B}-\mathsf{F}+\mathsf{f}}-\kappa_{\mathsf{B}}^2=\sum\limits_{\mathsf{g}=1}^{\mathsf{F}-\mathsf{f}}\big[\kappa_{\mathsf{B}-\mathsf{g}}^2-\kappa^2_{\mathsf{B}-\mathsf{g}+1}\big]\leq \frac{\phi(\mathsf{F}-\mathsf{f})}{\mathsf{F}}\,\big[\kappa^{2}_{\mathsf{A}}-\kappa^{2}_{\mathsf{B}}\big]\,.
\end{align}
Now, inequalities~\eqref{ineq-well-definedness-3},~\eqref{ineq-well-definedness-4} and $\phi < 1$ together with $\mathsf{f}\leq \mathsf{F}$ imply that
\begin{align*}
\begin{split}
\kappa_{\mathsf{J}_{\mathsf{f}-1}}^2-\kappa_{\mathsf{B}-\mathsf{F}+\mathsf{f}}^2&=\big[\kappa^{2}_{\mathsf{A}}-\kappa^{2}_{\mathsf{B}}\big]-\big[\kappa^2_{\mathsf{A}}-\kappa^2_{\mathsf{J}_{\mathsf{f}-1}}\big]-\big[\kappa^2_{\mathsf{B}-\mathsf{F}+\mathsf{f}}-\kappa_{\mathsf{B}}^2\big]\\&\geq\left[1-\frac{\mathsf{f}-1}{\mathsf{F}}-\frac{\phi(\mathsf{F}-\mathsf{f})}{\mathsf{F}}\right]\big[\kappa^{2}_{\mathsf{A}}-\kappa^{2}_{\mathsf{B}}\big]\\&\geq\frac{(\mathsf{F}-\mathsf{f})(1-\phi)+1}{\mathsf{F}}\,\big[\kappa^{2}_{\mathsf{A}}-\kappa^{2}_{\mathsf{B}}\big]-\frac{\phi}{\mathsf{F}}\,\big[{\kappa_{\mathsf{A}}^2-\kappa^2_{\mathsf{B}}}\big]\\
&\geq\frac{1-\phi}{\mathsf{F}}\,\big[{\kappa_{\mathsf{A}}^2-\kappa^2_{\mathsf{B}}}\big]
\end{split}
\end{align*}
and, therefore,~\eqref{ineq-well-definedness-1} is satisfied so that the numbers $\mathsf{J}_1,\mathsf{J}_2,\dots,\mathsf{J}_{\mathsf{F}}$ are indeed well-defined.\hfill $\diamond$

\vspace{.2cm}

Next, we define the numbers
\begin{align}\label{def-sub-partition-bis}
\mathsf{I}_{\mathsf{f}} := \left\{
\begin{array}{ll}
\mathsf{J}_{\mathsf{f}}\,,\quad & \mathsf{f}\in\{0,\dots,\mathsf{F}-1\}\,,\\
\mathsf{B}\,, & \mathsf{f}=\mathsf{F}\,.\\
\end{array}
\right.
\end{align}

\noindent \underline{\textit{Step 2.}} \textit{The numbers $\mathsf{I}_1,\dots,\mathsf{I}_{\mathsf{F}}$ fulfill the condition~\eqref{ineq-subdivision-strict-partition}.}\\
\noindent The sequence of inequalities
\begin{align*}
\mathsf{A}=\mathsf{J}_0<\mathsf{J}_1<\dots<\mathsf{J}_{\mathsf{F}-1}<\mathsf{J}_{\mathsf{F}}\leq \mathsf{B}
\end{align*}
is obvious by construction. This and the definition~\eqref{def-sub-partition-bis} directly imply the condition~\eqref{ineq-subdivision-strict-partition}. \hfill $\diamond$

\vspace{.2cm}

\noindent \underline{\textit{Step 3.}} \textit{The inequality~\eqref{ineq-subdivision-result} is satisfied for all $\mathsf{f}\in\{1,\dots,\mathsf{F}\}$.}\\
\noindent Let $\mathsf{f}\in\{1,\dots,\mathsf{F}\}$. By construction, one has
$$
\kappa_{\mathsf{J}_{\mathsf{f}-1}}^2-\kappa_{\mathsf{J}_{\mathsf{f}}}^2\geq\frac{1-\phi}{\mathsf{F}}\,\big[\kappa^{2}_{\mathsf{A}}-\kappa^{2}_{\mathsf{B}}\big]\,,
$$
which implies
\begin{align}\label{ineq-subdivision-1}
\eta(\mathsf{J}_{\mathsf{f}-1},\mathsf{J}_{\mathsf{f}})=\kappa_{\mathsf{J}_{\mathsf{f}-1}}^{-2}\,\big[\kappa_{\mathsf{J}_{\mathsf{f}-1}}^2-\kappa_{\mathsf{J}_{\mathsf{f}}}^2\big]\geq \kappa_{\mathsf{J}_{\mathsf{f}-1}}^{-2}\,\frac{1-\phi}{\mathsf{F}}\,\big[\kappa^{2}_{\mathsf{A}}-\kappa^{2}_{\mathsf{B}}\big]=\kappa_{\mathsf{J}_{\mathsf{f}-1}}^{-2}\,\kappa_{\mathsf{A}}^{2}\,\frac{1-\phi}{\mathsf{F}}\,\eta({\mathsf{A}},{\mathsf{B}})\geq\frac{1-\phi}{\mathsf{F}}\,\eta({\mathsf{A}},{\mathsf{B}})\,.
\end{align}
If $\mathsf{f}\neq\mathsf{F}$, one has $\eta(\mathsf{I}_{\mathsf{f}-1},\mathsf{I}_{\mathsf{f}})=\eta(\mathsf{J}_{\mathsf{f}-1},\mathsf{J}_{\mathsf{f}})$ due to~\eqref{def-sub-partition-bis} so that~\eqref{ineq-subdivision-1} implies~\eqref{ineq-subdivision-result}.~Further,~one~has
\begin{align}\label{ineq-subdivision-2}
\eta(\mathsf{I}_{\mathsf{F}-1},\mathsf{I}_{\mathsf{F}})=\eta(\mathsf{J}_{\mathsf{F}-1},\mathsf{B})=1-\kappa^2_{\mathsf{B}}\kappa^{-2}_{\mathsf{J}_{\mathsf{F}-1}}\geq 1-\kappa^2_{\mathsf{J}_{\mathsf{F}}}\kappa^{-2}_{\mathsf{J}_{\mathsf{F}-1}}=\eta(\mathsf{J}_{\mathsf{F}-1},\mathsf{J}_{\mathsf{F}})
\end{align}
because $\mathsf{J}_{\mathsf{F}}\leq \mathsf{B}$. The inequality~\eqref{ineq-subdivision-result} for $\mathsf{f}=\mathsf{F}$ then follows from~\eqref{ineq-subdivision-1} and~\eqref{ineq-subdivision-2}.\hfill $\diamond$
\hfill $\square$

\vspace{.2cm}

\noindent\textbf{Proof of Lemma~\ref{lemma-auxiliary-action}.} Let $\Upsilon \in \mathbb{F}_{\mathsf{L},\mathsf{w}}$ and $\Upsilon^{\perp} \in \mathbb{F}_{\mathsf{L},\mathsf{L}-\mathsf{w}}$ such that $\Upsilon\Upsilon^* = W$ and $\Upsilon^{\perp}(\Upsilon^{\perp})^* = W^{\perp}$. Note that $\begin{pmatrix} \Upsilon & v \end{pmatrix} \in \mathbb{F}_{\mathsf{L},\mathsf{w}+1}$ then is a frame for the projection $W+vv^*$. From Lemma~\ref{lemma-complement-action} we have $(\mathcal{T} \cdot W)^{\perp} = (\mathcal{T}^{-1})^* \cdot W^{\perp}$. We now find
\begin{align}\label{ineq-auxiliary-action}
\begin{split}
\|(\mathcal{T} \cdot W)^{\perp}\mathcal{T}v\|^2 &= v^*\mathcal{T}^*((\mathcal{T}^{-1})^* \cdot W^{\perp})\mathcal{T}v\\
&= v^*\mathcal{T}^*(\mathcal{T}^*)^{-1}\Upsilon^{\perp}\left[(\Upsilon^{\perp})^*\mathcal{T}^{-1}(\mathcal{T}^{-1})^*\Upsilon^{\perp}\right]^{-1}(\Upsilon^{\perp})^*\mathcal{T}^{-1}\mathcal{T}v\\
&\geq v^*\Upsilon^{\perp}\left[(\Upsilon^{\perp})^*\|\mathcal{T}^{-1}(\mathcal{T}^{-1})^*\|\Upsilon^{\perp}\right]^{-1}(\Upsilon^{\perp})^*v\\
&= \|\mathcal{T}^{-1}(\mathcal{T}^{-1})^*\|^{-1} v^*\Upsilon^{\perp}(\Upsilon^{\perp})^*v\\
&= \|\mathcal{T}^{-1}\|^{-2} > 0\,,
\end{split}
\end{align}
since $v^*\Upsilon^{\perp}(\Upsilon^{\perp})^*v = v^*W^{\perp}v = v^*v = 1$ (by $Wv = \mathbf{0}$) and $\|\mathcal{T}^{-1}\|^{-2} > 0$ as $\mathcal{T} \in \textnormal{GL}(\mathsf{L},\mathbb{C})$. Therefore also $(\mathcal{T} \cdot W)^{\perp}\mathcal{T}v \neq \mathbf{0}$. In what follows, we will use the Schur complement formula (for matrices $X$, $Y$, $Z$ and $T$ of suitable sizes, with $\det(X) \neq 0 \neq \det(T - ZX^{-1}Y)$):
\[ \begin{pmatrix} X & Y \\ Z & T \end{pmatrix}^{-1} = \begin{pmatrix} X^{-1}+X^{-1}Y(T - ZX^{-1}Y)^{-1}ZX^{-1} & -X^{-1}Y(T - ZX^{-1}Y)^{-1} \\ -(T - ZX^{-1}Y)^{-1}ZX^{-1} & (T - ZX^{-1}Y)^{-1} \end{pmatrix}. \]
Since $\mathcal{T}$ and hence $\mathcal{T}^*\mathcal{T}$is invertible, the square of its smallest singular value $\mu_1(\mathcal{T}^*\mathcal{T})$ must be strictly positive. Now, as $\Upsilon^*\mathcal{T}^*\mathcal{T}\Upsilon \geq \mu_1(\mathcal{T}^*\mathcal{T})\Upsilon^*\Upsilon >\mathbf{0}$ so $\Upsilon^*\mathcal{T}^*\mathcal{T}\Upsilon$ is invertible, and
\[ v^*\mathcal{T}^*\mathcal{T}v - v^*\mathcal{T}^*\mathcal{T}\Upsilon\left(\Upsilon^*\mathcal{T}^*\mathcal{T}\Upsilon\right)^{-1}\Upsilon^*\mathcal{T}^*\mathcal{T}v= v^*\mathcal{T}^*\left[\mathbf{1}_L - (\mathcal{T} \cdot W)\right]\mathcal{T}v = \|(\mathcal{T} \cdot W)^{\perp}\mathcal{T}v\|^2 \]
was proven to be nonzero in~\eqref{ineq-auxiliary-action}, we can apply the Schur complement formula to obtain
\allowdisplaybreaks
\begin{align*}
\mathcal{T} \cdot &\left(W + vv^*\right) = \begin{pmatrix} \mathcal{T}\Upsilon & \mathcal{T}v \end{pmatrix}\begin{pmatrix} \Upsilon^*\mathcal{T}^*\mathcal{T}\Upsilon & \Upsilon^*\mathcal{T}^*\mathcal{T}v \\ v^*\mathcal{T}^*\mathcal{T}\Upsilon & v^*\mathcal{T}^*\mathcal{T}v \end{pmatrix}^{-1} \begin{pmatrix} \Upsilon^*\mathcal{T}^* \\ v^*\mathcal{T}^* \end{pmatrix}\\
&= \mathcal{T}\Upsilon\left(\Upsilon^*\mathcal{T}^*\mathcal{T}\Upsilon\right)^{-1}\left[\mathbf{1}+\Upsilon^*\mathcal{T}^*\mathcal{T}v\|(\mathcal{T} \cdot W)^{\perp}\mathcal{T}v\|^{-2}v^*\mathcal{T}^*\mathcal{T}\Upsilon\left(\Upsilon^*\mathcal{T}^*\mathcal{T}\Upsilon\right)^{-1}\right]\Upsilon^*\mathcal{T}^*\\
&\qquad-\mathcal{T}\Upsilon\left(\Upsilon^*\mathcal{T}^*\mathcal{T}\Upsilon\right)^{-1}\Upsilon^*\mathcal{T}^*\mathcal{T}v\|(\mathcal{T} \cdot W)^{\perp}\mathcal{T}v\|^{-2}v^*\mathcal{T}^*\\
&\qquad-\mathcal{T}v\|(\mathcal{T} \cdot W)^{\perp}\mathcal{T}v\|^{-2}v^*\mathcal{T}^*\mathcal{T}\Upsilon\left(\Upsilon^*\mathcal{T}^*\mathcal{T}\Upsilon\right)^{-1}\Upsilon^*\mathcal{T}^*\\
&\qquad+\mathcal{T}v\|(\mathcal{T} \cdot W)^{\perp}\mathcal{T}v\|^{-2}v^*\mathcal{T}^*\\
&= \mathcal{T} \cdot W + \left[\mathbf{1}_L - \left(\mathcal{T} \cdot W\right)\right]\mathcal{T}v\|(\mathcal{T} \cdot W)^{\perp}\mathcal{T}v\|^{-2}v^*\mathcal{T}^*\left[\mathbf{1}_L - \left(\mathcal{T} \cdot W\right)\right]\\
&= \mathcal{T}\cdot W+\left[((\mathcal{T}\cdot W)^{\perp}\mathcal{T})\circ v\right]\left[((\mathcal{T}\cdot W)^{\perp}\mathcal{T})\circ v\right]^*\,,
\end{align*}
which is the desired formula.
\hfill $\square$

\vspace{.2cm}

\noindent\textbf{Proof of Lemma~\ref{lemma-vector-expansion}.} Let $(W,v) \in \mathfrak{W}$ and $\mathtt{P} \in \mathfrak{P}$. Using the reversed triangle inequality,~\eqref{def-Z} in the third step,~\eqref{ineq-expansion} in the fourth step and~\eqref{ineq-norm-3} in the sixth step, the bound~\eqref{ineq-vector-expansion-1} follows:
\begin{align}\label{ineq-vector-expansion-proof}
\begin{split}
&\big|\|\mathfrak{d}([(e^{\lambda\mathtt{P}}\cdot W)^{\perp}e^{\lambda\mathtt{P}}]\cdot v)\|^2-\|\mathfrak{d}(v)\|^2\big|\\
&=\big|\|\hat{\Psi}^*\left(e^{\lambda\mathtt{P}}\cdot(W+vv^*)-e^{\lambda\mathtt{P}}\cdot W\right)\hat{\Psi}\|-\|\hat{\Psi}^*vv^*\hat{\Psi}\|\big|\\
&\leq\|\hat{\Psi}^*\left(e^{\lambda\mathtt{P}}\cdot(W+vv^*)-e^{\lambda\mathtt{P}}\cdot W-vv^*\right)\hat{\Psi}\|\\
&\leq\|\left[e^{\lambda\mathtt{P}}\cdot(W+vv^*)-(W+vv^*)\right] - \left[e^{\lambda\mathtt{P}}\cdot W-W\right]\|\\
&\leq\lambda\,\|\mathtt{X}(W+vv^*,\mathtt{P})-\mathtt{X}(W,\mathtt{P})\|+\lambda^2\,\|\mathtt{Y}(W+vv^*,\mathtt{P})\|+\lambda^2\,\|\mathtt{Y}(W,\mathtt{P})\|\\
&\quad+\lambda^3\,\|\mathtt{Z}^{(\lambda)}(W+vv^*,\mathtt{P})\|+\lambda^3\,\|\mathtt{Z}^{(\lambda)}(W,\mathtt{P})\|\\
&\leq\lambda\,\|\mathtt{X}(W+vv^*,\mathtt{P})-\mathtt{X}(W,\mathtt{P})\|+\left(\mbox{\small $\frac{3}{2}$}+\mbox{\small $\frac{3}{2}$}\right)\lambda^2+(20+20)\lambda^3\\
&=\lambda\,\|(vv^*)^{\perp}\left[(W+vv^*)^{\perp}\mathtt{P}-W\mathtt{P}^*\right]vv^*+\textnormal{h.c.}\|+3\,\lambda^2+40\,\lambda^3\\
&\leq \lambda\,\|(W+vv^*)^{\perp}\mathtt{P}-W\mathtt{P}^*\|+4\,\lambda^2\\
&= \lambda\,\|\left[(W+vv^*)^{\perp}\mathtt{P}-W\mathtt{P}^*\right]^*\left[(W+vv^*)^{\perp}\mathtt{P}-W\mathtt{P}^*\right]\|^{\frac{1}{2}}+4\,\lambda^2\\
&= \lambda\,\|\mathtt{P}^*(W+vv^*)^{\perp}\mathtt{P}+\mathtt{P}W\mathtt{P}^*\|^{\frac{1}{2}}+4\,\lambda^2\\
&\leq 2^{\frac{1}{2}}\lambda + 4\,\lambda^2\\
&\leq \mbox{\small $\frac{3}{2}$}\,\lambda\,.
\end{split}
\end{align}
As for~\eqref{ineq-vector-expansion-3}, using Lemma~\ref{lemma-auxiliary-action},~\eqref{def-Z} and~\eqref{ineq-expansion} we find
\begin{align*}
&\big|\|\mathfrak{d}([(e^{\lambda\mathtt{P}}\cdot W)^{\perp}e^{\lambda\mathtt{P}}]\circ v)\|^2-\|\mathfrak{d}(v)\|^2 - \lambda\,\mathbf{A}_{\mathfrak{d}}(W,v,\mathtt{P})\big|\\
&=\big|\|\hat{\Psi}^*([(e^{\lambda\mathtt{P}}\cdot W)^{\perp}e^{\lambda\mathtt{P}}]\circ v)([(e^{\lambda\mathtt{P}}\cdot W)^{\perp}e^{\lambda\mathtt{P}}]\circ v)^*\hat{\Psi}\|-\|\hat{\Psi}^*vv^*\hat{\Psi}\| - \lambda\,\mathbf{A}_{\mathfrak{d}}(W,v,\mathtt{P})\big|\\
&=\Big|\operatorname{tr}\left[\hat{\Psi}^*\left(\left[e^{\lambda\mathtt{P}}\cdot(W+vv^*)-(W+vv^*)-\lambda\,\mathtt{X}(W+vv^*,\mathtt{P})\right]-\left[e^{\lambda\mathtt{P}}\cdot W-W-\lambda\,\mathtt{X}(W,\mathtt{P})\right]\right)\hat{\Psi}\right]\Big|\\
&=\lambda^2\,\Big|\operatorname{tr}\Big[\hat{\Psi}^*\Big(\big[\mathtt{Y}(W+vv^*,\mathtt{P})-\mathtt{Y}(W,\mathtt{P})\big]+\lambda\big[\mathtt{Z}^{(\lambda)}(W+vv^*,\mathtt{P})-\mathtt{Z}^{(\lambda)}(W,\mathtt{P})\big]\Big)\hat{\Psi}\Big]\Big|\\
&\leq \lambda^2\|\mathtt{Y}(W+vv^*,\mathtt{P})-\mathtt{Y}(W,\mathtt{P})\|\operatorname{rk}\big(\mathtt{Y}(W+vv^*,\mathtt{P})-\mathtt{Y}(W,\mathtt{P})\big)\\
&\quad+\lambda^3\|\mathtt{Z}^{(\lambda)}(W+vv^*,\mathtt{P})-\mathtt{Z}^{(\lambda)}(W,\mathtt{P})\|\operatorname{rk}\big(\mathtt{Z}^{(\lambda)}(W+vv^*,\mathtt{P})-\mathtt{Z}^{(\lambda)}(W,\mathtt{P})\big)\\
&\leq 3\,\lambda^2\left(\|\mathtt{Y}(W+vv^*,\mathtt{P})\|+\|\mathtt{Y}(W,\mathtt{P})\|\right)+4\,\lambda^3\left(\|\mathtt{Z}^{(\lambda)}(W+vv^*,\mathtt{P})\|+\|\mathtt{Z}^{(\lambda)}(W,\mathtt{P})\|\right)\\
&\leq 2 \cdot \mbox{\small $\frac{3}{2}$} \cdot 3\,\lambda^2+2 \cdot 20 \cdot 4\,\lambda^3\,.
\end{align*}
This and~\eqref{ineq-vector-expansion-proof} up to the penultimate step then imply
\begin{align*}
&\lambda\,\big|\mathbf{A}_{\mathfrak{d}}(W,v,\mathtt{P})\big|\\
&\leq \big|\|\mathfrak{d}([(e^{\lambda\mathtt{P}}\cdot W)^{\perp}e^{\lambda\mathtt{P}}]\circ v)\|^2-\|\mathfrak{d}(v)\|^2 - \lambda\,\mathbf{A}_{\mathfrak{d}}(W,v,\mathtt{P})\big| + \big|\|\mathfrak{d}([(e^{\lambda\mathtt{P}}\cdot W)^{\perp}e^{\lambda\mathtt{P}}]\circ v)\|^2-\|\mathfrak{d}(v)\|^2\big|\\
&\leq 9\,\lambda^2+160\,\lambda^3+2^{\frac{1}{2}}\lambda+4\,\lambda^2\,.
\end{align*}
This means that $\big|\mathbf{A}_{\mathfrak{d}}(W,v,\mathtt{P})\big| \leq 2^{\frac{1}{2}} + 13\,\lambda+164\,\lambda^2$ holds for all $\lambda > 0$, which implies~\eqref{ineq-vector-expansion-2}.
\hfill $\square$

\vspace{.2cm}

\noindent\textbf{Proof of Lemma~\ref{lemma-vector-contraction}.} Let $\Upsilon\in\mathbb{F}_{\mathsf{L},\mathsf{w}}$ be such that $W=\Upsilon\Upsilon^*$.
As in~\eqref{ineq-trace-contraction-1} and~\eqref{ineq-trace-contraction-2}, one has
\begin{align}\label{ineq-trace-vector-contraction-1}
\operatorname{tr}\Big[\hat{\chi}_{\mathsf{m}}^*(\mathcal{R}\cdot (W+vv^*))\hat{\chi}_{\mathsf{m}}\Big]\leq\operatorname{tr}\left[\hat{\chi}_{\mathsf{m}}^* (W+vv^*)\left[\mathbf{1}_{\mathsf{L}}+\tau_{\mathsf{m}}\,\hat{\zeta}_{\mathsf{m}}\hat{\zeta}_{\mathsf{m}}^*\right](W+vv^*)\hat{\chi}_{\mathsf{m}}\right]
\end{align}
and
\begin{align}\label{ineq-trace-vector-contraction-2}
\operatorname{tr}\left[\hat{\zeta}_{\mathsf{m}}^*(\mathcal{R}\cdot (W+vv^*))\hat{\zeta}_{\mathsf{m}}\right]\geq\operatorname{tr}\left[\hat{\zeta}_{\mathsf{m}}^* (W+vv^*)\Big[\mathbf{1}_{\mathsf{L}}+\tau_{\mathsf{m}}\,\hat{\chi}_{\mathsf{m}}\hat{\chi}_{\mathsf{m}}^*\Big](W+vv^*)\hat{\zeta}_{\mathsf{m}}\right]\,.
\end{align}
Moreover, using again the notation $\mu_1(T)$ for the smallest eigenvalue of a matrix $T$, we estimate
\begin{align}\label{ineq-trace-vector-contraction-3}
\begin{split}
\tau_{\mathsf{m}}\,\operatorname{tr}\left[\hat{\chi}_{\mathsf{m}}^*(\Upsilon,v)\,(\Upsilon,v)^*\hat{\zeta}_{\mathsf{m}}\hat{\zeta}_{\mathsf{m}}^* (\Upsilon,v)\,(\Upsilon,v)^*\hat{\chi}_{\mathsf{m}}\right]&\geq\tau_{\mathsf{m}}\,\mu_1\big((\Upsilon,v)^*\hat{\zeta}_{\mathsf{m}}\hat{\zeta}_{\mathsf{m}}^* (\Upsilon,v)\big)\,\operatorname{tr}\left[\hat{\chi}_{\mathsf{m}}^*(\Upsilon,v)(\Upsilon,v)^*\hat{\chi}_{\mathsf{m}}\right]\\
&\geq \tau_{\mathsf{m}}\,\left(\|\llc_{\mathsf{m}}(v)\|^2-\operatorname{tr}\left[(\hat{\zeta}_{\mathsf{m}}^{\perp})^*W\hat{\zeta}_{\mathsf{m}}^{\perp} \right]\right)\|\ulc_{\mathsf{m}}(v)\|^2\\
&\geq \tau_{\mathsf{m}}\,\|\ulc_{\mathsf{m}}(v)\|^2\,\|\llc_{\mathsf{m}}(v)\|^2-\operatorname{tr}\left[(\hat{\zeta}_{\mathsf{m}}^{\perp})^*W\hat{\zeta}_{\mathsf{m}}^{\perp} \right]
\end{split}
\end{align}
by using
\begin{align*}
\begin{split}
\mu_1\big((\Upsilon,v)^*\hat{\zeta}_{\mathsf{m}}\hat{\zeta}_{\mathsf{m}}^* (\Upsilon,v)\big)&\geq \operatorname{tr}\left[(\Upsilon,v)^*\hat{\zeta}_{\mathsf{m}}\hat{\zeta}_{\mathsf{m}}^* (\Upsilon,v)\right]-\mathsf{w}\\
&=\operatorname{tr}\left[v^*\hat{\zeta}_{\mathsf{m}}^* \hat{\zeta}_{\mathsf{m}}v\right]+\operatorname{tr}\left[\Upsilon^*\hat{\zeta}_{\mathsf{m}}^* \hat{\zeta}_{\mathsf{m}}\Upsilon\right]-\operatorname{tr}\left[\Upsilon^*\Upsilon\right]\\
&=\|\hat{\zeta}_{\mathsf{m}}^*v\|^2-\operatorname{tr}\left[\Upsilon^*\hat{\zeta}_{\mathsf{m}}^{\perp}(\hat{\zeta}_{\mathsf{m}}^{\perp})^* \Upsilon\right]\\
&=\|\llc_{\mathsf{m}}(v)\|^2-\operatorname{tr}\left[(\hat{\zeta}_{\mathsf{m}}^{\perp})^*W\hat{\zeta}_{\mathsf{m}}^{\perp} \right]
\end{split}
\end{align*}
and
\begin{align*}
\operatorname{tr}\left[\hat{\chi}_{\mathsf{m}}^*(\Upsilon,v)(\Upsilon,v)^*\hat{\chi}_{\mathsf{m}}\right]=\operatorname{tr}\left[\hat{\chi}_{\mathsf{m}}^*(W+vv^*)\hat{\chi}_{\mathsf{m}}\right]\geq\operatorname{tr}\left[\hat{\chi}_{\mathsf{m}}^*vv^*\hat{\chi}_{\mathsf{m}}\right]=\|\hat{\chi}_{\mathsf{m}}^*v\|^2=\|\ulc_{\mathsf{m}}(v)\|^2\,.
\end{align*}
Now for the proof of~\eqref{ineq-vector-contraction-1}, we use~\eqref{eq-auxiliary-action},~\eqref{ineq-trace-vector-contraction-1}, $\hat{\chi}_{\mathsf{m}}\hat{\chi}_{\mathsf{m}}^*\leq \hat{\zeta}_{\mathsf{m}}^{\perp}(\hat{\zeta}_{\mathsf{m}}^{\perp})^*$ and~\eqref{ineq-trace-vector-contraction-3} in the second, fourth, seventh and eighth step when estimating 
\begin{align*}
&\| \ulc_{\mathsf{m}}\left(\left((\mathcal{R} \cdot W)^{\perp}\mathcal{R}\right) \circ v\right) \|^2\\
&=\operatorname{tr}\left[ \hat{\chi}_{\mathsf{m}}^*\left[\left((\mathcal{R} \cdot W)^{\perp}\mathcal{R}\right) \circ v\right]\left[\left((\mathcal{R} \cdot W)^{\perp}\mathcal{R}\right) \circ v\right]^*\hat{\chi}_{\mathsf{m}}\right]\\
& = \operatorname{tr}\left[\hat{\chi}_{\mathsf{m}}^*\left(\mathcal{R}\cdot(W+vv^*)\right)\hat{\chi}_{\mathsf{m}}\right] - \operatorname{tr}\left[\hat{\chi}_{\mathsf{m}}^*\left(\mathcal{R}\cdot W\right)\hat{\chi}_{\mathsf{m}}\right]\\
& \leq \operatorname{tr}\left[\hat{\chi}_{\mathsf{m}}^*\left(\mathcal{R}\cdot(W+vv^*)\right)\hat{\chi}_{\mathsf{m}}\right]\\
&\leq\operatorname{tr}\left[\hat{\chi}_{\mathsf{m}}^* (W+vv^*)\left[\mathbf{1}_{\mathsf{L}}-\tau_{\mathsf{m}}\,\hat{\zeta}_{\mathsf{m}}\hat{\zeta}_{\mathsf{m}}^*\right](W+vv^*)\hat{\chi}_{\mathsf{m}}\right]\\
&= \|\hat{\chi}_{\mathsf{m}}^*v\|^2+ \operatorname{tr}\left[\hat{\chi}_{\mathsf{m}}^* W\hat{\chi}_{\mathsf{m}}\right]-\tau_{\mathsf{m}}\,\operatorname{tr}\left[\hat{\chi}_{\mathsf{m}}^*(W+vv^*)\hat{\zeta}_{\mathsf{m}}\hat{\zeta}_{\mathsf{m}}^* (W+vv^*)\hat{\chi}_{\mathsf{m}}\right]\\
&=\|\ulc_{\mathsf{m}}(v)\|^2+\operatorname{tr}\left[\Upsilon^*\hat{\chi}_{\mathsf{m}}\hat{\chi}_{\mathsf{m}}^*\Upsilon\right]-\tau_{\mathsf{m}}\,\operatorname{tr}\left[\hat{\chi}_{\mathsf{m}}^*(\Upsilon,v)\,(\Upsilon,v)^*\hat{\zeta}_{\mathsf{m}}\hat{\zeta}_{\mathsf{m}}^* (\Upsilon,v)\,(\Upsilon,v)^*\hat{\chi}_{\mathsf{m}}\right]\\
&\leq \|\ulc_{\mathsf{m}}(v)\|^2+\operatorname{tr}\left[\Upsilon^*\hat{\zeta}_{\mathsf{m}}^{\perp}(\hat{\zeta}_{\mathsf{m}}^{\perp})^*\Upsilon\right]-\tau_{\mathsf{m}}\,\operatorname{tr}\left[\hat{\chi}_{\mathsf{m}}^*(\Upsilon,v)\,(\Upsilon,v)^*\hat{\zeta}_{\mathsf{m}}\hat{\zeta}_{\mathsf{m}}^* (\Upsilon,v)\,(\Upsilon,v)^*\hat{\chi}_{\mathsf{m}}\right]\\
&\leq\|\ulc_{\mathsf{m}}(v)\|^2\left[1-\tau_{\mathsf{m}}\,\|\llc_{\mathsf{m}}(v)\|^2\right]+2\,\operatorname{tr}\left[(\hat{\zeta}^{\perp}_{\mathsf{m}})^*W\hat{\zeta}^{\perp}_{\mathsf{m}}\right]\,.
\end{align*}
As for~\eqref{ineq-vector-contraction-2}, let us use~\eqref{eq-auxiliary-action},~\eqref{ineq-trace-vector-contraction-2} and~\eqref{ineq-trace-vector-contraction-3} in the second, third and seventh step to show
\begin{align*}
&\| \llc_{\mathsf{m}}\left(\left((\mathcal{R} \cdot W)^{\perp}\mathcal{R}\right) \circ v\right) \|^2\\
&=\operatorname{tr}\left[ \hat{\zeta}_{\mathsf{m}}^*\left[\left((\mathcal{R} \cdot W)^{\perp}\mathcal{R}\right) \circ v\right]\left[\left((\mathcal{R} \cdot W)^{\perp}\mathcal{R}\right) \circ v\right]^*\hat{\zeta}_{\mathsf{m}}\right]\\
& = \operatorname{tr}\left[\hat{\zeta}_{\mathsf{m}}^*\left(\mathcal{R}\cdot(W+vv^*)\right)\hat{\zeta}_{\mathsf{m}}\right] - \operatorname{tr}\left[\hat{\zeta}_{\mathsf{m}}^*\left(\mathcal{R}\cdot W\right)\hat{\zeta}_{\mathsf{m}}\right]\\
&\geq\operatorname{tr}\left[\hat{\zeta}_{\mathsf{m}}^* (W+vv^*)\left[\mathbf{1}_{\mathsf{L}}+\tau_{\mathsf{m}}\,\hat{\chi}_{\mathsf{m}}\hat{\chi}_{\mathsf{m}}^*\right](W+vv^*)\hat{\zeta}_{\mathsf{m}}\right]-\operatorname{tr}\left[\left(\mathcal{R}\cdot W\right)\hat{\zeta}_{\mathsf{m}}\hat{\zeta}_{\mathsf{m}}^*\left(\mathcal{R}\cdot W\right)\right]\\
&\geq\operatorname{tr}\left[\hat{\zeta}_{\mathsf{m}}^* (W+vv^*)\hat{\zeta}_{\mathsf{m}}\right]+\tau_{\mathsf{m}}\,\operatorname{tr}\left[\hat{\zeta}_{\mathsf{m}}^*(W+vv^*)\hat{\chi}_{\mathsf{m}}\hat{\chi}_{\mathsf{m}}^* (W+vv^*)\hat{\zeta}_{\mathsf{m}}\right]-\mathsf{w}\\
&=\|\hat{\zeta}_{\mathsf{m}}^*v\|^2+\operatorname{tr}\left[\Upsilon^*\hat{\zeta}_{\mathsf{m}}\hat{\zeta}_{\mathsf{m}}^*\Upsilon\right]+\tau_{\mathsf{m}}\,\operatorname{tr}\left[\hat{\chi}_{\mathsf{m}}^*(\Upsilon,v)\,(\Upsilon,v)^*\hat{\zeta}_{\mathsf{m}}\hat{\zeta}_{\mathsf{m}}^* (\Upsilon,v)\,(\Upsilon,v)^*\hat{\chi}_{\mathsf{m}}\right]-\operatorname{tr}\left[\Upsilon^*\Upsilon\right]\\
&=\|\llc_{\mathsf{m}}(v)\|^2+\tau_{\mathsf{m}}\,\operatorname{tr}\left[\hat{\chi}_{\mathsf{m}}^*(\Upsilon,v)\,(\Upsilon,v)^*\hat{\zeta}_{\mathsf{m}}\hat{\zeta}_{\mathsf{m}}^* (\Upsilon,v)\,(\Upsilon,v)^*\hat{\chi}_{\mathsf{m}}\right]-\operatorname{tr}\left[\Upsilon^*\hat{\zeta}_{\mathsf{m}}^{\perp}(\hat{\zeta}_{\mathsf{m}}^{\perp})^*\Upsilon\right]\\
&\geq\|\llc_{\mathsf{m}}(v)\|^2\left[1+\tau_{\mathsf{m}}\,\|\ulc_{\mathsf{m}}(v)\|^2\right]-2\,\operatorname{tr}\left[(\hat{\zeta}^{\perp}_{\mathsf{m}})^*W\hat{\zeta}^{\perp}_{\mathsf{m}}\right]\,,
\end{align*}
concluding the proof.
\hfill $\square$

\vspace{.2cm}

\noindent\textbf{Proof of Corollary~\ref{lemma-deterministic-vector}.} 
Inequality~\eqref{ineq-deterministic-vector-1} follows from~\eqref{ineq-vector-expansion-1},~\eqref{ineq-vector-contraction-1} and~\eqref{ineq-trace-order-lambda}:
\begin{align*}
\| \ulc_{\mathsf{m}}\left(\left(((e^{\lambda\mathtt{P}}\mathcal{R}) \cdot W)^{\perp}(e^{\lambda\mathtt{P}}\mathcal{R})\right) \circ v\right) \|^2& \leq \| \ulc_{\mathsf{m}}\left(\left((\mathcal{R} \cdot W)^{\perp}\mathcal{R}\right) \circ v\right) \|^2+\frac{3}{2}\lambda\\
&\leq \|\ulc_{\mathsf{m}}(v)\|^2\left[1-{\tau}_{\mathsf{m}}\,\|\llc_{\mathsf{m}}(v)\|^2\right]+2\,\operatorname{tr}\left[(\hat{\zeta}^{\perp}_{\mathsf{m}})^*W\hat{\zeta}^{\perp}_{\mathsf{m}}\right]+\frac{3}{2}\lambda\\
&\leq \|\ulc_{\mathsf{m}}(v)\|^2+2\,\operatorname{tr}\left[(\hat{\zeta}^{\perp}_{\mathsf{m}})^*W\hat{\zeta}^{\perp}_{\mathsf{m}}\right]+\frac{3}{2}\lambda\\
&\leq \|\ulc_{\mathsf{m}}(v)\|^2+\frac{1}{4}\lambda+\frac{3}{2}\lambda\\
&= \|\ulc_{\mathsf{m}}(v)\|^2+\frac{7}{4}\lambda\,.
\end{align*}
Moreover,  from~\eqref{ineq-vector-expansion-1},~\eqref{ineq-vector-contraction-2} and~\eqref{ineq-trace-order-lambda}:
\begin{align*}
\| \llc_{\mathsf{m}}\left(\left(((e^{\lambda\mathtt{P}}\mathcal{R}) \cdot W)^{\perp}(e^{\lambda\mathtt{P}}\mathcal{R})\right) \circ v\right) \|^2& \geq \| \llc_{\mathsf{m}}\left(\left((\mathcal{R} \cdot W)^{\perp}\mathcal{R}\right) \circ v\right) \|^2-\frac{3}{2}\lambda\\
&\geq \|\llc_{\mathsf{m}}(v)\|^2\left[1+{\tau}_{\mathsf{m}}\,\|\ulc_{\mathsf{m}}(v)\|^2\right]-2\,\operatorname{tr}\left[(\hat{\zeta}^{\perp}_{\mathsf{m}})^*W\hat{\zeta}^{\perp}_{\mathsf{m}}\right]-\frac{3}{2}\lambda\\
&\geq \|\llc_{\mathsf{m}}(v)\|^2-2\,\operatorname{tr}\left[(\hat{\zeta}^{\perp}_{\mathsf{m}})^*W\hat{\zeta}^{\perp}_{\mathsf{m}}\right]-\frac{3}{2}\lambda\\
&\geq \|\llc_{\mathsf{m}}(v)\|^2-\frac{1}{4}\lambda-\frac{3}{2}\lambda\\
&= \|\llc_{\mathsf{m}}(v)\|^2-\frac{7}{4}\lambda\,,
\end{align*}
which is inequality~\eqref{ineq-deterministic-vector-2}.
\hfill $\square$

\vspace{.2cm}

\noindent\textbf{Proof of Corollary~\ref{coro-ladder}.} 
Inequality~\eqref{ineq-diminish-strictly} follows from~\eqref{ineq-vector-expansion-1},~\eqref{ineq-vector-contraction-1},~\eqref{ineq-trace-order-lambda} and~\eqref{ineq-ladder}:
\begin{align*}
\| \ulc_{\mathsf{m}}\left(\left(((e^{\lambda\mathtt{P}}\mathcal{R}) \cdot W)^{\perp}(e^{\lambda\mathtt{P}}\mathcal{R})\right) \circ v\right) \|^2& \leq \| \ulc_{\mathsf{m}}\left(\left((\mathcal{R} \cdot W)^{\perp}\mathcal{R}\right) \circ v\right) \|^2+\frac{3}{2}\lambda\\
&\leq \|\ulc_{\mathsf{m}}(v)\|^2\left[1-{\tau}_{\mathsf{m}}\,\|\llc_{\mathsf{m}}(v)\|^2\right]+2\,\operatorname{tr}\left[(\hat{\zeta}^{\perp}_{\mathsf{m}})^*W\hat{\zeta}^{\perp}_{\mathsf{m}}\right]+\frac{3}{2}\lambda\\
&= \|\ulc_{\mathsf{m}}(v)\|^2-2^{-2}\lambda-\left[\tau_{\mathsf{m}}\,\|\ulc_{\mathsf{m}}(v)\|^2\|\llc_{\mathsf{m}}(v)\|^2-2\lambda\right]\\
&\qquad-2\left(2^{-3}\lambda-\operatorname{tr}\left[(\hat{\zeta}^{\perp}_{\mathsf{m}})^*W\hat{\zeta}^{\perp}_{\mathsf{m}}\right]\right)\\
&\leq \|\ulc_{\mathsf{m}}(v)\|^2-2^{-2}\lambda\,.
\end{align*}
Moreover,  from~\eqref{ineq-vector-expansion-1},~\eqref{ineq-vector-contraction-2},~\eqref{ineq-trace-order-lambda} and~\eqref{ineq-ladder},
\begin{align*}
\| \llc_{\mathsf{m}}\left(\left(((e^{\lambda\mathtt{P}}\mathcal{R}) \cdot W)^{\perp}(e^{\lambda\mathtt{P}}\mathcal{R})\right) \circ v\right) \|^2& \geq \| \llc_{\mathsf{m}}\left(\left((\mathcal{R} \cdot W)^{\perp}\mathcal{R}\right) \circ v\right) \|^2-\frac{3}{2}\lambda\\
&\geq \|\llc_{\mathsf{m}}(v)\|^2\left[1+{\tau}_{\mathsf{m}}\,\|\ulc_{\mathsf{m}}(v)\|^2\right]-2\,\operatorname{tr}\left[(\hat{\zeta}^{\perp}_{\mathsf{m}})^*W\hat{\zeta}^{\perp}_{\mathsf{m}}\right]-\frac{3}{2}\lambda\\
&= \|\llc_{\mathsf{m}}(v)\|^2+2^{-2}\lambda+\left[\tau_{\mathsf{m}}\,\|\ulc_{\mathsf{m}}(v)\|^2\|\llc_{\mathsf{m}}(v)\|^2-2\lambda\right]\\
&\qquad+2\left(2^{-3}\lambda-\operatorname{tr}\left[(\hat{\zeta}^{\perp}_{\mathsf{m}})^*W\hat{\zeta}^{\perp}_{\mathsf{m}}\right]\right)\\
&\geq \|\llc_{\mathsf{m}}(v)\|^2+2^{-2}\lambda\,,
\end{align*}
which is inequality~\eqref{ineq-enlarge-strictly}. 
\hfill $\square$

\vspace{.2cm}

\noindent \textbf{Proof of Lemma~\ref{lemma-allowed-movements}.} Let $\mathtt{P}\in\mathfrak{P}$. Recall the assumption that the dynamics lies in $\mathfrak{Q}$ persistently. The satisfaction of~\eqref{ineq-ladder} is then sufficient for the validity of~\eqref{ineq-diminish-strictly} and~\eqref{ineq-enlarge-strictly} by Corollary~\ref{coro-ladder}.
\begin{enumerate}
\item Let $(W,v)\in\mathfrak{W}\setminus(\mathfrak{C}^{\sigma}_{\mathsf{m}}\cup\mathfrak{A}^{\sigma,\tau}_{\mathsf{m}})$. Then,~\eqref{ineq-ladder-1} and~\eqref{ineq-ladder-2} are satisfied and imply~\eqref{ineq-ladder}. Thus, by Corollary~\ref{coro-ladder},~\eqref{ineq-diminish-strictly} and~\eqref{ineq-enlarge-strictly} hold. Therefore, $\|\llc_{\mathsf{m}}(v)\|^2$ will be increased at least by $2^{-2}\lambda$ per time step until~\eqref{ineq-ladder} is violated. The increase of $\|\llc_{\mathsf{m}}(v)\|^2$ guarantees that~\eqref{ineq-ladder-2} remains valid so that~\eqref{ineq-ladder} can only be violated if~\eqref{ineq-ladder-1} is violated. In conclusion, $\|\llc_{\mathsf{m}}(v)\|^2$ will be increased strictly at least until~\eqref{ineq-ladder-1} is violated. In other words, there is a deterministic run into $\mathfrak{C}^{\sigma}_{\mathsf{m}}$ and the dynamics will not enter $\mathfrak{A}^{\sigma,\tau}_{\mathsf{m}}$ before having entered $\mathfrak{C}^{\sigma}_{\mathsf{m}}$.
\item If $(W,v)\hspace{-0.55mm}\in\hspace{-0.55mm}\mathfrak{C}^{\sigma}_{\mathsf{m}}\hspace{-0.55mm}\setminus\hspace{-0.55mm}\mathfrak{A}^{\sigma,\tau}_{\mathsf{m}}$, so $\|\ulc_{\mathsf{m}}(v)\|^2\hspace{-0.55mm}\leq\hspace{-0.55mm} \sigma$ and $\|\llc_{\mathsf{m}}(v)\|^2\hspace{-0.55mm}>\hspace{-0.55mm}\frac{2}{\sigma}\frac{\lambda}{\tau}$, then~\eqref{ineq-vector-expansion-1},~\eqref{ineq-vector-contraction-1} and~\eqref{ineq-trace-order-lambda}~imply
\begin{align*}
\| \ulc_{\mathsf{m}}\left(\left(((e^{\lambda\mathtt{P}}\mathcal{R}) \cdot W)^{\perp}(e^{\lambda\mathtt{P}}\mathcal{R})\right) \circ v\right) \|^2& \leq \| \ulc_{\mathsf{m}}\left(\left((\mathcal{R} \cdot W)^{\perp}\mathcal{R}\right) \circ v\right) \|^2+\frac{3}{2}\lambda\\
&\leq \|\ulc_{\mathsf{m}}(v)\|^2\left[1-{\tau}_{\mathsf{m}}\,\|\llc_{\mathsf{m}}(v)\|^2\right]+2\,\operatorname{tr}\left[(\hat{\zeta}^{\perp}_{\mathsf{m}})^*W\hat{\zeta}^{\perp}_{\mathsf{m}}\right]+\frac{3}{2}\lambda\\
&\leq \sigma\left[1-{\tau}\,\frac{2}{\sigma}\frac{\lambda}{\tau}\right]+2\cdot2^{-3}\lambda+\frac{3}{2}\lambda\\
&\leq \sigma-2^{-2}\lambda\\&<\sigma\,,
\end{align*}
\textit{i.e.}, $e^{\lambda\mathtt{P}}\mathcal{R}\star (W,v) \in\mathfrak{C}^{\sigma}_{\mathsf{m}}$.
\item Let $(W,v)\in\mathfrak{W}\setminus\mathfrak{A}^{\sigma,\tau}_{\mathsf{m}+1}$ such that $e^{\lambda\mathtt{P}}\mathcal{R}\star (W,v)\in\mathfrak{A}^{\sigma,\tau}_{\mathsf{m}+1}$ and suppose that $(W,v)\not\in\mathfrak{I}_{\mathsf{m}+1/2}^{\sigma,\tau}$,~\textit{i.e.},
$$
\neg\left(\|\ulc_{\mathsf{m}}(v)\|^2> \sigma\geq\|\ulc_{\mathsf{m}+1}(v)\|^2\qquad\wedge\qquad \|\llc_{\mathsf{m}}(v)\|^2\leq \frac{2}{\sigma}\frac{\lambda}{\tau}<\|\llc_{\mathsf{m}+1}(v)\|^2 \right)\,.
$$
Therefore, at least one of the following holds:
\begin{enumerate}
\item $\|\ulc_{\mathsf{m}}(v)\|^2\leq \sigma\,,$
\item $\|\ulc_{\mathsf{m}+1}(v)\|^2> \sigma\,,$
\item $\|\llc_{\mathsf{m}}(v)\|^2> \frac{3}{\sigma}\frac{\lambda}{\tau}\,,$
\item $\|\llc_{\mathsf{m}+1}(v)\|^2\leq \frac{3}{\sigma}\frac{\lambda}{\tau}\,.$
\end{enumerate}
We find a contradiction in each case.
\begin{enumerate}
\item The inequalities~\eqref{ineq-vector-expansion-1},~\eqref{ineq-vector-contraction-1},~\eqref{ineq-trace-order-lambda} and~\eqref{ineq-sigma-tau} allow for the estimate
\begin{align*}
\| \ulc_{\mathsf{m}}\left(\left(((e^{\lambda\mathtt{P}}\mathcal{R}) \cdot W)^{\perp}(e^{\lambda\mathtt{P}}\mathcal{R})\right) \circ v\right) \|^2 &\leq \| \ulc_{\mathsf{m}}\left(\left((\mathcal{R} \cdot W)^{\perp}\mathcal{R}\right) \circ v\right) \|^2+\frac{3}{2}\lambda\\
&\leq \|\ulc_{\mathsf{m}}(v)\|^2+2\,\operatorname{tr}\left[(\hat{\zeta}^{\perp}_{\mathsf{m}})^*W\hat{\zeta}^{\perp}_{\mathsf{m}}\right]+\frac{3}{2}\lambda\\&\leq \sigma+2\cdot 2^{-3}\lambda+\frac{3}{2}\lambda\\&=\sigma+\frac{7}{4}\lambda\\&<1-\frac{2}{\sigma}\frac{\lambda}{\tau}\,,
\end{align*}
hence $\| \llc_{\mathsf{m}+1}\hspace{-1mm}\left(\left(((e^{\lambda\mathtt{P}}\mathcal{R}) \hspace{-1.1mm}\cdot\hspace{-1.2mm} W)^{\perp}\hspace{-0.3mm}(e^{\lambda\mathtt{P}}\mathcal{R})\right) \hspace{-1.5mm}\circ \hspace{-1.1mm}v\right) \|^2\hspace{-1.2mm}>\hspace{-1.2mm}\frac{2}{\sigma}\frac{\lambda}{\tau}$, equivalent to $e^{\lambda\mathtt{P}}\mathcal{R}\star (W,v)\hspace{-0.8mm}\not\in\hspace{-0.8mm}\mathfrak{A}^{\sigma,\tau}_{\mathsf{m}+1}$.
\item If $\|\ulc_{\mathsf{m}+1}(v)\|^2>\sigma$ and $(W,v)\not\in\mathfrak{A}^{\sigma,\tau}_{\mathsf{m}+1}$, one has
$$
\|\ulc_{\mathsf{m}+1}(v)\|^2\|\llc_{\mathsf{m}+1}(v)\|^2\geq 2\lambda\tau^{-1}\geq 2 {\lambda}{\tau_{\mathsf{m}+1}^{-1}}\,.
$$
Due to~\eqref{ineq-trace-order-lambda}, the statement~\eqref{ineq-enlarge-strictly} of Corollary~\ref{coro-ladder} (and $(W,v)\not\in\mathfrak{A}^{\sigma,\tau}_{\mathsf{m}+1}$) then implies 
$$
\| \llc_{\mathsf{m}+1}\left(\left(((e^{\lambda\mathtt{P}}\mathcal{R}) \cdot W)^{\perp}(e^{\lambda\mathtt{P}}\mathcal{R})\right) \circ v\right) \|^2\geq\|\llc_{\mathsf{m}+1}(v)\|^2+2^{-2}\lambda>\|\llc_{\mathsf{m}+1}(v)\|^2>\frac{2}{\sigma}\frac{\lambda}{\tau}\,.
$$
Therefore, one has $e^{\lambda\mathtt{P}}\mathcal{R}\star (W,v)\not\in\mathfrak{A}^{\sigma,\tau}_{\mathsf{m}+1}$.
\item The assumptions $(W,v)\in\mathfrak{W}\setminus\mathfrak{A}^{\sigma,\tau}_{\mathsf{m}+1}$ and $e^{\lambda\mathtt{P}}\mathcal{R}\star (W,v)\in\mathfrak{A}^{\sigma,\tau}_{\mathsf{m}+1}$ imply that
$$
\|\llc_{\mathsf{m}+1}(v)\|^2>\frac{2}{\sigma}\frac{\lambda}{\tau}\geq \| \llc_{\mathsf{m}+1}\left(\left(((e^{\lambda\mathtt{P}}\mathcal{R}) \cdot W)^{\perp}(e^{\lambda\mathtt{P}}\mathcal{R})\right) \circ v\right) \|^2\,,
$$
hence, in particular, that
$$
\|\llc_{\mathsf{m}+1}(v)\|^2>\| \llc_{\mathsf{m}+1}\left(\left(((e^{\lambda\mathtt{P}}\mathcal{R}) \cdot W)^{\perp}(e^{\lambda\mathtt{P}}\mathcal{R})\right) \circ v\right) \|^2\,,
$$
or, equivalently,
$$
\|\ulc_{\mathsf{m}}(v)\|^2<\| \ulc_{\mathsf{m}}\left(\left(((e^{\lambda\mathtt{P}}\mathcal{R}) \cdot W)^{\perp}(e^{\lambda\mathtt{P}}\mathcal{R})\right) \circ v\right) \|^2\,,
$$
\textit{i.e.},~\eqref{ineq-diminish-strictly} is violated. In view of Corollary~\ref{coro-ladder}, it follows that~\eqref{ineq-ladder} is violated, as $(W,v)\in\mathfrak{Q}$ is guaranteed by assumption. Thus, either~\eqref{ineq-ladder-1} or~\eqref{ineq-ladder-2} is violated, \textit{i.e.}, at least one of the following holds:
\begin{enumerate}
\item $\|\ulc_{\mathsf{m}}(v)\|^2<\sigma\,,$
\item $\|\llc_{\mathsf{m}}(v)\|^2<\frac{2}{\sigma}\frac{\lambda}{\tau_{\mathsf{m}}}\,.$
\end{enumerate}
We find a contradiction in each case. 
\begin{enumerate}
\item In this case, the inequality $\|\llc_{\mathsf{m}+1}(v)\|^2>1-\sigma$ holds and allows to infer
\begin{align*}
\| \llc_{\mathsf{m}+1}\left(\left(((e^{\lambda\mathtt{P}}\mathcal{R}) \cdot W)^{\perp}(e^{\lambda\mathtt{P}}\mathcal{R})\right) \circ v\right) \|^2&\geq \| \llc_{\mathsf{m}+1}\left(\left((\mathcal{R} \cdot W)^{\perp}\mathcal{R}\right) \circ v\right) \|^2-\frac{3}{2}\lambda\\
&\geq\|\llc_{\mathsf{m}+1}(v)\|^2-2\,\operatorname{tr}\left[(\hat{\zeta}^{\perp}_{\mathsf{m}})^*W\hat{\zeta}^{\perp}_{\mathsf{m}}\right]-\frac{3}{2}\lambda\\
&> 1-\sigma-2\cdot 2^{-3}\lambda-\frac{3}{2}\,\lambda\\
&=1-\sigma-\frac{7}{4}\lambda\\&\geq \frac{2}{\sigma}\frac{\lambda}{\tau}
\end{align*}
by using~\eqref{ineq-vector-expansion-1},~\eqref{ineq-vector-contraction-2},~\eqref{ineq-trace-order-lambda} and~\eqref{ineq-sigma-tau}. Therefore, $e^{\lambda\mathtt{P}}\mathcal{R}\circ v\not\in\mathfrak{A}^{\sigma,\tau}_{\mathsf{m}+1}$.
\item In this case $\|\llc_{\mathsf{m}}(v)\|^2< \frac{2}{\sigma}\frac{\lambda}{\tau_{\mathsf{m}}}$ holds, which contradicts $\|\llc_{\mathsf{m}}(w)\|^2> \frac{2}{\sigma}\frac{\lambda}{\tau}$, as $\tau\leq \tau_{\mathsf{m}}$.
\end{enumerate}
\item In this case $(W,v)\in\mathfrak{A}^{\sigma,\tau}_{\mathsf{m}+1}$ holds, which contradicts the assumption $(W,v)\in\mathfrak{W}\setminus\mathfrak{A}^{\sigma,\tau}_{\mathsf{m}+1}$.
\end{enumerate}
\item Let $(W,v)\in\mathfrak{S}^{\sigma,\tau}_{\mathsf{m}}$ and assume that $e^{\lambda\mathtt{P}}\mathcal{R}\star (W,v)\in\mathfrak{A}^{\sigma,\tau}_{\mathsf{m}}\setminus(\mathfrak{C}^{\sigma}_{\mathsf{m}}\cup\mathfrak{I}^{\sigma,\tau}_{\mathsf{m}+1/2})=(\mathfrak{A}_{\mathsf{m}}^{\sigma,\tau}\setminus\mathfrak{C}^{\sigma}_{\mathsf{m}+1})\cup\mathfrak{S}^{\sigma,\tau}_{\mathsf{m}+1}$, \textit{i.e.}, that one of the following holds:
\begin{enumerate}
\item $(W,v)\hspace{-0.5mm}\in\hspace{-0.5mm}\mathfrak{S}^{\sigma,\tau}_{\mathsf{m}}\subset\mathfrak{C}^{\sigma}_{\mathsf{m}+1}\,\wedge\, e^{\lambda\mathtt{P}}\mathcal{R}\star (W,v)\hspace{-0.5mm}\in\hspace{-0.5mm}\mathfrak{A}_{\mathsf{m}}^{\sigma,\tau}\setminus\mathfrak{C}^{\sigma}_{\mathsf{m}+1}$, \textit{i.e.}, the dynamics is leaving $\mathfrak{C}^{\sigma}_{\mathsf{m}+1}$.
\item $(W,v)\in\mathfrak{S}^{\sigma,\tau}_{\mathsf{m}}\subset\mathfrak{C}^{\sigma}_{\mathsf{m}}\,\wedge\, e^{\lambda\mathtt{P}}\mathcal{R}\star (W,v)\in\mathfrak{S}^{\sigma,\tau}_{\mathsf{m}+1}\subset\mathfrak{A}^{\sigma,\tau}_{\mathsf{m}+1}$, \textit{i.e.}, one has the inequalities
$$
\|\ulc_{\mathsf{m}}(v)\|^2\leq \sigma\qquad\wedge\qquad\| \llc_{\mathsf{m}+1}\left(\left(((e^{\lambda\mathtt{P}}\mathcal{R}) \cdot W)^{\perp}(e^{\lambda\mathtt{P}}\mathcal{R})\right) \circ v\right) \|^2\leq \frac{2\lambda}{\sigma\tau}\,.
$$
\end{enumerate}
We find a contradiction in each case. 
\begin{enumerate}
\item The statement \textit{2} of the present Lemma guarantees the following:
If the dynamics is leaving $\mathfrak{C}^{\sigma}_{\mathsf{m}+1}$, it is leaving $\mathfrak{S}^{\sigma,\tau}_{\mathsf{m}+1}$. But $(W,v)\not\in\mathfrak{S}^{\sigma,\tau}_{\mathsf{m}+1}$, \textit{i.e.}, it is not being in $\mathfrak{S}^{\sigma,\tau}_{\mathsf{m}+1}$.
\item Due to $\|\llc_{\mathsf{m}+1}(v)\|^2=1-\|\ulc_{\mathsf{m}}(v)\|^2$ and~\eqref{ineq-vector-expansion-1},~\eqref{ineq-vector-contraction-2},~\eqref{ineq-trace-order-lambda} and~\eqref{ineq-sigma-tau}, one has
\begin{align*}
\| \llc_{\mathsf{m}+1}\left(\left(((e^{\lambda\mathtt{P}}\mathcal{R}) \cdot W)^{\perp}(e^{\lambda\mathtt{P}}\mathcal{R})\right) \circ v\right) \|^2&\geq \| \llc_{\mathsf{m}+1}\left(\left((\mathcal{R} \cdot W)^{\perp}\mathcal{R}\right) \circ v\right) \|^2-\frac{3}{2}\lambda\\
&\geq\|\llc_{\mathsf{m}+1}(v)\|^2-2\,\operatorname{tr}\left[(\hat{\zeta}^{\perp}_{\mathsf{m}})^*W\hat{\zeta}^{\perp}_{\mathsf{m}}\right]-\frac{3}{2}\lambda\\
&\geq \|\llc_{\mathsf{m}+1}(v)\|^2-2\cdot 2^{-3}\lambda-\frac{3}{2}\,\lambda\\
&=1-\frac{7}{4}\lambda-\|\ulc_{\mathsf{m}}(v)\|^2\\
&> \sigma+\frac{2}{\sigma}\frac{\lambda}{\tau}-\|\ulc_{\mathsf{m}}(v)\|^2
\end{align*}
or, equivalently,
$$
\|\ulc_{\mathsf{m}}(v)\|^2+\| \llc_{\mathsf{m}+1}\left(\left(((e^{\lambda\mathtt{P}}\mathcal{R}) \cdot W)^{\perp}(e^{\lambda\mathtt{P}}\mathcal{R})\right) \circ v\right) \|^2> \sigma+\frac{2}{\sigma}\frac{\lambda}{\tau}\,,
$$
which is contradictory to the statement of this case.
\end{enumerate}
\item Let $(W,v)\in\mathfrak{I}_{\mathsf{m}-1/2}^{\sigma,\tau}\subset \mathfrak{C}^{\sigma}_{\mathsf{m}}$ and assume that
$$
e^{\lambda\mathtt{P}}\mathcal{R}\star (W,v) \in \mathfrak{A}^{\sigma,\tau}_{\mathsf{m}-1}\setminus\big(\mathfrak{I}^{\sigma,\tau}_{\mathsf{m}-1/2}\cup\mathfrak{S}^{\sigma,\tau}_{\mathsf{m}-1}\cup\mathfrak{S}^{\sigma,\tau}_{\mathsf{m}}\big)=\mathfrak{A}_{\mathsf{m}-1}^{\sigma,\tau}\setminus \mathfrak{C}^{\sigma}_{\mathsf{m}}\,.
$$
That means, the dynamics is leaving $\mathfrak{C}^{\sigma}_{\mathsf{m}}$. The statement \textit{2} of the present Lemma guarantees the following:
If the dynamics is leaving $\mathfrak{C}^{\sigma}_{\mathsf{m}}$, it is leaving $\mathfrak{S}^{\sigma,\tau}_{\mathsf{m}}$. But $(W,v)\not\in\mathfrak{S}^{\sigma,\tau}_{\mathsf{m}}$, \textit{i.e.}, it is not being in $\mathfrak{S}^{\sigma,\tau}_{\mathsf{m}}$, which is a contradiction.\hfill $\square$
\end{enumerate}

\vspace{.2cm}

\noindent\textbf{Proof of Lemma~\ref{lemma-deterministic-path}.} The argument starts with two preparatory steps.

\vspace{.2cm}

\noindent \underline{\textit{Step 1.}} \textit{Let $(W_0,v_0)\not\in\mathfrak{A}^{\overline{\sigma},\tau_0}_0$. Then there is an integer $N_{\downarrow}\in [0,4\,\lambda^{-1}]$ satisfying $(W_{N_{\downarrow}},v_{N_{\downarrow}})\in\mathfrak{C}^{\overline{\sigma}}_0$.}\\
\noindent We may assume that $(W_0,v_0)\not\in\mathfrak{C}_0^{\overline{\sigma}}$, as the claim holds trivially with $N_{\downarrow}=0$ otherwise. Therefore,
\begin{align}\label{ineq-deterministic-path-1}
(W_n,v_n)\not\in\mathfrak{C}^{\overline{\sigma}}_0\qquad\wedge\qquad (W_n,v_n)\not\in\mathfrak{A}^{{\overline{\sigma}},\tau_0}_0\qquad\wedge\qquad\|\ulc_0(v_n)\|^2\leq\|\ulc_0(v_0)\|^2-2^{-2}\,n\,\lambda
\end{align}
holds with $n=0$. Now if~\eqref{ineq-deterministic-path-1} holds for some $n\in\mathbb{N}_0$, then~\eqref{ineq-ladder-1},~\eqref{ineq-ladder-2} and thus~\eqref{ineq-ladder} hold with $v=v_n$ and $\mathsf{m}=0$. Moreover, all $n\in \{0,\dots,\lfloor 4\,\lambda^{-1}\rfloor-1\}$ fulfill $(W_n,v_n)\in\mathfrak{Q}$ by assumption. By Corollary~\ref{coro-ladder}, one then has $\left\|\llc_0(v_{n+1})\right\|^2\geq \left\|\llc_0(v_n)\right\|^2+2^{-2}\,\lambda\geq 2\,\lambda (\overline{\sigma}\tau_0)^{-1}$ and $\left\|\ulc_0(v_{n+1})\right\|^2\leq \left\|\ulc_0(v_n)\right\|^2-2^{-2}\,\lambda\leq \left\|\ulc_0(v_0)\right\|^2-2^{-2}\,(n+1)\,\lambda$ and, therefore, it holds that 
$$
(W_{n+1},v_{n+1})\not\in\mathfrak{A}^{\overline{\sigma},\tau_0}_0\qquad\wedge\qquad\|\ulc_0(v_{n+1})\|^2\leq\|\ulc_0(v_0)\|^2-2^{-2}\,(n+1)\,\lambda\,.
$$
In conclusion, the condition~\eqref{ineq-deterministic-path-1} for some $n\in\mathbb{N}_0$ implies the condition
$$
(W_{n+1},v_{n+1})\in\mathfrak{C}^{\overline{\sigma}}_0\quad\vee\quad \bigg(\begin{array}{l}
(W_{n+1},v_{n+1})\not\in\mathfrak{C}^{\overline{\sigma}}_0\qquad\wedge\qquad (W_{n+1},v_{n+1})\not\in\mathfrak{A}^{\overline{\sigma},\tau_0}_0\quad\wedge\\
\qquad\wedge\qquad\|\ulc_0(v_{n+1})\|^2\leq\|\ulc_0(v_0)\|^2-2^{-2}\,(n+1)\,\lambda
\end{array} \bigg)\,.
$$
Hence, if $(W_n,v_n)\in\mathfrak{C}^{\overline{\sigma}}_0$ were false for all integers $n\in (0,4\,\lambda^{-1}]$, then all these integers $n\in (0,4\,\lambda^{-1}]$ would satisfy the condition~\eqref{ineq-deterministic-path-1}. In particular,~\eqref{ineq-deterministic-path-1} would hold with $n=\lfloor 4\,\lambda^{-1}\rfloor$ and therefore
\begin{align}\label{ineq-deterministic-path-2}
\begin{split}
0&\leq \|\ulc_0(v_n)\|^2\\
&\leq\|\ulc_0(v_0)\|^2-2^{-2}\,\lfloor 4\,\lambda^{-1}\rfloor \,\lambda\\
&\leq 1-\|\llc_0(v_0)\|^2-2^{-2}\,\lfloor 4\,\lambda^{-1}\rfloor \,\lambda\\
&\leq 1-2\,\lambda\, [\overline{\sigma}\,\tau_0]^{-1}-2^{-2}\,\lfloor 4\,\lambda^{-1}\rfloor \,\lambda\\
&\leq 1-2^{\frac{5}{2}}\lambda-2^{-2}\,\lfloor 4\,\lambda^{-1}\rfloor \,\lambda
\end{split}
\end{align}
were satisfied (as $\tau_0 \leq 1$). But~\eqref{ineq-deterministic-path-2} is equivalent to the contradiction $4\,\lambda^{-1}-\lfloor 4\,\lambda^{-1}\rfloor\geq 2^{\frac{9}{2}}$.
\hfill $\diamond$

\vspace{.2cm}

\noindent \underline{\textit{Step 2.}} \textit{Let $(W_0,v_0)\in\mathfrak{C}^{\overline{\sigma}}_0$. Then all integer $N_{\uparrow}\in [0,4\,\lambda^{-1}]$ satisfy $(W_{N_{\uparrow}},v_{N_{\uparrow}})\not\in\mathfrak{A}^{\overline{\sigma},\overline{\tau}}_{24}$.}\\
\noindent Assume that there were some integer $N_{\uparrow}\in [0,4\,\lambda^{-1}]$ satisfying $(W_{N_{\uparrow}},v_{N_{\uparrow}})\in\mathfrak{A}^{\overline{\sigma},\overline{\tau}}_{24}$. According to Lemma~\ref{lemma-allowed-movements}, a run from $\mathfrak{C}^{\overline{\sigma}}_0$ to $\mathfrak{A}^{\overline{\sigma},\overline{\tau}}_{24}$ is only possible via a path of the form
$$
\mathfrak{C}^{\overline{\sigma}}_0\supset\mathfrak{S}_0^{\overline{\sigma},\overline{\tau}}\rightarrow\mathfrak{S}^{\overline{\sigma},\overline{\tau}}_1\rightarrow\mathfrak{S}_2^{\overline{\sigma},\overline{\tau}}\rightarrow\dots\rightarrow\mathfrak{S}^{\overline{\sigma},\overline{\tau}}_{23}
\rightarrow\mathfrak{S}^{\overline{\sigma},\overline{\tau}}_{24}\subset\mathfrak{A}^{\overline{\sigma},\overline{\tau}}_{24}
\,,$$
where all steps of the ladder $\mathfrak{S}_{0}^{\overline{\sigma},\overline{\tau}},\mathfrak{S}_{1}^{\overline{\sigma},\overline{\tau}},\dots,\mathfrak{S}_{24}^{\overline{\sigma},\overline{\tau}}$ have to be entered at least once. Thus, there would be integers $s_0,s_1,\dots,s_{24}$ obeying
$0\leq s_0<s_1<s_2<\dots<s_{23}<s_{24}\leq N_{\uparrow}$
and $v_{s_{\mathsf{m}}}\in\mathfrak{S}_{\mathsf{m}}^{\overline{\sigma},\overline{\tau}}$
for all $\mathsf{m}\in\{0,\dots,24\}$. This and equation~\eqref{eq-sigma-tau} would imply that all $\mathsf{m}\in\{0,\dots,23\}$ satisfy
\begin{align}\label{ineq-deterministic-path-3}
\left\|\ulc_{\mathsf{m}}(v_{s_{\mathsf{m}+1}})\right\|^2-\left\|\ulc_{\mathsf{m}}(v_{s_{\mathsf{m}}})\right\|^2=1-\left\|\llc_{\mathsf{m}+1}(v_{s_{\mathsf{m}+1}})\right\|^2-\left\|\ulc_{\mathsf{m}}(v_{s_{\mathsf{m}}})\right\|^2\geq 1-\frac{2}{\overline{\sigma}}\frac{\lambda}{\overline{\tau}}-\overline{\sigma}=1-2^{-\frac{1}{2}}> \frac{7}{24}\,.
\end{align}
Now due to inequality~\eqref{ineq-deterministic-vector-1}, one has $\left\|\ulc_{\mathsf{m}}(v_{n+1})\right\|^2-\left\|\ulc_{\mathsf{m}}(v_n)\right\|^2\leq \frac{7}{4}\lambda$ for all $n\in\mathbb{N}_0$ and therefore
\begin{align}\label{ineq-deterministic-path-4}
\left\|\ulc_{\mathsf{m}}(v_{s_{\mathsf{m}+1}})\right\|^2-\left\|\ulc_{\mathsf{m}}(v_{s_{\mathsf{m}}})\right\|^2=\sum\limits_{n=s_{\mathsf{m}}+1}^{s_{\mathsf{m}+1}}\Big[\left\|\ulc_{\mathsf{m}}(v_{n+1})\right\|^2-\left\|\ulc_{\mathsf{m}}(v_n)\right\|^2\Big]\leq \frac{7}{4}\lambda\,\big[s_{\mathsf{m}+1}-s_{\mathsf{m}}\big]
\end{align}
Combining~\eqref{ineq-deterministic-path-3} with~\eqref{ineq-deterministic-path-4} and summing from $\mathsf{m}=0$ to $23$ would yield
$$
N_{\uparrow}\geq s_{24}-s_0=\sum\limits_{\mathsf{m}=0}^{23}\big[s_{\mathsf{m}+1}-s_{\mathsf{m}}\big]\geq \frac{4}{7}\,\lambda^{-1}\sum\limits_{\mathsf{m}=0}^{23}\Big[\left\|\ulc_{\mathsf{m}}(v_{s_{\mathsf{m}+1}})\right\|^2-\left\|\ulc_{\mathsf{m}}(v_{s_{\mathsf{m}}})\right\|^2\Big]> \frac{4}{7}\,\lambda^{-1}\sum\limits_{\mathsf{m}=0}^{23}\frac{7}{24}=4\,\lambda^{-1},
$$
which is a contradiction.
\hfill $\diamond$

\vspace{.2cm}

\noindent \underline{\textit{Conclusion.}}  By the assumption $(W_0,v_0)\not\in\mathfrak{A}_0^{\overline{\sigma},\tau_0}$ and \textit{Step 1}, there is some integer $N_{\downarrow}\in [0,4\,\lambda^{-1}]$ satisfying $(W_{N_{\downarrow}},v_{N_{\downarrow}})\in\mathfrak{C}^{\overline{\sigma}}_0$. We now set $N_{\uparrow}=\lfloor 4\,\lambda^{-1}\rfloor-N_{\downarrow}$. Clearly, one then has $N_{\uparrow}\in[0,4\,\lambda^{-1}]$.  By \textit{Step 2}, this implies that $(W_{\lfloor 4\,\lambda^{-1}\rfloor},v_{\lfloor 4\,\lambda^{-1}\rfloor})=(W_{N_{\downarrow}+N_{\uparrow}},v_{N_{\downarrow}+N_{\uparrow}})\not\in\mathfrak{A}_{24}^{\overline{\sigma},\overline{\tau}}$.
\hfill $\square$

\vspace{.2cm}

For the proof of Lemma~\ref{lemma-diffusion-probability}, we first formulate and prove a quantitative perturbative bound. 

\begin{lemma}\label{lemma-Hamiltonian-bound}
	Let $\mathtt{P}\in\mathfrak{P}$ and $Q\in\mathbb{G}_{\mathsf{L},\mathsf{w}+1}$. All
	$\Phi,\widetilde{\Phi}\in\mathbb{F}_{\mathsf{L},\mathsf{w}+1}$ with $Q=\Phi\Phi^*$ and $e^{\lambda\mathtt{P}}\cdot Q=\widetilde{\Phi}\widetilde{\Phi}^*$~obey
	\begin{align}\label{ineq-Hamiltonian-bound-1}
	\mu_1\big(\widetilde{\Phi}^*\hat{P}_{\llc}\widetilde{\Phi}\big)\geq \mu_1\big(\mathtt{H}\big)-2^5\lambda^3\,,\qquad\qquad \mathtt{H}=\mathtt{H}_0+\lambda\,\mathtt{H}_1+\lambda^2\mathtt{H}_2
	\end{align}
	with
	$$
	\mathtt{H}_0=\Phi^*\hat{P}_{\llc}\Phi\,,\qquad\qquad\mathtt{H}_1=\Phi^*\big[\hat{P}_{\llc}\,\mathtt{x}+\mathtt{x}^*\hat{P}_{\llc}\big]\Phi\,,\qquad\qquad\mathtt{H}_2=\Phi^*\big[\mathtt{x}^*\hat{P}_{\llc}\,\mathtt{x}+\hat{P}_{\llc}\,\mathtt{y}+\mathtt{y}^*\hat{P}_{\llc}\big]\Phi\,,
	$$
	where
	$$
	\mathtt{x}=Q^{\perp}\mathtt{P}\,,\qquad\qquad\mathtt{y}=\mbox{\small $\frac{1}{2}$}\left[Q^{\perp}\mathtt{P}\left[Q^{\perp}-Q\right]-Q\mathtt{P}^*Q^{\perp}\right]\mathtt{P}\,.
	$$
	Moreover, it holds that
	\begin{align}\label{ineq-Hamiltonian-bound-2}
	\|\mathtt{x}\| \leq 1, \qquad \|\mathtt{y}\| \leq 2^{-\frac{1}{2}}, \qquad \|\mathtt{H}_0\| \leq 1, \qquad \|\mathtt{H}_1\| \leq 2, \qquad \|\mathtt{H}_2\| \leq 1 + 2^{\frac{1}{2}}\,.
	\end{align}
	For $\psi \in \mathbb{S}_{\mathbb{C}}^{\mathsf{w}}$ for which we set $\varphi := \hat{P}_{\llc}^{\perp}\Phi\psi$, there exists a $\widetilde{W} \in \mathbb{G}_{\mathsf{L},\mathsf{L}_{\mathfrak{c}}-\mathsf{q}+1}$ with $\widetilde{W} \leq \hat{P}_{\mathfrak{c}}$ such that
	\begin{align}\label{ineq-Hamiltonian-bound-3}
	\langle \psi| (\mathtt{H}_2-\mathtt{H}_1^2)\psi \rangle \geq \|\widetilde{W}\mathtt{P}\varphi\|^2 - \frac{\beta}{15} - \frac{242}{\beta}\langle \psi| \mathtt{H}_0\psi \rangle\,.
	\end{align}
\end{lemma}

\noindent\textbf{Proof of Lemma~\ref{lemma-Hamiltonian-bound}.} The bounds indicated in~\eqref{ineq-Hamiltonian-bound-2} all follow from the triangle inequality for norms and the direct observation that $\|\mathtt{x}\| \leq 1$ by $\|\mathtt{P}\| \leq 1$. To bound $\|\mathtt{y}\|$, applying~\eqref{ineq-norm-1} yields
\begin{align}\label{ineq-Hamiltonian-bound-y}
\|\mathtt{y}\|=\|\mathtt{y}^*\mathtt{y}\|^{\frac{1}{2}}=\mbox{\small $\frac{1}{2}$}\left\|\left[Q^{\perp}-Q\right]\mathtt{P}^*Q^{\perp}\mathtt{P}\left[Q^{\perp}-Q\right]+Q^{\perp}\mathtt{P}^*Q\mathtt{P}Q^{\perp}\right\|^{\frac{1}{2}}\leq\mbox{\small $\frac{1}{2}$}\,(1+1)^{\frac{1}{2}}=2^{-\frac{1}{2}}\,.
\end{align}
Now let $\mathtt{X}(Q,\mathtt{P})$, $\mathtt{Y}(Q,\mathtt{P})$ and $\mathtt{Z}^{(\lambda)}(Q,\mathtt{P})$ be as in Lemma~\ref{lemma-expansion}. Then,
$$
\mathtt{x}\,Q+Q\,\mathtt{x}^*-\mathtt{X}(Q,\mathtt{P})=0\,,\qquad\qquad\qquad \mathtt{x}\,Q\,\mathtt{x}^*+Q\,\mathtt{y}^*+\mathtt{y}\,Q-\mathtt{Y}(Q,\mathtt{P})=0
$$
which with the bounds of~\eqref{ineq-expansion} implies
\begin{align}\label{ineq-Hamiltonian-bound-expansion}
\begin{split}
&\left\|\left[\mathbf{1}+\lambda\,\mathtt{x}+\lambda^2\,\mathtt{y}\right]Q\left[\mathbf{1}+\lambda\,\mathtt{x}+\lambda^2\,\mathtt{y}\right]^*-e^{\lambda\mathtt{P}}\cdot Q\right\|\\
&=\lambda^3\left\|\mathtt{x}\,Q\,\mathtt{y}^*+\mathtt{y}\,Q\,\mathtt{x}^*-\mathtt{Z}^{(\lambda)}(Q,\mathtt{P})+\lambda\,\mathtt{y}Q\mathtt{y}^*\right\|\\
&\leq \lambda^3\left[2\|\mathtt{x}\|\|\mathtt{y}\| +  \|\mathtt{Z}^{(\lambda)}(Q,\mathtt{P})\| + \lambda\,\|\mathtt{y}\|^2\right]\\
&\leq (20+2^{\frac{1}{2}}+2^{-1}\lambda)\lambda^3
\end{split}
\end{align}
If we denote $\mu_{\mathsf{D}-\mathsf{w}}(\cdot)$ for the $(\mathsf{D}-\mathsf{w})$th smallest eigenvalue of its argument (a self-adjoint matrix), we find $\mu_1(\mathtt{A}^*\mathtt{A}) = \mu_{\mathsf{D}-\mathsf{w}}(\mathtt{A}\mathtt{A}^*)$ for all $\mathtt{A} \in \mathbb{C}^{\mathsf{D} \times (\mathsf{w}+1)}$. This, $\|\mathtt{x}\|\leq 1$,~\eqref{ineq-Hamiltonian-bound-y} and~\eqref{ineq-Hamiltonian-bound-expansion} yield~\eqref{ineq-Hamiltonian-bound-1}:
\begin{align*}
\mu_1\big((\Phi^{\prime})^*\hat{P}_{\llc}\Phi^{\prime}\big)&=\mu_{\mathsf{D}-\mathsf{w}}\big(\zeta^*(e^{\lambda\mathtt{P}}\cdot Q)\zeta\big)\\
&\geq \mu_{\mathsf{D}-\mathsf{w}}\big(\zeta^*\left[\mathbf{1}+\lambda\,\mathtt{x}+\lambda^2\,\mathtt{y}\right]Q\left[\mathbf{1}+\lambda\,\mathtt{x}+\lambda^2\,\mathtt{y}\right]^*\zeta\big)-(20+2^{\frac{1}{2}}+2^{-1}\lambda)\lambda^3\\
&= \mu_{1}\big(\Phi^*\left[\mathbf{1}+\lambda\,\mathtt{x}+\lambda^2\,\mathtt{y}\right]^*\hat{P}_{\llc}\left[\mathbf{1}+\lambda\,\mathtt{x}+\lambda^2\,\mathtt{y}\right]\Phi\big)-(20+2^{\frac{1}{2}}+2^{-1}\lambda)\lambda^3\\
&=\mu_{1}\big(\mathtt{H}+\lambda^3\,\Phi^*\big[\mathtt{x}^*\hat{P}_{\llc}\mathtt{y}+\mathtt{y}^*\hat{P}_{\llc}\mathtt{x}\big]\Phi+\lambda^4\,\Phi^*\mathtt{y}^*\hat{P}_{\llc}\mathtt{y}\Phi\big)-(20+2^{\frac{1}{2}}+2^{-1}\lambda)\lambda^3\\
&\geq\mu_{1}(\mathtt{H})-\lambda^3\,\|\mathtt{x}^*\hat{P}_{\llc}\mathtt{y}+\mathtt{y}^*\hat{P}_{\llc}\mathtt{x}\|-(20+2^{\frac{1}{2}}+2^{-1}\lambda)\lambda^3\\
&\geq \mu_{1}(\mathtt{H})-2\|\mathtt{x}\|\|\mathtt{y}\|\,\lambda^3-(2^5-2^{\frac{1}{2}})\lambda^3\\
&\geq \mu_{1}(\mathtt{H})-2^5\,\lambda^3\,.
\end{align*}
Now we concentrate on the quantity $\langle \psi| (\mathtt{H}_2-\mathtt{H}_1^2)\psi \rangle$. First we formulate the following bound, which will be used several times. For all $\alpha > 0$, $\mathtt{A} \in \mathbb{C}^{\mathsf{L}\times\mathsf{L}}$ and $\mathtt{B} \in \mathbb{C}^{\mathsf{L}\times(\mathsf{w}+1)}$, it holds that
\begin{align}\label{ineq-inner-product-norms}
\langle \psi| \mathtt{B}^*(\mathtt{A} + \mathtt{A}^*)\mathtt{B}\psi \rangle \geq - \alpha\|\mathtt{B}\psi\|^2 - \alpha^{-1}\|\mathtt{A}\mathtt{B}\psi\|^2\,,
\end{align}
since adding the positive quantity $\|(\alpha^{\frac{1}{2}}\mathbf{1}_{\mathsf{L}}+\alpha^{-\frac{1}{2}}\mathtt{A})\mathtt{B}\psi\|^2$ on the right makes it an equality.

Applying this three times (taking $\mathtt{A}$ equal to $\mathtt{x}^*\hat{P}_{\llc}Q\mathtt{x}^*\hat{P}_{\llc}$, $\mathtt{y}^*\hat{P}_{\llc}$ and $\hat{P}_{\llc}^{\perp}\mathtt{x}^*\hat{P}_{\llc}Q^{\perp}\hat{P}_{\llc}\mathtt{x}\hat{P}_{\llc}$, respectively, $\alpha = \frac{\beta}{60}$ and $\mathtt{B} = \Phi$) and using~\eqref{ineq-Hamiltonian-bound-2} as shown earlier yields
\begin{align}\label{ineq-Hamiltonian-bound-4}
\begin{split}
\langle \psi| (\mathtt{H}_2-\mathtt{H}_1^2)\psi \rangle &= \langle \psi| \Phi^*\left[\mathtt{x}^*\hat{P}_{\llc}Q^{\perp}\hat{P}_{\llc}\mathtt{x} - \hat{P}_{\llc}\mathtt{x}Q\hat{P}_{\llc}\mathtt{x} - \mathtt{x}^*\hat{P}_{\llc}Q\mathtt{x}^*\hat{P}_{\llc} + \hat{P}_{\llc}\mathtt{y} + \mathtt{y}^*\hat{P}_{\llc} - \hat{P}_{\llc}\mathtt{x}Q\mathtt{x}^*\hat{P}_{\llc}\right]\Phi\psi \rangle\\
&\geq \langle \psi| \Phi^*(\hat{P}_{\llc}+\hat{P}_{\llc}^{\perp})\mathtt{x}^*\hat{P}_{\llc}Q^{\perp}\hat{P}_{\llc}\mathtt{x}(\hat{P}_{\llc}+\hat{P}_{\llc}^{\perp})\Phi\psi \rangle - \frac{\beta}{60}\|\Phi\psi\|^2 - \frac{60}{\beta}\|\mathtt{x}^*\hat{P}_{\llc}Q\mathtt{x}^*\hat{P}_{\llc}\Phi\psi\|^2\\
&\qquad- \frac{\beta}{60}\|\Phi\psi\|^2 - \frac{60}{\beta}\|\mathtt{y}^*\hat{P}_{\llc}\Phi\psi\|^2 - \|Q\mathtt{x}^*\hat{P}_{\llc}\Phi\psi\|^2\\
&\geq \|Q^{\perp}\hat{P}_{\llc}\mathtt{x}\hat{P}_{\llc}^{\perp}\Phi\psi\|^2 + \|Q^{\perp}\hat{P}_{\llc}\mathtt{x}\hat{P}_{\llc}\Phi\psi\|^2 - \frac{\beta}{60}\|\Phi\psi\|^2 - \frac{60}{\beta}\|\hat{P}_{\llc}^{\perp}\mathtt{x}^*\hat{P}_{\llc}Q^{\perp}\hat{P}_{\llc}\mathtt{x}\hat{P}_{\llc}\Phi\psi\|^2\\
&\qquad- \frac{\beta}{30} - \left(\frac{180}{\beta}+1\right)\|\hat{P}_{\llc}\Phi\psi\|^2\\
&\geq \langle \psi|\Phi^*\hat{P}_{\llc}^{\perp}\mathtt{x}^*\hat{P}_{\llc}Q^{\perp}\hat{P}_{\llc}\mathtt{x}\hat{P}_{\llc}^{\perp}\Phi\psi \rangle - \frac{\beta}{20} - \left(\frac{240}{\beta}+1\right)\|\hat{P}_{\llc}\Phi\psi\|^2\,.
\end{split}
\end{align}
Since $\psi \in \mathbb{S}_{\mathbb{C}}^{\mathsf{w}}$, we have $\Phi\psi\psi^*\Phi^* \in \mathbb{G}_{\mathsf{L},1}$ and then $\Phi\psi\psi^*\Phi^* \leq \Phi\Phi^* = Q$ follows from $\psi\psi^* \leq \mathbf{1}_{\mathsf{w}}$. This also implies $Q^{\perp} + \Phi\psi\psi^*\Phi^* = (Q - \Phi\psi\psi^*\Phi^*)^{\perp} \in \mathbb{G}_{\mathsf{L},\mathsf{L}-\mathsf{w}}$. Now we observe that
\begin{align*}
	\dim\left[\operatorname{Ran}(Q^{\perp} + \Phi\psi\psi^*\Phi^*) \cap \operatorname{Ran}(\hat{P}_{\mathfrak{c}})\right] &= \dim\left[\operatorname{Ran}(Q^{\perp} + \Phi\psi\psi^*\Phi^*)\right] + \dim\left[\operatorname{Ran}(\hat{P}_{\mathfrak{c}})\right]\\
	&\quad- \dim\left[\operatorname{Ran}(Q^{\perp} + \Phi\psi\psi^*\Phi^*) + \operatorname{Ran}(\hat{P}_{\mathfrak{c}})\right]\\
	&\geq \dim\left[\operatorname{Ran}(Q^{\perp} + \Phi\psi\psi^*\Phi^*)\right] + \dim\left[\operatorname{Ran}(\hat{P}_{\mathfrak{c}})\right] - \mathsf{L}\\
	&= (\mathsf{L}-\mathsf{w}) + \mathsf{L}_{\mathfrak{c}} - \mathsf{L}\\
	&\geq \mathsf{L}_{\mathfrak{c}} - (\mathsf{q} - 1)\,.
\end{align*}
Therefore there exists a projection $\widetilde{W} \in \mathbb{G}_{\mathsf{L},\mathsf{L}_{\mathfrak{c}}-\mathsf{q}+1}$ such that $\widetilde{W} \leq Q^{\perp} + \Phi\psi\psi^*\Phi^*$ and $\widetilde{W} \leq \hat{P}_{\mathfrak{c}} \leq \hat{P}_{\llc}$. Now as $\mathtt{x} = Q^{\perp}\mathtt{P}$, using~\eqref{ineq-inner-product-norms} with $\mathtt{A} = -Q^{\perp}\hat{P}_{\llc}\widetilde{W}\hat{P}_{\llc}\Phi\psi\psi^*\Phi^*$, $\alpha = \frac{\beta}{60}$ and $\mathtt{B} = \mathtt{P}\hat{P}_{\llc}^{\perp}\Phi$, we find
\begin{align}\label{ineq-Hamiltonian-bound-5}
\begin{split}
&\langle \psi|\Phi^*\hat{P}_{\llc}^{\perp}\mathtt{P}^*Q^{\perp}\hat{P}_{\llc}Q^{\perp}\hat{P}_{\llc}Q^{\perp}\mathtt{P}\hat{P}_{\llc}^{\perp}\Phi\psi \rangle\\
&\geq \langle \psi|\Phi^*\hat{P}_{\llc}^{\perp}\mathtt{P}^*Q^{\perp}\hat{P}_{\llc}\widetilde{W}\hat{P}_{\llc}Q^{\perp}\mathtt{P}\hat{P}_{\llc}^{\perp}\Phi\psi \rangle - \langle \psi|\Phi^*\hat{P}_{\llc}^{\perp}\mathtt{P}^*Q^{\perp}\hat{P}_{\llc}\Phi\psi\psi^*\Phi^*\hat{P}_{\llc}Q^{\perp}\mathtt{P}\hat{P}_{\llc}^{\perp}\Phi\psi \rangle\\
&= \langle \varphi|\mathtt{P}^*\widetilde{W}\mathtt{P}\varphi \rangle - \langle \varphi|\mathtt{P}^*\left[Q^{\perp}\hat{P}_{\llc}\widetilde{W}\hat{P}_{\llc}\Phi\psi\psi^*\Phi^* + \Phi\psi\psi^*\Phi^*\hat{P}_{\llc}\widetilde{W}\hat{P}_{\llc}Q^{\perp}\right]\mathtt{P}\varphi \rangle\\
&\quad+ \|\widetilde{W}\hat{P}_{\llc}\Phi\psi\psi^*\Phi^*\mathtt{P}\varphi\|^2 - \|\psi^*\Phi^*\hat{P}_{\llc}Q^{\perp}\mathtt{P}\varphi\|^2\\
&\geq \langle \varphi|\mathtt{P}^*\widetilde{W}\mathtt{P}\varphi \rangle - \frac{\beta}{60}\|\mathtt{P}\varphi\|^2 - \frac{60}{\beta}\|Q^{\perp}\hat{P}_{\llc}\widetilde{W}\hat{P}_{\llc}\Phi\psi\psi^*\Phi^*\mathtt{P}\varphi\|^2 - \|\varphi^*\mathtt{P}^*Q^{\perp}\hat{P}_{\llc}\Phi\psi\|^2\\
&\geq \|\widetilde{W}\mathtt{P}\varphi\|^2 - \frac{\beta}{60} - \left(\frac{60}{\beta}+1\right)\|\hat{P}_{\llc}\Phi\psi\|^2\,.
\end{split}
\end{align}
Finally, combining~\eqref{ineq-Hamiltonian-bound-4},~\eqref{ineq-Hamiltonian-bound-5}, $\|\hat{P}_{\llc}\Phi\psi\|^2 = \langle \psi| \mathtt{H}_0 \psi \rangle$ and $1 \leq \frac{1}{\beta}$ yields~\eqref{ineq-Hamiltonian-bound-3}.
\hfill $\square$

\vspace{.2cm}

\noindent\textbf{Proof of Lemma~\ref{lemma-diffusion-probability}.} A central quantity in this proof is $\mu_1(\Phi^*\hat{P}_{\llc}\Phi)$ for $(W,v) \in \mathfrak{W}$. Here we used the usual abbreviation $\Phi = \begin{pmatrix}\Upsilon & v\end{pmatrix} \in \mathbb{F}_{\mathsf{L},\mathsf{w}+1}$ with $\Upsilon \in \mathbb{F}_{\mathsf{L},\mathsf{w}}$ such that $W = \Upsilon\Upsilon^*$. Then $\|\llc(v)\|^2$ is a diagonal element of the matrix $\Phi^*\hat{P}_{\llc}\Phi$, which implies
\begin{align}\label{ineq-eigenvalue-norm}
\mu_1(\Phi^*\hat{P}_{\llc}\Phi) \leq \|\llc(v)\|^2\,.
\end{align}
Another bound follows from the matrix inequalities $\mathbf{0} \leq \hat{\zeta}^*(W+vv^*)\hat{\zeta} \leq \hat{\zeta}^*\hat{\zeta} = \mathbf{1}_{\mathsf{D}}$:
\begin{align}\label{ineq-norm-eigenvalue}
\begin{split}
\|\llc(v)\|^2 &= \operatorname{tr}\left(\hat{\zeta}^*(W+vv^*)\hat{\zeta}\right) - \operatorname{tr}\left(\hat{\zeta}^*W\hat{\zeta}\right)\\
&\leq \mu_1(\Phi^*\hat{P}_{\llc}\Phi) + \mathsf{w} - \operatorname{tr}\left(\Upsilon^*\hat{P}_{\llc}\Upsilon\right) = \mu_1(\Phi^*\hat{P}_{\llc}\Phi) + \operatorname{tr}\left(\Upsilon^*(\mathbf{1}_{\mathsf{L}}-\hat{P}_{\llc})\Upsilon\right)\\
&= \mu_1(\Phi^*\hat{P}_{\llc}\Phi) + \operatorname{tr}\left((\hat{\zeta}^{\perp})^*W\hat{\zeta}^{\perp}\right)\,,
\end{split}
\end{align}
The fact that the action~\eqref{def-star-action} is well-defined implies that $\mathcal{R} \star (W,v) \in \mathfrak{W}$, so analogously there exists a $\Phi^{\prime} \in \mathbb{F}_{\mathsf{L},\mathsf{w}+1}$ such that $\Phi^{\prime}(\Phi^{\prime})^* = \mathcal{R} \cdot (W + vv^*)$ according to Lemma~\ref{lemma-auxiliary-action}. Now we note that Lemma~\ref{lemma-norm-contraction} implies that $\|\hat{\alpha}^*(\mathcal{R} \cdot Q)\hat{\alpha}\| \leq \|\hat{\alpha}^*Q\hat{\alpha}\|$. Completely analogously one can show that $\|(\hat{\zeta}^{\perp})^*(\mathcal{R} \cdot (W+vv^*))\hat{\zeta}^{\perp}\| \leq \|(\hat{\zeta}^{\perp})^*(W+vv^*)\hat{\zeta}^{\perp}\|$, and then it follows that
\begin{align}\label{ineq-eigenvalue-growth}
\begin{split}
\mu_1((\Phi^{\prime})^*\hat{P}_{\llc}\Phi^{\prime}) &= \mu_1(\mathbf{1}_{\mathsf{w}+1} - (\Phi^{\prime})^*\hat{P}_{\llc}^{\perp}\Phi^{\prime}) = 1 - \|(\Phi^{\prime})^*\hat{P}_{\llc}^{\perp}\Phi^{\prime}\|\\
&= 1 - \|(\hat{\zeta}^{\perp})^*(\mathcal{R} \cdot (W+vv^*))\hat{\zeta}^{\perp}\|\\
&\geq 1 - \|(\hat{\zeta}^{\perp})^*(W+vv^*)\hat{\zeta}^{\perp}\|\\
&= \mu_1(\Phi^*\hat{P}_{\llc}\Phi)\,.
\end{split}
\end{align}
Let $n \in \lbrace 1,\dots,N \rbrace$. If we write $\Phi_n^{\prime} = \begin{pmatrix}\Upsilon_n^{\prime} & v_n^{\prime}\end{pmatrix} \in \mathbb{F}_{\mathsf{L},\mathsf{w}+1}$ with $v_n^{\prime}:=(\mathcal{R} \cdot W_{n-1})^{\perp}\mathcal{R} \circ v_{n-1}$ and ${\Upsilon_n^{\prime} \in \mathbb{F}_{\mathsf{L},\mathsf{w}}}$ such that $\Upsilon_n^{\prime}(\Upsilon_n^{\prime})^* = \mathcal{R} \cdot W_{n-1}$, then using Lemma~\ref{lemma-auxiliary-action} and~\eqref{ineq-eigenvalue-growth} one finds $\mu_1((\Phi_n^{\prime})^*\hat{P}_{\llc}\Phi_n^{\prime}) \geq \mu_1(\Phi_{n-1}^*\hat{P}_{\llc}\Phi_{n-1})$. 

\vspace{.2cm}

In the first step of this proof, we show an enlargement inequality for $\mathbb{E}\,\left[\mu_1(\Phi_n^*\hat{P}_{\llc}\Phi_n)\,|\,\mathfrak{O}_{0,n-1}\right]$. In the second step we then apply the reverse Markov inequality and~\eqref{ineq-eigenvalue-norm} to show~\eqref{ineq-diffusion-probability}.

\vspace{.2cm}

\noindent \underline{\textit{Step 1.}} \textit{If $(W_0,v_0) \in \mathfrak{Q}$, then all $n \in \mathbb{N}$ satisfy}
\begin{align}\label{ineq-eigenvalue-enlargment}
\mathbb{E}\,\left[\mu_1(\Phi_n^*\hat{P}_{\llc}\Phi_n)\,|\,\mathfrak{O}_{0,n-1}\right] \geq \left[1-2^{-\frac{6}{5}}\beta^{\frac{3}{5}}\pmb{\eta}^{-\frac{1}{5}}\vartheta^{-\frac{3}{5}}\lambda^{\frac{7}{5}}\right]\mathbb{E}\,\left[\mu_1((\Phi_{n-1})^*\hat{P}_{\llc}\Phi_{n-1})\,|\,\mathfrak{O}_{0,n-1}\right] + \frac{2\beta}{3}\,\lambda^2\,.
\end{align}
To prove this, we will make a further distinction, for which we introduce the event
$$
\mathfrak{V}_n := \left\lbrace\mu_1((\Phi^{\prime}_n)^*\hat{P}_{\llc}\Phi^{\prime}_n) \leq 2^{-2}\right\rbrace\,.
$$
In \textit{Step 1.a)} we will assume that $\mathfrak{V}_n$ holds, and the contrary case is treated in \textit{Step 1.b)}. We also note that $\mathfrak{V}_n$, $\neg\,\mathfrak{V}_n$ and $\mathfrak{O}_{0,n-1}$ only depend on $\mathcal{P}_1,\dots,\mathcal{P}_{n-1}$. Therefore all $n \in \mathbb{N}$ satisfy
\begin{align}\label{eq-P-centered}
\mathbb{E}\,\left[\mathcal{P}_n\,|\,\mathfrak{V}_n\wedge\mathfrak{O}_{0,n-1}\right] = \mathbf{0} = \mathbb{E}\,\left[\mathcal{P}_n\,|\,\neg\,\mathfrak{V}_n\wedge\mathfrak{O}_{0,n-1}\right]\,.
\end{align}

\vspace{.2cm}

\noindent \underline{\textit{Step 1.a)}} \textit{If $(W_0,v_0) \in \mathfrak{Q}$, then all $n \in \mathbb{N}$ satisfy}
\begin{align}
\label{ineq-eigenvalue-enlargment-V}
\mathbb{E}\,& 
\left[\mu_1(\Phi_n^*\hat{P}_{\llc}\Phi_n)\,|\,\mathfrak{V}_n\wedge\mathfrak{O}_{0,n-1}\right] 
\nonumber
\\
& \geq \left[1-2^{-\frac{6}{5}}\beta^{\frac{3}{5}}\pmb{\eta}^{-\frac{1}{5}}\vartheta^{-\frac{3}{5}}\lambda^{\frac{7}{5}}\right]\mathbb{E}\,\left[\mu_1((\Phi_{n-1})^*\hat{P}_{\llc}\Phi_{n-1})\,|\,\mathfrak{V}_n\wedge\mathfrak{O}_{0,n-1}\right] + \frac{2\beta}{3}\,\lambda^2\,.
\end{align}
Let us now use the notation of Lemma~\ref{lemma-Hamiltonian-bound} by setting $\mathtt{P} = \mathcal{P}_n$, $\Phi = \Phi_n^{\prime}$ and $\widetilde{\Phi} = \Phi_n$. Then it indeed holds that $e^{\lambda\mathcal{P}_n} \cdot \Phi_n^{\prime}(\Phi_n^{\prime})^* = \Phi_n\Phi_n^*$. Moreover, let us identify $E_0^{(\lambda)} = \mu_1(\mathtt{H})$ and $E_0 = \mu_1(\mathtt{H}_0)$, and write $\psi$ instead of $\psi_0$, hence $\mu_1((\Phi^{\prime}_n)^*\hat{P}_{\llc}\Phi^{\prime}_n) = \langle \psi| \mathtt{H}_0\psi \rangle$.

We now show that assuming $\mathfrak{V}_n$, one can apply Lemma~\ref{lemma-Hamiltonian} (see Appendix~\ref{app-perturbative-bounds}) to the $\mathtt{H}$, $\mathtt{H}_0$, $\mathtt{H}_1$ and $\mathtt{H}_2$ of Lemma~\ref{lemma-Hamiltonian-bound}. For this, one has to find $G > 0$ and $g > 0$ with $\frac{G}{2} > g$ that fulfill~\eqref{ineq-radius}. Under the assumption $\mathfrak{O}_{0,n-1}$, we verify that this is the case for $G = \frac{1}{2}$ and $g = \frac{1}{8}$. First note that for all $k \in \lbrace 1,\dots,\mathsf{w} \rbrace$, $E_k$ is not the smallest eigenvalue of $\mathtt{H}_0 = (\Phi^{\prime}_n)^*\hat{P}_{\llc}\Phi^{\prime}_n$ (as this is $E_0$). Since $(\Upsilon^{\prime}_n)^*\hat{P}_{\llc}\Upsilon^{\prime}_n$ is a compression of $\mathtt{H}_0$ to a $\mathsf{w}$-dimensional subspace, it follows from Cauchy's interlacing theorem that
\begin{align}\label{ineq-other-eigenvalues}
\begin{split}
E_k &\geq \mu_1((\Upsilon^{\prime}_n)^*\hat{P}_{\llc}\Upsilon^{\prime}_n)\\
&\geq \mu_1(\Upsilon_{n-1}^*\hat{P}_{\llc}\Upsilon_{n-1}) = 1 - \|\mathbf{1}_{\mathsf{w}} - \Upsilon_{n-1}^*\hat{P}_{\llc}\Upsilon_{n-1}\| = 1 - \|\Upsilon_{n-1}^*\hat{P}_{\llc}^{\perp}\Upsilon_{n-1}\|\\
&\geq 1 - \operatorname{tr}(\Upsilon_{n-1}^*\hat{P}_{\llc}^{\perp}\Upsilon_{n-1}) = 1 - \operatorname{tr}((\hat{\zeta}^{\perp})^*W_{n-1}\hat{\zeta}^{\perp})\,,
\end{split}
\end{align}
in which the second inequality is fully analogous to~\eqref{ineq-eigenvalue-growth}. By $\mathfrak{V}_n$ and $\mathfrak{O}_{0,n-1}$, with the latter implying~\eqref{ineq-trace-order-lambda}, we indeed find for all $k \in \lbrace 1,\dots,\mathsf{w} \rbrace$ that
$$
|E_0-E_k| = E_k-E_0 \geq 1 - \operatorname{tr}((\hat{\zeta}^{\perp})^*W_{n-1}\hat{\zeta}^{\perp}) - \mu_1((\Phi^{\prime}_n)^*\hat{P}_{\llc}\Phi^{\prime}_n) \geq 1 - \frac{1}{4} - \frac{1}{4} = \frac{1}{2} = G\,.
$$
As $\|\mathtt{H}_1\| \leq 2$ and $\|\mathtt{H}_2\| \leq 1+2^{\frac{1}{2}}$, it holds that $\|\mathtt{H}-\mathtt{H}_0\| \leq 2\,\lambda + (1+2^{\frac{1}{2}})\,\lambda^2$. If the eigenvalues of $\mathtt{H}$ and $\mathtt{H}_0$ are ordered according to their indices, one also finds $|E_k-E_k^{(\lambda)}| \leq \|\mathtt{H}-\mathtt{H}_0\|$ for all $k \in \lbrace 0,\dots,\mathsf{w} \rbrace$. In particular, $|E_0^{(\lambda)}-E_0| \leq 2\,\lambda + (1+2^{\frac{1}{2}})\,\lambda^2 \leq \frac{1}{8} = g$, and for all $k \in \lbrace 1,\dots,\mathsf{w} \rbrace$
$$|E_0-E_k^{(\lambda)}| \geq |E_0-E_k| - |E_k-E_k^{(\lambda)}| \geq G - 2\,\lambda - (1+2^{\frac{1}{2}})\,\lambda^2 \geq G - \frac{1}{8} = G - g\,.$$
Therefore the assumptions~\eqref{ineq-radius} of Lemma~\ref{lemma-Hamiltonian} are satisfied. By estimating $\|\mathtt{H}_2\| \leq 2$ and using the bound $\lambda \leq 2^{-6}$ from Hypothesis~\ref{hyp-lambda}, we can simplify~\eqref{ineq-Hamiltonian} to
\begin{align}\label{ineq-E-lambda-bound}
\begin{split}
|E^{(\lambda)}| &\leq \left[1+2^5\lambda\left(1+2^3(1+2\,\lambda+2\,\lambda^2)\right)(2+2\,\lambda)\right]\left[2^6+2^4\right]\\
&\qquad\; +2^4\lambda \left[1+2^3(1+2\,\lambda+2\,\lambda^2)\right]\left[2^2+2^5\right]\\
&\leq 2^{10}-1\,.
\end{split}
\end{align}
Furthermore, $\mathfrak{O}_{0,n-1}$ and $\mathfrak{V}_n$ imply that $0 \leq x := \operatorname{tr}((\hat{\zeta}^{\perp})^*W_{n-1}\hat{\zeta}^{\perp}) + \mu_1((\Phi^{\prime}_n)^*\hat{P}_{\llc}\Phi^{\prime}_n) \leq \frac{1}{4} + \frac{1}{4} = \frac{1}{2}$. Therefore, using~\eqref{ineq-other-eigenvalues} and the fact that $(1-x)^{-1} \leq 1+2x$ for all $x \in \left[0,\frac{1}{2}\right]$, for all $k \in \lbrace 1,\dots,\mathsf{w} \rbrace$
\begin{align}\label{ineq-eigenvalue-difference}
\begin{split}
\left[E_k-E_0\right]^{-1} - 1 &\leq \left[1 - \operatorname{tr}((\hat{\zeta}^{\perp})^*W_{n-1}\hat{\zeta}^{\perp}) - \mu_1((\Phi^{\prime}_n)^*\hat{P}_{\llc}\Phi^{\prime}_n)\right]^{-1} - 1\\
&\leq 2\operatorname{tr}((\hat{\zeta}^{\perp})^*W_{n-1}\hat{\zeta}^{\perp}) + 2\mu_1((\Phi^{\prime}_n)^*\hat{P}_{\llc}\Phi^{\prime}_n)\\
&\leq 2^{-\frac{16}{5}}\beta^{\frac{3}{5}}\pmb{\eta}^{-\frac{1}{5}}\vartheta^{-\frac{3}{5}}\lambda^{\frac{7}{5}} + 2\mu_1((\Phi^{\prime}_n)^*\hat{P}_{\llc}\Phi^{\prime}_n)
\end{split}
\end{align}
holds. Combining~\eqref{ineq-E-lambda-bound},~\eqref{ineq-eigenvalue-difference}, the result~\eqref{def-E-lambda} from Lemma~\ref{lemma-Hamiltonian}, $\|\mathtt{H}_1\| \leq 2$ and~\eqref{ineq-Hamiltonian-bound-1} from Lemma~\ref{lemma-Hamiltonian-bound} under the assumptions $\mathfrak{O}_{0,n-1}$ and $\mathfrak{V}_n$, we find
\begin{align}\label{ineq-Hamiltonian-applied}
\mu_1(\Phi_n^*\hat{P}_{\llc}\Phi_n) \geq \left[1-2^3\lambda^2\right]\mu_1((\Phi^{\prime}_n)^*\hat{P}_{\llc}\Phi^{\prime}_n) + \lambda\langle \psi|\mathtt{H}_1\psi \rangle + \lambda^2\langle \psi|(\mathtt{H}_2-\mathtt{H}_1^2)\psi \rangle - \frac{\beta}{60}\,\lambda^2\,,
\end{align}
for which used the following estimate, obtained from~\eqref{ineq-trace-order-lambda} and Hypothesis~\ref{hyp-q}:
$$
2^{-\frac{6}{5}}\beta^{\frac{3}{5}}\pmb{\eta}^{-\frac{1}{5}}\vartheta^{-\frac{3}{5}}\lambda^{\frac{7}{5}} + (2^{10}-1)\lambda \leq 2^{10}\lambda \leq \frac{2^{36}}{60}\lambda\,\pmb{\eta}^{-3}\mathsf{q}^5\vartheta \leq \frac{2^{36}}{60}\pmb{\eta}^{-3}\vartheta\lambda(2^{-\frac{36}{5}}\beta^{\frac{1}{5}}\pmb{\eta}^{\frac{3}{5}}\vartheta^{-\frac{1}{5}}\lambda^{-\frac{1}{5}})^5 = \frac{\beta}{60}\,.
$$
Recall that $\|\hat{P}_{\llc}\Phi^{\prime}_n\psi\|^2 = \mu_1((\Phi^{\prime}_n)^*\hat{P}_{\llc}\Phi^{\prime}_n) \leq \|\llc(v_n^{\prime})\|^2$ from~\eqref{ineq-eigenvalue-norm}. Moreover, the assumption $\mathfrak{V}_n$ implies $\|\varphi\|^2 = 1 - \|\hat{P}_{\llc}\Phi^{\prime}_n\psi\|^2 \geq \frac{3}{4}$. Combining this with~\eqref{ineq-Hamiltonian-bound-3} for some $\widetilde{W} \in \mathbb{G}_{\mathsf{L},\mathsf{L}_{\mathfrak{c}}-\mathsf{q}+1}$ yields
\begin{align}\label{ineq-Hamiltonian-bound-applied}
\begin{split}
\langle \psi| (\mathtt{H}_2-\mathtt{H}_1^2)\psi \rangle &\geq \|\widetilde{W}\mathcal{P}_n\varphi\|^2\frac{\|\varphi\|^2}{\|\varphi\|^2} - \frac{\beta}{12} - \frac{242}{\beta}\mu_1((\Phi^{\prime}_n)^*\hat{P}_{\llc}\Phi^{\prime}_n)\\
&\geq \frac{3\|\widetilde{W}\mathcal{P}_n\varphi\|^2}{4\|\varphi\|^2} - \frac{\beta}{12} - \frac{2^9-2^3}{\beta}\mu_1((\Phi^{\prime}_n)^*\hat{P}_{\llc}\Phi^{\prime}_n)\,.
\end{split}
\end{align}
Combining~\eqref{ineq-Hamiltonian-applied} and~\eqref{ineq-Hamiltonian-bound-applied} under the assumptions $\mathfrak{V}_n$ and $\mathfrak{O}_{0,n-1}$, and Hypothesis~\ref{hyp-lambda}, we obtain
\begin{align}\label{ineq-eigenvalue-enlargment-V-1}
\begin{split}
\mu_1(\Phi_n^*\hat{P}_{\llc}\Phi_n) &\geq \left[1-\left(2^3+\frac{2^9-2^3}{\beta}\right)\lambda^{\frac{3}{5}}\lambda^{\frac{7}{5}}\right]\mu_1((\Phi^{\prime}_n)^*\hat{P}_{\llc}\Phi^{\prime}_n) + \lambda\langle \psi|\mathtt{H}_1\psi \rangle + \frac{3\|\widetilde{W}\mathcal{P}_n\varphi\|^2}{4\|\varphi\|^2}\,\lambda^2 - \frac{\beta}{12}\,\lambda^2\\
&\geq \left[1 - 2^{-\frac{6}{5}}\beta^{\frac{3}{5}}\pmb{\eta}^{-\frac{1}{5}}\vartheta^{-\frac{3}{5}}\lambda^{\frac{7}{5}}\right]\mu_1((\Phi^{\prime}_n)^*\hat{P}_{\llc}\Phi^{\prime}_n) + \lambda\langle \psi|\mathtt{H}_1\psi \rangle + \frac{3\|\widetilde{W}\mathcal{P}_n\varphi\|^2}{4\|\varphi\|^2}\,\lambda^2 - \frac{\beta}{12}\,\lambda^2\,.
\end{split}
\end{align}
Clearly, one has $\frac{\varphi}{\|\varphi\|} \in \mathbb{S}_{\mathbb{C}}^{\mathsf{L}-1}$ and $\lgl(\frac{\varphi}{\|\varphi\|})=0$. From Lemma~\ref{lemma-Hamiltonian-bound} we also have $\widetilde{W} \leq \hat{P}_{\mathfrak{c}}$, and then Hypothesis~\ref{hyp-P} implies that
\begin{align}\label{ineq-eigenvalue-enlargment-V-2}
\mathbb{E}\,\left[\lambda\langle \psi|\mathtt{H}_1\psi \rangle + \frac{3\|\widetilde{W}\mathcal{P}_n\varphi\|^2}{4\|\varphi\|^2}\,\lambda^2\,|\,\mathfrak{V}_n\wedge\mathfrak{O}_{0,n-1}\right] \geq \frac{2\,\beta}{3}\,\lambda^2 + \frac{\beta}{12}\,\lambda^2\,.
\end{align}
in which the identity $\mathtt{H}_1 = (\Phi^{\prime}_n)^*\left[\hat{P}_{\llc}(\mathcal{R} \cdot W_{n-1} + v_n^{\prime}(v_n^{\prime})^*)^{\perp}\mathcal{P}_n + \mathcal{P}_n^*(\mathcal{R} \cdot W_{n-1} + v_n^{\prime}(v_n^{\prime})^*)^{\perp}\hat{P}_{\llc}\right]\Phi^{\prime}_n$ was used as well as~\eqref{eq-P-centered}. Finally, combining~\eqref{ineq-eigenvalue-growth},~\eqref{ineq-eigenvalue-enlargment-V-1} and~\eqref{ineq-eigenvalue-enlargment-V-2} yields~\eqref{ineq-eigenvalue-enlargment-V}.
\hfill $\diamond$

\vspace{.2cm}

\noindent \underline{\textit{Step 1.b)}} \textit{If $(W_0,v_0) \in \mathfrak{Q}$, then all $n \in \mathbb{N}$ satisfy}
\begin{align}
\label{ineq-eigenvalue-enlargment-neg-V}
\mathbb{E}\,& \left[\mu_1(\Phi_n^*\hat{P}_{\llc}\Phi_n)\,|\,\neg\,\mathfrak{V}_n\wedge\mathfrak{O}_{0,n-1}\right] 
\nonumber
\\
&
\geq \left[1-2^{-\frac{6}{5}}\beta^{\frac{3}{5}}\pmb{\eta}^{-\frac{1}{5}}\vartheta^{-\frac{3}{5}}\lambda^{\frac{7}{5}}\right]\mathbb{E}\,\left[\mu_1((\Phi_{n-1})^*\hat{P}_{\llc}\Phi_{n-1})\,|\,\neg\,\mathfrak{V}_n\wedge\mathfrak{O}_{0,n-1}\right] + \frac{2\beta}{3}\,\lambda^2\,.
\end{align}
To prove~\eqref{ineq-eigenvalue-enlargment-neg-V}, assume $\neg\,\mathfrak{V}_n$, so $\mu_1((\Phi^{\prime}_n)^*\hat{P}_{\llc}\Phi^{\prime}_n) > 2^{-2}$. Recall that ${v_n^{\prime} = (\mathcal{R} \cdot W_{n-1})^{\perp}\mathcal{R} \circ v_{n-1}}$. Combining this with Lemma~\ref{lemma-auxiliary-action} implies that ${v_n = (e^{\lambda\mathcal{P}} \cdot (\mathcal{R} \cdot W_{n-1}))^{\perp}e^{\lambda\mathcal{P}_n} \circ v_n^{\prime} = W_n^{\perp}e^{\lambda\mathcal{P}_n} \circ v_n^{\prime}}$. From Lemma~\ref{lemma-vector-expansion} and~\eqref{eq-P-centered}, we find that $\mathbb{E}\,\left[\mathbf{A}_{\llc}(\mathcal{R} \cdot W_{n-1}, v_n^{\prime}, \mathcal{P}_n)\,|\,\neg\,\mathfrak{V}_n\wedge\mathfrak{O}_{0,n-1}\right] = 0$, and $\mathbb{E}\,\left[\mathtt{X}(\mathcal{R} \cdot W_{n-1}, \mathcal{P}_n)\,|\,\neg\,\mathfrak{V}_n\wedge\mathfrak{O}_{0,n-1}\right] = \mathbf{0}$ by Lemma~\ref{lemma-expansion}. All this, the bounds from Lemmata~\ref{lemma-expansion} and~\ref{lemma-vector-expansion},~\eqref{ineq-eigenvalue-norm},~\eqref{ineq-norm-eigenvalue},~\eqref{ineq-eigenvalue-growth} and an estimate similar to~\eqref{ineq-trace-contraction-1} in the fourth step then imply
\begin{align*}
&\mathbb{E}\,\left[\mu_1(\Phi_n^*\hat{P}_{\llc}\Phi_n)\,|\,\neg\,\mathfrak{V}_n\wedge\mathfrak{O}_{0,n-1}\right]\\
&\geq \mathbb{E}\,\left[\|\llc(v_n)\|^2-\operatorname{tr}\left((\hat{\zeta}^{\perp})^*W_n\hat{\zeta}^{\perp}\right)\,|\,\neg\,\mathfrak{V}_n\wedge\mathfrak{O}_{0,n-1}\right]\\
&= \mathbb{E}\,\left[\|\llc(W_n^{\perp}e^{\lambda\mathcal{P}_n} \circ v_n^{\prime})\|^2-\operatorname{tr}\left((\hat{\zeta}^{\perp})^*\left[e^{\lambda\mathcal{P}_n} \cdot (\mathcal{R} \cdot W_{n-1})\right]\hat{\zeta}^{\perp}\right)\,|\,\neg\,\mathfrak{V}_n\wedge\mathfrak{O}_{0,n-1}\right]\\
&\geq \mathbb{E}\,\left[\|\llc(v_n^{\prime})\|^2 + \lambda\,\mathbf{A}_{\llc}(\mathcal{R} \cdot W_{n-1}, v_n^{\prime}, \mathcal{P}_n)\,|\,\neg\,\mathfrak{V}_n\wedge\mathfrak{O}_{0,n-1}\right] - (9+160\,\lambda)\lambda^2  - (\mbox{$\frac{3}{2}$}+20\,\lambda)\mathsf{w}\,\lambda^2\\
&\qquad- \mathbb{E}\,\left[\operatorname{tr}\left((\hat{\zeta}^{\perp})^*(\mathcal{R} \cdot W_{n-1})\hat{\zeta}^{\perp}\right) + \operatorname{tr}\left((\hat{\zeta}^{\perp})^*\mathtt{X}(\mathcal{R} \cdot W_{n-1}, \mathcal{P}_n)\hat{\zeta}^{\perp}\right)\,|\,\neg\,\mathfrak{V}_n\wedge\mathfrak{O}_{0,n-1}\right]\\
&\geq \mathbb{E}\,\left[\mu_1((\Phi_n^{\prime})^*\hat{P}_{\llc}\Phi_n^{\prime}) - \operatorname{tr}\left((\hat{\zeta}^{\perp})^*W_{n-1}\hat{\zeta}^{\perp}\right)\,|\,\neg\,\mathfrak{V}_n\wedge\mathfrak{O}_{0,n-1}\right] - \frac{2^6}{3}(\mathsf{w}+1)\lambda^2\\
&\geq \left[1-2^{-\frac{6}{5}}\beta^{\frac{3}{5}}\pmb{\eta}^{-\frac{1}{5}}\vartheta^{-\frac{3}{5}}\lambda^{\frac{7}{5}}\right]\mathbb{E}\,\left[\mu_1((\Phi_{n-1})^*\hat{P}_{\llc}\Phi_{n-1})\,|\,\neg\,\mathfrak{V}_n\wedge\mathfrak{O}_{0,n-1}\right] - \frac{2^6}{3}\mathsf{q}\,\lambda^2\\
&\qquad+\mathbb{E}\,\left[2^{-\frac{6}{5}}\beta^{\frac{3}{5}}\pmb{\eta}^{-\frac{1}{5}}\vartheta^{-\frac{3}{5}}\lambda^{\frac{7}{5}}\mu_1((\Phi_{n-1})^*\hat{P}_{\llc}\Phi_{n-1}) - \operatorname{tr}\left((\hat{\zeta}^{\perp})^*W_{n-1}\hat{\zeta}^{\perp}\right)\,|\,\neg\,\mathfrak{V}_n\wedge\mathfrak{O}_{0,n-1}\right]\\
&\geq \left[1-2^{-\frac{6}{5}}\beta^{\frac{3}{5}}\pmb{\eta}^{-\frac{1}{5}}\vartheta^{-\frac{3}{5}}\lambda^{\frac{7}{5}}\right]\mathbb{E}\,\left[\mu_1((\Phi_{n-1})^*\hat{P}_{\llc}\Phi_{n-1})\,|\,\neg\,\mathfrak{V}_n\wedge\mathfrak{O}_{0,n-1}\right] + 2^{-\frac{21}{5}}\beta^{\frac{3}{5}}\pmb{\eta}^{-\frac{1}{5}}\vartheta^{-\frac{3}{5}}\lambda^{\frac{7}{5}} - \frac{2^6}{3}\mathsf{q}\,\lambda^2
\end{align*}
in which the final step incorporated $\neg\,\mathfrak{V}_n$ and $\mathfrak{O}_{0,n-1}$. Then one finds~\eqref{ineq-eigenvalue-enlargment-neg-V} by combining this with the following estimate, which follows from Hypothesis~\ref{hyp-q}:
\begin{align*}
2^{-\frac{21}{5}}\beta^{\frac{3}{5}}\pmb{\eta}^{-\frac{1}{5}}\vartheta^{-\frac{3}{5}}\lambda^{\frac{7}{5}} - \frac{2^6}{3}\mathsf{q}\,\lambda^2 &\geq 2^{-\frac{78}{5}}\beta^{\frac{3}{5}}\pmb{\eta}^{-\frac{1}{5}}\vartheta^{-\frac{3}{5}}\lambda^{\frac{7}{5}}\pmb{\eta}^2\mathsf{q}^{-2} - \frac{2^6}{3}\mathsf{q}\,\lambda^2\\
&\geq \frac{2^{-\frac{1}{5}}}{3}\beta^{\frac{1}{5}}\pmb{\eta}^{\frac{3}{5}}\vartheta^{-\frac{1}{5}}\lambda^{\frac{9}{5}} \geq \frac{2^{-\frac{1}{5}}}{3}\beta^{\frac{1}{5}}\pmb{\eta}^{\frac{3}{5}}\vartheta^{-\frac{1}{5}}\lambda^{\frac{9}{5}}2^{-6}\mathsf{q}^{-1}\beta \geq \frac{2\beta}{3}\,\lambda^2\,.
\end{align*}
\hfill $\diamond$

\vspace{.2cm}

\noindent \underline{\textit{Step 1.c)}} We now use the results~\eqref{ineq-eigenvalue-enlargment-V} and~\eqref{ineq-eigenvalue-enlargment-neg-V} from \textit{Step 1.a)} and \textit{1.b)} to prove~\eqref{ineq-eigenvalue-enlargment}:
\begin{align}\label{ineq-eigenvalue-enlargment-proof}
\begin{split}
\mathbb{E}\,&\left[\mu_1(\Phi_n^*\hat{P}_{\llc}\Phi_n)\,\Big|\,\mathfrak{O}_{0,n-1}\right]\mathbb{P}\left[\mathfrak{O}_{0,n-1}\right] = \mathbb{E}\,\left[\mu_1(\Phi_n^*\hat{P}_{\llc}\Phi_n)\,\Big|\,\neg\,\mathfrak{V}_n\wedge\mathfrak{O}_{0,n-1}\right]\mathbb{P}\left[\neg\,\mathfrak{V}_n\wedge\mathfrak{O}_{0,n-1}\right]\\
&\phantom{\left[\mu_1(\Phi_n^*\hat{P}_{\llc}\Phi_n)\,\Big|\,\mathfrak{O}_{0,n-1}\right]\mathbb{P}\left[\mathfrak{O}_{0,n-1}\right] =}\,+ \mathbb{E}\,\left[\mu_1(\Phi_n^*\hat{P}_{\llc}\Phi_n)\,\Big|\,\mathfrak{V}_n\wedge\mathfrak{O}_{0,n-1}\right]\mathbb{P}\left[\mathfrak{V}_n\wedge\mathfrak{O}_{0,n-1}\right]\\
&\geq \left[1-2^{-\frac{6}{5}}\beta^{\frac{3}{5}}\pmb{\eta}^{-\frac{1}{5}}\vartheta^{-\frac{3}{5}}\lambda^{\frac{7}{5}}\right]\mathbb{E}\,\left[\mu_1((\Phi_{n-1})^*\hat{P}_{\llc}\Phi_{n-1})\,\Big|\,\mathfrak{O}_{0,n-1}\right]\mathbb{P}\left[\mathfrak{O}_{0,n-1}\right] + \frac{2\beta}{3}\mathbb{P}\left[\mathfrak{O}_{0,n-1}\right]\,\lambda^2\,.
\end{split}
\end{align}
If either $\mathbb{P}\left[\mathfrak{V}_n\wedge\mathfrak{O}_{0,n-1}\right] = 0$ or $\mathbb{P}\left[\neg\,\mathfrak{V}_n\wedge\mathfrak{O}_{0,n-1}\right] = 0$, one simply omits the corresponding terms the calculation of~\eqref{ineq-eigenvalue-enlargment-proof}, so there is no need to treat these cases separately. As we assumed that $\mathbb{P}\left[\mathfrak{O}_{0,n-1}\right] > 0$, dividing both sides of~\eqref{ineq-eigenvalue-enlargment-proof} by this quantity yields~\eqref{ineq-eigenvalue-enlargment}.
\hfill $\diamond$

\vspace{.2cm}

\noindent\underline{\textit{Step 2.}} \textit{The inequality~\eqref{ineq-eigenvalue-enlargment} of {\rm Step 1} allows to conclude the statement~\eqref{ineq-diffusion-probability}.}\\
\noindent For the proof of the statement~\eqref{ineq-diffusion-probability}, we first note that for all $n, N \in \mathbb{N}$ it holds that
\begin{align}\label{eq-overwhelming-events}
\mathfrak{O}_{0,n-1}=\mathfrak{O}_{0,n}\sqcup\big(\mathfrak{O}_{0,n-1}\setminus\mathfrak{O}_{n-1,n}\big)\,,\qquad\neg\mathfrak{O}_{0,N}=\bigsqcup\limits_{k=1}^{N}\,\left(\mathfrak{O}_{0,k-1}\setminus\mathfrak{O}_{k-1,k}\right)\,.
\end{align}
This, the deterministic bound $\mu_1(\Phi^*_n\hat{P}_{\llc}\Phi_n)\leq 1$ and the inequality~\eqref{ineq-eigenvalue-enlargment} imply that for all $n\in\mathbb{N}$
\begin{align*}
&\mathbb{E}\left(\mu_1(\Phi_n^*\hat{P}_{\llc}\Phi_n)\,\Big|\,\mathfrak{O}_{0,n}\right)\mathbb{P}(\mathfrak{O}_{0,n})\\
&
\!\!=\mathbb{E}\left(\mu_1(\Phi_n^*\hat{P}_{\llc}\Phi_n)\,\Big|\,\mathfrak{O}_{0,n-1}\right)\mathbb{P}(\mathfrak{O}_{0,n-1})-\mathbb{E}\left(\mu_1(\Phi_n^*\hat{P}_{\llc}\Phi_n)\,\Big|\,\mathfrak{O}_{0,n-1}\wedge \neg \mathfrak{O}_{n-1,n}\right)\mathbb{P}(\mathfrak{O}_{0,n-1}\wedge \neg \mathfrak{O}_{n-1,n})\\
&
\!\!
\geq \mathbb{E}\left(\mu_1(\Phi_n^*\hat{P}_{\llc}\Phi_n)\,\Big|\,\mathfrak{O}_{0,n-1}\right)\mathbb{P}(\mathfrak{O}_{0,n-1})-\mathbb{P}(\mathfrak{O}_{0,n-1}\wedge \neg \mathfrak{O}_{n-1,n})\\
&
\!\!
\geq \left[\left[1-2^{-\frac{6}{5}}\beta^{\frac{3}{5}}\pmb{\eta}^{-\frac{1}{5}}\vartheta^{-\frac{3}{5}}\lambda^{\frac{7}{5}}\right]\,\mathbb{E}\left(\mu_1(\Phi_{n-1}^*\hat{P}_{\llc}\Phi_{n-1})\,\Big|\,\mathfrak{O}_{0,n-1}\right)+\frac{2\,\beta}{3}\,\lambda^2\right]\mathbb{P}(\mathfrak{O}_{0,n-1})
\\
&\qquad -\mathbb{P}(\mathfrak{O}_{0,n-1}\wedge \neg \mathfrak{O}_{n-1,n})
\end{align*}
holds. Due to $\mathfrak{O}_{0,N}\subset\mathfrak{O}_{0,n}\subset \mathfrak{O}_{0,n-1}$ and hence $\mathbb{P}\left(\mathfrak{O}_{0,N}\right)\leq\mathbb{P}\left(\mathfrak{O}_{0,n}\right)\leq \mathbb{P}\left(\mathfrak{O}_{0,n-1}\right)$, this implies
\begin{align*}
\mathbb{E} & \left(\mu_1(\Phi_n^*\hat{P}_{\llc}\Phi_n)\,\Big|\,\mathfrak{O}_{0,n}\right)\\
&\geq \left[\left[1-2^{-\frac{6}{5}}\beta^{\frac{3}{5}}\pmb{\eta}^{-\frac{1}{5}}\lambda^{\frac{7}{5}}\vartheta^{-\frac{3}{5}}\right]\,\mathbb{E}\left(\mu_1(\Phi_{n-1}^*\hat{P}_{\llc}\Phi_{n-1})\,\Big|\,\mathfrak{O}_{0,n-1}\right)+\frac{2\,\beta}{3}\,\lambda^2\right]\frac{\mathbb{P}(\mathfrak{O}_{0,n-1})}{\mathbb{P}\left(\mathfrak{O}_{0,n}\right)}
\\
&
\qquad
-\frac{\mathbb{P}(\mathfrak{O}_{0,n-1}\wedge \neg \mathfrak{O}_{n-1,n})}{\mathbb{P}\left(\mathfrak{O}_{0,n}\right)}\\
&\geq \left[1-2^{-\frac{6}{5}}\beta^{\frac{3}{5}}\pmb{\eta}^{-\frac{1}{5}}\vartheta^{-\frac{3}{5}}\lambda^{\frac{7}{5}}\right]\,\mathbb{E}\left(\mu_1(\Phi_{n-1}^*\hat{P}_{\llc}\Phi_{n-1})\,\Big|\,\mathfrak{O}_{0,n-1}\right)+\frac{2\,\beta}{3}\,\lambda^2
\\
&
\qquad
-{\mathbb{P}(\mathfrak{O}_{0,n-1}\wedge \neg \mathfrak{O}_{n-1,n})}\,{\mathbb{P}\left(\mathfrak{O}_{0,N}\right)}^{-1}\,.
\end{align*}
Iterating the foregoing for $n=1$ to $N$ yields
\begin{align}\label{ineq-diffusion-probability-proof}
\begin{split}
\mathbb{E} & \left(\mu_1(\Phi_N^*\hat{P}_{\llc}\Phi_N)\,\Big|\,\mathfrak{O}_{0,N}\right)\\
&\geq \frac{2\,\beta}{3}\,\lambda^2\sum\limits_{n=1}^{N}\left[1-2^{-\frac{6}{5}}\beta^{\frac{3}{5}}\pmb{\eta}^{-\frac{1}{5}}\vartheta^{-\frac{3}{5}}\lambda^{\frac{7}{5}}\right]^{N-n}\\
&\qquad-{\mathbb{P}\left(\mathfrak{O}_{0,N}\right)}^{-1}\sum\limits_{n=1}^{N}\left[1-2^{-\frac{6}{5}}\beta^{\frac{3}{5}}\pmb{\eta}^{-\frac{1}{5}}\vartheta^{-\frac{3}{5}}\lambda^{\frac{7}{5}}\right]^{N-n}\,{\mathbb{P}(\mathfrak{O}_{0,n-1}\wedge \neg \mathfrak{O}_{n-1,n})}\\
&\geq \frac{2\,\beta}{3}\,\lambda^2\sum\limits_{n=1}^{N}\left[1-2^{-\frac{6}{5}}\beta^{\frac{3}{5}}\pmb{\eta}^{-\frac{1}{5}}\vartheta^{-\frac{3}{5}}\lambda^{\frac{7}{5}}\right]^{N-n}-{\mathbb{P}\left(\mathfrak{O}_{0,N}\right)}^{-1}\sum\limits_{n=1}^{N}{\mathbb{P}(\mathfrak{O}_{0,n-1}\wedge \neg \mathfrak{O}_{n-1,n})}\\
&\geq \frac{2^{\frac{11}{5}}}{3}\beta^{\frac{2}{5}}\pmb{\eta}^{\frac{1}{5}}\vartheta^{\frac{3}{5}}\lambda^{\frac{3}{5}}\,\Big(1-\exp\big[-2^{-\frac{6}{5}}\beta^{\frac{3}{5}}\pmb{\eta}^{-\frac{1}{5}}\vartheta^{-\frac{3}{5}}\lambda^{\frac{7}{5}}N\big]\Big)-{\mathbb{P}\left(\neg\mathfrak{O}_{0,N}\right)}\,{\mathbb{P}\left(\mathfrak{O}_{0,N}\right)}^{-1}\,,
\end{split}
\end{align}
where the last step incorporates the identity~\eqref{eq-overwhelming-events} and the estimate
\begin{align*}
2^{-\frac{6}{5}}\beta^{\frac{3}{5}}\pmb{\eta}^{-\frac{1}{5}}\vartheta^{-\frac{3}{5}}\lambda^{\frac{7}{5}}\,\sum\limits_{n=1}^{N}\left[1-2^{-\frac{6}{5}}\beta^{\frac{3}{5}}\pmb{\eta}^{-\frac{1}{5}}\vartheta^{-\frac{3}{5}}\lambda^{\frac{7}{5}}\right]^{N-n}&=1-\left[1-2^{-\frac{6}{5}}\beta^{\frac{3}{5}}\pmb{\eta}^{-\frac{1}{5}}\vartheta^{-\frac{3}{5}}\lambda^{\frac{7}{5}}\right]^N\\
&\geq 1-\exp\big[-2^{-\frac{6}{5}}\beta^{\frac{3}{5}}\pmb{\eta}^{-\frac{1}{5}}\vartheta^{-\frac{3}{5}}\lambda^{\frac{7}{5}}N\big]\,.
\end{align*}
As $0\leq \mu_1(\Phi_N^*\hat{P}_{\llc}\Phi_N)\leq 1$, applying~\eqref{ineq-eigenvalue-norm} and the reverse Markov inequality to~\eqref{ineq-diffusion-probability-proof} yields
\begin{align*}
\mathbb{P} & \left((W_N,v_N)\not\in\mathfrak{A}^{\sigma,\tau}_0\,\big|\,\mathfrak{O}_{0,N}\right)
\\
&=\mathbb{P}\left(\|\llc(v_N)\|^2>\frac{2}{\sigma}\frac{\lambda}{\tau}\,\bigg|\,\mathfrak{O}_{0,N}\right)\\
&\geq\mathbb{P}\left(\mu_1(\Phi_N^*\hat{P}_{\llc}\Phi_N)>\frac{2}{\sigma}\frac{\lambda}{\tau}\,\bigg|\,\mathfrak{O}_{0,N}\right)\\
&\geq \left[\mathbb{E}\left(\mu_1(\Phi_N^*\hat{P}_{\llc}\Phi_N)\,\Big|\,\mathfrak{O}_{0,N}\right)-\frac{2}{\sigma}\frac{\lambda}{\tau}\right]\,\left[1-\frac{2}{\sigma}\frac{\lambda}{\tau}\right]^{-1}\\
&\geq \mathbb{E}\left(\mu_1(\Phi_N^*\hat{P}_{\llc}\Phi_N)\,\Big|\,\mathfrak{O}_{0,N}\right)-\frac{2}{\sigma}\frac{\lambda}{\tau}\\
&\geq \frac{2^{\frac{11}{5}}}{3}\beta^{\frac{2}{5}}\pmb{\eta}^{\frac{1}{5}}\vartheta^{\frac{3}{5}}\lambda^{\frac{3}{5}}\,\Big(1-\exp\big[-2^{-\frac{6}{5}}\beta^{\frac{3}{5}}\pmb{\eta}^{-\frac{1}{5}}\vartheta^{-\frac{3}{5}}\lambda^{\frac{7}{5}}N\big]\Big)-\frac{\mathbb{P}\left(\neg\mathfrak{O}_{0,N}\right)}{\mathbb{P}\left(\mathfrak{O}_{0,N}\right)}-\frac{2}{\sigma}\frac{\lambda}{\tau}\,.
\end{align*}
This, in turn, implies
\begin{align*}
\mathbb{P}&\big((W_N,v_N)\in\mathfrak{A}^{\sigma,\tau}_0\,\wedge\,\mathfrak{O}_{0,N}\big)\\
&=\mathbb{P}\left((W_N,v_N)\in\mathfrak{A}^{\sigma,\tau}_0\,\big|\,\mathfrak{O}_{0,N}\right)\,\mathbb{P}\left(\mathfrak{O}_{0,N}\right)\\
&=\left[1-\mathbb{P}\left((W_N,v_N)\not\in\mathfrak{A}^{\sigma,\tau}_0\,\big|\,\mathfrak{O}_{0,N}\right)\right]\,\mathbb{P}\left(\mathfrak{O}_{0,N}\right)\\
&\leq\left[1- \frac{2^{\frac{11}{5}}}{3}\beta^{\frac{2}{5}}\pmb{\eta}^{\frac{1}{5}}\vartheta^{\frac{3}{5}}\lambda^{\frac{3}{5}}\,\Big(1-\exp\big[-2^{-\frac{6}{5}}\beta^{\frac{3}{5}}\pmb{\eta}^{-\frac{1}{5}}\vartheta^{-\frac{3}{5}}\lambda^{\frac{7}{5}}N\big]\Big)+\frac{\mathbb{P}\left(\neg\mathfrak{O}_{0,N}\right)}{\mathbb{P}\left(\mathfrak{O}_{0,N}\right)}+\frac{2}{\sigma}\frac{\lambda}{\tau}\right]\,\mathbb{P}\left(\mathfrak{O}_{0,N}\right)\\
&=1-\left[\frac{2^{\frac{11}{5}}}{3}\beta^{\frac{2}{5}}\pmb{\eta}^{\frac{1}{5}}\vartheta^{\frac{3}{5}}\lambda^{\frac{3}{5}}\,\Big(1-\exp\big[-2^{-\frac{6}{5}}\beta^{\frac{3}{5}}\pmb{\eta}^{-\frac{1}{5}}\vartheta^{-\frac{3}{5}}\lambda^{\frac{7}{5}}N\big]\Big)-\frac{2}{\sigma}\frac{\lambda}{\tau}\right]\,\mathbb{P}\left(\mathfrak{O}_{0,N}\right)\,.
\end{align*}
Now if $\mathbb{P}(\mathfrak{O}_{0,N})\geq \frac{1}{2}$, then inequality~\eqref{ineq-diffusion-probability} directly follows from this. But if $\mathbb{P}(\mathfrak{O}_{0,N})\leq \frac{1}{2}$, inequality~\eqref{ineq-diffusion-probability} is trivially satisfied, as due to Hypothesis~\ref{hyp-lambda} it holds that
\begin{align*}
&1-\left[\frac{2^{\frac{6}{5}}}{3}\beta^{\frac{2}{5}}\pmb{\eta}^{\frac{1}{5}}\vartheta^{\frac{3}{5}}\lambda^{\frac{3}{5}}\,\Big(1-\exp\big[-2^{-\frac{6}{5}}\beta^{\frac{3}{5}}\pmb{\eta}^{-\frac{1}{5}}\vartheta^{-\frac{3}{5}}\lambda^{\frac{7}{5}}N\big]\Big)-\frac{1}{\sigma}\frac{\lambda}{\tau}\right]\\
&\geq 1-\frac{2^{\frac{6}{5}}}{3}\beta^{\frac{2}{5}}\pmb{\eta}^{\frac{1}{5}}\vartheta^{\frac{3}{5}}\lambda^{\frac{3}{5}} \geq 1-\frac{2^{-9}\beta^2}{3} \geq \frac{1}{2}\,.
\end{align*}
\hfill $\diamond$ $\Box$

\vspace{.2cm}

\noindent\textbf{Proof of Corollary~\ref{coro-diffusion-consequence}.}
Recall that $\overline{\sigma} = 2^{-\frac{3}{2}}$. From~\eqref{ineq-mesoscopic-gaps}, we have $\tau_0\geq 2^{-5}\mathsf{q}^{-1}\pmb{\eta}$. Combining these two statements in the last step, using the inequality $1-e^{-a}\geq a\left[1-\frac{a}{2}\right]$ for all $a\geq 0$ in the first step and the assumption~\eqref{ineq-diffusion-N} in the fourth and sixth step, we obtain
\begin{align*}
&\frac{2^{\frac{6}{5}}}{3}\beta^{\frac{2}{5}}\pmb{\eta}^{\frac{1}{5}}\vartheta^{\frac{3}{5}}\lambda^{\frac{3}{5}}\,\Big(1-\exp\big[-2^{-\frac{6}{5}}\beta^{\frac{3}{5}}\pmb{\eta}^{-\frac{1}{5}}\lambda^{\frac{7}{5}}\vartheta^{-\frac{3}{5}}(N-\lfloor 4\,\lambda^{-1}\rfloor)\big]\Big)\\
&\geq 3^{-1}\beta\,\lambda^2(N-\lfloor 4\,\lambda^{-1}\rfloor)\,\left[1-2^{-\frac{11}{5}}\beta^{\frac{3}{5}}\pmb{\eta}^{-\frac{1}{5}}\vartheta^{-\frac{3}{5}}\lambda^{\frac{7}{5}}(N-\lfloor 4\,\lambda^{-1}\rfloor)\right]\\
&\geq 3^{-1}\beta\,\lambda^2(N-\lfloor 4\,\lambda^{-1}\rfloor)-\frac{2^{-\frac{11}{5}}}{3}\beta^{\frac{8}{5}}\pmb{\eta}^{-\frac{1}{5}}\vartheta^{-\frac{3}{5}}\lambda^{\frac{17}{5}}N^2\\
&\geq 3^{-1}\beta\,\lambda^2N(1-2^{-\frac{11}{5}}\beta^{\frac{3}{5}}\pmb{\eta}^{-\frac{1}{5}}\vartheta^{-\frac{3}{5}}\lambda^{\frac{7}{5}}N)-\frac{4\beta}{3}\,\lambda\\
&\geq 3^{-1}\beta\,\lambda^2N(1-2^{-\frac{11}{5}}\beta^{\frac{3}{5}}\pmb{\eta}^{-\frac{1}{5}}\vartheta^{-\frac{3}{5}}\lambda^{\frac{7}{5}}N)-(2^7-2^{\frac{13}{2}})\pmb{\eta}^{-1}\mathsf{q}\,\lambda\\
&\qquad-3^{-1}\beta\,\lambda^2\left[2^{-3}-2^{-\frac{11}{5}}\beta^{\frac{3}{5}}\pmb{\eta}^{-\frac{1}{5}}\vartheta^{-\frac{3}{5}}\lambda^{\frac{7}{5}}2^{-\frac{4}{5}}\beta^{-\frac{3}{5}}\pmb{\eta}^{\frac{1}{5}}\vartheta^{\frac{3}{5}}\lambda^{-\frac{7}{5}}\right]N\\
&\geq 3^{-1}\beta\lambda^2N(1-2^{-\frac{11}{5}}\beta^{\frac{3}{5}}\pmb{\eta}^{-\frac{1}{5}}\vartheta^{-\frac{3}{5}}\lambda^{\frac{7}{5}}N)-(2^7-2^{\frac{13}{2}})\pmb{\eta}^{-1}\mathsf{q}\,\lambda\\
&\qquad-2^{-3}\beta\,\lambda^2\left[3^{-1}-2^{\frac{4}{5}}\beta^{\frac{3}{5}}\pmb{\eta}^{-\frac{1}{5}}\vartheta^{-\frac{3}{5}}\lambda^{\frac{7}{5}}2^{10}\beta^{-1}\pmb{\eta}^{-1}\mathsf{q}\,\lambda^{-1}\right]N\\
&= -\frac{2^{-\frac{11}{5}}}{3}\beta^{\frac{8}{5}}\pmb{\eta}^{-\frac{1}{5}}\vartheta^{-\frac{3}{5}}\lambda^{\frac{17}{5}}\left(N-2^{-\frac{4}{5}}\beta^{-\frac{3}{5}}\pmb{\eta}^{\frac{1}{5}}\vartheta^{\frac{3}{5}}\lambda^{-\frac{7}{5}}\right)\left(N-3 \cdot 2^{10}\beta^{-1}\pmb{\eta}^{-1}\mathsf{q}\,\lambda^{-1}\right)
\\
& \qquad
+2^{-2}N\,\beta\,\lambda^2+2^{\frac{13}{2}}\pmb{\eta}^{-1}\mathsf{q}\,\lambda\\
&\geq 2^{-2}N\,\beta\,\lambda^2+\frac{1}{\overline{\sigma}}\frac{\lambda}{\tau_0}\,.
\end{align*}
Then, the statement of Corollary~\ref{coro-diffusion-consequence} follows from inequality~\eqref{ineq-diffusion-probability} with $(\sigma,\tau)=(\overline{\sigma},\tau_0)$ and $N-\lfloor 4\,\lambda^{-1}\rfloor$ instead of $N$.
\hfill $\square$

\vspace{.2cm}

\noindent \textbf{Proof of Lemma~\ref{lemma-diffusion-conclusion}.} As $N$ fulfills~\eqref{ineq-diffusion-N}, we can apply Corollary~\ref{coro-diffusion-consequence} and certainly $N > 4\,\lambda^{-1}$ holds. Therefore Lemma~\ref{lemma-deterministic-path} implies the time-shifted and time-reversed deterministic implication
\begin{align*}
(W_{N},v_{N})\in\mathfrak{A}_{24}^{\overline{\sigma},\overline{\tau}}
\,
\wedge
\,
\mathfrak{O}_{N-\lfloor 4\,\lambda^{-1}\rfloor-1,N}
\,
\qquad
\Longrightarrow
\qquad
\,
\big(W_{N-\lfloor 4\,\lambda^{-1}\rfloor},v_{N-\lfloor 4\,\lambda^{-1}\rfloor}\big)\in\mathfrak{A}_0^{\overline{\sigma},\tau_0}\,.
\end{align*}
Using this implication in the second step and Corollary~\ref{coro-diffusion-consequence} in the third step, we estimate
\begin{align*}
\mathbb{P}\,\Big((W_N,v_N) \in \mathfrak{A}^{\overline{\sigma},\overline{\tau}}_{24}\,\wedge\,\mathfrak{O}_{0,N}\Big) &= \mathbb{P}\,\Big(\left[(W_N,v_N) \in \mathfrak{A}^{\overline{\sigma},\overline{\tau}}_{24}\wedge\mathfrak{O}_{N-\lfloor 4\,\lambda^{-1}\rfloor-1,N}\right]\wedge\mathfrak{O}_{0,N-\lfloor 4\,\lambda^{-1}\rfloor}\Big)\\
&\leq \mathbb{P}\,\Big((W_{N-\lfloor 4\,\lambda^{-1}\rfloor},v_{N-\lfloor 4\,\lambda^{-1}\rfloor}) \in \mathfrak{A}^{\overline{\sigma},\tau_0}_{0}\,\wedge\,\mathfrak{O}_{0,N-\lfloor 4\,\lambda^{-1}\rfloor}\Big)\\
&\leq 1 - 2^{-2}\,N\,\beta\,\lambda^2\,,
\end{align*}
which shows the claim.
\hfill $\square$

\vspace{.2cm}

\noindent \textbf{Proof of Lemma~\ref{lemma-ascension}.}
According to Lemma~\ref{lemma-allowed-movements}, a run from $\mathfrak{W}\setminus\mathfrak{A}_{\mathsf{m}}^{\sigma,\tau}$ to $\mathfrak{A}^{\sigma,\tau}_{\mathsf{m}+1}$ is~\textemdash~provided one has $(W_0,v_0)\in\mathfrak{Q}$ and $\mathfrak{O}_{0,N}$~\textemdash~only possible via
$$
\mathfrak{W}\setminus\mathfrak{A}_{\mathsf{m}}^{\sigma,\tau}\supset\mathfrak{C}_{\mathsf{m}}^{\sigma}\setminus\mathfrak{S}_{\mathsf{m}}^{\sigma,\tau}\rightarrow\mathfrak{S}_{\mathsf{m}}^{\sigma,\tau}\rightarrow\mathfrak{I}_{\mathsf{m}+1/2}^{\sigma,\tau}\rightarrow\mathfrak{S}_{\mathsf{m}+1}^{\sigma,\tau}\subset\mathfrak{A}^{\sigma,\tau}_{\mathsf{m}+1}
\,,$$
where both steps of the ladder $\mathfrak{S}_{\mathsf{m}}^{\sigma,\tau}$ and $\mathfrak{S}_{\mathsf{m}+1}^{\sigma,\tau}$ must be entered at least once. In particular, the event $(W_N,v_N)\in\mathfrak{A}_{\mathsf{m}+1}^{\sigma,\tau}$ requires the existence of an  $n\in \mathbb{N}\cap [1,N)$ for which $(W_n,v_n)\in\mathfrak{S}_{\mathsf{m}}^{\sigma,\tau}$ holds. 
Hence, the event of interest $(W_N,v_N)\in\mathfrak{A}_{\mathsf{m}+1}^{\sigma,\tau}\,\wedge\,\mathfrak{O}_{0,N}$ equals the following disjoint union:
\begin{align}\label{eq-ascension-disjoint-union}
\begin{split}
(W_N,v_N)\in\mathfrak{A}_{\mathsf{m}+1}^{\sigma,\tau}\,\wedge\,\mathfrak{O}_{0,N}=\bigsqcup\limits_{n=1}^{N-1}\Big[\,(W_N,v_N)\in\mathfrak{A}_{\mathsf{m}+1}^{\sigma,\tau}\,\wedge\,\mathfrak{O}_{0,N}\,\wedge\,n=n_{\textnormal{min}}\,\Big]\,.
\end{split}
\end{align}
Here, $n_{\textnormal{min}}$ denotes the random time of first presence in $\mathfrak{S}_{\mathsf{m}}^{\sigma,\tau}$:
$$
n_{\textnormal{min}}:=\min\left(\left\{n^{\prime}\in \mathbb{N}:(W_{n^{\prime}},v_{n^{\prime}})\in\mathfrak{S}_{\mathsf{m}}^{\sigma,\tau}\right\}\cup\{\infty\}\right)\,.
$$
\noindent\underline{\textit{Step 1.}} \textit{For all $n\in\{1,\dots,N-1\}$, it holds that}
\begin{align*}
&\qquad\qquad\qquad(W_N,v_N)\in\mathfrak{A}_{\mathsf{m}+1}^{\sigma,\tau}\,\wedge\,\mathfrak{O}_{0,N}\,\wedge\,n=n_{\textnormal{min}}\\
&\qquad\qquad\qquad\qquad\qquad\qquad\qquad\Downarrow 
\\
&\quad 2^{-\frac{1}{2}}\sum\limits_{n^{\prime}=n+1}^{N}\mathbf{A}_{\llc_{\mathsf{m}+1}}\left(\mathcal{R}\cdot W_{n^{\prime}-1},\left((\mathcal{R} \cdot W_{n^{\prime}-1})^{\perp}\mathcal{R}\right) \circ v_{n^{\prime}-1},\mathcal{P}_{n^{\prime}}\right)\geq 2^{-\frac{5}{2}}\lambda^{-1}\,.
\end{align*}
\noindent In order to prove this, let us assume that the first line holds. Now we use~\eqref{ineq-vector-expansion-3} and~\eqref{ineq-vector-contraction-2} to show that one has~\textemdash~on condition of $\mathfrak{O}_{0,N}$~\textemdash~for all $n^{\prime}\in\{n+1,\dots, N\}$ the inequality
\begin{align*}
&\|\llc_{\mathsf{m}+1}(v_{n^{\prime}-1})\|^2-\|\llc_{\mathsf{m}+1}(v_{n^{\prime}})\|^2\\
&=\|\llc_{\mathsf{m}+1}(v_{n^{\prime}-1})\|^2-\| \llc_{\mathsf{m}+1}\left(\left(((e^{\lambda\mathcal{P}_{n^{\prime}}}\mathcal{R}) \cdot W_{n^{\prime}-1})^{\perp}(e^{\lambda\mathcal{P}_{n^{\prime}}}\mathcal{R})\right) \circ v_{n^{\prime}-1}\right) \|^2\\
&\leq 2^{-\frac{21}{5}}\beta^{\frac{3}{5}}\pmb{\eta}^{-\frac{1}{5}}\vartheta^{-\frac{3}{5}}\lambda^{\frac{7}{5}}+\| \llc_{\mathsf{m}+1}\left(\left((\mathcal{R} \cdot W_{n^{\prime}-1})^{\perp}\mathcal{R}\right) \circ v_{n^{\prime}-1}\right) \|^2
\\
&
\qquad
-\| \llc_{\mathsf{m}+1}\left(\left(((e^{\lambda\mathcal{P}_{n^{\prime}}}\mathcal{R}) \cdot W_{n^{\prime}-1})^{\perp}(e^{\lambda\mathcal{P}_{n^{\prime}}}\mathcal{R})\right) \circ v_{n^{\prime}-1}\right) \|^2\\
&\leq 2^{-\frac{21}{5}}\beta^{\frac{3}{5}}\pmb{\eta}^{-\frac{1}{5}}\vartheta^{-\frac{3}{5}}\lambda^{\frac{7}{5}}+\lambda\,\mathbf{A}_{\llc_{\mathsf{m}+1}}\left(\mathcal{R}\cdot W_{n^{\prime}-1},\left((\mathcal{R} \cdot W_{n^{\prime}-1})^{\perp}\mathcal{R}\right) \circ v_{n^{\prime}-1},\mathcal{P}_{n^{\prime}}\right)+(9+160\,\lambda)\lambda^2\,.
\end{align*}
Combining this with the assumptions~\eqref{ineq-ascension-assumption} and Hypothesis~\ref{hyp-lambda} yields
\begin{align}\label{ineq-ascension-1}
\begin{split}
&\|\llc_{\mathsf{m}+1}(v_n)\|^2-\|\llc_{\mathsf{m}+1}(v_{N})\|^2\\&=\sum_{n^{\prime}=n+1}^{N}\Big[\|\llc_{\mathsf{m}+1}(v_{n^{\prime}-1})\|^2-\|\llc_{\mathsf{m}+1}(v_{n^{\prime}})\|^2\Big]\\
&\leq \lambda\sum_{n^{\prime}=n+1}^{N}\mathbf{A}_{\llc_{\mathsf{m}+1}}\left(\mathcal{R}\cdot W_{n^{\prime}-1},\left((\mathcal{R} \cdot W_{n^{\prime}-1})^{\perp}\mathcal{R}\right) \circ v_{n^{\prime}-1},\mathcal{P}_{n^{\prime}}\right)\\
&\qquad+2^{-\frac{21}{5}}\beta^{\frac{3}{5}}\pmb{\eta}^{-\frac{1}{5}}\vartheta^{-\frac{3}{5}}\lambda^{\frac{7}{5}}N[1+2^{\frac{21}{5}}(9+160\,\lambda)\beta^{-\frac{3}{5}}\pmb{\eta}^{\frac{1}{5}}\vartheta^{\frac{3}{5}}\lambda^{\frac{3}{5}}]\\
&\leq \lambda\sum_{n^{\prime}=n+1}^{N}\mathbf{A}_{\llc_{\mathsf{m}+1}}\left(\mathcal{R}\cdot W_{n^{\prime}-1},\left((\mathcal{R} \cdot W_{n^{\prime}-1})^{\perp}\mathcal{R}\right) \circ v_{n^{\prime}-1},\mathcal{P}_{n^{\prime}}\right)
\\
&
\qquad
+2^{-\frac{21}{5}}\beta^{\frac{3}{5}}\pmb{\eta}^{-\frac{1}{5}}\vartheta^{-\frac{3}{5}}\lambda^{\frac{7}{5}}N[1+2^{-6}(9+160\,\lambda)\beta]\\
&\leq \lambda\sum_{n^{\prime}=n+1}^{N}\mathbf{A}_{\llc_{\mathsf{m}+1}}\left(\mathcal{R}\cdot W_{n^{\prime}-1},\left((\mathcal{R} \cdot W_{n^{\prime}-1})^{\perp}\mathcal{R}\right) \circ v_{n^{\prime}-1},\mathcal{P}_{n^{\prime}}\right)+1-2^{-\frac{1}{2}}-2^{-2}\,.
\end{split}
\end{align}
According to the first and third statement of assumption of the implication that needs to be shown, it holds that $(W_N,v_N)\in\mathfrak{A}_{\mathsf{m}+1}^{\sigma,\tau}$ and $(W_n,v_n)\in\mathfrak{S}_{\mathsf{m}}^{\sigma,\tau}\subset\mathfrak{C}^{\sigma}_{\mathsf{m}}$, \textit{i.e.}, $\|\llc_{\mathsf{m}+1}(v_N)\|^2\leq \frac{2}{\overline{\sigma}}\frac{\lambda}{\overline{\tau}}$ and $\|\ulc_{\mathsf{m}}(v_n)\|^2\leq {\overline{\sigma}}$, which implies~\textemdash~due to $\|\llc_{\mathsf{m}+1}(v_n)\|^2=1-\|\ulc_{\mathsf{m}}(v_n)\|^2$ and~\eqref{eq-sigma-tau}~\textemdash~the bound
\begin{align}\label{ineq-ascension-2}
\|\llc_{\mathsf{m}+1}(v_n)\|^2-\|\llc_{\mathsf{m}+1}(v_{N})\|^2=1-\|\ulc_{\mathsf{m}}(v_n)\|^2-\|\llc_{\mathsf{m}+1}(v_{N})\|^2\geq 1-\overline{\sigma}-\frac{2}{\overline{\sigma}}\frac{\lambda}{\overline{\tau}}=1-2^{-\frac{1}{2}}\,.
\end{align}
Now combining inequalities~\eqref{ineq-ascension-1} and~\eqref{ineq-ascension-2} yields
$$
\lambda\sum_{n^{\prime}=n+1}^{N}\mathbf{A}_{\llc_{\mathsf{m}+1}}\left(\mathcal{R}\cdot W_{n^{\prime}-1},\left((\mathcal{R} \cdot W_{n^{\prime}-1})^{\perp}\mathcal{R}\right) \circ v_{n^{\prime}-1},\mathcal{P}_{n^{\prime}}\right)+1-2^{-\frac{1}{2}}-2^{-2} \geq 1-2^{-\frac{1}{2}}\,,
$$
and then the implication indeed holds.\hfill $\diamond$

\vspace{.2cm}

\noindent\underline{\textit{Step 2.}} \textit{The implication shown in {\rm Step 1} allows to conclude the statement~\eqref{ineq-ascension-statement}.}\\
\noindent Let us define the set
$$
X_N:=\{n\in\{1,\dots,N-1\}: \mathbb{P}(n=n_{\textnormal{min}})>0\}\,.
$$
We use this definition, the identity~\eqref{eq-ascension-disjoint-union} and the implication of \textit{Step 1} to estimate
\begin{align}\label{ineq-ascension-3}
\begin{split}
\mathbb{P}&\big((W_N,v_N)\in\mathfrak{A}_{\mathsf{m}+1}^{\sigma,\tau}\,\wedge\,\mathfrak{O}_{0,N}\big)\\
&=\sum\limits_{n=1}^{N-1}\,\mathbb{P}\Big((W_N,v_N)\in\mathfrak{A}^{\sigma,\tau}_{\mathsf{m}+1}\,\wedge\,\mathfrak{O}_{0,N}\,\wedge\,n=n_{\textnormal{min}}\Big)\\
&\leq\sum\limits_{n=1}^{N-1}\,\mathbb{P}\left(2^{-\frac{1}{2}}\sum\limits_{n^{\prime}=n+1}^{N}\mathbf{A}_{\llc_{\mathsf{m}+1}}\left(\mathcal{R}\cdot W_{n^{\prime}-1},\left((\mathcal{R} \cdot W_{n^{\prime}-1})^{\perp}\mathcal{R}\right) \circ v_{n^{\prime}-1},\mathcal{P}_{n^{\prime}}\right)\geq 2^{-\frac{5}{2}}\lambda^{-1}\,\wedge\,n=n_{\textnormal{min}}\right)\\
&=\sum\limits_{n\in X_N}\hspace{-0.75mm}\mathbb{P}\left(2^{-\frac{1}{2}}\sum\limits_{n^{\prime}=n+1}^{N}\mathbf{A}_{\llc_{\mathsf{m}+1}}\left(\mathcal{R}\cdot W_{n^{\prime}-1},\left((\mathcal{R} \cdot W_{n^{\prime}-1})^{\perp}\mathcal{R}\right) \circ v_{n^{\prime}-1},\mathcal{P}_{n^{\prime}}\right)\geq 2^{-\frac{5}{2}}\lambda^{-1}\,\wedge\,n=n_{\textnormal{min}}\right)\\
&=\sum\limits_{n\in X_N}\mathbb{P}\left(n=n_{\textnormal{min}}\right)\times\\
&\qquad\times\mathbb{P}\left(2^{-\frac{1}{2}}\sum\limits_{n^{\prime}=n+1}^{N}\mathbf{A}_{\llc_{\mathsf{m}+1}}\left(\mathcal{R}\cdot W_{n^{\prime}-1},\left((\mathcal{R} \cdot W_{n^{\prime}-1})^{\perp}\mathcal{R}\right) \circ v_{n^{\prime}-1},\mathcal{P}_{n^{\prime}}\right)\geq 2^{-\frac{5}{2}}\lambda^{-1}\,\bigg|\,n=n_{\textnormal{min}}\right)\\
&\leq\max\limits_{n\in X_N}\,\mathbb{P}\left(2^{-\frac{1}{2}}\sum\limits_{n^{\prime}=n+1}^{N}\mathbf{A}_{\llc_{\mathsf{m}+1}}\left(\mathcal{R}\cdot W_{n^{\prime}-1},\left((\mathcal{R} \cdot W_{n^{\prime}-1})^{\perp}\mathcal{R}\right) \circ v_{n^{\prime}-1},\mathcal{P}_{n^{\prime}}\right)\geq 2^{-\frac{5}{2}}\lambda^{-1}\,\bigg|\,t=t_{\textnormal{min}}\right)\\
&=\max\limits_{n\in X_N}\,\mathbb{P}\left(\aleph_N-\aleph_n\geq 2^{-\frac{5}{2}}\lambda^{-1}\,\Big|\,n=n_{\textnormal{min}}\right)\,,
\end{split}
\end{align}
where we introduced the abbreviation 
\begin{align}\label{def-martingale}
\aleph_k:=2^{-\frac{1}{2}}\sum\limits_{n^{\prime}=1}^{k}\mathbf{A}_{\llc_{\mathsf{m}+1}}\left(\mathcal{R}\cdot W_{n^{\prime}-1},\left((\mathcal{R} \cdot W_{n^{\prime}-1})^{\perp}\mathcal{R}\right) \circ v_{n^{\prime}-1},\mathcal{P}_{n^{\prime}}\right)\,,\qquad k\in\mathbb{N}_0\,.
\end{align}
To tackle the r.h.s. of~\eqref{ineq-ascension-3}, let us use the definition of $n_{\textnormal{min}}$ to estimate for all ${n\in X_N}$
\begin{align}\label{ineq-ascension-4}
\begin{split}
&\mathbb{P}\left(\aleph_N-\aleph_n\geq 2^{-\frac{5}{2}}\lambda^{-1}\,\Big|\,n=n_{\textnormal{min}}\right)\\
&=\mathbb{P}\left(\aleph_N-\aleph_n\geq 2^{-\frac{5}{2}}\lambda^{-1}\,\Big|\quad (W_n,v_n)\in\mathfrak{S}_{\mathsf{m}}^{\sigma,\tau}\quad\wedge\quad(W_{n^{\prime}},v_{n^{\prime}})\not\in\mathfrak{S}^{\sigma,\tau}_{\mathsf{m}} \quad\forall\,n^{\prime}\in\{0,\dots,n-1\}\right)\\
&=\mathbb{P}\Big(((W_1,v_1),\dots, (W_n,v_n))\in(\mathfrak{W}\setminus\mathfrak{S}_{\mathsf{m}}^{\sigma,\tau})^{\times(n-1)}\times\mathfrak{S}_{\mathsf{m}}^{\sigma,\tau}\Big)^{-1}\times\\
&\qquad\times\int_{(\mathfrak{W}\setminus\mathfrak{S}^{\sigma,\tau}_{\mathsf{m}})^{\times(n-1)}\times\mathfrak{S}^{\sigma,\tau}_{\mathsf{m}}}\textnormal{d}\mathbb{P}(((W_1,v_1),\dots, (W_n,v_n))\in\bullet)(x_1,\dots,x_n)\times\\
&\qquad\qquad\qquad\qquad\qquad\qquad
\times\mathbb{P}_{((W_1,v_1),\dots, (W_n,v_n))=(x_1,\dots,x_n)}\left(\aleph_N-\aleph_n\geq 2^{-\frac{5}{2}}\lambda^{-1}\right)\\
&\leq \sup\limits_{(x_1,\dots,x_n)\in(\mathfrak{W}\setminus\mathfrak{S}^{\sigma,\tau}_{\mathsf{m}})^{\times(n-1)}\times\mathfrak{S}^{\sigma,\tau}_{\mathsf{m}}}\mathbb{P}_{((W_1,v_1),\dots, (W_n,v_n))=(x_1,\dots,x_n)}\left(\aleph_N-\aleph_n\geq 2^{-\frac{5}{2}}\lambda^{-1}\right)
\,,
\end{split}
\end{align}
where the transition kernel, which fulfills the second step of~\eqref{ineq-ascension-4} by definition, satisfies
\begin{align}\label{eq-elementary-Markov-property}
\mathbb{P}_{((W_1,v_1),\dots, (W_n,v_n))=(x_1,\dots,x_n)}\left(\aleph_N-\aleph_n\geq 2^{-\frac{5}{2}}\lambda^{-1}\right)=\mathbb{P}_{(W_0,v_0)=x_t}\left(\aleph_{N-n}-\aleph_0\geq 2^{-\frac{5}{2}}\lambda^{-1}\right)
\end{align}
for all $(x_1,\dots,x_n)\in(\mathfrak{W}\setminus\mathfrak{S}^{\sigma,\tau}_{\mathsf{m}})^{\times(n-1)}\times\mathfrak{S}_{\mathsf{m}}^{\sigma,\tau}$ due to the elementary Markov property. Moreover, the inequality~\eqref{ineq-ascension-4} and the identity~\eqref{eq-elementary-Markov-property} imply
\begin{align}\label{ineq-ascension-5}
\begin{split}
\mathbb{P}\left(\aleph_N-\aleph_n\geq 2^{-\frac{5}{2}}\lambda^{-1}\,\Big|\,n=n_{\textnormal{min}}\right)&\leq \sup\limits_{x_n\in\mathfrak{S}_{\mathsf{m}}^{\sigma,\tau}}\mathbb{P}_{(W_0,v_0)=x_n}\left(\aleph_{N-n}-\aleph_0\geq 2^{-\frac{5}{2}}\lambda^{-1}\right)\\
&\leq \sup\limits_{(W,v)\in\mathfrak{W}}\mathbb{P}_{(W_0,v_0)=(W,v)}\left(\aleph_{N-n}-\aleph_0\geq 2^{-\frac{5}{2}}\lambda^{-1}\right)\,.
\end{split}
\end{align}
Now the stochastic process $(\aleph_k)_{k\in\mathbb{N}_0}$ is a martingale as obvious by its definition~\eqref{def-martingale} and it satisfies $|\aleph_{k+1}-\aleph_k|\leq 1$ for all $k\in\mathbb{N}_0$ due to~\eqref{ineq-vector-expansion-2}. Therefore, Azuma's inequality implies
\begin{align}\label{ineq-ascension-6}
\mathbb{P}_{(W_0,v_0)=(W,v)}\left(\aleph_{N-n}-\aleph_{0}\geq 2^{-\frac{5}{2}}\lambda^{-1}\right)\leq\exp\left[-\frac{(2^{-\frac{5}{2}}\lambda^{-1})^2}{2(N-n)}\right]\leq\exp\left[-2^{-6}N^{-1}\lambda^{-2}\right]
\end{align}
holds for all $(W,v)\in\mathfrak{W}$. Finally, combining~\eqref{ineq-ascension-3},~\eqref{ineq-ascension-5} and~\eqref{ineq-ascension-6} implies~\eqref{ineq-ascension-statement}.
\hfill $\diamond$ $\square$

\vspace{.2cm}

\noindent \textbf{Proof of Lemma~\ref{lemma-anticone-probability}.} We begin with the elementary identity for all $\mathsf{m}\in\{0,\dots,\mathsf{M}-1\}$
\begin{align}\label{eq-anticone-probability-0}
\begin{split}
\mathbb{P} & \Big(\big(W_{N(\mathsf{m}+1)},v_{N(\mathsf{m}+1)}\big)\in\mathfrak{A}_{\mathsf{m}+2}^{\overline{\sigma},\overline{\tau}}\,\wedge\,\mathfrak{O}_{0,N(\mathsf{m}+1)}\Big)\\
&= \mathbb{P}\,\Big(\big(W_{N(\mathsf{m}+1)},v_{N(\mathsf{m}+1)}\big)\in\mathfrak{A}_{\mathsf{m}+2}^{\overline{\sigma},\overline{\tau}}\,\wedge\,\mathfrak{O}_{N\mathsf{m},N(\mathsf{m}+1)}\,\wedge\,\big(W_{N\mathsf{m}},v_{N\mathsf{m}}\big)\in\mathfrak{A}_{\mathsf{m}+1}^{\overline{\sigma},\overline{\tau}}\,\wedge\,\mathfrak{O}_{0,N\mathsf{m}}\Big)\\
&\qquad+\mathbb{P}\,\Big(\big(W_{N(\mathsf{m}+1)},v_{N(\mathsf{m}+1)}\big)\in\mathfrak{A}_{\mathsf{m}+2}^{\overline{\sigma},\overline{\tau}}\,\wedge\,\mathfrak{O}_{N\mathsf{m},N(\mathsf{m}+1)}\,\wedge\,\big(W_{N\mathsf{m}},v_{N\mathsf{m}}\big)\not\in\mathfrak{A}_{\mathsf{m}+1}^{\overline{\sigma},\overline{\tau}}\,\wedge\,\mathfrak{O}_{0,N\mathsf{m}}\Big)\,.
\end{split}
\end{align}
We may assume w.l.o.g. that $\mathbb{P}(\mathfrak{O}_{0,N\mathsf{M}})>0$. For all $\mathsf{m}\in\{0,\dots,\mathsf{M}-1\}$ satisfying
\begin{align}\label{ineq-anticone-probability-condition}
\mathbb{P}\,\big(\big(W_{N\mathsf{m}},v_{N\mathsf{m}}\big)\in\mathfrak{A}_{\mathsf{m}+1}^{\overline{\sigma},\overline{\tau}}\,\wedge\,\mathfrak{O}_{0,N\mathsf{m}}\big)>0\,,
\end{align}
we estimate
\begin{align}\label{ineq-anticone-probability-1}
\begin{split}
\mathbb{P} & \Big(\big(W_{N(\mathsf{m}+1)},v_{N(\mathsf{m}+1)}\big)\in\mathfrak{A}_{\mathsf{m}+2}^{\overline{\sigma},\overline{\tau}}\,\wedge\,\mathfrak{O}_{{N\mathsf{m}},N(\mathsf{m}+1)}\,\Big|\,\big(W_{N\mathsf{m}},v_{N\mathsf{m}}\big)\in\mathfrak{A}_{\mathsf{m}+1}^{\overline{\sigma},\overline{\tau}}\,\wedge\,\mathfrak{O}_{0,N\mathsf{m}}\Big)\\
&=\mathbb{P}\,\Big(\big((W_{N\mathsf{m}+1},v_{N\mathsf{m}+1}),\dots,\big(W_{N(\mathsf{m}+1)},v_{N(\mathsf{m}+1)}\big)\big)\in\mathfrak{Q}^{\times (N-1)}\times\big(\mathfrak{Q}\cap\mathfrak{A}_{\mathsf{m}+2}^{\overline{\sigma},\overline{\tau}}\big)\,\Big|\\
&\qquad\qquad\qquad\Big|\,((W_1,v_1),\dots, (W_{N\mathsf{m}},v_{N\mathsf{m}}))\in \mathfrak{Q}^{\times (N\mathsf{m}-1)}\times\big(\mathfrak{Q}\cap\mathfrak{A}_{\mathsf{m}+1}^{\overline{\sigma},\overline{\tau}}\big)\Big)\\
&=\mathbb{P}\,\Big(((W_1,v_1),\dots, (W_{N\mathsf{m}},v_{N\mathsf{m}}))\in \mathfrak{Q}^{\times (N\mathsf{m}-1)}\times\big(\mathfrak{Q}\cap\mathfrak{A}_{\mathsf{m}+1}^{\overline{\sigma},\overline{\tau}}\big)\Big)^{-1}\times\\
&\quad\times\int_{\mathfrak{Q}^{\times (N\mathsf{m}-1)}\times(\mathfrak{Q}\,\cap\,\mathfrak{A}_{\mathsf{m}+1}^{\overline{\sigma},\overline{\tau}})}\textnormal{d}\mathbb{P}(((W_1,v_1),\dots, (W_{N\mathsf{m}},v_{N\mathsf{m}})))\in\bullet)(y_1,\dots,y_{N\mathsf{m}})\,\times\\
&\qquad\quad
\times\mathbb{P}_{((W_1,v_1),\dots, (W_{N\mathsf{m}},v_{N\mathsf{m}}))=(y_1,\dots,y_{N\mathsf{m}})}\Bigg(\begin{array}{l}
\vspace{-1mm}
\mbox{\footnotesize $\big((W_{N\mathsf{m}+1},v_{N\mathsf{m}+1}),\dots,\big(W_{N(\mathsf{m}+1)},v_{N(\mathsf{m}+1)}\big)\big)$}\\
\qquad\qquad \qquad\quad~\reflectbox{\rotatebox[origin=c]{-90}{$\in$}}\\
\qquad\quad\mbox{\small $\mathfrak{Q}^{\times (N-1)}\times\big(\mathfrak{Q}\cap\mathfrak{A}_{\mathsf{m}+2}^{\overline{\sigma},\overline{\tau}}\big)$}
\end{array}
\Bigg)\\
&\leq \sup\limits_{\vspace{-2mm}\begin{array}{l}
	\vspace{-2mm}
	\mbox{\scriptsize $\qquad (y_1,\dots,y_{N\mathsf{m}})$}\\ 
	\vspace{-2mm}\qquad\quad\mbox{\scriptsize $\reflectbox{\rotatebox[origin=c]{-90}{$\in$}}$}\\
	\mbox{\scriptsize $\mathfrak{Q}^{\times N\mathsf{m}-1}\times(\mathfrak{Q}\,\cap\,\mathfrak{A}_{\mathsf{m}+1}^{\overline{\sigma},\overline{\tau}})$}
	\end{array}}
\!\!\!\!\!\!\!\!
\mathbb{P}_{((W_1,v_1),\dots, (W_{N\mathsf{m}},v_{N\mathsf{m}}))\atop=(y_1,\dots,y_{N\mathsf{m}})}\,\Bigg(\begin{array}{l}
\vspace{-1mm}
\mbox{\footnotesize $\big((W_{N\mathsf{m}+1},v_{N\mathsf{m}+1}),\dots,\big(W_{N(\mathsf{m}+1)},v_{N(\mathsf{m}+1)}\big)\big)$}\\
\qquad\qquad \qquad\quad~\reflectbox{\rotatebox[origin=c]{-90}{$\in$}}\\
\qquad\quad\mbox{\small $\mathfrak{Q}^{\times (N-1)}\times\big(\mathfrak{Q}\cap\mathfrak{A}_{\mathsf{m}+2}^{\overline{\sigma},\overline{\tau}}\big)$}
\end{array}
\Bigg)\\
&= \sup\limits_{y\in \mathfrak{Q}\,\cap\,\mathfrak{A}_{\mathsf{m}+1}^{\overline{\sigma},\overline{\tau}}} \mathbb{P}_{(W_0,v_0)=y}\,\Big(\big((W_{1},v_{1}),\dots,(W_{N},v_{N})\big)\in\mathfrak{Q}^{\times (N-1)}\times\big(\mathfrak{Q}\cap\mathfrak{A}_{\mathsf{m}+2}^{\overline{\sigma},\overline{\tau}}\big)
\Big)\\
&= \sup\limits_{y\in \mathfrak{Q}\,\cap\,\mathfrak{A}_{\mathsf{m}+1}^{\overline{\sigma},\overline{\tau}}}\mathbb{P}_{(W_0,v_0)=y}\,\Big((W_{N},v_{N})\in\mathfrak{A}_{\mathsf{m}+2}^{\overline{\sigma},\overline{\tau}}\,\wedge\,\mathfrak{O}_{0,N}\Big)\,,
\end{split}
\end{align}
in which the second step incorporates the definition of the transition kernel and the fourth step follows for all $(y_1,\dots,y_{N\mathsf{m}})\in \mathfrak{W}^{\times N}$ by the elementary Markov property. Moreover,  we use the inclusions $\mathfrak{Q}\,\cap\,\mathfrak{A}_{\mathsf{m}+1}^{\overline{\sigma},\overline{\tau}}\subset\mathfrak{Q}$ and $\mathfrak{A}_{\mathsf{m}+2}^{\overline{\sigma},\overline{\tau}}\subset\mathfrak{A}_{24}^{\overline{\sigma},\overline{\tau}}$ for all $\mathsf{m}\geq 22$
to prove
\begin{align}\label{ineq-anticone-probability-2}
\begin{split}
\sup\limits_{y\in \mathfrak{Q}\,\cap\, \mathfrak{A}_{\mathsf{m}+1}^{\overline{\sigma},\overline{\tau}}}
&
\mathbb{P}_{(W_0,v_0)=y}\,\Big((W_{N},v_{N})\in\mathfrak{A}_{\mathsf{m}+2}^{\overline{\sigma},\overline{\tau}}\,\wedge\,\mathfrak{O}_{0,N}\Big)
\\
&\leq \sup\limits_{y\in \mathfrak{Q}}\mathbb{P}_{(W_0,v_0)=y}\,\Big((W_{N},v_{N})\in\mathfrak{A}_{\mathsf{m}+2}^{\overline{\sigma},\overline{\tau}}\,\wedge\,\mathfrak{O}_{0,N}\Big)\\
&\leq \sup\limits_{y\in \mathfrak{Q}}\mathbb{P}_{(W_0,v_0)=y}\,\Big((W_{N},v_{N})\in\mathfrak{A}_{24}^{\overline{\sigma},\overline{\tau}}\,\wedge\,\mathfrak{O}_{0,N}\Big)\,,
\end{split}
\end{align}
Therefore, in view of inequalities~\eqref{ineq-anticone-probability-1} and~\eqref{ineq-anticone-probability-2}, all $\mathsf{m}$ obeying~\eqref{ineq-anticone-probability-condition} satisfy
\begin{align*}
\mathbb{P}&\Big(\big(W_{N(\mathsf{m}+1)},v_{N(\mathsf{m}+1)}\big)\in\mathfrak{A}_{\mathsf{m}+2}^{\overline{\sigma},\overline{\tau}}\,\wedge\,\mathfrak{O}_{{N\mathsf{m}},N(\mathsf{m}+1)}\,\Big|\,\big(W_{N\mathsf{m}},v_{N\mathsf{m}}\big)\in\mathfrak{A}_{\mathsf{m}+1}^{\overline{\sigma},\overline{\tau}}\,\wedge\,\mathfrak{O}_{0,N\mathsf{m}}\Big)\\
&\leq \sup\limits_{y\in \mathfrak{Q}}\mathbb{P}_{(W_0,v_0)=y}\,\Big((W_{N},v_{N})\in\mathfrak{A}_{24}^{\overline{\sigma},\overline{\tau}}\,\wedge\,\mathfrak{O}_{0,N}\Big)\,.
\end{align*}
Therefore, for all $\mathsf{m}$ obeying~\eqref{ineq-anticone-probability-condition} Lemma~\ref{lemma-diffusion-conclusion} then implies
\begin{align}\label{ineq-anticone-probability-3}
\begin{split}
\mathbb{P} &\Big(\big(W_{N(\mathsf{m}+1)},v_{N(\mathsf{m}+1)}\big)\in\mathfrak{A}_{\mathsf{m}+2}^{\overline{\sigma},\overline{\tau}}\,\wedge\,\mathfrak{O}_{N\mathsf{m},N(\mathsf{m}+1)}\,\wedge\,\big(W_{N\mathsf{m}},v_{N\mathsf{m}}\big)\in\mathfrak{A}_{\mathsf{m}+1}^{\overline{\sigma},\overline{\tau}}\,\wedge\,\mathfrak{O}_{0,N\mathsf{m}}\Big)\\
&\leq \mathbb{P}\,\Big(\big(W_{N(\mathsf{m}+1)},v_{N(\mathsf{m}+1)}\big)\in\mathfrak{A}_{\mathsf{m}+2}^{\overline{\sigma},\overline{\tau}}\,\wedge\,\mathfrak{O}_{{N\mathsf{m}},N(\mathsf{m}+1)}\,\Big|\,\big(W_{N\mathsf{m}},v_{N\mathsf{m}}\big)\in\mathfrak{A}_{\mathsf{m}+1}^{\overline{\sigma},\overline{\tau}}\,\wedge\,\mathfrak{O}_{0,N\mathsf{m}}\Big)\,\times\\
&\qquad\times\,\mathbb{P}\,\Big(\big(W_{N\mathsf{m}},v_{N\mathsf{m}}\big)\in\mathfrak{A}_{\mathsf{m}+1}^{\overline{\sigma},\overline{\tau}}\,\wedge\,\mathfrak{O}_{0,N\mathsf{m}}\Big)\\
&\leq \bigg[\sup\limits_{y\in \mathfrak{Q}}\mathbb{P}_{(W_0,v_0)=y}\,\Big((W_{N},v_{N})\in\mathfrak{A}_{24}^{\overline{\sigma},\overline{\tau}}\,\wedge\,\mathfrak{O}_{0,N}\Big)\bigg]\,\, \mathbb{P}\,\Big(\big(W_{N\mathsf{m}},v_{N\mathsf{m}}\big)\in\mathfrak{A}_{\mathsf{m}+1}^{\overline{\sigma},\overline{\tau}}\,\wedge\,\mathfrak{O}_{0,N\mathsf{m}}\Big)\\
&\leq \left[1-2^{-2}N\,\beta\,\lambda^2\right]\,\,\mathbb{P}\,\Big(\big(W_{N\mathsf{m}},v_{N\mathsf{m}}\big)\in\mathfrak{A}_{\mathsf{m}+1}^{\overline{\sigma},\overline{\tau}}\,\wedge\,\mathfrak{O}_{0,N\mathsf{m}}\Big)\,.
\end{split}
\end{align}
Further, the overall bound in~\eqref{ineq-anticone-probability-3} holds trivially if $\mathsf{m}$ violates~\eqref{ineq-anticone-probability-condition}. Thus the overall bound in~\eqref{ineq-anticone-probability-3} holds for all $\mathsf{m}\in\{0,\dots,\mathsf{M}-1\}$. 

Now similarly to~\eqref{ineq-anticone-probability-1}, for all $\mathsf{m}\in\{0,\dots,\mathsf{M}-1\}$ satisfying
\begin{align}\label{ineq-anticone-probability-condition-bis}
\mathbb{P}\,\big(\big(W_{N\mathsf{m}},v_{N\mathsf{m}}\big)\notin\mathfrak{A}_{\mathsf{m}+1}^{\overline{\sigma},\overline{\tau}}\,\wedge\,\mathfrak{O}_{0,N\mathsf{m}}\big)>0\,,
\end{align}
we estimate
\begin{align*}
\mathbb{P}&\Big(\big(W_{N(\mathsf{m}+1)},v_{N(\mathsf{m}+1)}\big)\in\mathfrak{A}_{\mathsf{m}+2}^{\overline{\sigma},\overline{\tau}}\,\wedge\,\mathfrak{O}_{{N\mathsf{m}},N(\mathsf{m}+1)}\,\Big|\,\big(W_{N\mathsf{m}},v_{N\mathsf{m}}\big)\notin\mathfrak{A}_{\mathsf{m}+1}^{\overline{\sigma},\overline{\tau}}\,\wedge\,\mathfrak{O}_{0,N\mathsf{m}}\Big)\\
&\leq \sup\limits_{y\in \mathfrak{Q}\setminus \mathfrak{A}_{\mathsf{m}+1}^{\overline{\sigma},\overline{\tau}}}\mathbb{P}_{(W_0,v_0)=y}\,\Big((W_{N},v_{N})\in\mathfrak{A}_{\mathsf{m}+2}^{\overline{\sigma},\overline{\tau}}\,\wedge\,\mathfrak{O}_{0,N}\Big)\,.
\end{align*}
Combining this with Lemma~\ref{lemma-ascension} then implies for all $\mathsf{m}$ obeying~\eqref{ineq-anticone-probability-condition-bis} the bound
\begin{align*}
\mathbb{P}&\Big(\big(W_{N(\mathsf{m}+1)},v_{N(\mathsf{m}+1)}\big)\in\mathfrak{A}_{\mathsf{m}+2}^{\overline{\sigma},\overline{\tau}}\,\wedge\,\mathfrak{O}_{N\mathsf{m},N(\mathsf{m}+1)}\,\wedge\,\big(W_{N\mathsf{m}},v_{N\mathsf{m}}\big)\notin\mathfrak{A}_{\mathsf{m}+1}^{\overline{\sigma},\overline{\tau}}\,\wedge\,\mathfrak{O}_{0,N\mathsf{m}}\Big)\\
&\leq \mathbb{P}\,\Big(\big(W_{N(\mathsf{m}+1)},v_{N(\mathsf{m}+1)}\big)\in\mathfrak{A}_{\mathsf{m}+2}^{\overline{\sigma},\overline{\tau}}\,\wedge\,\mathfrak{O}_{{N\mathsf{m}},N(\mathsf{m}+1)}\,\Big|\,\big(W_{N\mathsf{m}},v_{N\mathsf{m}}\big)\notin\mathfrak{A}_{\mathsf{m}+1}^{\overline{\sigma},\overline{\tau}}\,\wedge\,\mathfrak{O}_{0,N\mathsf{m}}\Big)\,\times\\
&\qquad\times\,\mathbb{P}\,\Big(\big(W_{N\mathsf{m}},v_{N\mathsf{m}}\big)\notin\mathfrak{A}_{\mathsf{m}+1}^{\overline{\sigma},\overline{\tau}}\,\wedge\,\mathfrak{O}_{0,N\mathsf{m}}\Big)\\
&\leq \Bigg[\sup\limits_{y\in \mathfrak{Q}\setminus \mathfrak{A}_{\mathsf{m}+1}^{\overline{\sigma},\overline{\tau}}}\mathbb{P}_{(W_0,v_0)=y}\,\Big((W_{N},v_{N})\in\mathfrak{A}_{\mathsf{m}+2}^{\overline{\sigma},\overline{\tau}}\,\wedge\,\mathfrak{O}_{0,N}\Big)\Bigg]\,\, \mathbb{P}\,\Big(\big(W_{N\mathsf{m}},v_{N\mathsf{m}}\big)\notin\mathfrak{A}_{\mathsf{m}+1}^{\overline{\sigma},\overline{\tau}}\,\wedge\,\mathfrak{O}_{0,N\mathsf{m}}\Big)\\
&\leq \exp\big[-2^{-6}N^{-1}\lambda^{-2}\big]\,\,\mathbb{P}\,\Big(\big(W_{N\mathsf{m}},v_{N\mathsf{m}}\big)\notin\mathfrak{A}_{\mathsf{m}+1}^{\overline{\sigma},\overline{\tau}}\,\wedge\,\mathfrak{O}_{0,N\mathsf{m}}\Big)\,,
\end{align*}
which holds trivially if $\mathsf{m}$ violates~\eqref{ineq-anticone-probability-condition-bis}, so this overall bound is valid for all $\mathsf{m}\in\{0,\dots,\mathsf{M}-1\}$.

Now, the foregoing inequality,~\eqref{eq-anticone-probability-0} and~\eqref{ineq-anticone-probability-3} imply that all $\mathsf{m}\in\{1,\dots,\mathsf{M}\}$~obey
\begin{align*}
\mathbb{P}&\Big(\big(W_{N(\mathsf{m}+1)},v_{N(\mathsf{m}+1)}\big)\in\mathfrak{A}_{\mathsf{m}+2}^{\overline{\sigma},\overline{\tau}}\,\wedge\,\mathfrak{O}_{0,N(\mathsf{m}+1)}\Big)\\
&\leq \left[1-2^{-2}N\,\beta\,\lambda^2\right]\mathbb{P}\,\Big(\big(W_{N\mathsf{m}},v_{N\mathsf{m}}\big)\in\mathfrak{A}_{\mathsf{m}+1}^{\overline{\sigma},\overline{\tau}}\,\wedge\,\mathfrak{O}_{0,N\mathsf{m}}\Big)+\exp\big[-2^{-6}N^{-1}\lambda^{-2}\big]\,.
\end{align*}
Iterating this from $\mathsf{m}=22$ to $\mathsf{M}-1$ using~\eqref{ineq-iterative-estimate}, $x^2e^{-x} \leq e^{-\frac{x}{4}}$ and $\log(2^{-\frac{54}{5}}x^{-1})^{13} \leq 2^{\frac{93}{5}}x^{-1}$ for all $x \geq 0$ in the second and seventh step respectively, and inserting the lower bound~\eqref{def-M} yields
\begin{align*}
\mathbb{P} &\Big(\big(W_{N\mathsf{M}},v_{N\mathsf{M}}\big)\in\mathfrak{A}_{\mathsf{M}+1}^{\overline{\sigma},\overline{\tau}}\,\wedge\,\mathfrak{O}_{0,N\mathsf{M}}\Big)\\
&\leq \exp\left[-2^{-2}N\,\beta\,\lambda^2(\mathsf{M}-22)\right] + 2^2N^{-1}\beta^{-1}\lambda^{-2}\exp\big[-2^{-6}N^{-1}\lambda^{-2}\big]\\
&\leq \exp\left[-2^{-2}N\,\beta\,\lambda^2(2^{-10}\pmb{\eta}\,\mathsf{q}^{-1}\lambda^{-1}-23)\right] 
\\
&
\qquad
+ 2^{14}N\beta^{-1}\lambda^2(N^{-1}N)(\vartheta^{-1}\lambda^{-1}\vartheta\lambda)^{\frac{3}{5}}\exp\big[-2^{-8}N^{-1}\lambda^{-2}\big]\\
&= \exp\left[23 \cdot 2^{-2}N\beta\,\lambda^2\right]\exp\left[-2^{-12}N\beta\,\pmb{\eta}\,\mathsf{q}^{-1}\lambda\right]\\
&\qquad+ 2^{14}N\beta^{-1}\lambda^2(N^{-1}2^{-\frac{4}{5}}\beta^{-\frac{3}{5}}\pmb{\eta}^{\frac{1}{5}}\vartheta^{\frac{3}{5}}\lambda^{-\frac{7}{5}})(\vartheta^{-1}\lambda^{-1}2^{-17}\beta^{\frac{8}{3}}\pmb{\eta}^{-\frac{1}{3}})^{\frac{3}{5}}\exp\big[-2^{-8}N^{-1}\lambda^{-2}(N^{-1}N)^2\big]\\
&\leq \exp\left[23 \cdot 2^{-2}N\beta^{\frac{3}{5}}\pmb{\eta}^{-\frac{1}{5}}\lambda^2\right]\exp\left[-2^{-12}N\beta\,\pmb{\eta}\,\mathsf{q}^{-1}\lambda\right]\\
&\qquad+ 2^3\exp\big[-2^{-8}N^{-1}\lambda^{-2}(2^{\frac{4}{5}}\beta^{\frac{3}{5}}\pmb{\eta}^{-\frac{1}{5}}\vartheta^{-\frac{3}{5}}\lambda^{\frac{7}{5}}N)^2(\mathsf{q}^{-1}\mathsf{q})^{\frac{1}{3}}(\vartheta^{-1}\lambda^{-1}\vartheta\lambda)^{\frac{1}{20}}2^{-\frac{73}{10}}\pmb{\eta}^{\frac{19}{12}}\mathsf{q}^{-\frac{4}{3}}\big]\\
&\leq \exp\left[23 \cdot 2^{-4}\lambda^{\frac{3}{5}}\right]\exp\left[-2^{-12}N\beta\,\pmb{\eta}\,\mathsf{q}^{-1}\lambda\right]\\
&\qquad+ 2^3\exp\big[-2^{-\frac{137}{10}}N\beta^{\frac{6}{5}}\pmb{\eta}^{\frac{71}{60}}\mathsf{q}^{-\frac{4}{3}}\vartheta^{-\frac{6}{5}}\lambda^{\frac{4}{5}}(\mathsf{q}\,2^{\frac{36}{5}}\beta^{-\frac{1}{5}}\pmb{\eta}^{-\frac{3}{5}}\vartheta^{\frac{1}{5}}\lambda^{\frac{1}{5}})^{\frac{1}{3}}(\vartheta\lambda\,2^{17}\beta^{-\frac{8}{3}}\pmb{\eta}^{\frac{1}{3}})^{\frac{1}{20}}\big]\\
&\leq \exp\left[23 \cdot 2^{-\frac{56}{5}}\right]\exp\left[-2^{-12}N\beta\,\pmb{\eta}\,\mathsf{q}^{-1}\lambda\right] + 2^3\exp\big[-2^{-\frac{209}{20}}N\beta\,\pmb{\eta}\,\mathsf{q}^{-1}\vartheta^{-\frac{13}{12}}\lambda^{\frac{11}{12}}\big]\\
&\leq 2^3\exp\left[-2^{-12}N\beta\,\pmb{\eta}\,\mathsf{q}^{-1}\lambda\right] + 2^3\exp\big[-2^{-\frac{209}{20}}N\beta\,\pmb{\eta}\,\mathsf{q}^{-1}\lambda^{\frac{11}{12}}(2^{\frac{93}{5}}\lambda^{-1})^{-\frac{1}{12}}\big]\\
&= 2^4\exp\left[-2^{-12}N\beta\,\pmb{\eta}\,\mathsf{q}^{-1}\lambda\right]\,,
\end{align*}
which is the statement of Lemma~\ref{lemma-anticone-probability}.
\hfill $\square$

\vspace{.2cm}

\noindent\textbf{Proof of Lemma~\ref{lemma-outside-anticone}.}
Let us assume that the statements on the l.h.s. of the implication hold. It then follows that all $\tilde{n}\in\{1,\dots,S_0\}$ satisfy $(W_{\tilde{n}-1},v_{\tilde{n}-1})\in\mathfrak{Q}$, which implies
\begin{align}\label{ineq-vector-enlargment}
\|\llc_{\mathsf{M}+1}(v_{\tilde{n}})\|^2\geq \|\llc_{\mathsf{M}+1}(v_{\tilde{n}-1})\|^2-\frac{7}{4}\lambda
\end{align}
due to~\eqref{ineq-deterministic-vector-2}.
Further, the assumption $(W_{0},v_{0})\hspace{-0.25mm}\not\in\hspace{-0.25mm}\mathfrak{A}_{\mathsf{M}+1}^{\overline{\sigma},\overline{\tau}}$ is equivalent~to
$$
\|\llc_{\mathsf{M}+1}(v_0)\|^2>\frac{2}{\overline{\sigma}}\,\frac{\lambda}{\overline{\tau}}=2^{-\frac{3}{2}}\,.
$$
Using this and~\eqref{ineq-trace-order-lambda}, iterating~\eqref{ineq-vector-enlargment} for $\tilde{n}=1$ to $n-1$ for some $n\in\{1,\dots, S_0\}$ yields
\begin{align}\label{ineq-outside-anticone-1}
\begin{split}
\|\llc_{\mathsf{M}+1}(v_{n-1})\|^2&\geq \|\llc_{\mathsf{M}+1}(v_{0})\|^2-\frac{7}{4}\lambda\,(n-1)\geq 2^{-\frac{3}{2}}-\frac{7}{4}\lambda\,(S_0-1) \geq 2^{-\frac{3}{2}} - 7 \cdot 2^{-\frac{13}{2}}\\
&\geq 2^{-2} + 2^{-15} \geq 2^{-2} + 2^{-3}\lambda \geq 2^{-2} + 2^{-\frac{21}{5}}\beta^{\frac{3}{5}}\pmb{\eta}^{-\frac{1}{5}}\vartheta^{-\frac{3}{5}}\lambda^{\frac{7}{5}}
\end{split}
\end{align}
because $S_0-1\leq 2^{-\frac{9}{2}}\,\lambda^{-1}$ and by Hypothesis~\ref{hyp-lambda}.
We now use $\hat{\zeta}_{\mathsf{M}+1}^{\perp}(\hat{\zeta}_{\mathsf{M}+1}^{\perp})^{*}\leq \hat{\zeta}^{\perp}(\hat{\zeta}^{\perp})^{*}$ and $\hat{\zeta}_{\mathsf{M}+1}^{\perp}(\hat{\zeta}_{\mathsf{M}+1}^{\perp})^{*}=\mathbf{1}_{\mathsf{L}}-\hat{\zeta}_{\mathsf{M}+1}\hat{\zeta}_{\mathsf{M}+1}^{*}$ in the third step and $(W_{n-1},v_{n-1})\in\mathfrak{Q}$ and~\eqref{ineq-outside-anticone-1} in the fifth step to estimate
\begin{align}\label{ineq-outside-anticone-2}
\begin{split}
\big\|(\hat{\zeta}^{\perp}_{\mathsf{M}+1})^*(W_{n-1}+v_{n-1}v_{n-1}^*)\hat{\zeta}^{\perp}_{\mathsf{M}+1}\big\|&\leq \operatorname{tr}\left[(\hat{\zeta}^{\perp}_{\mathsf{M}+1})^*(W_{n-1}+v_{n-1}v_{n-1}^*)\hat{\zeta}^{\perp}_{\mathsf{M}+1}\right]\\
&=\operatorname{tr}\left[W_{n-1}\hat{\zeta}_{\mathsf{M}+1}^{\perp}(\hat{\zeta}_{\mathsf{M}+1}^{\perp})^{*}W_{n-1}\right]+\operatorname{tr}\left[v_{n-1}^*\hat{\zeta}_{\mathsf{M}+1}^{\perp}(\hat{\zeta}_{\mathsf{M}+1}^{\perp})^{*}v_{n-1}\right]\\
&\leq\operatorname{tr}\left[W_{n-1}\hat{\zeta}^{\perp}(\hat{\zeta}^{\perp})^{*}W_{n-1}\right]+\operatorname{tr}\left[v_{n-1}^*\big(\mathbf{1}_{\mathsf{L}}-\hat{\zeta}_{\mathsf{M}+1}\hat{\zeta}_{\mathsf{M}+1}^{*}\big)v_{n-1}\right]\\
&=\operatorname{tr}\left[(\hat{\zeta}^{\perp})^*W_{n-1}\hat{\zeta}^{\perp}\right]+1-\left\|\llc_{\mathsf{M}+1}(v_{n-1})\right\|^2\\
&\leq 2^{-\frac{21}{5}}\beta^{\frac{3}{5}}\pmb{\eta}^{-\frac{1}{5}}\vartheta^{-\frac{3}{5}}\lambda^{\frac{7}{5}}+1-2^{-2}-2^{-\frac{21}{5}}\beta^{\frac{3}{5}}\pmb{\eta}^{-\frac{1}{5}}\vartheta^{-\frac{3}{5}}\lambda^{\frac{7}{5}}\\
&= 1-2^{-2}
\end{split}
\end{align}
for all $n\in\{1,\dots, S_0\}$. Now inequality~\eqref{ineq-d-contraction-bis} with $W_{n-1}+v_{n-1}v_{n-1}^*$ instead of $Q$ and with $\hat{\chi}\equiv\hat{\chi}_{\mathsf{M}+1}$ and $\hat{\zeta}_{\mathsf{M}+1}^{\perp}$ instead of $\hat{\alpha}$ and $\hat{\gamma}^{\perp}$, respectively, and thus with $\tau_{\mathsf{M}+1}$ instead of $\pmb{\eta}$ reads
\begin{align}\label{ineq-outside-anticone-3}
\begin{split}
\operatorname{tr}& \left[\hat{\chi}^*(\mathcal{R}\cdot (W_{n-1}+v_{n-1}v_{n-1}^*))\hat{\chi}\right]\\
&\leq\left(1-\tau_{\mathsf{M}+1}\left[1-\big\|(\hat{\zeta}^{\perp}_{\mathsf{M}+1})^*(W_{n-1}+v_{n-1}v_{n-1}^*)\hat{\zeta}^{\perp}_{\mathsf{M}+1}\big\|\right]\right)
\, \operatorname{tr}\left[\hat{\chi}^*(W_{n-1}+v_{n-1}v_{n-1}^*)\hat{\chi}\right]\,.
\end{split}
\end{align}
Combining~\eqref{ineq-outside-anticone-2},~\eqref{ineq-outside-anticone-3} and the second item of~\eqref{ineq-mesoscopic-gaps} yields for all $n\in\{1,\dots,S_0\}$
\begin{align}\label{ineq-outside-anticone-4}
\operatorname{tr}\left[\hat{\chi}^*(\mathcal{R}\cdot (W_{n-1}+v_{n-1}v_{n-1}^*))\hat{\chi}\right]\leq\left(1-2^{-7}\mathsf{q}^{-1}\pmb{\eta}\right)
\, \operatorname{tr}\left[\hat{\chi}^*(W_{n-1}+v_{n-1}v_{n-1}^*)\hat{\chi}\right]\,.
\end{align}
We now use~\eqref{def-Z} and the inequalities~\eqref{ineq-expansion} with $Q=\mathcal{R}\cdot (W_{n-1}+v_{n-1}v_{n-1}^*)$, $Q =\mathbf{0}$ and $Q^{\prime}=\mathcal{R}\cdot (W_{n-1}+v_{n-1}v_{n-1}^*)$ in order to estimate
\begin{align}\label{ineq-outside-anticone-5}
\begin{split}
\operatorname{tr}&\left[\hat{\chi}^*(W_n+v_nv_n^*)\hat{\chi}\right]\\
&=\operatorname{tr}\left[\hat{\chi}^*(\mathcal{R}\cdot (W_{n-1}+v_{n-1}v_{n-1}^*))\hat{\chi}\right]+\lambda\,\operatorname{tr}\left[\hat{\chi}^*\,\mathtt{X}(\mathcal{R}\cdot (W_{n-1}+v_{n-1}v_{n-1}^*),\mathcal{P}_n)\,\hat{\chi}\right]\\
&\quad+\lambda^2\,\operatorname{tr}\left[\hat{\chi}^*\,\mathtt{Y}(\mathcal{R}\cdot (W_{n-1}+v_{n-1}v_{n-1}^*),\mathcal{P}_n)\,\hat{\chi}\right]+\lambda^3\,\operatorname{tr}\left[\hat{\chi}^*\,\mathtt{Z}^{(\lambda)}(\mathcal{R}\cdot (W_{n-1}+v_{n-1}v_{n-1}^*),\mathcal{P}_n)\,\hat{\chi}\right]\\
&\leq \operatorname{tr}\left[\hat{\chi}^*(\mathcal{R}\cdot (W_{n-1}+v_{n-1}v_{n-1}^*))\hat{\chi}\right]+\lambda\,\operatorname{tr}\left[\hat{\chi}^*\,\mathtt{X}(\mathcal{R}\cdot (W_{n-1}+v_{n-1}v_{n-1}^*),\mathcal{P}_n)\,\hat{\chi}\right]\\
&\quad+\lambda^2\,\operatorname{rk}\left[\mathtt{Y}(\mathcal{R}\cdot (W_{n-1}+v_{n-1}v_{n-1}^*),\mathcal{P}_n)\right]\,\left\|\mathtt{Y}(\mathcal{R}\cdot (W_{n-1}+v_{n-1}v_{n-1}^*),\mathcal{P}_n)\right\|\\
&\quad+\lambda^3\,\operatorname{rk}\left[\mathtt{Z}^{(\lambda)}(\mathcal{R}\cdot (W_{n-1}+v_{n-1}v_{n-1}^*),\mathcal{P}_n)\right]\,\left\|\mathtt{Z}^{(\lambda)}(\mathcal{R}\cdot (W_{n-1}+v_{n-1}v_{n-1}^*),\mathcal{P}_n)\right\|\\
&\leq \operatorname{tr}\left[\hat{\chi}^*(\mathcal{R}\cdot (W_{n-1}+v_{n-1}v_{n-1}^*))\hat{\chi}\right]+\lambda\,\operatorname{tr}\left[\hat{\chi}^*\,\mathtt{X}(\mathcal{R}\cdot (W_{n-1}+v_{n-1}v_{n-1}^*),\mathcal{P}_n)\,\hat{\chi}\right]\\
&\quad+\left[3 \cdot \mbox{$\frac{3}{2}$}+4 \cdot 20\,\lambda\right]\,(\mathsf{w}+1)\,\lambda^2\\
&\leq \operatorname{tr}\left[\hat{\chi}^*(\mathcal{R}\cdot (W_{n-1}+v_{n-1}v_{n-1}^*))\hat{\chi}\right]+\lambda\,\operatorname{tr}\left[\hat{\chi}^*\,\mathtt{X}(\mathcal{R}\cdot (W_{n-1}+v_{n-1}v_{n-1}^*),\mathcal{P}_n)\,\hat{\chi}\right] + 2^8\mathsf{q}^2\pmb{\eta}^{-1}\lambda^2\,.
\end{split}
\end{align}
Let $\Phi_n^{\prime} \in \mathbb{F}_{\mathsf{L},\mathsf{w}+1}$ such that $\Phi_n^{\prime}(\Phi_n^{\prime})^*=\mathcal{R}\cdot (W_{n-1}+v_{n-1}v_{n-1}^*)$. Now we apply the Cauchy-Schwarz inequality to the Hilbert-Schmidt inner product to prove the inequality
\begin{align}\label{ineq-outside-anticone-6}
\begin{split}
\operatorname{tr}&\left[\hat{\chi}^*\,\mathtt{X}(\mathcal{R}\cdot (W_{n-1}+v_{n-1}v_{n-1}^*),\mathcal{P}_n)\,\hat{\chi}\right]\\
&= \operatorname{tr}\left[\hat{\chi}^*[\Phi_n^{\prime}(\Phi_n^{\prime})^*]^{\perp}\mathcal{P}_n\Phi_n^{\prime}(\Phi_n^{\prime})^*\hat{\chi}+\hat{\chi}^*\Phi_n^{\prime}(\Phi_n^{\prime})^*\mathcal{P}_n^*[\Phi_n^{\prime}(\Phi_n^{\prime})^*]^{\perp}\hat{\chi}\right]\\
&= \operatorname{tr}\left[(\hat{\chi}^*\Phi_n^{\prime})^*(\hat{\chi}^*[\Phi_n^{\prime}(\Phi_n^{\prime})^*]^{\perp}\mathcal{P}_n\Phi_n^{\prime})\right]+\operatorname{tr}\left[(\hat{\chi}^*[\Phi_n^{\prime}(\Phi_n^{\prime})^*]^{\perp}\mathcal{P}_n\Phi_n^{\prime})^*(\hat{\chi}^*\Phi_n^{\prime})\right]\\
&\leq 2\left|\operatorname{tr}\left[(\hat{\chi}^*\Phi_n^{\prime})^*(\hat{\chi}^*[\Phi_n^{\prime}(\Phi_n^{\prime})^*]^{\perp}\mathcal{P}_n\Phi_n^{\prime})\right]\right|\\
&\leq 2\left(\operatorname{tr}\left[(\hat{\chi}^*\Phi_n^{\prime})^*\hat{\chi}^*\Phi_n^{\prime}\right]\right)^{\frac{1}{2}}\left(\operatorname{tr}\left[(\hat{\chi}^*[\Phi_n^{\prime}(\Phi_n^{\prime})^*]^{\perp}\mathcal{P}_n\Phi_n^{\prime})^*\hat{\chi}^*[\Phi_n^{\prime}(\Phi_n^{\prime})^*]^{\perp}\mathcal{P}_n\Phi_n^{\prime}\right]\right)^{\frac{1}{2}}\\
&= 2\left(\operatorname{tr}\left[(\Phi_n^{\prime})^*\hat{\chi}\hat{\chi}^*\Phi_n^{\prime}\right]\right)^{\frac{1}{2}}\left(\operatorname{tr}\left[(\Phi_n^{\prime})^*\mathcal{P}_n^*[\Phi_n^{\prime}(\Phi_n^{\prime})^*]^{\perp}\hat{\chi}\hat{\chi}^*[\Phi_n^{\prime}(\Phi_n^{\prime})^*]^{\perp}\mathcal{P}_n\Phi_n^{\prime}\right]\right)^{\frac{1}{2}}\\
&\leq 2\left(\operatorname{tr}\left[(\Phi_n^{\prime})^*\hat{\chi}\hat{\chi}^*\Phi_n^{\prime}\right]\right)^{\frac{1}{2}}\left(\operatorname{tr}\left[(\Phi_n^{\prime})^*\Phi_n^{\prime}\right]\right)^{\frac{1}{2}}\\
&= 2(\mathsf{w}+1)^{\frac{1}{2}}\left(\operatorname{tr}\left[\hat{\chi}^*\Phi_n^{\prime}(\Phi_n^{\prime})^*\hat{\chi}\right]\right)^{\frac{1}{2}}\\
&\leq 2\mathsf{q}^{\frac{1}{2}}\left(\operatorname{tr}\left[\hat{\chi}^*\Phi_n^{\prime}(\Phi_n^{\prime})^*\hat{\chi}\right]\right)^{\frac{1}{2}} + \mathsf{q}^{\frac{1}{2}}\left(2^{16}\pmb{\eta}^{-2}\mathsf{q}^3\lambda^2\right)^{-\frac{1}{2}}\left[\left(\operatorname{tr}\left[\hat{\chi}^*\Phi_n^{\prime}(\Phi_n^{\prime})^*\hat{\chi}\right]\right)^{\frac{1}{2}} - \left(2^{16}\pmb{\eta}^{-2}\mathsf{q}^3\lambda^2\right)^{\frac{1}{2}}\right]^2\\
&= 2^{-8}\pmb{\eta}\,\mathsf{q}^{-1}\lambda^{-1}\operatorname{tr}\left[\hat{\chi}^*(\mathcal{R}\cdot (W_{n-1}+v_{n-1}v_{n-1}^*))\hat{\chi}\right] + 2^8\pmb{\eta}^{-1}\mathsf{q}^2\lambda\\
&\leq 2^{-8}\pmb{\eta}\,\mathsf{q}^{-1}\lambda^{-1}\operatorname{tr}\left[\hat{\chi}^*(W_{n-1}+v_{n-1}v_{n-1}^*)\hat{\chi}\right] + 2^8\pmb{\eta}^{-1}\mathsf{q}^2\lambda\,,
\end{split}
\end{align}
in which the last step follows from~\eqref{ineq-outside-anticone-4}. Combining~\eqref{ineq-outside-anticone-4},~\eqref{ineq-outside-anticone-5} and~\eqref{ineq-outside-anticone-6} then yields
\begin{align*}
\operatorname{tr}\left[\hat{\chi}^*(W_n+v_nv_n^*)\hat{\chi}\right] \leq \left(1-2^{-8}\mathsf{q}^{-1}\pmb{\eta}\right)\operatorname{tr}\left[\hat{\chi}^*(W_{n-1}+v_{n-1}v_{n-1}^*)\hat{\chi}\right] + 2^9\pmb{\eta}^{-1}\mathsf{q}^2\lambda^2\,.
\end{align*}
Applying~\eqref{ineq-iterative-estimate}, using Hypothesis~\ref{hyp-lambda}, $S_0 \geq 2^{-\frac{9}{2}}\lambda^{-1}$ and that $\left(\frac{5e}{12}\right)^{\frac{12}{5}}e^{-x} \leq x^{-\frac{12}{5}}$ for all $x > 0$ yields
\begin{align*}
\operatorname{tr} & \left[\hat{\chi}^*(W_{S_0}+v_{S_0}v_{S_0}^*)\hat{\chi}\right] 
\\
&\leq \exp\left[-2^{-8}\mathsf{q}^{-1}\pmb{\eta}S_0\right]\operatorname{tr}\left[\hat{\chi}^*(W_0+v_0v_0^*)\hat{\chi}\right] + 2^{17}\pmb{\eta}^{-2}\mathsf{q}^3\lambda^2\\
&\leq \exp\left[-2^{-\frac{25}{2}}\pmb{\eta}\,\mathsf{q}^{-1}\lambda^{-1}\right]\left(\operatorname{tr}\left[(\hat{\zeta}^{\perp})^*W_0\hat{\zeta}^{\perp}\right]+1\right) + 2^{17}\pmb{\eta}^{-2}\mathsf{q}^3\lambda^2\\
&\leq \left(2^{-3}\lambda+1\right)\exp\left[-2^{-\frac{25}{2}}\pmb{\eta}\,\mathsf{q}^{-1}\lambda^{-1}\right] + 2^{-\frac{23}{5}}\beta^{\frac{3}{5}}\pmb{\eta}^{-\frac{1}{5}}\lambda^{\frac{7}{5}}\\
&\leq 2^{-\frac{2}{5}}\left(\frac{5e}{12}\right)^{\frac{12}{5}}\exp\left[-2^{-\frac{25}{2}}\pmb{\eta}\,\mathsf{q}^{-1}\lambda^{-1}\right]\beta^{-\frac{2}{5}}\pmb{\eta}^{-\frac{4}{5}}\mathsf{q}^{\frac{13}{5}}\vartheta^{\frac{2}{5}} + 2^{-\frac{23}{5}}\beta^{\frac{3}{5}}\pmb{\eta}^{-\frac{1}{5}}\lambda^{\frac{7}{5}}\\
&\leq 2^{\frac{148}{5}}\beta^{-\frac{2}{5}}\pmb{\eta}^{-\frac{16}{5}}\mathsf{q}^{5}\vartheta^{\frac{2}{5}}\lambda^{\frac{12}{5}} + 2^{-\frac{23}{5}}\beta^{\frac{3}{5}}\pmb{\eta}^{-\frac{1}{5}}\lambda^{\frac{7}{5}}\\
&\leq 2^{\frac{148}{5}}(\vartheta^{-\frac{1}{5}}\lambda^{-\frac{1}{5}}2^{-\frac{36}{5}}\beta^{\frac{1}{5}}\pmb{\eta}^{\frac{3}{5}})^{5}\beta^{-\frac{2}{5}}\pmb{\eta}^{-\frac{16}{5}}\lambda^{\frac{12}{5}}\vartheta^{\frac{2}{5}} + (2^{-\frac{21}{5}}-2^{-\frac{32}{5}})\beta^{\frac{3}{5}}\pmb{\eta}^{-\frac{1}{5}}\vartheta^{-\frac{3}{5}}\lambda^{\frac{7}{5}}\\
&=  2^{-\frac{21}{5}}\beta^{\frac{3}{5}}\pmb{\eta}^{-\frac{1}{5}}\vartheta^{-\frac{3}{5}}\lambda^{\frac{7}{5}}\,,
\end{align*}
where we used~\eqref{ineq-trace-order-lambda} in the third step. Then the r.h.s. of the implication indeed holds.
\hfill$\square$

\appendix

\section{Appendix: Quantitative analytic perturbation theory}
\label{app-perturbative-bounds}

Analytic perturbation theory is a well-known classical subject \cite{Kat}. Here we provide a quantitative bound on the third order error terms of the expansion of the fundamental. It is spelled out introducing some notation that fits the application for the proof of Lemma~\ref{lemma-diffusion-probability}.

\begin{lemma}\label{lemma-Hamiltonian}
Let $\mathtt{H}= \mathtt{H}_0 + \lambda\mathtt{H}_1 + \lambda^2\mathtt{H}_2$ be a self-adjoint operator on a $\mathtt{w}+1$-dimensional complex Hilbert space depending on a positive parameter $\lambda > 0$. Denote the eigenvalues of $\mathtt{H}$ by $E_0^{(\lambda)},E_1^{(\lambda)},\dots,E_{\mathtt{w}}^{(\lambda)}$, those of $\mathtt{H}_0$ by $E_0,E_1,\dots,E_{\mathtt{w}}$ and write $\psi_0,\psi_1,\dots,\psi_{\mathtt{w}}$ for corresponding normalized eigenvectors, so in particular $E_0 = \langle \psi_0| \mathtt{H}_0\psi_0 \rangle$. Assume that neither $E_0$ nor $E_0^{(\lambda)}$ is a degenerate eigenvalue. If there exist $G > 0$ and $g > 0$ with $\frac{G}{2} > g$ such that
	\begin{align}\label{ineq-radius}
	|E_0^{(\lambda)}-E_0| \leq g\,, \qquad |E_0-E_k| \geq G\qquad\text{and}\qquad |E_0-E^{(\lambda)}_k| \geq G-g
	\end{align}
	holds for all $k \in \lbrace 1,\dots,\mathtt{w} \rbrace$, then the third order error term $E^{(\lambda)}$ characterized by
	\begin{align}\label{def-E-lambda}
	E_0^{(\lambda)} = E_0 + \lambda\langle \psi_0|\mathtt{H}_1\psi_0 \rangle + \lambda^2\langle \psi_0|\mathtt{H}_2\psi_0 \rangle - \lambda^2\sum_{k=1}^{\mathtt{w}}\frac{|\langle \psi_0|\mathtt{H}_1\psi_k \rangle|^2}{E_k-E_0} + \lambda^3E^{(\lambda)}
	\end{align}
	fulfills
	\begin{align}\label{ineq-Hamiltonian}
	\begin{split}
	|E^{(\lambda)}| &\leq \left[1+8\lambda G^{-2}\left(1+2\|\mathtt{H}\|(G-2g)^{-1}\right)(\|\mathtt{H}_1\|+\lambda\|\mathtt{H}_2\|)\right]\left[2G^{-2}\|\mathtt{H}_1\|^3+2G^{-1}\|\mathtt{H}_1\|\|\mathtt{H}_2\|\right]\\
	&\qquad\; +4\,\lambda G^{-2}\left[1+2\|\mathtt{H}\|(G-2g)^{-1}\right]\left[\|\mathtt{H}_2\|^2+2G^{-1}\|\mathtt{H}_1\|^2\|\mathtt{H}_2\|\right]\,.
	\end{split}
	\end{align}
\end{lemma}

\noindent\textbf{Proof of Lemma~\ref{lemma-Hamiltonian}.} The conditions on $G$ and $g$ specify the separation of the eigenvalues $E_0$ and $E_0^{(\lambda)}$ and the rest of the spectrum of the operators $\mathtt{H}_0$ and $\mathtt{H}$, respectively. Concretely, the circle in the complex plane with middle point $E_0$ and radius $\frac{G}{2}$ only contains $E_0$ and $E_0^{(\lambda)}$, and no other eigenvalues of $\mathtt{H}_0$ and $\mathtt{H}$. Let us denote this (positively oriented) curve by $\Gamma$. Setting $\mathtt{R}_0(z) = (\mathtt{H}_0-z)^{-1}$ for all $z \in \Gamma$, one has
\begin{align}\label{eq-resolvent}
\frac{-1}{2\pi i}\oint_{\Gamma}(\mathtt{H}_0-z)^{-1}dz = \frac{-1}{2\pi i}\oint_{\Gamma}\mathtt{R}_0(z)dz = \psi_0\psi_0^*\,,
\end{align}
which is the projection on $\psi_0$. Moreover, from~\eqref{ineq-radius} it follows for all $z \in \Gamma$ that $\|\mathtt{R}_0(z)\| = 2G^{-1}$ and $\|(\mathtt{H}-z)^{-1}\| \leq \left(\frac{G}{2}-g\right)^{-1}$. Now let us verify the following: for all $z \in \Gamma$, it holds that
\begin{align}\label{eq-resolvent-expansion}
\begin{split}
(\mathtt{H}-z)^{-1} &= \mathtt{R}_0(z) - \lambda\mathtt{R}_0(z)\mathtt{H}_1\mathtt{R}_0(z) - \lambda^2\mathtt{R}_0(z)\left[\mathtt{H}_2-\mathtt{H}_1\mathtt{R}_0(z)\mathtt{H}_1\right]\mathtt{R}_0(z)\\
&\qquad- \lambda^3\mathtt{R}_0(z)\left[\mathtt{H}_1\mathtt{R}_0(z)\mathtt{H}_1\mathtt{R}_0(z)\mathtt{H}_1 - \mathtt{H}_2\mathtt{R}_0(z)\mathtt{H}_1 - \mathtt{H_1}\mathtt{R}_0(z)\mathtt{H}_2\right]\mathtt{R}_0(z) + \lambda^4\mathtt{R}^{(\lambda)}(z)\,,
\end{split}
\end{align}
in which
\begin{align*}
\begin{split}
\mathtt{R}^{(\lambda)}(z) &= (\mathtt{H}-z)^{-1}\big[\mathtt{H}_2\mathtt{R}_0(z)\mathtt{H}_2-\mathtt{H}_2\mathtt{R}_0(z)\mathtt{H}_1\mathtt{R}_0(z)\mathtt{H}_1\\
&\qquad+ \left(\mathtt{H}_1+\lambda\mathtt{H}_2\right)\mathtt{R}_0(z)\left(\mathtt{H}_1\mathtt{R}_0(z)\mathtt{H}_1\mathtt{R}_0(z)\mathtt{H}_1 - \mathtt{H}_2\mathtt{R}_0(z)\mathtt{H}_1 - \mathtt{H}_1\mathtt{R}_0(z)\mathtt{H}_2\right)\big]\mathtt{R}_0(z)\,.
\end{split}
\end{align*}
These results can be obtained from the identity
\begin{align*}
(\mathbf{1}+\lambda\mathtt{I}+\lambda^2\mathtt{J})^{-1} &= \mathbf{1}-\lambda\mathtt{I}-\lambda^2(\mathtt{J}-\mathtt{I}^2)-\lambda^3(\mathtt{I}^3-\mathtt{J}\mathtt{I}-\mathtt{I}\mathtt{J})\\
&\qquad+\lambda^4(\mathbf{1}+\lambda\mathtt{I}+\lambda^2\mathtt{J})^{-1}(\mathtt{J}^2-\mathtt{J}\mathtt{I}^2+(\mathtt{I}+\lambda\mathtt{J})(\mathtt{I}^3-\mathtt{J}\mathtt{I}-\mathtt{I}\mathtt{J}))
\end{align*}
which holds for all square matrices $\mathtt{I}$ and $\mathtt{J}$ with norms that are uniformly bounded in $\lambda$ (for $\lambda$ small enough such that the l.h.s. is well-defined), using $(H-z)^{-1} = (\mathbf{1}+\lambda\mathtt{I}+\lambda^2\mathtt{J})^{-1}\mathtt{R}_0(z)$ where we put $\mathtt{I} = \mathtt{R}_0(z)\mathtt{H}_1$ and $\mathtt{I} = \mathtt{R}_0(z)\mathtt{H}_2$. It now follows that, for all $z \in \Gamma$, 
\begin{align}\label{ineq-resolvent}
\|\mathtt{R}^{(\lambda)}(z)\| \leq \frac{8G^{-2}}{G-2g}\left[\|\mathtt{H}_2\|^2+2G^{-1}\|\mathtt{H}_2\|\|\mathtt{H}_1\|^2 + 2\left(\|\mathtt{H}_1\|+\lambda\|\mathtt{H}_2\|\right)\left(2G^{-2}\|\mathtt{H}_1\|^3+2G^{-1}\|\mathtt{H}_1\|\|\mathtt{H}_2\|\right)\right]\,.
\end{align}
Now the eigenvector for the eigenvalue $E_0^{(\lambda)}$ of $\mathtt{H}$ is required to be normalized. This corresponds to imposing that the trace of the projection onto this eigenspace equals one:
\begin{align*}
1 &= \frac{-1}{2\pi i}\oint_{\Gamma}\operatorname{tr}\left((\mathtt{H}-z)^{-1}\right)dz\\
&= \frac{-1}{2\pi i}\oint_{\Gamma}\operatorname{tr}\left(\mathtt{R}_0(z)\right)dz - \lambda\left(\frac{-1}{2\pi i}\right)\oint_{\Gamma}\operatorname{tr}\left(\mathtt{R}_0(z)\mathtt{H}_1\mathtt{R}_0(z)\right)dz\\
&\qquad- \lambda^2\left(\frac{-1}{2\pi i}\right)\oint_{\Gamma}\operatorname{tr}\left(\mathtt{R}_0(z)\left[\mathtt{H}_2-\mathtt{H}_1\mathtt{R}_0(z)\mathtt{H}_1\right]\mathtt{R}_0(z)\right)dz + \lambda^4\left(\frac{-1}{2\pi i}\right)\oint_{\Gamma}\operatorname{tr}\left(\mathtt{R}^{(\lambda)}(z)\right)dz\\
&\qquad- \lambda^3\left(\frac{-1}{2\pi i}\right)\oint_{\Gamma}\operatorname{tr}\left(\mathtt{R}_0(z)\left[\mathtt{H}_1\mathtt{R}_0(z)\mathtt{H}_1\mathtt{R}_0(z)\mathtt{H}_1 - \mathtt{H}_2\mathtt{R}_0(z)\mathtt{H}_1 - \mathtt{H_1}\mathtt{R}_0(z)\mathtt{H}_2\right]\mathtt{R}_0(z)\right)dz
\end{align*}
Since the argument of this trace is analytic in $\lambda$ (albeit for $\lambda$ small enough) \cite{Kat} and as the zeroth order in $\lambda$ equals $\langle \psi_0|\psi_0 \rangle = 1$ by~\eqref{eq-resolvent}, it follows that all other orders in $\lambda$ must vanish (independently of how small $\lambda$ is). By calculating residues with spectral calculus, the first order vanishes identically, and the second and third order yield
\begin{align}\label{eq-normalization}
\begin{split}
0 &= \frac{-1}{2\pi i}\oint_{\Gamma}\operatorname{tr}\left(\mathtt{R}_0(z)\mathtt{H}_1\mathtt{R}_0(z)\mathtt{H}_1\mathtt{R}_0(z)\right)\,,\\
0 &= \frac{-1}{2\pi i}\oint_{\Gamma}\operatorname{tr}\left(\mathtt{R}_0(z)\left[\mathtt{H}_1\mathtt{R}_0(z)\mathtt{H}_1\mathtt{R}_0(z)\mathtt{H}_1 - \mathtt{H}_2\mathtt{R}_0(z)\mathtt{H}_1 - \mathtt{H_1}\mathtt{R}_0(z)\mathtt{H}_2\right]\mathtt{R}_0(z)\right)\,.
\end{split}
\end{align}
Now, using $\mathtt{H} = \mathtt{H}_0 + \lambda\mathtt{H}_1 + \lambda^2\mathtt{H}_2$ and the expansion obtained in~\eqref{eq-resolvent-expansion}, one finds
\begin{align}\label{eq-eigenvalue-expansion}
\begin{split}
E_0^{(\lambda)} &= \frac{-1}{2\pi i}\oint_{\Gamma}\operatorname{tr}\left(\mathtt{H}(\mathtt{H}-z)^{-1}\right)dz\\
&= \frac{-1}{2\pi i}\oint_{\Gamma}\operatorname{tr}\big(\mathtt{H}_0\mathtt{R}_0(z) + \lambda\left[\mathtt{H}_1\mathtt{R}_0(z)-\mathtt{H}_0\mathtt{R}_0(z)\mathtt{H}_1\mathtt{R}_0(z)\right]\\
&\qquad\qquad\qquad+ \lambda^2\left[\mathtt{H}_2\mathtt{R}_0(z)-\mathtt{H}_0\mathtt{R}_0(z)\mathtt{H}_2\mathtt{R}_0(z)+\left(\mathtt{H}_0\mathtt{R}_0(z)-\mathbf{1}\right)\mathtt{H}_1\mathtt{R}_0(z)\mathtt{H}_1\mathtt{R}_0(z)\right]\big) + \lambda^3E^{(\lambda)}\,.
\end{split}
\end{align}
In order to prove that the $E^{(\lambda)}$ written here is indeed the one characterized in the statement, we must show that the first and second order of the above indeed coincide with the expressions in~\eqref{def-E-lambda} (it is immediate that the zeroth order yields $E_0$). The first order in $\lambda$ indeed equals
$$\operatorname{tr}\left(\mathtt{H}_1\frac{-1}{2\pi i}\oint_{\Gamma}\mathtt{R}_0(z)dz + \mathtt{H}_1\mathtt{H}_0\frac{-1}{2\pi i}\oint_{\Gamma}\mathtt{R}_0(z)^2dz\right) = \operatorname{tr}\left(\mathtt{H}_1\psi_0\psi_0^* + \mathbf{0}\right) = \langle \psi_0|\mathtt{H}_1\psi_0 \rangle\,.$$
For the second order, it turns out to be convenient to subtract $E_0$ multiplied with the first expression of~\eqref{eq-normalization}, as one can then simplify $(\mathtt{H}_0-E_0)\mathtt{R}_0(z) - \mathbf{1} = -(E_0-z)\mathtt{R}_0(z)$ for all $z \in \Gamma$:
\begin{align*}
&\operatorname{tr}\left(\mathtt{H}_2\frac{-1}{2\pi i}\oint_{\Gamma}\mathtt{R}_0(z)dz + \mathtt{H}_2\mathtt{H}_0\frac{-1}{2\pi i}\oint_{\Gamma}\mathtt{R}_0(z)^2dz + \frac{-1}{2\pi i}\oint_{\Gamma}\left[(\mathtt{H}_0-E_0)\mathtt{R}_0(z) - \mathbf{1}\right]\mathtt{H}_1\mathtt{R}_0(z)\mathtt{H}_1\mathtt{R}_0(z)dz\right)\\
&\quad= \operatorname{tr}\left(\mathtt{H}_2\psi_0\psi_0^* + \mathbf{0}\right) - \sum_{l=0}^{\mathtt{w}}\left(\frac{-1}{2\pi i}\right)\oint_{\Gamma}(E_0-z)\langle \psi_l|\mathtt{R}_0(z)\mathtt{H}_1\mathtt{R}_0(z)\mathtt{H}_1\mathtt{R}_0(z)\psi_l \rangle dz\\
&\quad= \langle \psi_0|\mathtt{H}_2\psi_0 \rangle - \sum_{l=1}^{\mathtt{w}}\sum_{k=0}^{\mathtt{w}}\left(\frac{-1}{2\pi i}\right)\oint_{\Gamma}\frac{E_0-z}{E_l-z}\langle \psi_l|\mathtt{H}_1\mathtt{R}_0(z)\psi_k \rangle \langle \psi_k|\mathtt{H}_1\psi_l \rangle \frac{dz}{E_l-z}\\
&\quad\qquad-\sum_{k=0}^{\mathtt{w}}\left(\frac{-1}{2\pi i}\right)\oint_{\Gamma}\frac{E_0-z}{E_0-z}\langle \psi_0|\mathtt{H}_1\mathtt{R}_0(z)\psi_k \rangle \langle \psi_k|\mathtt{H}_1\psi_0 \rangle \frac{dz}{E_0-z}\\
&\quad= \langle \psi_0|\mathtt{H}_2\psi_0 \rangle - 0 -\sum_{k=1}^{\mathtt{w}}\left(\frac{-1}{2\pi i}\right)\oint_{\Gamma}\frac{\langle \psi_0|\mathtt{H}_1\psi_k \rangle \langle \psi_k|\mathtt{H}_1\psi_0 \rangle dz}{(E_k-z)(E_0-z)}\\
&\quad= \langle \psi_0|\mathtt{H}_2\psi_0 \rangle - \sum_{k=1}^{\mathtt{w}}\frac{|\langle \psi_0|\mathtt{H}_1\psi_k \rangle|^2}{E_k-E_0}\,.
\end{align*}
Therefore the expansion in~\eqref{eq-eigenvalue-expansion} is correct. Using the expression~\eqref{eq-resolvent-expansion}, one obtains
\begin{align}\label{eq-E-lambda}
\begin{split}
E^{(\lambda)} &= \frac{-1}{2\pi i}\oint_{\Gamma}\operatorname{tr}\big(\left[\mathbf{1}-\left(\mathtt{H}_0+\lambda\mathtt{H}_1+\lambda^2\mathtt{H}_2\right)\mathtt{R}_0(z)\right]\left[\mathtt{H}_1\mathtt{R}_0(z)\mathtt{H}_1\mathtt{R}_0(z)\mathtt{H}_1 - \mathtt{H}_2\mathtt{R}_0(z)\mathtt{H}_1 - \mathtt{H}_1\mathtt{R}_0(z)\mathtt{H}_2\right]\mathtt{R}_0(z)\\
&\qquad\qquad\qquad\quad+\lambda\mathtt{H}_2\mathtt{R}_0(z)\left[\mathtt{H}_2-\mathtt{H}_1\mathtt{R}_0(z)\mathtt{H}_1\right]\mathtt{R}_0(z) + \lambda\left[\mathtt{H}_0+\lambda\mathtt{H}_1+\lambda^2\mathtt{H}_2\right]\mathtt{R}^{(\lambda)}(z)\big)dz\,.
\end{split}
\end{align}
Now note that when evaluating this integral, each non-vanishing contribution originates from a summand in which at least one resolvent has a pole inside of $\Gamma$. By the construction of $\Gamma$ and by performing this spectral calculus, this operator is then only of rank one, so the trace can be estimated by the norm of its argument. Therefore, all terms in the foregoing that are at least of order 1 in $\lambda$ can be bounded by
\begin{align}\label{ineq-higher-orders}
\lambda\|\mathtt{H}\|\|\mathtt{R}^{(\lambda)}\|+4\,\lambda G^{-2}\left[\|\mathtt{H}_2\|^2+2G^{-1}\|\mathtt{H}_1\|^2\|\mathtt{H}_2\|+2\left(\|\mathtt{H}_1\|+\lambda\|\mathtt{H}_2\|\right)\left(2G^{-2}\|\mathtt{H}_1\|^3+2G^{-1}\|\mathtt{H}_1\|\|\mathtt{H}_2\|\right)\right].
\end{align}
Inserting $E_0$ times the second formula of~\eqref{eq-normalization} and again using $\mathbf{1} - (\mathtt{H}_0-E_0)\mathtt{R}_0(z) = (E_0-z)\mathtt{R}_0(z)$ for all $z \in \Gamma$, one can explicitly calculate the zeroth order in $\lambda$ of~\eqref{eq-E-lambda}, which then equals
\begin{align*}
&\frac{-1}{2\pi i}\oint_{\Gamma}\operatorname{tr}\left(\left[\mathbf{1}-(\mathtt{H}_0-E_0)\mathtt{R}_0(z)\right]\left[\mathtt{H}_1\mathtt{R}_0(z)\mathtt{H}_1\mathtt{R}_0(z)\mathtt{H}_1 - \mathtt{H}_2\mathtt{R}_0(z)\mathtt{H}_1 - \mathtt{H}_1\mathtt{R}_0(z)\mathtt{H}_2\right]\mathtt{R}_0(z)\right)dz\\
&\quad= \sum_{k=1}^{\mathtt{w}}\left(\frac{-1}{2\pi i}\right)\oint_{\Gamma}\frac{E_0-z}{E_k-z}\langle \psi_k|\left[\mathtt{H}_1\mathtt{R}_0(z)\mathtt{H}_1\mathtt{R}_0(z)\mathtt{H}_1 - \mathtt{H}_2\mathtt{R}_0(z)\mathtt{H}_1 - \mathtt{H}_1\mathtt{R}_0(z)\mathtt{H}_2\right]\psi_k \rangle \frac{dz}{E_k-z}\\
&\quad\qquad+ \frac{-1}{2\pi i}\oint_{\Gamma}\frac{E_0-z}{E_0-z}\langle \psi_0|\left[\mathtt{H}_1\mathtt{R}_0(z)\mathtt{H}_1\mathtt{R}_0(z)\mathtt{H}_1 - \mathtt{H}_2\mathtt{R}_0(z)\mathtt{H}_1 - \mathtt{H}_1\mathtt{R}_0(z)\mathtt{H}_2\right]\psi_0 \rangle \frac{dz}{E_0-z}\\
&\quad= \sum_{k=1}^{\mathtt{w}}\left(\frac{-1}{2\pi i}\right)\oint_{\Gamma}\frac{E_0-z}{E_k-z}\frac{\langle \psi_k|\mathtt{H}_1\psi_0 \rangle \langle \psi_0|\mathtt{H}_1\psi_0 \rangle \langle \psi_0|\mathtt{H}_1\psi_k \rangle}{(E_0-z)^2} \frac{dz}{E_k-z}\\
&\quad\qquad+ \sum_{l=1}^{\mathtt{w}}\sum_{m=1}^{\mathtt{w}}\left(\frac{-1}{2\pi i}\right)\oint_{\Gamma}\frac{\langle \psi_0|\mathtt{H}_1\psi_l \rangle \langle \psi_l|\mathtt{H}_1\psi_m \rangle  \langle\psi_m|\mathtt{H}_1\psi_0 \rangle}{(E_l-z)(E_m-z)} \frac{dz}{E_0-z}\\
&\quad\qquad- \sum_{l=1}^{\mathtt{w}}\left(\frac{-1}{2\pi i}\right)\oint_{\Gamma}\frac{\langle \psi_0|\mathtt{H}_2\psi_l \rangle \langle \psi_l|\mathtt{H}_1\psi_0 \rangle + \langle \psi_0|\mathtt{H}_1\psi_l \rangle \langle \psi_l|\mathtt{H}_2\psi_0 \rangle}{E_l-z} \frac{dz}{E_0-z}\\
&\quad= \sum_{k=1}^{\mathtt{w}}\frac{|\langle \psi_k|\mathtt{H}_1\psi_0 \rangle|^2 \langle \psi_0|\mathtt{H}_1\psi_0 \rangle}{(E_k-E_0)^2} + \sum_{l=1}^{\mathtt{w}}\sum_{m=1}^{\mathtt{w}}\frac{\langle \psi_0|\mathtt{H}_1\psi_l \rangle \langle \psi_l|\mathtt{H}_1\psi_m \rangle  \langle\psi_m|\mathtt{H}_1\psi_0 \rangle}{(E_l-E_0)(E_m-E_0)}\\
&\quad\qquad- \sum_{l=1}^{\mathtt{w}}\frac{\langle \psi_0|\mathtt{H}_2\psi_l \rangle \langle \psi_l|\mathtt{H}_1\psi_0 \rangle + \langle \psi_0|\mathtt{H}_1\psi_l \rangle \langle \psi_l|\mathtt{H}_2\psi_0 \rangle}{E_l-E_0}\,,
\end{align*}
which is bounded by $2G^{-2}\|\mathtt{H}_1\|^3 + 2G^{-1}\|\mathtt{H}_1\|\|\mathtt{H}_2\|$. Combining this with~\eqref{ineq-resolvent} and~\eqref{ineq-higher-orders}, one finds that the expression~\eqref{eq-E-lambda} is indeed bounded as indicated in the statement.
\hfill $\square$

\section{Appendix: Lyapunov exponents}
\label{app-Lyapunov}

This technical appendix contains the proofs of the statements of Section~\ref{sec-Lyapunov-exponents}.

\vspace{.2cm}

\noindent\textbf{Proof of Lemma~\ref{lemma-Lyapunov-estimate}.} We use the abbreviation $\mathcal{R}\cdot \Phi:=\mathcal{R}\Phi\left(\Phi^*\mathcal{R}^2\Phi\right)^{\frac{1}{2}}\in\mathbb{F}_{\mathsf{L},\mathsf{q}}$ and observe that one has $\mathcal{R}\cdot Q=(\mathcal{R}\cdot \Phi)(\mathcal{R}\cdot\Phi)^*$.
Next, we use
\begin{align*}
\big\|e^{\lambda\mathtt{P}}-\mathbf{1}-\lambda\mathtt{P}-\mbox{\small $\frac{1}{2}$}\lambda^2\mathtt{P}^2\big\|\leq e^{\lambda}-1-\lambda-\mbox{\small $\frac{1}{2}$}\lambda^2\leq 2^{-2}\,e^{-\lambda}\,\lambda^3
\end{align*}
in order to prove the lower bound
\begin{align}\label{ineq-Lyapunov-estimate-1}
\begin{split}
e^{\lambda\mathtt{P}^*} & \,e^{\lambda\mathtt{P}}
\\
&=\mathbf{1}+\lambda\left[\mathtt{P}^*+\mathtt{P}\right]+\mbox{\small $\frac{1}{2}$}\,\lambda^2\left[\mathtt{P}^2+(\mathtt{P}^*)^2+2\mathtt{P}^*\mathtt{P}\right]+\left[\left(e^{\lambda\mathtt{P}}-\mathbf{1}-\lambda\mathtt{P}-\mbox{\small $\frac{1}{2}$}\lambda^2\mathtt{P}^2\right)^*\,e^{\lambda\mathtt{P}}+\textnormal{h.c.}\right]\\
&\quad+\mbox{\small $\frac{1}{2}$}\lambda^3\,\left[\mathtt{P}^*\mathtt{P}^2+(\mathtt{P}^*)^2\mathtt{P}\right]+\mbox{\small $\frac{1}{4}$}\lambda^4\,(\mathtt{P}^*)^2\mathtt{P}^2-\left(e^{\lambda\mathtt{P}}-\mathbf{1}-\lambda\mathtt{P}-\mbox{\small $\frac{1}{2}$}\lambda^2\mathtt{P}^2\right)^*\left(e^{\lambda\mathtt{P}}-\mathbf{1}-\lambda\mathtt{P}-\mbox{\small $\frac{1}{2}$}\lambda^2\mathtt{P}^2\right)\\
&\geq \mathbf{1}+\lambda\left[\mathtt{P}^*+\mathtt{P}\right]+\mbox{\small $\frac{1}{2}$}\,\lambda^2\left[\mathtt{P}^2+(\mathtt{P}^*)^2+2\mathtt{P}^*\mathtt{P}\right]-2\lambda^3\,\mathbf{1}\\
&=\mathbf{1}+\lambda\left[\mathtt{P}^*+\mathtt{P}\right]+\mbox{\small $\frac{1}{2}$}\,\lambda^2\left[(\mathtt{P}+\mathtt{P}^*)^2+\mathtt{P}^*\mathtt{P}-\mathtt{P}\mathtt{P}^*\right]-2\lambda^3\,\mathbf{1}\\
&\geq\mathbf{1}+\lambda\left[\mathtt{P}^*+\mathtt{P}\right]-\mbox{\small $\frac{1}{2}$}\lambda^2\,\mathtt{P}\mathtt{P}^*-2\lambda^3\,\mathbf{1}\\
&\geq (1-\mbox{\small $\frac{1}{2}$}\,\lambda^2-2\lambda^3)\,\mathbf{1}+\lambda\left[\mathtt{P}^*+\mathtt{P}\right]\\
&\geq e^{-\frac{3}{4}\lambda^2}\,\mathbf{1}+\lambda\left[\mathtt{P}^*+\mathtt{P}\right]\,.
\end{split}
\end{align}
In the following sequence of (in)equalities, we exploit the operator monotonicity of the logarithm when using~\eqref{ineq-Lyapunov-estimate-1} in the second  and $\log(1+x)\geq x-\mbox{\small $\frac{17}{32}$}x^2$ for $x>-2^{-4}$ in the 13th step to show
\begin{align}\label{ineq-Lyapunov-estimate-2}
\begin{split}
&\log\,\det\left(\Phi^*\mathcal{R}e^{\lambda\mathtt{P}^*}\,e^{\lambda\mathtt{P}}\mathcal{R}\Phi\right)\\
&=\operatorname{tr}\,\log\left(\Phi^*\mathcal{R}e^{\lambda\mathtt{P}^*}\,e^{\lambda\mathtt{P}}\mathcal{R}\Phi\right)\\
&\geq \operatorname{tr}\,\log\left(e^{-\frac{3}{4}\lambda^2}\,\Phi^*\mathcal{R}^2\Phi+\lambda\,\Phi^*\mathcal{R}(\mathtt{P}^*+\mathtt{P})\mathcal{R}\Phi\right)\\
&= \log\,\det\left(e^{-\frac{3}{4}\lambda^2}\,\mathbf{1}_{\mathsf{q}}\left[\Phi^*\mathcal{R}^2\Phi+\lambda\,e^{\frac{3}{4}\lambda^2}\,\Phi^*\mathcal{R}(\mathtt{P}^*+\mathtt{P})\mathcal{R}\Phi\right]\right)\\
&= \log\,\left[\det\left(e^{-\frac{3}{4}\lambda^2}\,\mathbf{1}_{\mathsf{q}}\right)\,\det\left(\Phi^*\mathcal{R}^2\Phi+\lambda\,e^{\frac{3}{4}\lambda^2}\,\Phi^*\mathcal{R}(\mathtt{P}^*+\mathtt{P})\mathcal{R}\Phi\right)\right]\\
&= \log\,\left[\exp\left(-\mbox{\small $\frac{3}{4}$}\lambda^2\,{\mathsf{q}}\right)\,\det\left(\Phi^*\mathcal{R}^2\Phi+\lambda\,e^{\frac{3}{4}\lambda^2}\,\Phi^*\mathcal{R}(\mathtt{P}^*+\mathtt{P})\mathcal{R}\Phi\right)\right]\\
&=\log\,\det\left(\Phi^*\mathcal{R}^2\Phi+\lambda\,e^{\frac{3}{4}\lambda^2}\,\Phi^*\mathcal{R}(\mathtt{P}^*+\mathtt{P})\mathcal{R}\Phi\right)-\mbox{\small $\frac{3}{4}$}\lambda^2\,\mathsf{q}\\
&=\operatorname{tr}\,\log\left(\Phi^*\mathcal{R}^2\Phi+\lambda\,e^{\frac{3}{4}\lambda^2}\,\Phi^*\mathcal{R}(\mathtt{P}^*+\mathtt{P})\mathcal{R}\Phi\right)-\mbox{\small $\frac{3}{4}$}\lambda^2\,\mathsf{q}\\
&=\operatorname{tr}\,\log\left(\left(\Phi^*\mathcal{R}^2\Phi\right)^{\frac{1}{2}}\left[\mathbf{1}+\lambda\,e^{\frac{3}{4}\lambda^2}\,\left(\Phi^*\mathcal{R}^2\Phi\right)^{-\frac{1}{2}}\,\Phi^*\mathcal{R}(\mathtt{P}^*\!+\!\mathtt{P})\mathcal{R}\Phi\,\left(\Phi^*\mathcal{R}^2\Phi\right)^{-\frac{1}{2}}\right]\left(\Phi^*\mathcal{R}^2\Phi\right)^{\frac{1}{2}}\right)\!-\!\mbox{\small $\frac{3}{4}$}\lambda^2\,\mathsf{q}\\
&=\operatorname{tr}\,\log\left(\left(\Phi^*\mathcal{R}^2\Phi\right)^{\frac{1}{2}}\left[\mathbf{1}+\lambda\,e^{\frac{3}{4}\lambda^2}\,(\mathcal{R}\cdot \Phi)^*(\mathtt{P}^*+\mathtt{P})(\mathcal{R}\cdot \Phi)\right]\left(\Phi^*\mathcal{R}^2\Phi\right)^{\frac{1}{2}}\right)-\mbox{\small $\frac{3}{4}$}\lambda^2\,\mathsf{q}\\
&=\log\,\det\left(\left(\Phi^*\mathcal{R}^2\Phi\right)^{\frac{1}{2}}\left[\mathbf{1}+\lambda\,e^{\frac{3}{4}\lambda^2}\,(\mathcal{R}\cdot \Phi)^*(\mathtt{P}^*+\mathtt{P})(\mathcal{R}\cdot \Phi)\right]\left(\Phi^*\mathcal{R}^2\Phi\right)^{\frac{1}{2}}\right)-\mbox{\small $\frac{3}{4}$}\lambda^2\,\mathsf{q}\\
&=\log\,\det\left(\Phi^*\mathcal{R}^2\Phi\right)+\log\,\det\left(\mathbf{1}+\lambda\,e^{\frac{3}{4}\lambda^2}\,(\mathcal{R}\cdot \Phi)^*(\mathtt{P}^*+\mathtt{P})(\mathcal{R}\cdot \Phi)\right)-\mbox{\small $\frac{3}{4}$}\lambda^2\,\mathsf{q}\\
&=\operatorname{tr}\,\log\left(\Phi^*\mathcal{R}^2\Phi\right)+\operatorname{tr}\,\log\left(\mathbf{1}+\lambda\,e^{\frac{3}{4}\lambda^2}\,(\mathcal{R}\cdot \Phi)^*(\mathtt{P}^*+\mathtt{P})(\mathcal{R}\cdot \Phi)\right)-\mbox{\small $\frac{3}{4}$}\lambda^2\,\mathsf{q}\\
&\geq \operatorname{tr}\,\log\left(\Phi^*\mathcal{R}^2\Phi\right)+\lambda\,e^{\frac{3}{4}\lambda^2}\,\operatorname{tr}\,\left[(\mathcal{R}\cdot \Phi)^*(\mathtt{P}^*+\mathtt{P})(\mathcal{R}\cdot \Phi)\right]\\
&\quad-\mbox{\small $\frac{5}{8}$}\,\lambda^2\,e^{\frac{3}{2}\lambda^2}\,\operatorname{tr}\,\left[(\mathcal{R}\cdot \Phi)^*(\mathtt{P}^*+\mathtt{P})(\mathcal{R}\cdot \Phi)(\mathcal{R}\cdot \Phi)^*(\mathtt{P}^*+\mathtt{P})(\mathcal{R}\cdot \Phi)\right]-\mbox{\small $\frac{3}{4}$}\lambda^2\,\mathsf{q}\\
&\geq \operatorname{tr}\,\log\left(\Phi^*\mathcal{R}^2\Phi\right)+\lambda\,e^{\frac{3}{4}\lambda^2}\,\operatorname{tr}\,\left[(\mathcal{R}\cdot \Phi)^*\,(\mathtt{P}+\mathtt{P}^*)\,(\mathcal{R}\cdot \Phi)\right]-\mbox{\small $\frac{17}{32}$}\,\lambda^2\,e^{\frac{3}{2}\lambda^2}\,\operatorname{tr}\,\left[2^2\,\mathbf{1}_{\mathsf{q}}\right]-\mbox{\small $\frac{3}{4}$}\lambda^2\,\mathsf{q}\\
&\geq \operatorname{tr}\,\log\left(\Phi^*\mathcal{R}^2\Phi\right)+\lambda\,e^{\frac{3}{4}\lambda^2}\,\operatorname{tr}\,\left[(\mathtt{P}+\mathtt{P}^*)(\mathcal{R}\cdot \Phi)(\mathcal{R}\cdot \Phi)^*\right]-\left[\mbox{\small $\frac{17}{8}$}\,e^{\frac{3}{2}\lambda^2}+\mbox{\small $\frac{3}{4}$}\right]\lambda^2\,\mathsf{q}\\
&\geq \operatorname{tr}\,\log\left(\Phi^*\mathcal{R}^2\Phi\right)+\lambda\,e^{\frac{3}{4}\lambda^2}\,\operatorname{tr}\,\left[(\mathtt{P}+\mathtt{P}^*)\,(\mathcal{R}\cdot Q)\right]-3\,\lambda^2\,\mathsf{q}\,.
\end{split}
\end{align}
In the next series of (in)equalities, we use the bound $\log[1-(1-y)x]\geq \log(y)\,x$ for  $x\in [0,1]$ and $y\in (0,1]$ in the seventh step and the bound
$$
\mathcal{R}^2\geq \kappa^2_{\mathsf{L}_{\mgl}+\mathsf{L}_{\lgl}}\,\hat{\alpha}^{\perp}(\hat{\alpha}^{\perp})^*+\kappa_{\mathsf{L}}^2\,\hat{\alpha}\hat{\alpha}=\kappa^2_{\mathsf{L}_{\mgl}+\mathsf{L}_{\lgl}}\left[\mathbf{1}_{\mathsf{L}}-(1-\kappa^2_{\mathsf{L}}\kappa^{-2}_{\mathsf{L}_{\mgl}+\mathsf{L}_{\lgl}})\hat{\alpha}\hat{\alpha}^*\right]
$$
in the first step to estimate
\begin{align}\label{ineq-Lyapunov-estimate-3}
\begin{split}
\operatorname{tr}\,\log\left(\Phi^*\mathcal{R}^2\Phi\right)&\geq \operatorname{tr}\,\log\left(\kappa^2_{\mathsf{L}_{\mgl}+\mathsf{L}_{\lgl}}\mathbf{1}_{\mathsf{q}}\left[\mathbf{1}_{\mathsf{q}}-(1-\kappa^2_{\mathsf{L}}\kappa^{-2}_{\mathsf{L}_{\mgl}+\mathsf{L}_{\lgl}})\Phi^*\hat{\alpha}\hat{\alpha}^*\Phi\right]\right)\\
&=\log\,\det\left(\kappa^2_{\mathsf{L}_{\mgl}+\mathsf{L}_{\lgl}}\mathbf{1}_{\mathsf{q}}\left[\mathbf{1}_{\mathsf{q}}-(1-\kappa^2_{\mathsf{L}}\kappa^{-2}_{\mathsf{L}_{\mgl}+\mathsf{L}_{\lgl}})\Phi^*\hat{\alpha}\hat{\alpha}^*\Phi\right]\right)\\
&=\log\,\left[\det\left(\kappa^2_{\mathsf{L}_{\mgl}+\mathsf{L}_{\lgl}}\mathbf{1}_{\mathsf{q}}\right)\,\det\left(\left[\mathbf{1}_{\mathsf{q}}-(1-\kappa^2_{\mathsf{L}}\kappa^{-2}_{\mathsf{L}_{\mgl}+\mathsf{L}_{\lgl}})\Phi^*\hat{\alpha}\hat{\alpha}^*\Phi\right]\right)\right]\\
&=\log\,\left[\kappa^{2\mathsf{q}}_{\mathsf{L}_{\mgl}+\mathsf{L}_{\lgl}}\,\det\left(\left[\mathbf{1}_{\mathsf{q}}-(1-\kappa^2_{\mathsf{L}}\kappa^{-2}_{\mathsf{L}_{\mgl}+\mathsf{L}_{\lgl}})\Phi^*\hat{\alpha}\hat{\alpha}^*\Phi\right]\right)\right]\\
&=2\,\mathsf{q}\,\log(\kappa_{\mathsf{L}_{\mgl}+\mathsf{L}_{\lgl}})+\log\,\det\left(\mathbf{1}_{\mathsf{q}}-(1-\kappa^2_{\mathsf{L}}\kappa^{-2}_{\mathsf{L}_{\mgl}+\mathsf{L}_{\lgl}})\Phi^*\hat{\alpha}\hat{\alpha}^*\Phi\right)\\
&=2\,\mathsf{q}\,\log(\kappa_{\mathsf{L}_{\mgl}+\mathsf{L}_{\lgl}})+\operatorname{tr}\,\log\left(\mathbf{1}_{\mathsf{q}}-(1-\kappa^2_{\mathsf{L}}\kappa^{-2}_{\mathsf{L}_{\mgl}+\mathsf{L}_{\lgl}})\Phi^*\hat{\alpha}\hat{\alpha}^*\Phi\right)\\
&\geq 2\,\mathsf{q}\,\log(\kappa_{\mathsf{L}_{\mgl}+\mathsf{L}_{\lgl}})+\log(\kappa^2_{\mathsf{L}}\kappa^{-2}_{\mathsf{L}_{\mgl}+\mathsf{L}_{\lgl}})\,\operatorname{tr}\left[\Phi^*\hat{\alpha}\hat{\alpha}^*\Phi\right]\\
&= 2\,\mathsf{q}\,\log(\kappa_{\mathsf{L}_{\mgl}+\mathsf{L}_{\lgl}})+2\,\log(\kappa_{\mathsf{L}}\kappa^{-1}_{\mathsf{L}_{\mgl}+\mathsf{L}_{\lgl}})\,\operatorname{tr}\left[\hat{\alpha}^*Q\hat{\alpha}\right]\\
&= 2\,\mathsf{q}\,\log(\kappa_{\mathsf{L}_{\mgl}+\mathsf{L}_{\lgl}})+2\,\log(\kappa_{\mathsf{L}}\kappa^{-1}_{\mathsf{L}_{\mgl}+\mathsf{L}_{\lgl}})\,\mathsf{d}(Q)\,.
\end{split}
\end{align}
Combining~\eqref{ineq-Lyapunov-estimate-2} and~\eqref{ineq-Lyapunov-estimate-3} then implies the stated bound.
\hfill $\square$

\vspace{.2cm}

\noindent \textbf{Proof of Proposition~\ref{proposi-Lyapunov-reflection}.} 
For $\mathsf{q}\in\{1,\dots,\mathsf{L}\}$, we consider the reversed random dynamics induced by the sequence  $((\mathcal{T}_{n}^{*})^{-1})_{n\in\mathbb{N}}$ given by
$$
Q_n^{\prime}:=(\mathcal{T}_n^*)^{-1}\cdot Q_{n-1}^{\prime}\,,\qquad Q_0^{\prime}\in\mathbb{G}_{\mathsf{L},\mathsf{L}-\mathsf{q}}\,, \qquad n\in\mathbb{N}\,,
$$
where we choose $Q_0^{\prime}=Q_0^{\perp}$. By Lemma~\ref{lemma-complement-action}, one then has $Q_n^{\prime}=Q_n^{\perp}$ for all $n\in\mathbb{N}$, where $Q_n$ is as in~\eqref{dyn-grassmanian}. Now let $\Phi_n\in\mathbb{F}_{\mathsf{L},\mathsf{q}}$ and $\Phi_n^{\perp}\in\mathbb{F}_{\mathsf{L},\mathsf{L}-\mathsf{q}}$ be such that $\Phi_n\Phi_n^*=Q_n$ and $\Phi^{\perp}_n(\Phi^{\perp}_n)^*=Q_n^{\perp}=Q_n^{\prime}$.

Similarly to the statement of Proposition~\ref{proposi-Lyapunov-rewriting}, one then has
\begin{align}\label{eq-Lyapunov-reflection-1}
\sum\limits_{\mathsf{w}=1}^{\mathsf{L}}\gamma^{\prime}_{\mathsf{w}}-\sum\limits_{\mathsf{w}=1}^{\mathsf{L}-\mathsf{q}}\gamma_{\mathsf{w}}^{\prime}=\frac{1}{2}\,\lim\limits_{N\rightarrow\infty}\frac{1}{N}\sum\limits_{n=0}^{N-1}\,\mathbb{E}\,\left[\log\,\det\left(\mathcal{T}_{n+1}^{-1}(\mathcal{T}_{n+1}^*)^{-1}\right)-\log\,\det\left((\Phi_n^{\perp})^*\mathcal{T}_{n+1}^{-1}(\mathcal{T}_{n+1}^*)^{-1}\Phi_n^{\perp}\right)\right]\,.
\end{align}

To tackle the r.h.s. of~\eqref{eq-Lyapunov-reflection-1}, we compute the following sequence of identities for all $\mathcal{T}\in\mathbb{C}^{\mathsf{L}\times\mathsf{L}}$ and all pairs $(\Phi,\Phi^{\perp})\in\mathbb{F}_{\mathsf{L},\mathsf{q}}\times\mathbb{F}_{\mathsf{L},\mathsf{L}-\mathsf{q}}$ with $\Phi^*\Phi^{\perp}=0$, which incorporates Lemma~\ref{lemma-complement-action} in the eight step:
\begin{align}\label{eq-Lyapunov-reflection-2}
\begin{split}
&\log\,\det\left(\mathcal{T}^{-1}(\mathcal{T}^*)^{-1}\right)-\log\,\det\left((\Phi^{\perp})^*\mathcal{T}^{-1}(\mathcal{T}^*)^{-1}\Phi^{\perp}\right)\\
&=\log\left[\det(\mathcal{T}^{-1}(\mathcal{T}^*)^{-1})\,\det\left((\Phi^{\perp})^*\mathcal{T}^{-1}(\mathcal{T}^*)^{-1}\Phi^{\perp}\right)^{-1}\right]\\
&=\log\left[\det((\Phi,\Phi^{\perp})^*\mathcal{T}^{-1}(\mathcal{T}^*)^{-1}(\Phi,\Phi^{\perp}))\,\det\big(\operatorname{diag}\big(\mathbf{1}_{\mathsf{q}},[(\Phi^{\perp})^*\mathcal{T}^{-1}(\mathcal{T}^*)^{-1}\Phi^{\perp}]^{-\frac{1}{2}}\big)\big)^2\right]\\
&=\log\Big[\det\Big(\operatorname{diag}\big(\mathbf{1}_{\mathsf{q}},[(\Phi^{\perp})^*\mathcal{T}^{-1}(\mathcal{T}^*)^{-1}\Phi^{\perp}]^{-\frac{1}{2}}\big)\,(\Phi,\Phi^{\perp})^*\mathcal{T}^{-1}(\mathcal{T}^*)^{-1}(\Phi,\Phi^{\perp})\,\times\\
&\qquad\qquad\qquad\times\,\operatorname{diag}\big(\mathbf{1}_{\mathsf{q}},[(\Phi^{\perp})^*\mathcal{T}^{-1}(\mathcal{T}^*)^{-1}\Phi^{\perp}]^{-\frac{1}{2}}\big)\Big)\Big]\\
&=\log\left[\det\mbox{\small $\begin{pmatrix}\Phi^*\mathcal{T}^{-1}(\mathcal{T}^*)^{-1}\Phi&
\!\!\!\!\!\!\!\!\!\!\!
\Phi^*\mathcal{T}^{-1}(\mathcal{T}^*)^{-1}\Phi^{\perp}[(\Phi^{\perp})^*\mathcal{T}^{-1}(\mathcal{T}^*)^{-1}\Phi^{\perp}]^{-\frac{1}{2}}\\ 
[(\Phi^{\perp})^*\mathcal{T}^{-1}(\mathcal{T}^*)^{-1}\Phi^{\perp}]^{-\frac{1}{2}}(\Phi^{\perp})^*\mathcal{T}^{-1}(\mathcal{T}^*)^{-1}\Phi &\mathbf{1}_{\mathsf{q}}\end{pmatrix}$}\right]\\
&=\log\left[\det\big(\Phi^*\mathcal{T}^{-1}(\mathcal{T}^*)^{-1}\Phi-\Phi^*\mathcal{T}^{-1}(\mathcal{T}^*)^{-1}\Phi^{\perp}[(\Phi^{\perp})^*\mathcal{T}^{-1}(\mathcal{T}^*)^{-1}\Phi^{\perp}]^{-1}(\Phi^{\perp})^*\mathcal{T}^{-1}(\mathcal{T}^*)^{-1}\Phi\big)\right]\\
&=\log\left[\det\big(\Phi^*\mathcal{T}^{-1}(\mathcal{T}^*)^{-1}\Phi-\Phi^*\mathcal{T}^{-1}\left[(\mathcal{T}^*)^{-1}\cdot \Phi^{\perp}(\Phi^{\perp})^*\right](\mathcal{T}^*)^{-1}\Phi\big)\right]\\
&=\log\left[\det\big(\Phi^*\mathcal{T}^{-1}\left[(\mathcal{T}^*)^{-1}\cdot \Phi^{\perp}(\Phi^{\perp})^*\right]^{\perp}(\mathcal{T}^*)^{-1}\Phi\big)\right]\\
&=\log\left[\det\Big(\Phi^*\mathcal{T}^{-1}\left[\mathcal{T}\cdot \Phi\Phi^*\right](\mathcal{T}^*)^{-1}\Phi\Big)\right]\\
&=\log\left[\det\Big(\Phi^*\mathcal{T}^{-1}\mathcal{T}\Phi\big(\Phi^*\mathcal{T}^*\mathcal{T}\Phi^*\big)^{-1}\Phi^*\mathcal{T}^*(\mathcal{T}^*)^{-1}\Phi\Big)\right]\\
&=\log\left[\det\Big(\big(\Phi^*\mathcal{T}^*\mathcal{T}\Phi^*\big)^{-1}\Big)\right]\\
&=-\log\left[\det\big(\Phi^*\mathcal{T}^*\mathcal{T}\Phi^*\big)\right]\,.
\end{split}
\end{align}
Inserting $\mathcal{T}=\mathcal{T}_n$ and $(\Phi,\Phi^{\perp})=(\Phi_{n+1},\Phi_{n+1}^{\perp})$  into~\eqref{eq-Lyapunov-reflection-2} then yields the identity
\begin{align}\label{eq-Lyapunov-reflection-3}
-\log\,\det\left(\Phi_n^*\mathcal{T}_{n+1}^*\mathcal{T}_{n+1}\Phi_n\right)=\log\,\det\left(\mathcal{T}_{n+1}^{-1}(\mathcal{T}_{n+1}^*)^{-1}\right)-\log\,\det\left((\Phi_n^{\perp})^*\mathcal{T}_{n+1}^{-1}(\mathcal{T}_{n+1}^*)^{-1}\Phi_n^{\perp}\right)
\end{align}
for all $n\in\mathbb{N}$.
Combining Proposition~\ref{proposi-Lyapunov-rewriting},~\eqref{eq-Lyapunov-reflection-1} and~\eqref{eq-Lyapunov-reflection-3} then yields
$$
-\sum\limits_{\mathsf{w}=1}^{\mathsf{q}}\gamma_{\mathsf{w}}=\sum\limits_{\mathsf{w}=\mathsf{L}-\mathsf{q}+1}^{\mathsf{L}}\gamma^{\prime}_{\mathsf{w}}=\sum\limits_{\mathsf{w}=1}^{\mathsf{q}}\gamma^{\prime}_{\mathsf{L}-\mathsf{w}+1}\,.
$$
As $\mathsf{q}\in\{1,\dots,\mathsf{L}\}$ was arbitrary, this implies the claim~\eqref{eq-Lyapunov-reflection}.
\hfill $\square$

\vspace{.2cm}

\noindent\textbf{Proof of Lemma~\ref{lemma-complement-action}.}
Let $\Phi \in \mathbb{F}_{\mathsf{L},\mathsf{q}}$ and $\Phi^{\perp} \in \mathbb{F}_{\mathsf{L},\mathsf{L}-\mathsf{q}}$ such that $\Phi\Phi^* = Q$ and $\Phi^{\perp}(\Phi^{\perp})^* = Q^{\perp}$. By $\Phi^*\Phi = \mathbf{1}_{\mathsf{q}}$ and $(\Phi^{\perp})^*\Phi^{\perp} = \mathbf{1}_{\mathsf{L}-\mathsf{q}}$, it follows that
\begin{align*}
(\mathcal{T} \cdot Q) & ((\mathcal{T}^{-1})^* \cdot Q^{\perp}) 
\\
&
\;= \;\mathcal{T}\Phi\left[\Phi^*\mathcal{T}^*\mathcal{T}\Phi\right]^{-1}\Phi^*\mathcal{T}^*(\mathcal{T}^*)^{-1}\Phi^{\perp}\left[(\Phi^{\perp})^*\mathcal{T}^{-1}(\mathcal{T}^{-1})^*\Phi^{\perp}\right]^{-1}(\Phi^{\perp})^*\mathcal{T}^{-1}\\
&
\;=\; \mathcal{T}\Phi\left[\Phi^*\mathcal{T}^*\mathcal{T}\Phi\right]^{-1}\Phi^*\Phi\Phi^*\Phi^{\perp}(\Phi^{\perp})^*\Phi^{\perp}\left[(\Phi^{\perp})^*\mathcal{T}^{-1}(\mathcal{T}^{-1})^*\Phi^{\perp}\right]^{-1}(\Phi^{\perp})^*\mathcal{T}^{-1} = \mathbf{0}\,,
\end{align*}
as this expression contains $\Phi\Phi^*\Phi^{\perp}(\Phi^{\perp})^* = QQ^{\perp} = \mathbf{0}$ as a factor. As a consequence, we find that $\mathcal{T} \cdot Q + (\mathcal{T}^{-1})^* \cdot Q^{\perp}$ is a projection. Recall that the rank of a projection is equal to its trace. Therefore $\operatorname{tr}(\mathcal{T} \cdot Q + (\mathcal{T}^{-1})^* \cdot Q^{\perp}) = \mathsf{q} + \mathsf{L} - \mathsf{q} = \mathsf{L}$ implies that $(\mathcal{T} \cdot Q) + ((\mathcal{T}^{-1})^* \cdot Q^{\perp}) = \mathbf{1}$, hence $(\mathcal{T} \cdot Q)^{\perp} = (\mathcal{T}^{-1})^* \cdot Q^{\perp}$.
\hfill $\square$


\let\section=\oldsection

\end{document}